\documentclass{elsart}
\usepackage[dvips]{graphicx}
\usepackage{pennames}
\usepackage{./axodraw}
\usepackage{xspace}
\usepackage{cite}
\setkeys{Gin}{width=0.95\linewidth}
\graphicspath{{../eps/}{../oth/}}
\DeclareGraphicsExtensions{.ps,.ps.gz,.eps,.eps.gz}
\def\loq{,\kern-0.080em,\kern+0.05em}

\def\footloq{,\kern-0.07em,\kern+0.03em}

\def\quabla{{\raise.7ex\hbox{\boxed{{}}}}}
\newsavebox{\fminibox}
\newlength{\fminilength}
    {\setlength{\fminilength}%
    {#1-2\fboxsep-2\fboxrule}%
    \begin{lrbox}{\fminibox}%
    \begin{minipage}{\fminilength}}
   {\end{minipage}\end{lrbox}%
    \noindent\fbox{\usebox{\fminibox}}}
\newcommand{\Descriptionlabel}[1]{%
   \mbox{\Descriptionfont #1}\hfil}
\newcommand\Descriptionfont{\itshape}
\newcommand\Descriptionmargin{}

\newcommand{\xdescription}[1][normal]{%
\list{}{\settowidth{\labelwidth}%
{\mbox{\Descriptionfont\Descriptionmargin}}%
 \setlength{\itemindent}{0pt}%
 \setlength{\leftmargin}{\labelwidth+\labelsep}%
 \let\makelabel\Descriptionlabel
 \ifthenelse{\equal{#1}{compact}}%
  {\setlength{\itemsep}{0pt}%
   \setlength{\topsep}{0.5\topsep}}{}%
 }}
 {\end{list}}
 \newcounter{enumct}
 \newenvironment{Enumerate}{\begin{list}{\arabic{enumct}.}%
 {\usecounter{enumct}\setlength{\topsep}{0.2mm}%
 \setlength{\partopsep}{0.2mm}\setlength{\itemsep}{0.2mm}%
 \setlength{\parsep}{0.2mm}}}{\end{list}}
\newcommand\smfrac[2]{{\textstyle{\frac{#1}{#2}}}}
\newcommand{\X      }[2]{\ensuremath{#1\,\pm #2}\xspace}
\newcommand{\Y      }[3]{\ensuremath{#1\,^{+\,#2}_{-\,#3}}\xspace}

\newcommand{\XZ}[3]{\ensuremath{#1\,\pm #2\,\pm #3}\xspace}

\newcommand{\chiq }{\ensuremath{\chi^2}\xspace}

\newcommand{\chiqm}{\ensuremath{\chi^2_\mathrm{min}}\xspace}
\newcommand{\lumi}{\ensuremath{{{\mathcal L}_{int}}}\xspace}
\newcommand{\lsim}{\ensuremath{\raisebox{-0.5mm}{$\stackrel{<}{\scriptstyle{\sim}}$}}\xspace}
\newcommand{\omfw }{\ensuremath{{\mathcal{O}}(\frac{\mf}{W})}\xspace}
\newcommand{\omfwq}{\ensuremath{{\mathcal{O}}(\frac{\mf^2}{\wsq})}\xspace}
\newcommand{\zn     }{\ensuremath{\PZz}\xspace}

\newcommand{\epem   }{\ensuremath{\Pep\,\Pem}\xspace}

\newcommand{\gsg    }{\ensuremath{\Pgg^\star\,\Pgg}\xspace}
\newcommand{\mumu   }{\ensuremath{\mu^+\mu^-}\xspace}
\newcommand{\mupmum }{\ensuremath{\mu^+\mu^-}\xspace}
\newcommand{\tauptaum}{\ensuremath{\tau^+\tau^-}\xspace}
\newcommand{\ffbar  }{\ensuremath{\mathrm{f\bar{f}}}\xspace}
\newcommand{\ff     }{\ensuremath{\mathrm{f}\mathrm{f}}\xspace}

\newcommand{\qqbar  }{\ensuremath{\mathrm{q\bar{q}}}\xspace}

\newcommand{\GG     }{\ensuremath{{\gamma\gamma}}\xspace}

\newcommand{\ggmm   }{\ensuremath{\gamma^{(\star)}\gamma^{(\star)}}\xspace}
\newcommand{\ggs    }{\ensuremath{\gamma\gamma^{\star}}\xspace}
\newcommand{\ggss   }{\ensuremath{\gamma^{\star}\gamma^{\star}}\xspace}

\newcommand{\ee     }{\ensuremath{\mathrm{e}{\mathrm e}}\xspace}
\newcommand{\eG     }{\ensuremath{{\mathrm{e}\gamma}}\xspace}

\newcommand{\lam    }{\ensuremath{\Lambda}\xspace}
\newcommand{\lamsq  }{\ensuremath{\Lambda^2}\xspace}

\newcommand{\lamv   }{\ensuremath{\Lambda_{\rm 4}^{\scriptstyle\overline{\rm MS}}}\xspace}
\newcommand{\gev    }{\ensuremath{\mathrm{GeV}}\xspace}
\newcommand{\gevsq  }{\ensuremath{\mathrm{GeV^2}}\xspace}
\newcommand{\aem    }{\ensuremath{\alpha}\xspace}
\newcommand{\aemsq  }{\ensuremath{\aem^2}\xspace}
\newcommand{\al     }{\ensuremath{\alpha_{s}}\xspace}

\newcommand{\nf     }{\ensuremath{\mathrm{n_f}}\xspace}
\newcommand{\nc     }{\ensuremath{N_\mathrm{c}}\xspace}
\newcommand{\xg     }{\ensuremath{x_\gamma}\xspace}
\newcommand{\xp     }{\ensuremath{x_\mathrm{p}}\xspace}
\newcommand{\xgpm   }{\ensuremath{x_\gamma^\pm}\xspace}
\newcommand{\qsq    }{\ensuremath{Q^{2}}\xspace}
\newcommand{\qnsq   }{\ensuremath{Q_0^{2}}\xspace}
\newcommand{\sqsq   }{\ensuremath{q^{2}}\xspace}
\newcommand{\qzm    }{\ensuremath{\langle \qsq \rangle}\xspace}
\newcommand{\qsqmin }{\ensuremath{Q^{2}_\mathrm{min}}\xspace}

\newcommand{\psq    }{\ensuremath{P^{2}}\xspace}
\newcommand{\pcsq   }{\ensuremath{P_\mathrm{c}^{2}}\xspace}
\newcommand{\spsq   }{\ensuremath{p^{2}}\xspace}
\newcommand{\pzm    }{\ensuremath{\langle \psq \rangle}\xspace}
\newcommand{\psqmin }{\ensuremath{P^{2}_{\rm{min}}}\xspace}
\newcommand{\psqmax }{\ensuremath{P^{2}_{\rm{max}}}\xspace}

\newcommand{\wsq    }{\ensuremath{W^{2}}\xspace}

\newcommand{\xmin   }{\ensuremath{x_{\rm{min}}}\xspace}

\newcommand{\ssee   }{\ensuremath{\sqrt{s_{\rm e e}}}\xspace}
\newcommand{\sega   }{\ensuremath{\sqrt{s_{\rm e \gamma}}}\xspace}

\newcommand{\sgagaq }{\ensuremath{s_{\gamma\gamma}}\xspace}
\newcommand{\segaq  }{\ensuremath{s_{\rm e \gamma}}\xspace}

\newcommand{\eecms  }{\ensuremath{s_{\rm e e}}\xspace}

\newcommand{\Mses   }{\ensuremath{M_\mathrm{SES}}\xspace}
\newcommand{\pdfe   }{\ensuremath{\tilde{f}_{\gamma}}\xspace}
\newcommand{\pdfel  }{\ensuremath{\tilde{f}^{\rm L}_{\gamma}}\xspace}
\newcommand{\pdfet  }{\ensuremath{\tilde{f}^{\rm T}_{\gamma}}\xspace}
\newcommand{\pdfexq }{\ensuremath{\tilde{f}_{\gamma}(\xg,\qsq)}\xspace}
\newcommand{\pdfexqp}{\ensuremath{\tilde{f}_{\gamma}(\xg,\qsq,\psq)}\xspace}
\newcommand{\ft     }{\ensuremath{F_{2}^{\gamma}}\xspace}
\newcommand{\fti    }{\ensuremath{F_{2,i}^{\gamma}}\xspace}
\newcommand{\ftc    }{\ensuremath{F_{2,\mathrm{c}}^{\gamma}}\xspace}
\newcommand{\ftb    }{\ensuremath{F_{2,\mathrm{b}}^{\gamma}}\xspace}
\newcommand{\fth    }{\ensuremath{F_{2,\mathrm{h}}^{\gamma}}\xspace}
\newcommand{\ftn    }{\ensuremath{F_{2}^{\gamma}/\aem}\xspace}

\newcommand{\ftqd   }{\ensuremath{\ft(\qsq,uds)}\xspace}

\newcommand{\ftxq   }{\ensuremath{\ft(x,\qsq)}\xspace}
\newcommand{\ftqn   }{\ensuremath{\ft(\qsq)/\aem}\xspace}
\newcommand{\ftxqp  }{\ensuremath{\ft(x,\qsq,\psq)}\xspace}

\newcommand{\fthad  }{\ensuremath{F_{2,\mathrm{had}}^{\gamma}}\xspace}
\newcommand{\ftpl   }{\ensuremath{F_{2,\mathrm{pl}}^{\gamma}}\xspace}
\newcommand{\ftasy  }{\ensuremath{F_{2,\mathrm{asy}}^{\gamma}}\xspace}
\newcommand{\ftqpm  }{\ensuremath{F_{2,\mathrm{QPM}}^{\gamma}}\xspace}

\newcommand{\fl     }{\ensuremath{F_\mathrm{L}^{\gamma}}\xspace}
\newcommand{\fa     }{\ensuremath{F_\mathrm{A}^{\gamma}}\xspace}
\newcommand{\fb     }{\ensuremath{F_\mathrm{B}^{\gamma}}\xspace}
\newcommand{\fT     }{\ensuremath{F_\mathrm{T}^{\gamma}}\xspace}
\newcommand{\fTxq   }{\ensuremath{\fT(x,\qsq)}\xspace}
\newcommand{\fTxqp  }{\ensuremath{\fT(x,\qsq,\psq)}\xspace}
\newcommand{\flxq   }{\ensuremath{\fl(x,\qsq)}\xspace}
\newcommand{\flxqp  }{\ensuremath{\fl(x,\qsq,\psq)}\xspace}
\newcommand{\ftqed  }{\ensuremath{F_\mathrm{2,QED}^{\gamma}}\xspace}

\newcommand{\faqed  }{\ensuremath{F_\mathrm{A,QED}^{\gamma}}\xspace}
\newcommand{\fbqed  }{\ensuremath{F_\mathrm{B,QED}^{\gamma}}\xspace}

\newcommand{\faoft  }{\ensuremath{\faqed/\ftqed}\xspace}
\newcommand{\fboft  }{\ensuremath{\fbqed/\ftqed}\xspace}
\newcommand{\feff   }{\ensuremath{F_\mathrm{eff}^{\gamma}}\xspace}

\newcommand{\ftapr  }{\ensuremath{F_\mathrm{2,apr}^{\gamma}}\xspace}
\newcommand{\ftpion }{\ensuremath{F_\mathrm{2}^{\pi}}\xspace}
\newcommand{\FtA   }{\ensuremath{\widetilde{F}_\mathrm{A}^\gamma}\xspace}

\newcommand{\FtAqed}{\ensuremath{\widetilde{F}_\mathrm{A,QED}^\gamma}\xspace}
\newcommand{\FtBqed}{\ensuremath{\widetilde{F}_\mathrm{B,QED}^\gamma}\xspace}
\newcommand{\FtLqed}{\ensuremath{\widetilde{F}_\mathrm{L,QED}^\gamma}\xspace}
\newcommand{\FtTqed}{\ensuremath{\widetilde{F}_\mathrm{T,QED}^\gamma}\xspace}
\newcommand{\Fttqed}{\ensuremath{\widetilde{F}_\mathrm{2,QED}^\gamma}\xspace}
\newcommand{\slop   }{\ensuremath{\der(\ftn)/\der\ln\qsq}\xspace}

\newcommand{\ftp    }{\ensuremath{F_\mathrm{2}^\mathrm{p}}\xspace}
\newcommand{\ftpxq  }{\ensuremath{F_\mathrm{2}^\mathrm{p}(x,\qsq)}\xspace}

\newcommand{\Ngtt   }{\ensuremath{\frac{\der^2N^{\rm T}_\gamma}{\der z\der\psq}}\xspace}
\newcommand{\Ngot   }{\ensuremath{\frac{\der N^{\rm T}_\gamma}{\der z}}\xspace}
\newcommand{\Ngtl   }{\ensuremath{\frac{\der^2N^{\rm L}_\gamma}{\der z\der\psq}}\xspace}

\newcommand{\msbl   }{\ensuremath{\overline{\rm MS}}\xspace}

\newcommand{\disg   }{\ensuremath{{\rm DIS_\gamma}}\xspace}

\newcommand{\qg     }{\ensuremath{q^\gamma}\xspace}
\newcommand{\qbg    }{\ensuremath{\bar{q}^\gamma}\xspace}
\newcommand{\Gg     }{\ensuremath{g^\gamma}\xspace}
\newcommand{\frhoq  }{\ensuremath{f^2_\rho}\xspace}
\newcommand{\ffp    }{\ensuremath{f_\pi}\xspace}

\newcommand{\dsigdx}{\ensuremath{\der\sigma/\der x}\xspace}
\newcommand{\stt  }{\ensuremath{\sigma_\mathrm{TT}}\xspace}
\newcommand{\slt  }{\ensuremath{\sigma_\mathrm{LT}}\xspace}
\newcommand{\stl  }{\ensuremath{\sigma_\mathrm{TL}}\xspace}
\newcommand{\sll  }{\ensuremath{\sigma_\mathrm{LL}}\xspace}
\newcommand{\ttt  }{\ensuremath{\tau_\mathrm{TT}}\xspace}
\newcommand{\ttl  }{\ensuremath{\tau_\mathrm{TL}}\xspace}

\newcommand{\siggsg}{\ensuremath{\sigma_{\gamma^\star\gamma}}\xspace}
\newcommand{\siggsgs}{\ensuremath{\sigma_{\gamma^\star\gamma^\star}}\xspace}
\newcommand{\siggg}{\ensuremath{\sigma_{\gamma\gamma}}\xspace}
\newcommand{\siggp}{\ensuremath{\sigma_\mathrm{\gamma p}}\xspace}
\newcommand{\siggv}{\ensuremath{\sigma_\mathrm{\gamma V}}\xspace}
\newcommand{\sigpp}{\ensuremath{\sigma_\mathrm{pp}}\xspace}
\newcommand{\sighh}{\ensuremath{\sigma_\mathrm{hh}}\xspace}
\newcommand{\Mv   }{\ensuremath{M_\mathrm{V}}\xspace}
\newcommand{\barph}{\ensuremath{\bar{\phi}}\xspace}
\newcommand{\cosph}{\ensuremath{\cos\barph}\xspace}
\newcommand{\costph}{\ensuremath{\cos 2\barph}\xspace}
\newcommand{\ripp }{\ensuremath{\rho_{\rm i}^{++}}\xspace}
\newcommand{\ripm }{\ensuremath{\rho_{\rm i}^{+-}}\xspace}
\newcommand{\rinn }{\ensuremath{\rho_{\rm i}^{00}}\xspace}
\newcommand{\ripn }{\ensuremath{\rho_{\rm i}^{+0}}\xspace}
\newcommand{\ropp }{\ensuremath{\rho_1^{++}}\xspace}
\newcommand{\ropm }{\ensuremath{\rho_1^{+-}}\xspace}
\newcommand{\ronn }{\ensuremath{\rho_1^{00}}\xspace}
\newcommand{\ropn }{\ensuremath{\rho_1^{+0}}\xspace}
\newcommand{\rtpp }{\ensuremath{\rho_2^{++}}\xspace}
\newcommand{\rtpm }{\ensuremath{\rho_2^{+-}}\xspace}
\newcommand{\rtnn }{\ensuremath{\rho_2^{00}}\xspace}
\newcommand{\rtpn }{\ensuremath{\rho_2^{+0}}\xspace}
\newcommand{\ehad   }{\ensuremath{E_\mathrm{had}}\xspace}
\newcommand{\pzhad  }{\ensuremath{p_\mathrm{z,had}}\xspace}

\newcommand{\pthadq }{\ensuremath{p^2_\mathrm{t,had}}\xspace}

\newcommand{\eg     }{\ensuremath{E_\gamma}\xspace}

\newcommand{\etout  }{\ensuremath{E_\mathrm{t,out}}\xspace}

\newcommand{\eonep  }{\ensuremath{E_1^\prime}\xspace}

\newcommand{\etwop  }{\ensuremath{E_2^\prime}\xspace}

\newcommand{\etjet  }{\ensuremath{E^{\rm jet}_{\rm T}}\xspace}

\newcommand{\etjavq }{\ensuremath{\overline{E_{\rm T}}^2}\xspace}
\newcommand{\etjeto }{\ensuremath{E^{\rm jet1}_{\rm T}}\xspace}
\newcommand{\etjett }{\ensuremath{E^{\rm jet2}_{\rm T}}\xspace}
\newcommand{\etajet }{\ensuremath{\eta^{\rm jet}}\xspace}
\newcommand{\etajeto}{\ensuremath{\eta^{\rm jet1}}\xspace}
\newcommand{\etajett}{\ensuremath{\eta^{\rm jet2}}\xspace}

\newcommand{\pt     }{\ensuremath{p_{\mathrm{t}}}\xspace}
\newcommand{\ptq    }{\ensuremath{p^2_{\mathrm{t}}}\xspace}

\newcommand{\pthq   }{\ensuremath{\hat{p}^2_{\mathrm{t}}}\xspace}

\newcommand{\ptnq   }{\ensuremath{p^{2}_{\mathrm{t,0}}}\xspace}

\newcommand{\flow   }{\ensuremath{1/N~{\rm d}E/{\rm d}\eta}\xspace}
\newcommand{\zzp    }{\ensuremath{z_{\mathrm{p}}}\xspace}

\newcommand{\tonep  }{\ensuremath{\theta_1^\prime}\xspace}

\newcommand{\ttwop  }{\ensuremath{\theta_2^\prime}\xspace}
\newcommand{\ttwopmax}{\ensuremath{\theta_{\rm 2,max}^\prime}\xspace}

\newcommand{\phonep }{\ensuremath{\phi_1^\prime}\xspace}

\newcommand{\phtwop }{\ensuremath{\phi_2^\prime}\xspace}

\newcommand{\pone   }{\ensuremath{p_1}\xspace}
\newcommand{\ponep  }{\ensuremath{p_1^\prime}\xspace}
\newcommand{\ptwo   }{\ensuremath{p_2}\xspace}
\newcommand{\ptwop  }{\ensuremath{p_2^\prime}\xspace}

\newcommand{\vponep }{\ensuremath{\vec{p}_1^\prime}\xspace}

\newcommand{\vptwop }{\ensuremath{\vec{p}_2^\prime}\xspace}
\newcommand{\ptonepq}{\ensuremath{p_{1,\rm t}^{\prime 2}}\xspace}
\newcommand{\pzonep }{\ensuremath{p_{1,\rm z}^\prime}\xspace}
\newcommand{\pztwop }{\ensuremath{p_{2,\rm z}^\prime}\xspace}
\newcommand{\pfone  }{\ensuremath{p_{\rm f_1}}\xspace}
\newcommand{\pftwo  }{\ensuremath{p_{\rm f_2}}\xspace}

\newcommand{\vpfone }{\ensuremath{\vec{p}_{\rm f_1}}\xspace}
\newcommand{\vpftwo }{\ensuremath{\vec{p}_{\rm f_2}}\xspace}
\newcommand{\efone  }{\ensuremath{E_{\rm f_1}}\xspace}
\newcommand{\eftwo  }{\ensuremath{E_{\rm f_2}}\xspace}
\newcommand{\tfone  }{\ensuremath{\theta_{\rm f_1}}\xspace}
\newcommand{\tftwo  }{\ensuremath{\theta_{\rm f_2}}\xspace}

\newcommand{\act    }{\ensuremath{\vert\cos\theta\vert}\xspace}
\newcommand{\ts     }{\ensuremath{\theta^\star}\xspace}
\newcommand{\cts    }{\ensuremath{\cos\theta^\star}\xspace}
\newcommand{\ctsq   }{\ensuremath{\cos^2\theta^\star}\xspace}

\newcommand{\az     }{\ensuremath{\chi}\xspace}
\newcommand{\kt     }{\ensuremath{k_\mathrm{t}}\xspace}
\newcommand{\ktsq   }{\ensuremath{k^2_\mathrm{t}}\xspace}
\newcommand{\ktsqm  }{\ensuremath{k^2_\mathrm{t,max}}\xspace}
\newcommand{\kn     }{\ensuremath{k_0}\xspace}
\newcommand{\Wvis   }{\ensuremath{W_{\mathrm{vis}}}\xspace}

\newcommand{\xvis   }{\ensuremath{x_{\mathrm{vis}}}\xspace}
\newcommand{\vvis   }{\ensuremath{v_{\mathrm{vis}}}\xspace}
\newcommand{\Wrec   }{\ensuremath{W_{\mathrm{rec}}}\xspace}
\newcommand{\xrec   }{\ensuremath{x_{\mathrm{rec}}}\xspace}
\newcommand{\mc     }{\ensuremath{m_{\mathrm{c}}}\xspace}
\newcommand{\mb     }{\ensuremath{m_{\mathrm{b}}}\xspace}
\newcommand{\mh     }{\ensuremath{m_{\mathrm{h}}}\xspace}
\newcommand{\me     }{\ensuremath{m_{\mathrm{e}}}\xspace}
\newcommand{\ef     }{\ensuremath{{\rm e_{\rm f}}}\xspace}
\newcommand{\efv    }{\ensuremath{{\rm e^4_{\rm f}}}\xspace}
\newcommand{\mf     }{\ensuremath{m_{\rm f}}\xspace}

\newcommand{\eqv    }{\ensuremath{{\rm e^4_{\rm q}}}\xspace}

\newcommand{\eqit   }{\ensuremath{{\rm e^2_{\rm q_i}}}\xspace}

\newcommand{\eqkt   }{\ensuremath{{\rm e^2_{\rm q_k}}}\xspace}
\newcommand{\eqkv   }{\ensuremath{{\rm e^4_{\rm q_k}}}\xspace}
\newcommand{\eqht   }{\ensuremath{{\rm e^2_{\rm q_h}}}\xspace}
\newcommand{\eqhv   }{\ensuremath{{\rm e^4_{\rm q_h}}}\xspace}

\newcommand{\avet   }{\ensuremath{\langle {\rm e}^2\rangle}\xspace}
\newcommand{\gdet   }{\ensuremath{g^{\mathrm{det}}}\xspace}
\newcommand{\fpar   }{\ensuremath{f^{\mathrm{part}}}\xspace}
\newcommand{\fmult  }{\ensuremath{f_{\mathrm{mult}}(x)}\xspace}
\newcommand{\chidof }{\ensuremath{\chi^2/\mathrm{dof}}\xspace}
\newcommand{\der    }{\ensuremath{\mathrm{d}}\xspace}

\newcommand{\invpb  }{\ensuremath{\mathrm{pb}^{-1}}\xspace}
\newcommand{\pb     }{\ensuremath{\mathrm{pb}}\xspace}
\newcommand{\pz     }{\ensuremath{\phantom{0}}\xspace}

\begin{document}
%
%
 \begin{frontmatter}
 \title{The Photon Structure from Deep Inelastic Electron-Photon
        Scattering}
 \author{Richard Nisius}
 \thanks[MAIL]{E-mail: Richard.Nisius@cern.ch}
 \address{CERN, CH-1211 Gen\`eve 23, Switzerland}
 \begin{abstract}
 The present knowledge of the structure of the photon is presented
 with emphasis on measurements of the photon structure
 obtained from deep inelastic electron-photon scattering at \epem
 colliders.
 This review covers the leptonic and hadronic structure
 of quasi-real and also of highly virtual photons, based on
 measurements of structure functions and differential cross-sections.
 Future prospects of the investigation of the photon structure
 in view of the ongoing LEP2 programme and of a possible linear
 collider are addressed.
 The most relevant results in the context of measurements of the photon
 structure from photon-photon scattering at LEP and from photon-proton
 and electron-proton scattering at HERA are summarised.
 \end{abstract}
 \end{frontmatter}

\pagestyle{headings}
\pagenumbering{roman}
\tableofcontents

%
%
%
\section{Introduction}
\label{sec:intro}
\pagenumbering{arabic}
 The photon is a fundamental ingredient of our
 present understanding of the interactions of quarks and leptons.
 These interactions are successfully described in the framework of the
 standard model, a theory which consists of a combination of
 gauge theories.
 Being the gauge boson of the theory of Quantum Electro Dynamics, QED,
 the photon mediates the electromagnetic force between charged objects.
 In these interactions the photon can be regarded as a structureless object,
 called the \emph{direct}, or the \emph{bare\/} photon.
 Since QED is an abelian gauge theory, the photon has no self-couplings
 and to the best of our knowledge the photon is a massless particle.
 \par
 Due to the Heisenberg uncertainty
 principle\footnote{The units used are $c=\hbar=1$.},
 written as $\Delta E\Delta t>1$, the photon, denoted with $\gamma$,
 is allowed to violate the rule of conservation of energy
 by an amount of energy $\Delta E$ for a short period of time $\Delta t$
 and to fluctuate into a charged fermion anti-fermion, \ffbar, system
 carrying the same quantum numbers as the photon,
 $\gamma \rightarrow\ffbar\rightarrow \gamma$.
 If, during such a fluctuation, one of the
 fermions\footnote{Fermions and anti-fermions are not distinguished,
 for example, electrons and positrons are referred to as electrons.}
 interacts via a gauge boson with another object, then the parton
 content of the photon is resolved and the photon reveals its structure.
 In such interactions the photon can be regarded as an extended
 object consisting of charged fermions and also gluons,
 the so called \emph{resolved\/} photon.
 This possibility for the photon to interact either directly or in a
 resolved manner
 is another dual nature of the photon, which is the cause of a variety
 of phenomena and makes the photon a very interesting object to investigate.
 One possible description of the structure of the photon is given by the
 concept of photon structure functions, which is the main subject of this
 review.
 \par
 Today the main results on the structure of the photon are obtained from
 the electron-positron collider LEP
 and
 the electron-proton collider HERA.
 The largest part of this review is devoted to the discussion of
 deep inelastic electron-photon scattering and to the measurements
 of QED and hadronic structure functions of the photon.
 Other LEP results on the structure of the photon apart from those
 obtained from deep inelastic electron-photon scattering, as well as the
 measurements in photoproduction and deep inelastic electron-proton
 scattering at HERA, are summarised briefly.
 \par
 The review is organised in the following way.
 In Section~\ref{sec:kinem}
 the kinematical quantities are introduced and in Section~\ref{sec:detec}
 the capabilities of the detectors to measure the
 deep inelastic electron-photon scattering process are discussed.
 The theoretical formalism needed to measure the photon structure
 is outlined in Section~\ref{sec:theo}, with special emphasis on the QED
 and hadronic structure functions of the photon in Section~\ref{sec:QED}
 and Section~\ref{sec:QCD} respectively.
 A review of the available parametrisations of parton distribution
 functions of the photon is given in Section~\ref{sec:PDF}.
 The most important tools used to measure the photon structure
 are described in Section~\ref{sec:tools}, concentrating
 on event generators and unfolding methods.
 The measurements of the QED and hadronic structure of the photon
 obtained from leptonic and hadronic final states are discussed in
 Section~\ref{sec:qedres}, and Section~\ref{sec:qcdres} respectively.
 The prospects of future determinations of the structure of
 the photon are outlined.
 The measurements expected to be performed at LEP, using the high
 statistics, high energy data still expected within the ongoing
 LEP2 programme, are discussed in
 Section~\ref{sec:futlep}, followed by the discussion of measurements
 to be performed at a possible future linear collider in
 Section~\ref{sec:nlc}.
 Complementary investigations of the photon structure from LEP
 and selected HERA results are addressed in Section~\ref{sec:probe}.
 \par
 Additional information is presented in the appendices.
 The relation between the cross-section picture and the structure
 function picture is outlined in Appendix~\ref{sec:SIGTOSF},
 followed by a discussion of the general
 relation between the photon structure function, the parton distribution
 functions and the evolution equations given in Appendix~\ref{sec:PDFTH}.
 The Appendices~\ref{sec:tabqed} and~\ref{sec:tabqcd} contain
 a collection of numerical results on measurements of the photon
 structure.
%
%
\subsection{Theoretical description of photon interactions}
\label{sec:parton}
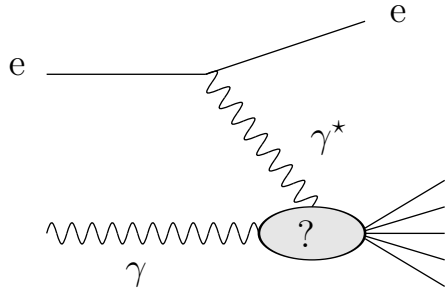
\begin{figure}[tbp]\unitlength 1pt
\begin{center}
\begin{picture}(300,110)(0,0)
 \Photon(50,20)(130,20){4}{10.5}
 \GOval(150,20)(10,20)(0){0.9}
 \Line(170,22)(200,40)
 \Line(170,21)(200,30)
 \Line(170,20)(200,20)
 \Line(170,19)(200,10)
 \Line(170,18)(200,00)
 \Text(080,00)[lb]{\mbox{\large $\gamma$}}
 \Text(145,15)[lb]{\mbox{\large ?}}
 \Line(50,80)(110,80)
 \Photon(110,80)(150,30){4}{8.5}
 \Line(110,80)(170,100)
 \Text(150,050)[lb]{\mbox{\large $\gamma^\star$}}
 \Text(036,080)[lb]{\mbox{\large $\rm e$}}
 \Text(180,100)[lb]{\mbox{\large $\rm e$}}
\end{picture}
\caption{
         Probing the structure of quasi-real photons, $\gamma$, by highly
         virtual photons, $\gamma^\star$.
        }\label{fig:chap1_01}
\end{center}
\end{figure}

 In deep inelastic electron-photon scattering the structure of
 a quasi-real photon, $\gamma$, is probed by a highly virtual photon,
 $\gamma^\star$, emitted by a deeply inelastically scattered electron,
 as sketched in Figure~\ref{fig:chap1_01}.
 \par
%
%
 The photon, as the mediator of the electromagnetic force, couples
 to charged objects.
 The fundamental coupling of the photon as described in the framework of
 QED is the coupling to charged fermions, $\rm f$, which can be either
 quarks, $\rm q$, or leptons, $\ell$, with $\ell = \rm e \mu\tau$.
 For the case of lepton pair production, the process can be calculated in QED.
 The relevant formulae are listed in Section~\ref{sec:theo} and the results
 on the QED structure of the photon are discussed in Section~\ref{sec:qedres}.
 \par
%
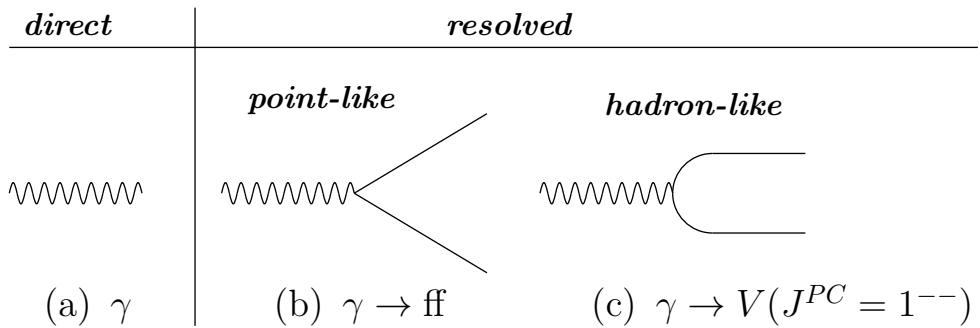
\begin{figure}[tbp]\unitlength 1pt
\begin{center}
\begin{picture}(300,120)(24,0)
 \Photon(0,50)(50,50){4}{8.5}
 \Photon(80,50)(130,50){4}{8.5}
 \Line(130,50)(180,80)
 \Line(130,50)(180,20)
 \Photon(200,50)(250,50){4}{8.5}
 \CArc(265,50)(15, 90,180)
 \CArc(265,50)(15,180,270)
 \Line(265,65)(300,65)
 \Line(265,35)(300,35)
 \Text( 13,0)[lb]{\mbox{\large (a)$\,\,\,\gamma$}}
 \Text(100,0)[lb]{\mbox{\large (b)$\,\,\,\gamma\rightarrow\ff$}}
 \Text(220,0)[lb]{\mbox{\large (c)$\,\,\,\gamma\rightarrow V(J^{PC}=1^{--})$}}
 \Text( 05,110)[lb]{\textbf{\emph{direct}}}
 \Text(165,110)[lb]{\textbf{\emph{resolved}}}
 \Text( 90,80)[lb]{\textbf{\emph{point-like}}}
 \Text(225,80)[lb]{\textbf{\emph{hadron-like}}}
 \Line( 0,105)(365,105)
 \Line(70,120)( 70,  0)
\end{picture}
\caption[
         The different appearances of the photon.
        ]
        {
         The different appearances of the photon.
         Shown are (a) the direct or bare photon, and (b,c) the resolved
         photon, which can be either point-like, (b), or hadron-like, (c).
        }\label{fig:chap1_02}
\end{center}
\end{figure}
%
 For the production of quark pairs the situation is more complex,
 since the spectrum of fluctuations is richer, and QCD
 corrections have to be taken into account.
 Therefore, the photon interactions receive several contributions
 shown in Figure~\ref{fig:chap1_02}.
 The leading order contributions are discussed in detail.
 The reactions of the photon are usually classified depending
 on the object which takes part in the hard interaction.
 If the photon directly, as a whole, takes part in the hard
 interaction, as shown in Figure~\ref{fig:chap1_02}(a),
 then it does not reveal a structure.
 These reactions are called direct interactions and the
 photon is named the \emph{direct}, or the \emph{bare\/} photon.
 If the photon first fluctuates into a hadronic state which
 subsequently interacts, the processes are called resolved photon
 processes and structure functions of the photon can be defined.
 The resolved photon processes are further subdivided into two parts.
 The first part, shown in Figure~\ref{fig:chap1_02}(b), is perturbatively
 calculable, as explained in Section~\ref{sec:QCD}, and called the
 contribution of the \emph{point-like}, or the \emph{anomalous\/} photon.
 Here the photon perturbatively splits into a quark pair of a certain
 relative transverse momentum and subsequently one of the quarks takes part
 in the hard interaction, which for deep inelastic electron-photon scattering
 in leading order is the process $\gamma^\star\rm q\rightarrow\rm q$.
 The second part, where the photon fluctuates into a hadronic state with
 the same quantum numbers as the photon, as shown in
 Figure~\ref{fig:chap1_02}(c), is usually called the \emph{hadron-like}, or
 \emph{hadronic\/} contribution\footnote{In this review the two parts of the
 resolved photon will be called point-like and hadron-like to avoid
 confusion with the term hadronic structure function of the photon
 which is used for the full \ft.}.
 The photon behaves like a hadron, and the hadron-like part of the
 hadronic photon structure function \ft can successfully be described
 by the vector meson dominance model, VMD, considering the low mass
 vector mesons $\rho,\omega$ and $\phi$, as outlined in
 Section~\ref{sec:VMD}.
 \par
 The leading order contributions are subject to QCD corrections due
 to the coupling of quarks to gluons.
 The hadronic photon structure function \ft receives contributions
 both from the point-like part and from the hadron-like part
 of the photon structure, discussed in detail in Section~\ref{sec:QCD}.
%
%

%
%
\section{Deep inelastic electron-photon scattering (DIS)}
\label{sec:egam}
 The classical way to investigate the structure of the photon at
 \epem colliders is the measurement of photon structure functions
 using the process
%
\begin{equation}
 \ee \rightarrow \ee \gsg \rightarrow \ee X.
 \label{eqn:react}
\end{equation}
%
 In this section the kinematical variables used to describe the reaction
 are introduced in Section~\ref{sec:kinem} and experimental aspects
 are discussed in Section~\ref{sec:detec}.
%
%
\subsection{Kinematics}
\label{sec:kinem}
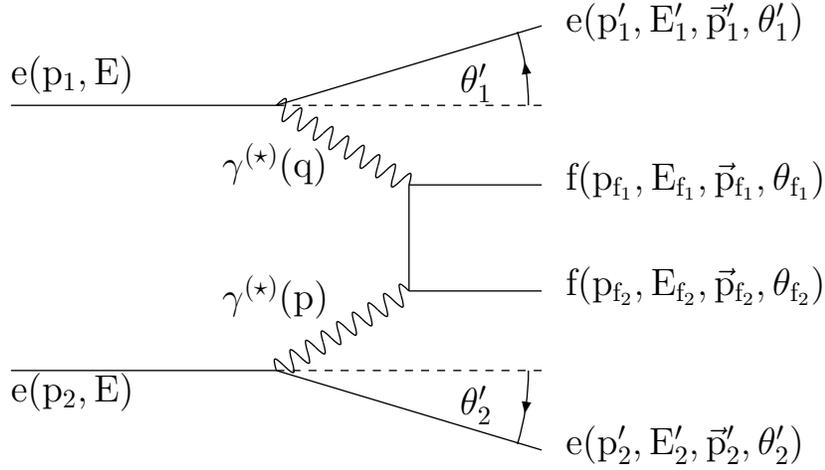
\begin{figure}[tbp]\unitlength 1pt
\begin{center}
\begin{picture}(300,200)(0,0)
 \Line(0,150)(100,150)
 \Line(100,150)(200,180)
 \Photon(100,150)(150,120){4}{8.5}   
 \Line(150,120)(150,80)
 \Line(150,120)(200,120)
 \DashLine(100,150)(200,150){3}
 \Text(170,152)[lb]{\mbox{\large $\tonep$}}
 \ArrowArc(100,150)(95,0,17)
 \Text(  0,155)[lb]{\mbox{\large $\rm e(\pone,E)$}}
 \Text(210,175)[lb]{\mbox{\large $\rm e(\ponep,\eonep,\vponep,\tonep)$}}
 \Text(  0, 35)[lb]{\mbox{\large $\rm e(\ptwo,E)$}}
 \Text(210, 15)[lb]{\mbox{\large $\rm e(\ptwop,\etwop,\vptwop,\ttwop)$}}
 \Text(210,115)[lb]{\mbox{\large $\rm f(\pfone,\efone,\vpfone,\tfone)$}}
 \Text(210, 75)[lb]{\mbox{\large $\rm f(\pftwo,\eftwo,\vpftwo,\tftwo)$}}
 \Text( 80,120)[lb]{\mbox{\large $\rm \gamma^{(\star)}(q)$}}
 \Text( 80, 70)[lb]{\mbox{\large $\rm \gamma^{(\star)}(p)$}}
 \Line(0,50)(100,50)
 \Line(100,50)(200,20)
 \Photon(100,50)(150,80){4}{8.5}   
 \DashLine(100,50)(200,50){3}
 \Text(170,30)[lb]{\mbox{\large $\ttwop$}}
 \ArrowArc(100,50)(-95,163,180)
 \Line(150,80)(200,80)
\end{picture}
\caption[
         A diagram of the reaction  $\ee \rightarrow \ee\ff$,
         proceeding via the exchange of two photons.
        ]
        {
         A diagram of the reaction  $\ee \rightarrow \ee\ff$,
         proceeding via the exchange of two photons.
        }\label{fig:chap2_01}
\end{center}
\end{figure}

 Figure~\ref{fig:chap2_01} shows a diagram of the scattering of two
 electrons, proceeding via the exchange of two photons of arbitrary
 virtualities, in the case of leading order fermion pair production,
 $X=\ff$.
 \par
%
%
 The reaction is described in the following notation
%
\begin{equation}
 \mathrm{e}(\pone) \mathrm{e}(\ptwo)
 \rightarrow
 \mathrm{e}(\ponep)\mathrm{e}(\ptwop)
 \gamma^{(\star)}(q)\gamma^{(\star)}(p)
 \rightarrow
 \mathrm{e}(\ponep)\mathrm{e}(\ptwop)
 \mathrm{f}(\pfone)\mathrm{f}(\pftwo).
 \label{eqn:reactkin}
\end{equation}
%
 The terms in brackets denote the four-vectors of the respective
 particles, and $E$ is the energy of the electrons of the beams.
 In addition the energies, momentum vectors and polar scattering angles
 of the outgoing particles are introduced in Figure~\ref{fig:chap2_01}.
 The symbol $(\star)$ indicates that the photons can be either
 quasi-real, $\gamma$, or virtual, $\gamma^\star$.
 The virtual photons have negative virtualities $\sqsq,\spsq \le 0$.
 For simplicity, the definitions $\qsq=-\sqsq\ge 0$ and $\psq=-\spsq\ge 0$
 are used, and the particles are ordered such that $\qsq>\psq$.
 A list of commonly used variables, which are valid for arbitrary
 virtualities and for any final state $X$, is given below
%
\begin{eqnarray}
 \eecms   &\equiv&(\pone+\ptwo)^2 = 2\pone\cdot\ptwo = 4E^2,  \label{eqn:see}\\
 \segaq   &\equiv&(\pone + p)^2,                              \label{eqn:seg}\\
 \sgagaq  &\equiv&\wsq\equiv (q+p)^2 = \qsq\frac{1-x}{x}-\psq,\label{eqn:wsq}\\
 x        &\equiv&\frac{\qsq}{2p\cdot q}=
                  \frac{\qsq}{\wsq+\qsq+\psq},                \label{eqn:x}\\
 y        &\equiv&\frac{p\cdot q}{\pone\cdot p},              \label{eqn:y}\\
 r        &\equiv&\frac{p\cdot q}{\ptwo\cdot q},              \label{eqn:r}\\
 \qsq     &=&x y(\segaq+\psq)=2E\eonep (1 - \cos\tonep),      \label{eqn:q2}\\
 \psq     &=&2E\etwop (1 - \cos\ttwop).                       \label{eqn:p2}
\end{eqnarray}
%
 Here
 \eecms is the invariant mass squared of the electron-electron system,
 \segaq the invariant mass squared of the electron-photon system,
 \sgagaq the invariant mass squared of the photon-photon system,
 and the mass of the electron has been neglected.
 \par
 Deep inelastic electron-photon scattering is characterised in the limit
 where one photon is highly virtual, $\qsq \gg 0$, and the other is
 quasi-real, $\psq \approx 0$.
 In this case \psq is neglected in the equations above and
 some simplified expressions are found
%
\begin{eqnarray}
 \segaq   &=&(\pone + p)^2= 2\pone\cdot p= 4E\eg,         \label{eqn:segpn}\\
  y       &=&1 - \frac{\eonep}{2E}(1 + \cos\tonep),       \label{eqn:ye}\\
  r       &=&\frac{E_\gamma}{E}\equiv z.                  \label{eqn:z}
\end{eqnarray}
%
 Here \eg is the energy of the quasi-real photon.
 \par
%
\begin{figure}[tbp]\unitlength 1pt
\begin{center}
\begin{picture}(360,290)(0,0)
%
%
 \Line(10,280)(130,280)
 \Line(10,180)(130,180)
 \Photon( 70,280)( 70,240){4}{6.5}
 \Line(70,240)( 70,220)
 \Photon( 70,180)( 70,220){4}{6.5}
 \Line( 70,240)(120,240)
 \Line( 70,220)(120,220)
 \Text(  0,275)[lb]{$\rm e$}
 \Text(140,275)[lb]{$\rm e$}
 \Text(  0,175)[lb]{$\rm e$}
 \Text(140,175)[lb]{$\rm e$}
 \Text(130,215)[lb]{$\rm f$}
 \Text(130,235)[lb]{$\rm f$}
 \Text( 44,250)[lb]{$\rm \gamma^{(\star)}$}
 \Text( 44,200)[lb]{$\rm \gamma^{(\star)}$}
 \Text(  0,225)[lb]{(a)}
%
%
 \Line(200,280)(330,280)
 \Photon(260,280)(260,180){4}{16.5}
 \Line(200,180)(330,180)
 \Photon(290,280)(310,250){4}{6.5}
 \Line(310,250)(330,240)
 \Line(310,250)(330,220)
 \Text(190,275)[lb]{$\rm e$}
 \Text(340,275)[lb]{$\rm e$}
 \Text(190,175)[lb]{$\rm e$}
 \Text(340,175)[lb]{$\rm e$}
 \Text(340,235)[lb]{$\rm f$}
 \Text(340,215)[lb]{$\rm f$}
 \Text(225,225)[lb]{$\rm \gamma^{(\star)}$}
 \Text(310,265)[lb]{$\rm \gamma^{(\star)}$}
 \Text(190,225)[lb]{(b)}
%
%
 \Line(10,130)(50,75)
 \Line(10, 20)(50,75)
 \Photon(50,75)(90,75){4}{6.5}
 \Line(90,75)(130,130)
 \Line(90,75)(130, 20)
 \Photon(30,102.5)(60,112.5){4}{6.5}
 \Line(60,112.5)( 90,127.5)
 \Line(60,112.5)( 90, 97.5)
 \Text(  0,130)[lb]{$\rm e$}
 \Text(  0, 10)[lb]{$\rm e$}
 \Text(135,130)[lb]{$\rm e$}
 \Text(135, 10)[lb]{$\rm e$}
 \Text( 95,120)[lb]{$\rm f$}
 \Text( 95, 95)[lb]{$\rm f$}
 \Text( 65, 55)[lb]{$\rm \gamma^{(\star)}$}
 \Text( 40,115)[lb]{$\rm \gamma^{(\star)}$}
 \Text(  0, 67)[lb]{(c)}
\end{picture}
\caption[The different contributions to the reaction
         $\ee\rightarrow\ee\ggmm\rightarrow\ee\ff$.
        ]
        {
         The different contributions to the reaction
         $\ee\rightarrow\ee\ggmm\rightarrow\ee\ff$.
         Shown are
         (a) the multipheripheral diagram,
         (b) the t-channel bremsstrahlung diagram
         and
         (c) the s-channel bremsstrahlung diagram.
         In all cases only one possible leading order diagram is shown.
        }\label{fig:chap2_02}
\end{center}
\end{figure}
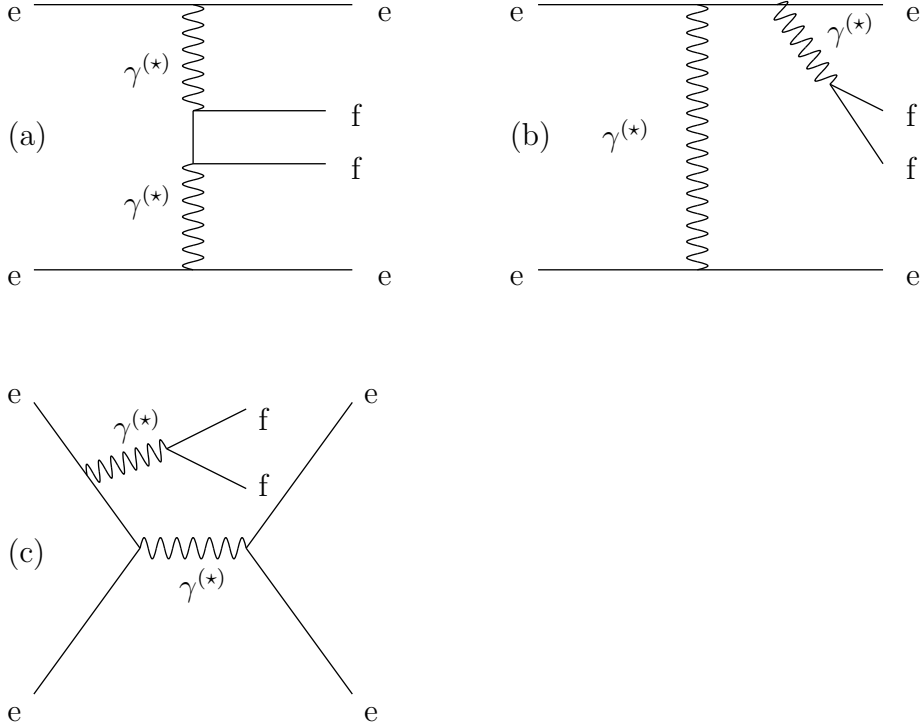
%
 The reaction receives contributions in leading order from the
 different Feynman diagrams shown in Figure~\ref{fig:chap2_02}.
 The relative sizes of the contributions of the different Feynman
 diagrams depend on the kinematical situation.
 In the region of deep inelastic scattering, $\qsq \gg \psq\approx 0$,
 and for moderate values of \qsq the dominant contribution stems from
 the multipheripheral diagram, Figure~\ref{fig:chap2_02}(a).
 The t-channel bremsstrahlung diagram, Figure~\ref{fig:chap2_02}(b), and
 the s-channel bremsstrahlung diagram,
 Figure~\ref{fig:chap2_02}(c), contribute much less to the total
 cross-section, as explained in
 Ref.~\cite{BHA-7701}\footnote{The contributions to the
 s-channel bremsstrahlung diagrams are sometimes called the annihilation
 and the conversion diagram, reserving the term bremsstrahlung only for
 the t-channel bremsstrahlung diagram.}.
 However, at large values of \qsq both bremsstrahlung diagrams
 have to be taken into account, especially the t-channel
 bremsstrahlung diagram can be important, predominantly at low
 invariant masses of the photon-photon system.
 \par
 The structure functions of the photon, introduced in
 Sections~\ref{sec:QED} and~\ref{sec:QCD}, are extracted from a
 measurement of the differential cross-sections of this reaction.
 For the measurement of the structure function \ft it is sufficient
 to describe the reaction in terms of $x$, \qsq and \psq.
 For the measurement of other structure functions like \fa and \fb
 further variables are necessary.
 For example, the measurement of \fa and \fb in deep inelastic
 electron-photon scattering involves the measurement of the azimuthal
 angle \az between the plane defined by the momentum vectors of the
 fermion and anti-fermion, called the fermion anti-fermion plane, and
 the plane defined by the momentum vectors of the incoming and the
 deeply inelastically scattered electron, called the electron
 scattering plane.
 \par
%
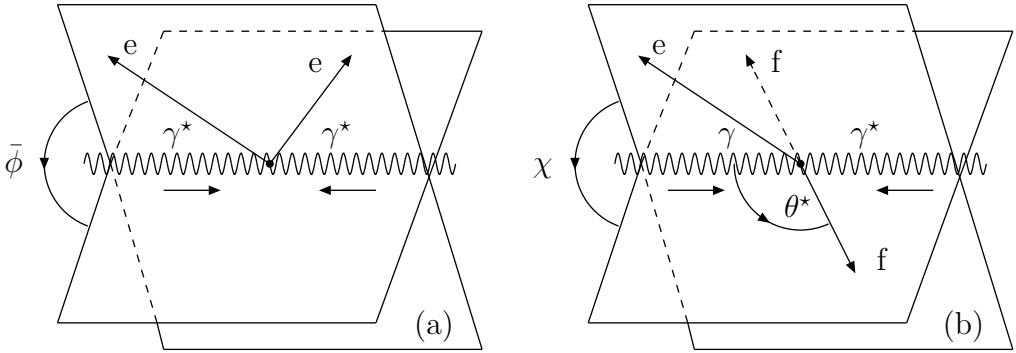
\begin{figure}[tbp]\unitlength 1pt
\begin{center}
\begin{picture}(400,150)(0,0)
 \Text(165,13)[lb]{(a)}
 \Line( 30,140)(150,140)
 \Line(155,130)(190,130)
 \DashLine(70,130)(155,130){3}
 \Line(190,10)(150,140)
 \Line(150,20)(190,130)
 \Line(30,20)(150,20)
 \Line(70,10)(190,10)
 \Line(30,140)(50,80)
 \DashLine(50,80)(67.5,20){3}
 \Line(67.5,20)(70,10)
 \DashLine(70,130)(50,80){3}
 \Line(50,80)(30,20)
 \Photon(40,80)(110,80){4}{14.5}
 \Vertex(110,80){1.5}
 \Photon(180,80)(110,80){4}{14.5}
 \LongArrow( 70,70)( 90,70)
 \LongArrow(150,70)(130,70)
 \LongArrow(110,80)( 50,120)
 \LongArrow(110,80)(140,120)
 \Text(130,86)[lb]{$\rm \gamma^{\star}$}
 \Text( 70,86)[lb]{$\rm \gamma^{\star}$}
 \Text(125,115)[lb]{${\rm e}$}
 \Text(55,122)[lb]{${\rm e}$}
 \Text(10,75)[lb]{$\barph$}
 \ArrowArc(50,80)(25,110,250)
%
%
 \Text(365,13)[lb]{(b)}
 \Line(230,140)(350,140)
 \Line(355,130)(390,130)
 \DashLine(270,130)(355,130){3}
 \Line(390,10)(350,140)
 \Line(350,20)(390,130)
 \Line(230,20)(350,20)
 \Line(270,10)(390,10)
 \Line(230,140)(250,80)
 \DashLine(250,80)(267.5,20){3}
 \Line(267.5,20)(270,10)
 \DashLine(270,130)(250,80){3}
 \Line(250,80)(230,20)
 \Photon(240,80)(310,80){4}{14.5}
 \Vertex(310,80){1.5}
 \Photon(380,80)(310,80){4}{14.5}
 \LongArrow(260,70)(280,70)
 \LongArrow(360,70)(340,70)
 \LongArrow(310,80)(250,120)
 \DashLine(310,80)(290,120){3}
 \LongArrow(292,116)(290,120)
 \LongArrow(310,80)(330,40)
 \Text(330,86)[lb]{$\rm \gamma^{\star}$}
 \Text(280,86)[lb]{$\rm \gamma$}
 \Text(300,115)[lb]{${\rm f}$}
 \Text(340, 40)[lb]{${\rm f}$}
 \Text(255,122)[lb]{${\rm e}$}
 \Text(305,60)[lb]{$\theta^\star$}
 \ArrowArc(310,80)(25,180,295)
 \Text(210,75)[lb]{$\chi$}
 \ArrowArc(250,80)(25,110,250)
\end{picture}
\caption[Illustrations of the scattering angles \barph, \ts and \az
         in the photon-photon centre-of-mass system.
        ]
        {
         Illustrations of the scattering angles \barph, \ts and \az
         in the photon-photon centre-of-mass system.
         Shown are
         (a) the angle between the scattering planes of the
         two scattered electrons, \barph,
         and
         (b) the scattering angle \ts of the fermion or anti-fermion
         with respect to the photon-photon axis, as well as the
         the azimuthal angle \az, defined as the angle between
         the observed electron and the fermion which, in the photon-photon
         centre-of-mass frame, is scattered at $\cts<0$.
        }\label{fig:chap2_03}
\end{center}
\end{figure}
%
 The experimentally exploited angles \barph, \ts and \az are
 introduced in Figure~\ref{fig:chap2_03}.
 The azimuthal angle \barph is defined as the angle between
 the scattering planes of the two electrons in the photon-photon
 centre-of-mass frame, as shown in Figure~\ref{fig:chap2_03}(a).
 The polar angle \ts is defined as the scattering angle of the
 fermion or anti-fermion with respect to the photon-photon axis
 in the photon-photon centre-of-mass frame,
 as shown in Figure~\ref{fig:chap2_03}(b).
 In this report, the azimuthal angle \az is defined, as in
 Ref.~\cite{SEY-9801}, as the angle between the observed electron
 and the fermion which, in the photon-photon centre-of-mass frame,
 is scattered at  $\cts<0$, as shown in Figure~\ref{fig:chap2_03}(b).
 There exist slightly different definitions of \az in the literature.
 The different definitions are due to the different
 choices made to accommodate the fact that the unintegrated
 structure function \FtA is antisymmetric in
 \cts if the angle \az is chosen to be the angle between the electron
 and the fermion or anti-fermion.
 There are several ways to redefine \az in such a way that the integration of
 \FtA with respect to \cts does not vanish, see Section~\ref{sec:QED}
 for details.
%
%
\subsection{Experimental considerations}
\label{sec:detec}
%
\begin{figure}[tbp]
\begin{center}
{\includegraphics[width=0.70\linewidth]{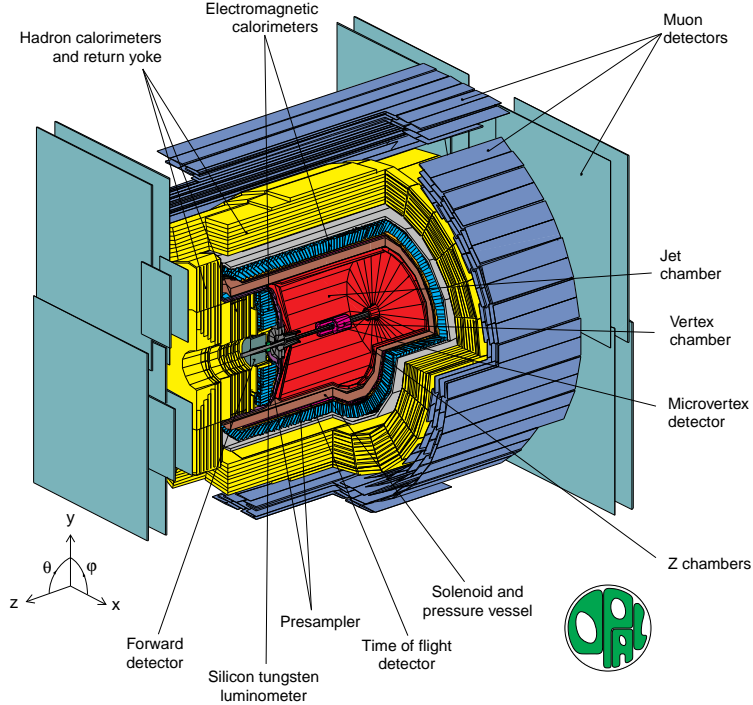}}
\caption[
         A schematic view of the OPAL detector.
        ]
        {
         A schematic view of the OPAL detector.
        }\label{fig:chap2_04}
\end{center}
\end{figure}
%
 The measurement of structure functions involves the determination
 of $x$, \qsq, \psq and \az.
 The capabilities of the different LEP detectors are very similar
 and they have only slightly different acceptances for the
 scattered electrons and the final state $X$.
 As an example, the acceptance of the OPAL detector, shown
 in Figure~\ref{fig:chap2_04}, is discussed.
 The scattered electrons are detected by various electromagnetic
 calorimeters. The final state $X$ is measured with tracking
 devices and calorimeters which are sensitive to electromagnetic as well
 as hadronic energy deposits, supplemented by muon detectors.
 The acceptance ranges of the various components of the OPAL detector
 are listed in Table~\ref{tab:chap2_01}.
 \par
\begin{figure}[tbp]
\begin{center}
\mbox{  }\\
\mbox{  }\\
{\includegraphics[width=1.0\linewidth]{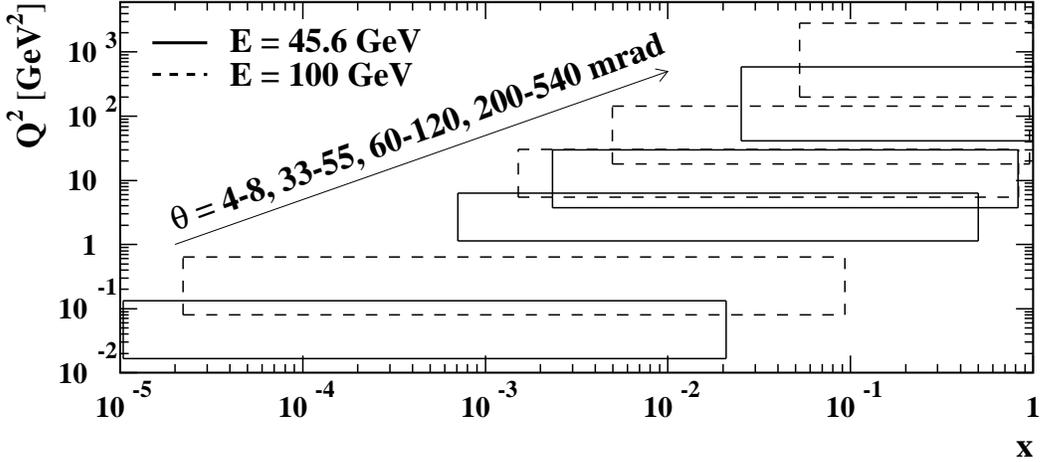}}
\caption[
         The kinematical coverage of the OPAL detector.
        ]
        {
         The kinematical coverage of the OPAL detector.
         Shown are the accepted ranges in $x$ and \qsq, for $\psq=0$ and
         for two
         values of the energy of the beam electrons, 45.6 and 100~\gev.
         The numbers are obtained for specific ranges in $W$ and
         a minimum energy required for the scattered electron,
         explained in the text.
        }\label{fig:chap2_05}
\end{center}
\end{figure}

 For two values of the energy of the beam electrons $E = 45.6$ and 100~\gev
 the covered phase space in terms of $x$ and \qsq, for $\psq=0$ is
 schematically shown in Figure~\ref{fig:chap2_05}.
 The values of $x$ and \qsq are obtained from the kinematical relations
 listed above, using a range in photon-photon invariant mass of
 $2.5<W<40/60$~\gev, for $E = 45.6/100$~\gev and assuming that the
 observed electrons carry at least 50$\%$ of the energy of the beam
 electrons.
 The kinematical coverage in principle extends from $10^{-5}$ to about $1$
 in $x$ and from $10^{-2}$ to $3\cdot10^3$~\gevsq in \qsq, but
 measurements of the photon structure cover only the approximate ranges
 of $10^{-3}<x<1$ and $1<\qsq<10^3$~\gevsq.
 This is because at large \qsq the statistics are small, and at very low
 \qsq the background conditions are severe.
%
\renewcommand{\arraystretch}{1.0}
\begin{table}[tbp]
\caption[
         The main parameters of the OPAL detector relevant
         for measurements of the photon structure.
        ]
        {
         The main parameters of the OPAL detector relevant
         for measurements of the photon structure.
         Shown are the acceptance ranges in polar angle $\theta$ for
         the scattered electrons and the final state $X$.
         The number 33~mrad reflects the electron acceptance for beam energies
         within the LEP2 programme, at LEP1 energies the clean acceptance
         already started at approximately 27~mrad.
         \\
         }\label{tab:chap2_01}
\begin{center}\begin{tabular}{cc}\hline
 \multicolumn{2}{c}{scattered electrons}                   \\\hline
 electromagnetic cluster   & 4-8, 33-55, 60-120, $>200$ [mrad]\\\hline
 \multicolumn{2}{c}{}                    \\
 \multicolumn{2}{c}{}                    \\\hline
 \multicolumn{2}{c}{final state $X$}   \\\hline
 charged particles         & $\act<0.96$ \\
 electromagnetic cluster   & $\act<0.98$ \\
 hadronic cluster          & $\act<0.99$ \\
 muons                     & $\act<0.98$ \\\hline
\end{tabular}
\end{center}\end{table}
%
 Therefore the present measurements of the photon structure are limited to
 $\tonep>33$~mrad, which means $\qsq>\qsqmin \approx 1.1 / 5.5$~\gevsq,
 for $E = 45.6/100$~\gev, as shown in Figure~\ref{fig:chap2_05}.
 Here \qsqmin, calculated from Eq.~(\ref{eqn:q2}) for
 $\eonep=0.5E$ and $\tonep=33$~mrad, is the minimum photon
 virtuality at which an electron can be observed.
 \par
%
%
 The considerations for the \qsq acceptance also apply to the
 acceptance in \psq for the second photon.
 Due to the limited coverage of the detector close to the beam direction the
 scattered electrons radiating the quasi-real photons cannot be detected
 up to $\ttwopmax=33$~mrad with the exception of a small region of
 4 to 8~mrad.
 Consequently, for deep inelastic electron-photon scattering the
 experiments effectively integrate over the invisible part of the
 \psq range up to a value \psqmax.
 Because \psq depends on energy and angle of the electron, \psqmax is
 not a fixed number, but depends on the minimum angle and energy
 required to observe an electron.
 Approximations of \psqmax are \qsqmin and the \psq value corresponding
 to an electron carrying the energy of the beam electrons and escaping
 at \ttwopmax, which is two times \qsqmin in the example from above.
 \par
 As a consequence of the limited acceptance the measured structure
 functions depend on the \psq distribution of the not observed
 quasi-real photon, and this dependence increases for increasing energy
 of the beam electrons.
%
%

%
%
\section{Theoretical framework}
\label{sec:theo}
 In this section the formalism needed for the interpretation of
 the measurements performed by the various experiments is outlined.
 The discussion is not complete, but focuses on general considerations
 and on the formulae relevant for the understanding of the
 experimental results.
 These results concern measurements of structure functions and of
 differential cross-sections, related to the QED and hadronic
 structure of quasi-real and virtual photons.
 The structure of quasi-real photons is investigated for
 $\qsq\gg\psq\approx 0$, and the structure of virtual photons
 for the regions $\qsq,\psq\gg\me^2$, or $\qsq,\psq\gg\lamsq$,
 where \me is the mass of the electron and \lam is the QCD
 scale\footnote{In principle one has to specify the number of
 flavours to which \lam corresponds, and in next-to-leading order
 also the factorisation scheme in which \lam is expressed.
 For example, \lamv means four active flavours and the \msbl
 factorisation scheme, see Ref.~\cite{PDG-9801} for details.
 However, when constructing parton distribution functions
 of the photon, in some cases \lam is taken as a fixed number independent
 of the number of flavours used, because, given the number of free
 parameters, there is no sensitivity to \lam.
 For simplicity, unless explicitly stated otherwise, here \lam is
 used either to denote a fixed number or as a shorthand for $\lam_4$.
 Numerically, in leading order, $\lam_4 = 0.2$~\gev corresponds to
 $\lam_3 = 0.232$~\gev.}.
 Since either the cross-section picture, or the structure function
 picture is relevant for the different measurements, both are
 discussed in detail, starting with the differential cross-section.
%
%
\subsection{Individual cross-sections}
\label{sec:cross}
 The general form of the differential cross-section for the scattering of
 two electrons via the exchange of two photons, Eq.~(\ref{eqn:reactkin}),
 using the multipheripheral diagram integrated over all angles
 except \barph is given as
%
\begin{eqnarray}
  \der^6\sigma
  &=&
  \der^6\sigma(\ee\rightarrow\ee X)
  \nonumber\\  &=&
  \frac{\der^3\ponep\der^3\ptwop}{\eonep\etwop}
  \frac{\aemsq}{16\pi^4\qsq\psq}
  \left[
  \frac{(p\cdot q)^2-\qsq\psq}{(\pone\cdot \ptwo)^2
        -m_{\mathrm e}^2m_{\mathrm e}^2}
  \right]^{1/2}
  \nonumber
  \\ &&
  \left(
  4\ropp\rtpp\stt +
  2 \vert \ropm\rtpm \vert \ttt\costph +
  2\ropp\rtnn\stl
  \right.
  \nonumber
  \\
  &&
  \left.
  + 2\ronn\rtpp\slt
  + \ronn\rtnn\sll -
  8 \vert\ropn\rtpn \vert \ttl\cosph
  \right)\, ,
\label{eqn:true}
\end{eqnarray}
%
 taken from Ref.~\cite[Eq.5.12]{BUD-7501}.
 The four-vectors and kinematic variables are defined
 in Section~\ref{sec:kinem}.
 The total cross-sections \stt, \stl, \slt and \sll
 and the interference terms \ttt and \ttl
 correspond to specific helicity states of the photons
 (T=transverse and L=longitudinal).
 Since a real photon can have only transverse polarisation, the terms where
 at least one photon has longitudinal polarisation have to vanish in the
 corresponding limit $\qsq \rightarrow 0$ or $\psq\rightarrow 0$.
 These terms have the following functional form:
 $\slt\propto\qsq$,
 $\stl\propto\psq$,
 $\sll\propto\qsq\psq$
 and
 $\ttl\propto\sqrt{\qsq\psq}$.
 The terms $\rho_1^{jk}$ and $\rho_2^{jk}$,
 where $j,k\in(+,-,0)$ denote the photon helicities,
 are elements of the photon density matrix
 which depend only on the four-vectors $q$, $p$, \pone and \ptwo and
 on \me.
 They are taken from Ref.~\cite[Eq.5.13]{BUD-7501} and have the following
 form
%
\begin{eqnarray}
 2\ropp          &=& \frac{\left(2\pone\cdot p-p\cdot q\right)^2}
                     {(p\cdot q)^2 -\qsq\psq} + 1 - 4\frac{\me^2}{\qsq},
                     \nonumber\\
 2\rtpp          &=& \frac{\left(2\ptwo\cdot q-p\cdot q\right)^2}
                     {(p\cdot q)^2 -\qsq\psq} + 1 - 4\frac{\me^2}{\psq},
                     \nonumber\\
 \ronn           &=& 2\ropp - 2 + 4\frac{\me^2}{\qsq}, \nonumber\\
 \rtnn           &=& 2\rtpp - 2 + 4\frac{\me^2}{\psq}, \nonumber\\
 \vert\ripm\vert &=& \ripp - 1,\nonumber\\
 \vert\ripn\vert &=& \sqrt{\left(\rinn+1\right)\vert\ripm\vert}.
\label{eqn:rhos}
\end{eqnarray}
%
 \par
 Experimentally two kinematical limits are studied for leptonic and
 hadronic final states.
 Firstly the situation where both photons are highly virtual and
 secondly the situation where one photon is quasi-real and the other
 highly virtual: the situation of deep inelastic
 electron-photon scattering.
 The corresponding limits of Eq.~(\ref{eqn:true}) are discussed next.
 \par
 If both photons are highly virtual the differential cross-section reduces
 to a much more compact form, because Eq.~(\ref{eqn:true}) can be evaluated
 in the limit $\qsq, \psq \gg m_{\mathrm e}^2$.
 In this limit the following relations can be obtained
 between the $\rho_i^{jk}$ given in Eq.~(\ref{eqn:rhos})
%
\begin{eqnarray}
 \rinn           &=& 2(\ripp - 1), \nonumber\\
 \vert\ripm\vert &=& \frac{\rinn}{2},\nonumber\\
 \frac{\vert\ripn\vert}{\ripp} &=&
 \sqrt{\frac{\rinn+1}{\ripp}\cdot\frac{\vert\ripm\vert}{\ripp}}=
 \sqrt{\frac{2(\rinn-\rinn/2+\ripp)}{2\ripp}\cdot\frac{\rinn}{2\ripp}}
 \nonumber\\
 &=&\sqrt{\left(\frac{\rinn}{2\ripp}+1\right)\cdot\frac{\rinn}{2\ripp}}.
\label{eqn:rhoslq}
\end{eqnarray}
%
 Defining $\rinn/2\ripp\equiv\epsilon_i$ Eq.~(\ref{eqn:true}) reads
%
 \begin{eqnarray}
 \mathrm{d}^6\sigma
  &=&
  \frac{\der^3\ponep\der^3\ptwop}{\eonep\etwop}
  \frac{\aemsq}{16\pi^4\qsq\psq}
  \left[
  \frac{(p\cdot q)^2-\qsq\psq}{(\pone\cdot\ptwo)^2-
         m_{\mathrm e}^2m_{\mathrm e}^2}
  \right]^{1/2} 4\ropp\rtpp
  \cdot
  \nonumber  \\  &&
  \left(
  \stt +
  \epsilon_2 \stl +
  \epsilon_1 \slt +
  \epsilon_1\epsilon_2 \sll +
  \half\epsilon_1\epsilon_2\ttt\costph
  \right. \nonumber  \\  && \left.
  -\sqrt{2(\epsilon_1+1)\epsilon_1}\sqrt{2(\epsilon_2+1)\epsilon_2}\,\ttl\cosph
  \right)\, .
 \label{eqn:truedbe}
 \end{eqnarray}
%
 Finally for $\epsilon_i\approx 1$, which is fulfilled by selecting
 events at low values of $y$ and $r$, the differential cross-section
 can be written as
%
 \begin{eqnarray}
 \mathrm{d}^6\sigma
  &=&
  \frac{\der^3\ponep\der^3\ptwop}{\eonep\etwop}
  \frac{\aemsq}{16\pi^4\qsq\psq}
  \left[
  \frac{(p\cdot q)^2-\qsq\psq}{(\pone\cdot\ptwo)^2-
         m_{\mathrm e}^2m_{\mathrm e}^2}
  \right]^{1/2} 4\ropp\rtpp
  \cdot
  \nonumber  \\  &&
  \left(
  \stt + \stl + \slt + \sll + \half\ttt\costph -  4\ttl\cosph
  \right)\, .
 \label{eqn:truedb}
 \end{eqnarray}
%
 This equation can be used to define an effective structure function
 $\feff\propto \stt + \stl + \slt + \sll + \half\ttt\costph - 4\ttl\cosph$.
 This effective structure function can be measured by experiments.
 However, to relate \feff to the structure functions \ft and \fl
 discussed below, further assumptions are needed.
 By assuming that the interference terms do not
 contribute that \sll is negligible and also using $\stl=\slt$,
 the effective structure function can be expressed
 by means of Eq.~(\ref{eqn:struc}), as $\feff = \ft + 3/2 \fl$.
 \par
 If the interference terms \ttt and \ttl are independent of \barph,
 the integration
 over \barph of the terms containing \cosph and \costph vanishes, and
 the cross-section is proportional to \stt + \stl + \slt + \sll.
 The total cross-sections and interference terms can be expressed
 using \qsq, \psq, \wsq, and the mass of the produced fermion, \mf.
 However, there is a kinematical correlation between these variables and
 \barph, which leads to the fact that in several kinematical regions
 \ttt and \ttl are not independent of \barph.
 Consequently, the terms proportional to \cosph and \costph do not vanish,
 even when integrated over the full range in \barph,
 as explained in Ref.~\cite{ART-9501}.
 The resulting contributions can be very large, depending on the ratios
 $\qsq/\psq$, $\qsq/\wsq$ and $\psq/\wsq$.
 Due to the large interference terms in some regions of phase space,
 cancellations occur in Eq.~(\ref{eqn:truedb}) between the
 cross-section and interference terms, and therefore no clear relation
 between a structure function and the cross-section terms can be found.
 In this situation the cleanest experimentally accessible measurement
 is the differential cross-section as defined by Eq.~(\ref{eqn:truedb}).
 \par
%
\renewcommand{\arraystretch}{1.1}
\begin{table}[tbp]
\caption[
         The individual contributions to the differential cross-section
         for muon pair production.
        ]
        {
         The individual contributions to the differential cross-section
         for muon pair production.
         The numbers given are for several kinematical situations
         explained in the text, and correspond to the integrals of
         the distributions shown in Figures~\ref{fig:chap3_01}
         and~\ref{fig:chap3_02}.\\
        }\label{tab:chap3_01}
\begin{center}
\begin{tabular}{cccccccccc}\cline{7-10}
 \multicolumn{6}{c}{}&
 \multicolumn{4}{c}{$\sigma$ [pb]}\\\hline
 \multicolumn{6}{c}{used terms in Eq.~(\protect\ref{eqn:true})}
 & \protect\ref{fig:chap3_01}(a)
 & \protect\ref{fig:chap3_01}(b)
 & \protect\ref{fig:chap3_02}(a)
 & \protect\ref{fig:chap3_02}(b) \\
\hline
\stt &      &      &      &      &      & 2.20& 1.26& 2.29& 285\\
\stt & \stl &      &      &      &      & 3.24& 1.68& 2.88& 349\\
\stt & \stl & \slt &      &      &      & 4.20& 2.07& 3.36& 360\\
\stt & \stl & \slt & \sll &      &      & 4.30& 2.12& 3.38& 362\\
\stt & \stl & \slt & \sll & \ttt &      & 4.17& 2.11& 3.37& 360\\
\stt & \stl & \slt & \sll & \ttt & \ttl & 3.35& 1.93& 3.24& 350\\
\hline\\
\end{tabular}
\end{center}\end{table}
%
\begin{figure}
\begin{center}
{\includegraphics[width=1.0\linewidth]{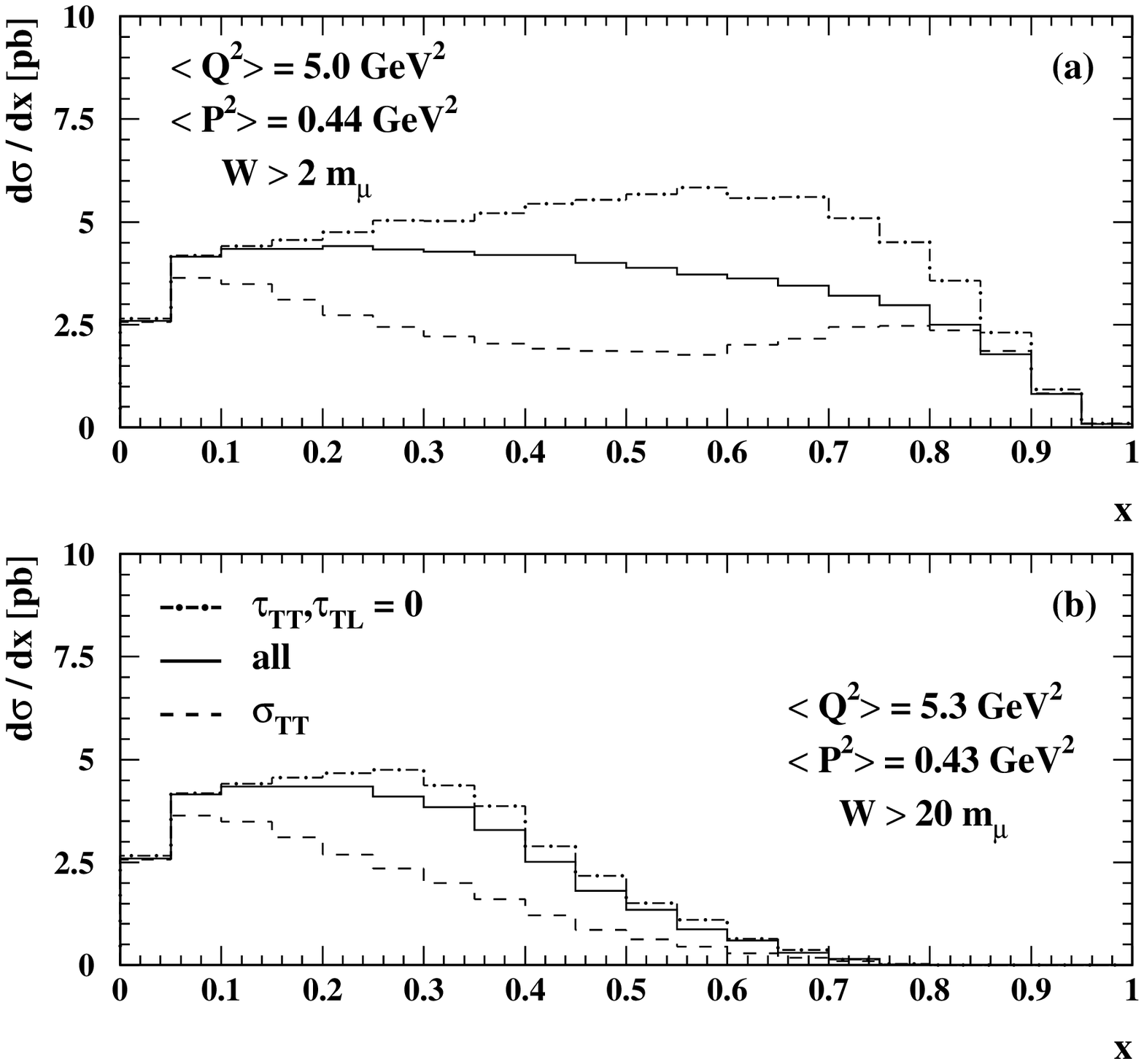}}
\caption[
         The predicted differential cross-section \dsigdx
         for the reaction $\ee\rightarrow\ee\mupmum$ for the acceptance
         of the PLUTO experiment, and for two different lower limits on $W$.
        ]
        {
         The predicted differential cross-section \dsigdx
         for the reaction $\ee\rightarrow\ee\mupmum$ for the acceptance
         of the PLUTO experiment, and for two different lower limits on $W$.
         Shown are the differential cross sections for $W>2 m_\mu$ in (a),
         and for $W>20 m_\mu$ in (b). See text for further details.
         The three different histograms correspond to the differential
         cross-section \dsigdx using all terms of Eq.~(\ref{eqn:true})
         (full), using only \stt (dash) or neglecting only the interference
         terms \ttt and \ttl (dot-dash).
        }\label{fig:chap3_01}
\end{center}
\end{figure}
%
 For the case of leading order QED fermion pair production the relevance
 of the individual terms for different kinematical regions can be studied.
 For example, Figure~\ref{fig:chap3_01} shows the differential
 cross-section \dsigdx for muon pair production in the kinematical
 acceptance range of the PLUTO experiment, Ref.~\cite{PLU-8405},
 and for two different lower limits on $W$.
 The kinematical requirements are, $\eonep,\etwop>0.35 E$ for
 $E=17.3$~\gev, $100<\tonep<250$~mrad, $31<\ttwop<55$~mrad, and
 in addition $W>2 m_\mu$ in Figure~\ref{fig:chap3_01}(a),
 and  $W>20 m_\mu$ in Figure~\ref{fig:chap3_01}(b).
 This leads to average values of \psq and \qsq of
 $\pzm\approx 0.44$~\gevsq and  $\qzm\approx 5.3$~\gevsq.
 The individual contributions are listed in Table~\ref{tab:chap3_01}.
 Shown in Figure~\ref{fig:chap3_01} are the differential cross
 sections \dsigdx for
 three different scenarios: \dsigdx using all terms of Eq.~(\ref{eqn:true}),
 using only \stt, or neglecting the interference terms \ttt and \ttl,
 all as predicted by the GALUGA program, Ref.~\cite{SCH-9801}, which is
 described in Section~\ref{sec:gener}.
 The difference between \dsigdx using only \stt and
 \dsigdx by neglecting only the interference terms, shows that
 there are large contributions from the cross-sections
 containing at least one longitudinal photon, \stl, \slt and
 \sll.
 But also the interference terms themselves give large negative
 contributions, as shown by the difference between the \dsigdx using all
 terms and \dsigdx by neglecting the interference terms.
 The importance of the interference terms decreases
 for increasing \wsq, as shown in Figure~\ref{fig:chap3_01}(b).
 However, this comes at the expense of a significant reduction in
 the acceptance at high values of $x$.
 \par
%
\begin{figure}
\begin{center}
{\includegraphics[width=1.0\linewidth]{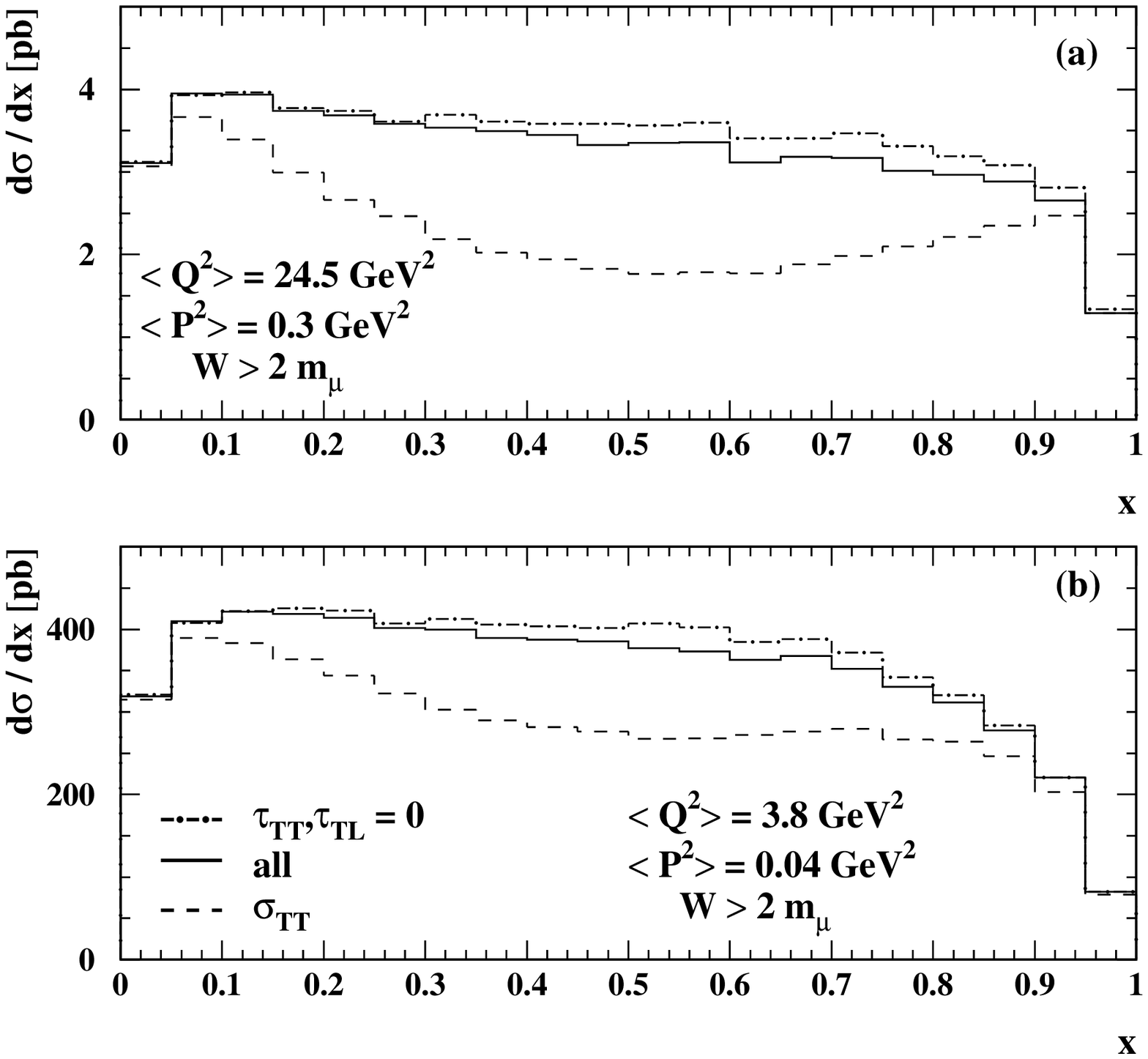}}
\caption[
         The predicted differential cross-section \dsigdx
         for the reaction $\ee\rightarrow\ee\mupmum$ for some typical
         acceptances of a LEP experiment.
        ]
        {
         The predicted differential cross-section \dsigdx
         for the reaction $\ee\rightarrow\ee\mupmum$ for some typical
         acceptances of a LEP experiment.
         Shown are the differential cross sections for $W>2 m_\mu$.
         In (a)
         \dsigdx is shown for a typical acceptance for the
         exchange of two virtual photons with two detected electrons
         and for $E=94.5$~\gev,
         and in
         (b)
         \dsigdx is shown for a typical acceptance for
         deep inelastic electron-photon scattering for $E=45.6$~\gev.
         See text for further details.
         The three different histograms correspond to the differential
         cross-section \dsigdx using all terms of Eq.~(\ref{eqn:true})
         (full), using only \stt (dash) or neglecting only the interference
         terms \ttt and \ttl (dot-dash).
        }\label{fig:chap3_02}
\end{center}
\end{figure}
%
 Figure~\ref{fig:chap3_02}(a) shows the same quantities for
 the typical acceptance of a LEP detector at $E=94.5$~\gev,
 when using the very low angle electromagnetic calorimeters and the
 calorimeters used for the high precision luminosity measurement.
 In this case, the kinematical requirements are,
 $\eonep,\etwop>0.5 E$ for $E=94.5$~\gev, $33<\tonep<120$~mrad,
 $4<\ttwop<8$~mrad, and $W>2 m_\mu$.
 The increase in the energy of the beam electrons is compensated by
 a smaller value of \ttwop, resulting in an average value
 of $\pzm= 0.3$~\gevsq, similar to the PLUTO acceptance.
 However, the average value of \qsq is increased to
 $\qzm= 24.5$~\gevsq, which results in increased ratios
 \qsq/\psq and \qsq/\wsq.
 The result is that the total contribution of the interference terms
 decreases the differential cross section by less than 4$\%$,
 compared to 28(10)$\%$ in the case of the PLUTO
 acceptance for $W>2(20)m_\mu$.
 This shows that the importance of the interference terms varies
 strongly as a function of the kinematical range.
 In the kinematical region of the LEP high energy programme the
 importance of the interference terms is smaller than for the PLUTO
 region.
 \par
 Unfortunately, no general statement of the importance of these terms
 can be made for the case of quark pair production in the framework
 of QCD.
 However, in the regions of phase space where the leading order
 point-like \qqbar production process dominates, the cross-section
 for quark pair production, in the quark parton model, is exactly the
 same as for muon pair production, except for the different masses of muons
 and quarks, and the above considerations can be applied.
 \par
 For deep inelastic electron-photon scattering, $\qsq\gg\psq\approx 0$,
 the terms \stl, \sll and \ttl vanish due to their \psq dependence as
 \psq approaches zero.
 This means that all contributions from longitudinal quasi-real photons
 can be neglected because the longitudinal polarisation state vanishes for
 $\psq=0$.
 Also the term proportional to \ttt vanishes, although \ttt itself
 does not vanish, because for $\psq=0$ the angle \barph is undefined.
 Consequently, for deep inelastic electron-photon scattering,
 the differential cross-section Eq.~(\ref{eqn:true}) reduces to
%
 \begin{eqnarray}
 \mathrm{d}^6\sigma
  &=&
  \frac{\der^3\ponep\der^3\ptwop}{\eonep\etwop}
  \frac{\aemsq}{16\pi^4\qsq\psq}
  \left[
  \frac{(p\cdot q)^2-\qsq\psq}{(\pone\cdot\ptwo)^2-
         m_{\mathrm e}^2m_{\mathrm e}^2}
  \right]^{1/2}
  \cdot
  \nonumber  \\  &&
  4\ropp\rtpp\phantom{\frac{}{}}
  \left[\stt+\frac{\ronn}{2\ropp}\slt\right]\, .
 \label{eqn:truesing}
 \end{eqnarray}
%
 This means that only the terms \stt and \slt contribute. They correspond
 to the situation where the structure of a transverse target photon,
 $p$, is probed by a transverse or longitudinal virtual photon, $q$,
 respectively.
 \par
 Experimentally, due to the limited acceptance discussed in
 Section~\ref{sec:detec}, \psq can only be kept small, but it is not
 exactly zero.
 The numerical effect of the various contributions due to the finite
 \psq are shown in Figure~\ref{fig:chap3_02}(b) for a typical
 acceptance of a LEP detector for $E=45.6$~\gev.
 The kinematical requirements are $\eonep>0.5 E$,
 $27<\tonep<120$~mrad, $\ttwop<27$~mrad, and $W>2 m_\mu$.
 The importance of the reduction of the cross-section by the
 interference terms is further decreased to
 around 3$\%$, and the contribution of \stl and \sll to the cross
 section is also around 3$\%$ and positive, such that the two
 almost cancel each other.
 In this situation the total cross-section is accurately described
 by \stt and \slt only.
 \par
 In the case of muon pair production, $\ff=\mupmum$, the cross-section is
 determined by QED. Equation~(\ref{eqn:true}) and consequently also the
 limits discussed above, Eqs.~(\ref{eqn:truedb},\ref{eqn:truesing}),
 contain the full information, and it is sufficient to describe the
 reaction in terms of cross-sections.
 However, most of the experimental results are expressed in terms of
 structure functions, since in the case of quark pair production the
 cross-section cannot be calculated in QCD and has to be parametrised
 by structure functions.
 The relations between the cross-sections and the structure functions
 are defined as
%
 \begin{eqnarray}
  2x\fTxqp &=&
  \frac{\qsq}{4\pi^2\aem}\frac{\sqrt{(p\cdot q)^2-\qsq\psq}}{p\cdot q}
  \nonumber \\& &
  \biggl[\stt(x,\qsq,\psq) -\half\stl(x,\qsq,\psq)\biggr],
  \nonumber \\
  \ftxqp &=&
  \frac{\qsq}{4\pi^2\aem}\frac{p\cdot q}{\sqrt{(p\cdot q)^2-\qsq\psq}}
  \biggr.
 \nonumber \\& &
  \biggl[\stt(x,\qsq,\psq) +\slt(x,\qsq,\psq)\biggr.
  \nonumber \\& &
  \biggl.
  -\half\sll(x,\qsq,\psq) -\half\stl(x,\qsq,\psq)\biggr],
  \nonumber \\
  \flxqp &=& \ftxqp - 2x\fT(x,\qsq,\psq),
\label{eqn:struc}
\end{eqnarray}
%
 as given, for example, in Ref.~\cite{BER-8701}.
 These equations can be used for the definition of both QED and hadronic
 structure functions. In the limit $\psq=0$ the relations
%
 \begin{eqnarray}
  2x\fTxq
       &=& \frac{\qsq}{4\pi^2\aem}\stt(x,\qsq) \nonumber \\
 \ftxq &=& \frac{\qsq}{4\pi^2\aem}\,
          \left[\stt(x,\qsq) +\slt(x,\qsq)\right] \nonumber \\
 \flxq &=& \frac{\qsq}{4\pi^2\aem}\, \slt(x,\qsq)
\label{eqn:strucnull}
\end{eqnarray}
%
 are obtained.
 In the QED case the structure functions can be calculated as
 discussed in Section~\ref{sec:QED}, whereas for the hadronic
 structure functions model assumptions have to be made which are
 discussed in detail in Section~\ref{sec:QCD}.
%
%
\subsection{Equivalent photon approximation}
\label{sec:EPA}
 In many experimental analyses of deep inelastic electron-photon
 scattering the differential cross-section for the reaction is not described
 in terms of cross-sections corresponding to specific helicity states
 of the photons, as outlined in Section~\ref{sec:cross}, but in terms
 of structure functions of the transverse quasi-real photon times a flux
 factor for the incoming quasi-real photons of transverse polarisation.
 \par
 In this notation the differential cross-section,
 Eq.~(\ref{eqn:truesing}), can be written in a factorised form as
%
 \begin{eqnarray}
 \frac{\der^4\sigma}
 {\der x\,\der\qsq\,\der z\,\der\psq}
 &=&
 \Ngtt \cdot \frac{2\pi\aemsq}{x\,Q^{4}} \cdot
 \left[1+(1-y)^2\right]\cdot\nonumber\\&&
 \left[2x\fTxq + \frac{2(1-y)}{1+(1-y)^2} \flxq\right]\, .
 \label{eqn:approxlt}
 \end{eqnarray}
%
 In Appendix~\ref{sec:SIGTOSF} this equation is derived from
 Eq.~(\ref{eqn:truesing}) using the limit $\psq\to 0$.
 \par
 By using in addition $\ft=2x\fT+\fl$ the widely used formula
%
 \begin{eqnarray}
 \frac{\der^4\sigma}
 {\der x\,\der\qsq\,\der z\,\der\psq}
 &=&
 \Ngtt \cdot \frac{2\pi\aemsq}{x\,Q^{4}} \cdot\nonumber\\&&
 \left[
 \left(1+(1-y)^2\right) \ftxq  - y^2 \flxq \right]\, ,
 \label{eqn:approx}
 \end{eqnarray}
%
 is obtained.
 Sometimes this formula is also used to study the \psq dependence of
 \ft by using \ftxqp instead of \ftxq. It should be kept in mind
 that the main approximation made in calculating Eq.~(\ref{eqn:approx})
 is $(p\cdot q)^2 - \qsq\psq \approx (p\cdot q)^2$ and that only
 results in the same limit of \ft are meaningful, see
 Appendix~\ref{sec:SIGTOSF} for details.
 To avoid this complication Eq.~(\ref{eqn:truesing}) should be used
 instead.
 \par
 The factor $\der^2N^{\rm T}_{\gamma}/\der z\der\psq$ describing the flux of
 incoming transversely polarised quasi-real photons of finite
 virtuality is the equivalent photon approximation, EPA,
 which was first derived in Ref.~\cite{KES-6001}.
 The EPA is given by
%
 \begin{equation}
  \Ngtt = \frac{\aem}{2\pi}\left[\frac{1+(1-z)^2}{z}
         \frac{1}{\psq} - \frac{2\,\me^2\,z}{P^4}\right],
 \label{eqn:epa}
 \end{equation}
%
 where the first term is dominant.
 The flux of longitudinal photons is
%
 \begin{equation}
  \Ngtl = \frac{\aem}{2\pi}\left[\frac{2(1-z)}{z}
          \frac{1}{\psq}\right],
 \label{eqn:epal}
 \end{equation}
%
 such that the ratio is given by
%
 \begin{equation}
  \Ngtl/\Ngtt \approx \frac{2(1-z)}{1+(1-z)^2} \equiv \epsilon(z).
 \label{eqn:eparat}
 \end{equation}
%
 Comparing this functional form to Eq.~(\ref{eqn:approxlt}) shows that
 the term in front of \fl corresponds to the ratio of the transverse and
 longitudinal flux of virtual photons.
 \par
 For the experimental situation where the electron which radiates
 the quasi-real photon is not detected, the EPA is often used integrated
 over the invisible part of the \psq range.
 The integration boundary \psqmin is given by four-vector conservation
 and \psqmax is determined by the experimental acceptance.
 The experimental values of \psqmax strongly depend on the detector
 acceptance and the energy of the beam electrons, as has been discussed
 in Section~\ref{sec:detec}.
 The integration of the EPA leads to the
 Weizs\"acker-Williams approximation~\cite{WEI-3401,WIL-3401},
 which is a formula for the flux of collinear real photons.
%
\begin{eqnarray}
 \Ngot &=& \int_{\psqmin}^{\psqmax}\der\psq\,\Ngtt
 \nonumber \\\label{eqn:weiz}
       &=&\frac{\aem}{2\pi}\left[
         \frac{1+(1-z)^2}{z}\ln\frac{\psqmax}{\psqmin}-
         2\,\me^2\,z\left(\frac{1}{\psqmin}-\frac{1}{\psqmax}\right)
         \right],\\
\mbox{where}&&\,\,\,\psqmin = \frac{\me^2\,z^2}{1-z}, \quad\,\,
\mbox{and}  \,\,   \psqmax  = (1-z)E^2\ttwopmax\nonumber.
\end{eqnarray}
%
 \par
%
\begin{figure}
\begin{center}
{\includegraphics[width=1.0\linewidth]{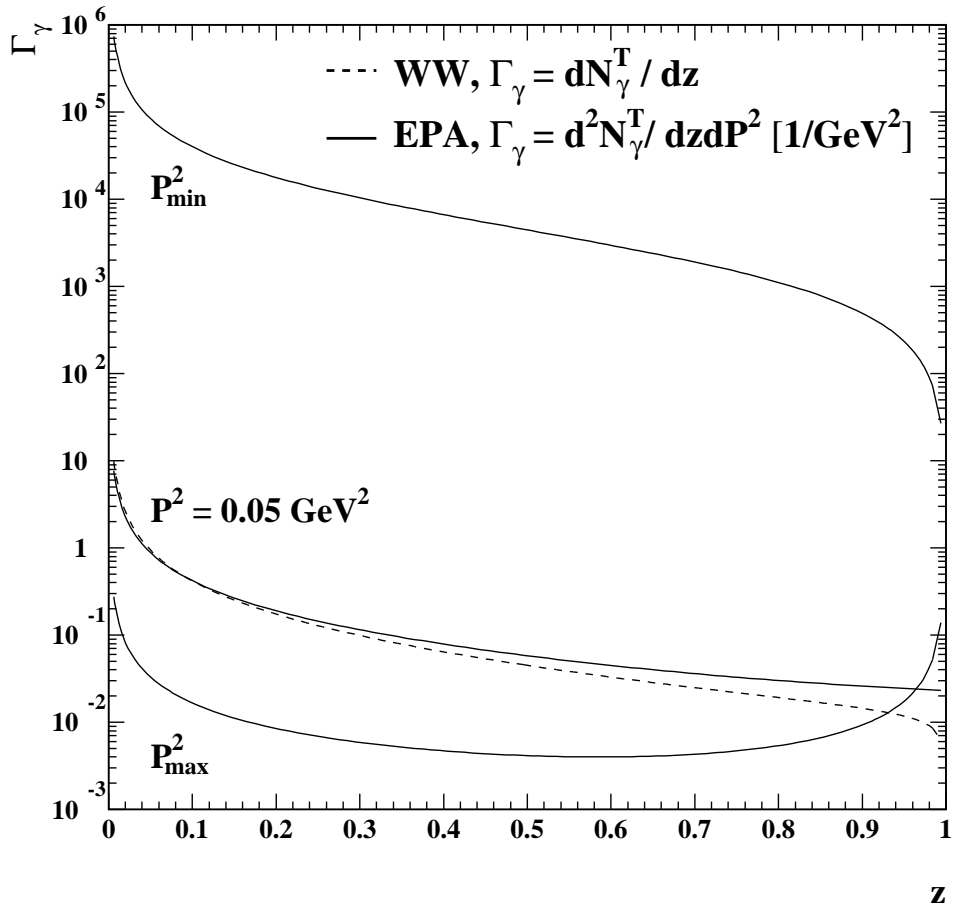}}
\caption[
         Comparison of the equivalent photon
         and the Weizs\"acker-Williams approximations.
        ]
        {
         Comparison of the equivalent photon
         and the Weizs\"acker-Williams approximations.
         The EPA is shown for three choices of \psq: \psqmin, \psqmax with
         $\psqmax(z=0)=1.4$~\gevsq and a fixed value of $\psq=0.05$~\gevsq.
         The EPA is compared to the Weizs\"acker-Williams approximation (WW),
         obtained by integrating the EPA, using the same values of
         \psqmin and \psqmax.
        }\label{fig:chap3_03}
\end{center}
\end{figure}
%
 The strong dependence of the EPA on the virtuality of the quasi-real
 photon is demonstrated in Figure~\ref{fig:chap3_03}, where the
 EPA is shown for three values of \psq.
 Shown are, firstly $\psq= \psqmin$ the smallest value possible,
 secondly  $\psq=\psqmax(z=0) = 1.4$~\gevsq, a typical value for a
 LEP detector for an \epem centre-of-mass energy of the mass
 of the $Z$ boson, $\ssee = m_Z$, and
 thirdly a typical value of an average \psq observed
 in an analysis of the QED structure of the photon, $\psq=0.05$~\gevsq.
 In the range \psqmin to \psqmax the EPA is reduced by about six orders
 of magnitude. In addition, the EPA is compared to the Weizs\"acker-Williams
 approximation, Eq.~(\ref{eqn:weiz}), using \psqmin and the same value of
 \psqmax.
 In this specific case the result of the integration
 is rather close to the EPA at the average \psq.
 \par
 It is clear that for different levels of accuracy different
 formulae have to be chosen for adequate comparisons to the theoretical
 predictions, and special care has to be taken when the \psq dependence
 is studied.
 \par
 Several improvements of the EPA have been suggested in the literature
 for different applications in electron-positron and electron-proton
 collisions. The discussion of these improvements is beyond the scope
 of this review and the reader is referred to the original
 publications, Refs.~\cite{HAG-9101,FRI-9301,DRE-9401,SCH-9602}.
 \par
%
%
\subsection{QED structure functions}
\label{sec:QED}
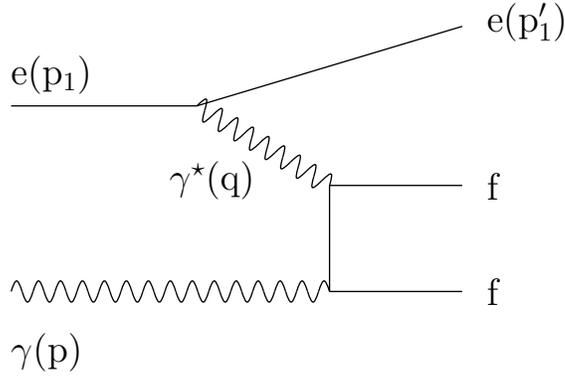
\begin{figure}[tbp]\unitlength 1pt
\begin{center}
\begin{picture}(200,150)(0,0)
 \Line(0,100)(70,100)
 \Line(70,100)(170,130)
 \Photon(70,100)(120,70){4}{8.5}
 \Line(120,70)(120,30)
 \Line(120,70)(170,70)
 \Text(  0,105)[lb]{\mbox{\large $\rm e(\pone)$}}
 \Text(180,125)[lb]{\mbox{\large $\rm e(\ponep)$}}
 \Text(180, 65)[lb]{\mbox{\large $\rm f$}}
 \Text(180, 25)[lb]{\mbox{\large $\rm f$}}
 \Text( 60, 65)[lb]{\mbox{\large $\rm \gamma^{\star}(q)$}}
 \Text(  0, 00)[lb]{\mbox{\large $\rm \gamma(p)$}}
 \Photon(0,30)(120,30){4}{14.5}
 \Line(120,30)(170,30)
\end{picture}
\caption[
         A diagram of the reaction ${\rm e}\gamma\to{\rm e}\gsg\to{\rm e}\ff$.
        ]
        {
         A diagram of the reaction ${\rm e}\gamma\to{\rm e}\gsg\to{\rm e}\ff$.
        }\label{fig:chap3_04}
\end{center}
\end{figure}

 Two topics concerning QED structure functions have been experimentally
 addressed by studying the deep inelastic electron-photon scattering
 reaction, shown in Figure~\ref{fig:chap3_04}.
 First the \az distribution has been measured, leading to the
 determination of the structure functions \faqed and
 \fbqed, which are obtained for real photons, $\psq=0$.
 Second the structure function \ftqed, and its dependence on
 \psq has been measured. The theoretical framework of these two
 topics is discussed here in turn.
 \par
%
%
 The starting point for the measurement of \faqed and \fbqed
 is the full differential cross-section for deep inelastic
 electron-photon scattering for real photons at $\psq=0$
%
\begin{eqnarray}
  \frac{\der^4\sigma_{\e\gamma\rightarrow\e\ff}}
       {\der x\,\der \qsq\, \der \zzp\, \der \az/2\pi} &=&
  \frac{2\pi\alpha^2}{xQ^4}
  \left[1+(1-y)^2\right]
  \left\{
  \left( 2x\FtTqed + \epsilon(y)\FtLqed \right)
  \right.
  \nonumber \\
  &&
  \left.
  -\rho(y)\FtAqed\cos\az
  +\smfrac12\epsilon(y)\FtBqed\cos2\az
  \right\},
 \label{eqn:crossaz}
\end{eqnarray}
%
 where the functions $\epsilon(y)$ and $\rho(y)$ are both of the
 form $1-{\mathcal{O}}(y^2)$
%
\begin{equation}
 \epsilon(y) = \frac{2(1-y)}{1+(1-y)^2},\quad\,
 \rho(y)     = \frac{(2-y)\sqrt{1-y}}{1+(1-y)^2}
             = \sqrt{2\left[\epsilon(y)+1\right]\epsilon(y)}\,.
\end{equation}
%
 The function $\epsilon(y)$, already defined in Eq.~(\ref{eqn:eparat}),
 is obtained from $\epsilon_1$ in the limit $\psq=0$, see
 Appendix~\ref{sec:SIGTOSF}.
 The function $\rho(y)$ stems from $\sqrt{2}\vert\ropn\vert/\ropp$
 evaluated in the same limit, as can be seen from Eq.~(\ref{eqn:rhoslq}).
 In leading order QED, the differential cross-section depends
 on four non-zero unintegrated structure functions, namely
 \FtTqed, \FtLqed,\FtAqed and \FtBqed.
 They are functions only of $x$, $\beta$ and \zzp, but do not
 depend on \az.
 The kinematic variables are defined from the four-vectors
 in Figure~\ref{fig:chap2_01} and listed in Section~\ref{sec:kinem}.
 The variable \zzp is related to the fermion scattering angle \ts in the
 photon-photon centre-of-mass frame, via $\zzp=\smfrac12(1+\beta\cts)$, with
 $\beta=\sqrt{1-4\mf^2/W^2}$, where \mf denotes the mass of the fermion.
 \par
\begin{figure}
\begin{center}
{\includegraphics[width=1.0\linewidth]{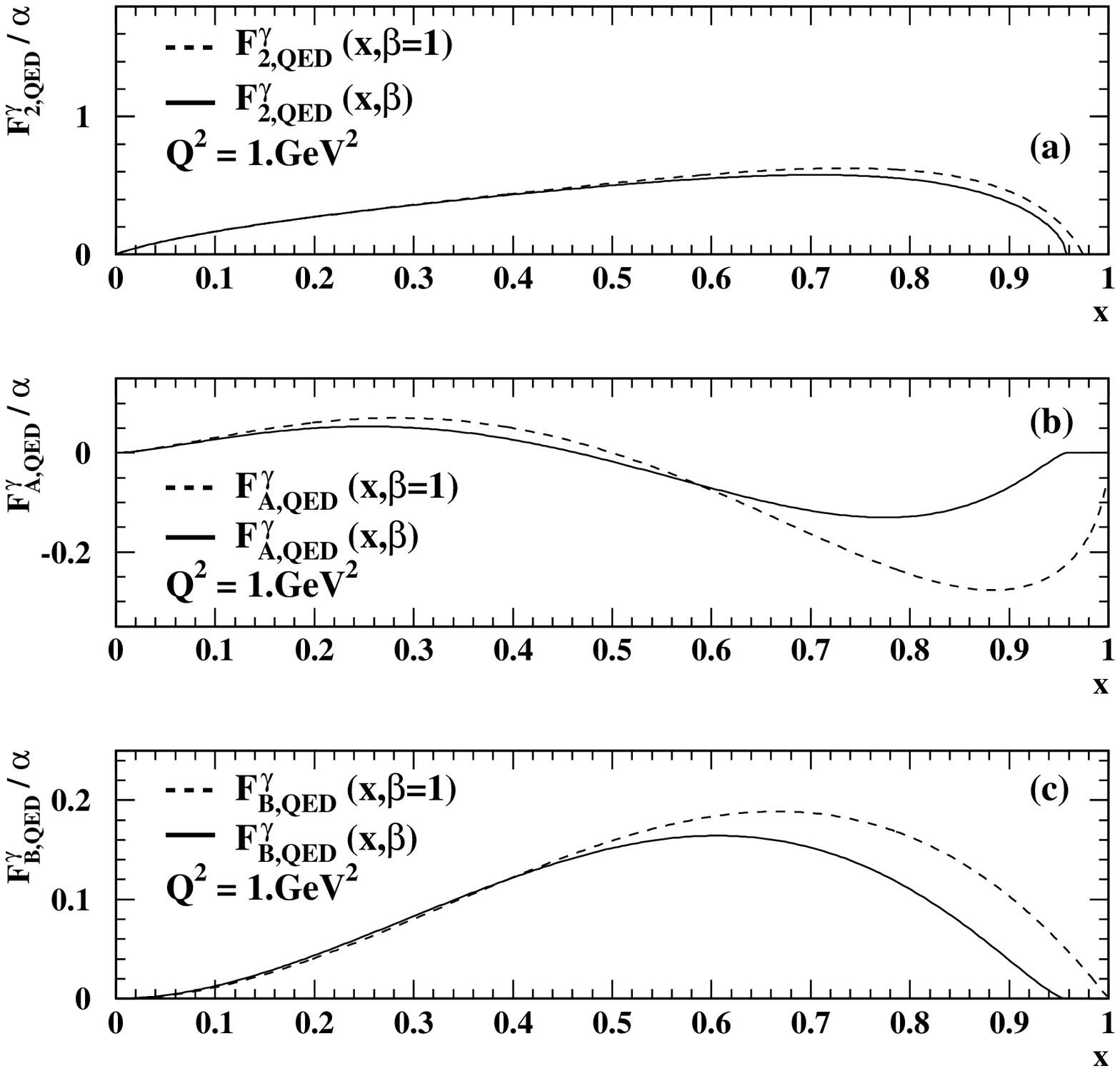}}
\caption[
         The structure functions \ftqed, \faqed and \fbqed
         for \mumu final states at $\qsq=1$~\gevsq.
        ]
        {
         The structure functions \ftqed, \faqed and \fbqed
         for \mumu final states at $\qsq=1$~\gevsq.
         The structure functions are shown with the full mass dependence
         (full) and in the leading logarithmic approximation (dash).
         Shown are
         (a) \ftqed,
         (b) \faqed,
         and
         (c) \fbqed.
        }\label{fig:chap3_05}
\end{center}
\end{figure}
\begin{figure}
\begin{center}
{\includegraphics[width=1.0\linewidth]{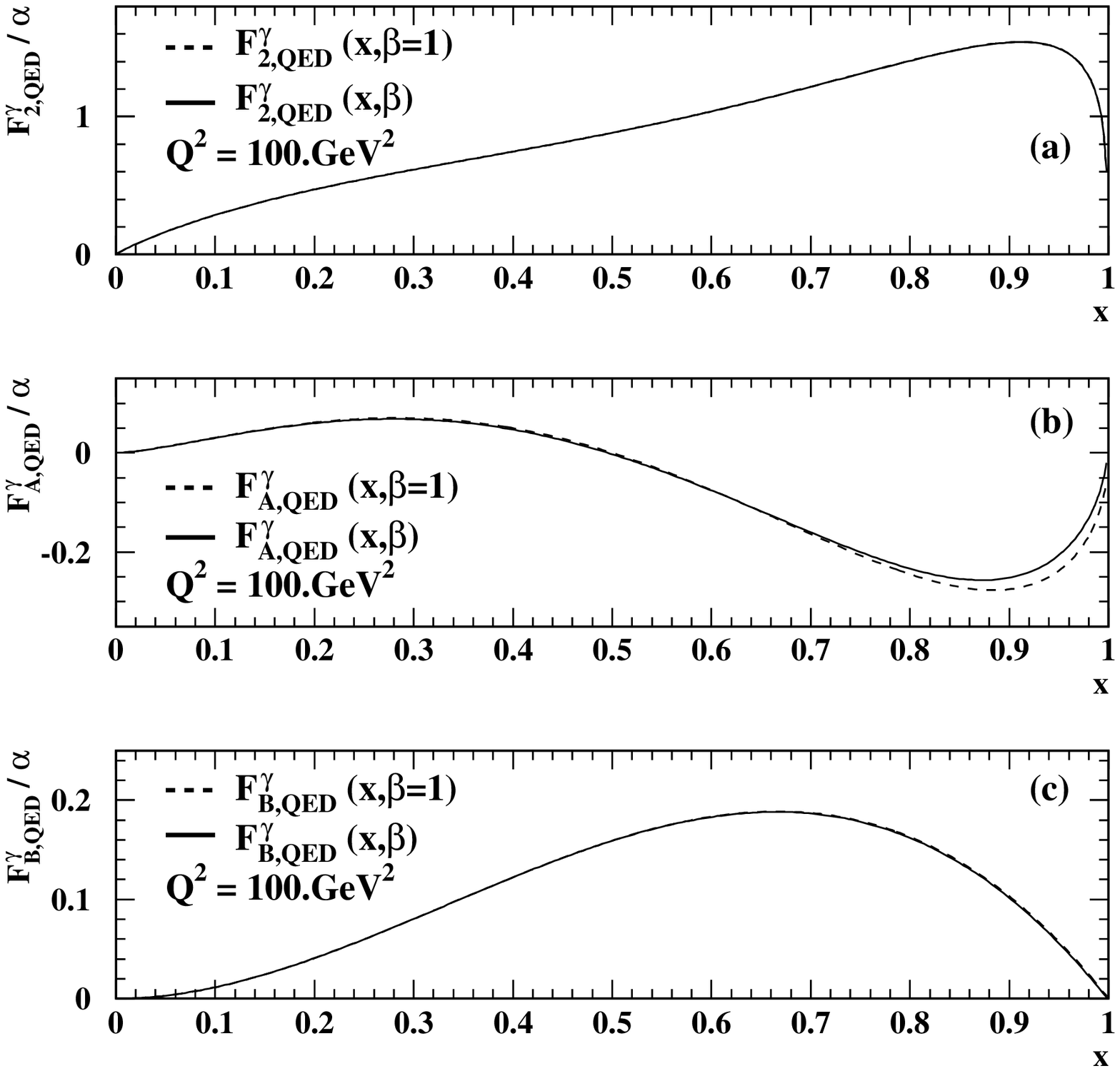}}
\caption[
         The structure functions \ftqed, \faqed and \fbqed
         for \mumu final states at $\qsq=100$~\gevsq.
        ]
        {
         The structure functions \ftqed, \faqed and \fbqed
         for \mumu final states at $\qsq=100$~\gevsq.
         The same quantities as in Figure\protect~\ref{fig:chap3_05}
         are shown.
        }\label{fig:chap3_06}
\end{center}
\end{figure}

 For real photons, $\psq=0$, the unintegrated structure functions,
 \FtTqed, \FtLqed, \FtAqed and \FtBqed have been calculated
 in the leading logarithmic approximation and
 can be found, for example, in Ref.~\cite{AUR-9601}.
 Only recently, in Ref.~\cite{SEY-9801}, the calculation
 has been extended beyond the leading logarithmic approximation,
 for all four unintegrated structure functions, by retaining
 the full dependence on the mass of the produced fermion up to terms of
 the order of \omfwq. But the limitation to real photons,
 $\psq=0$, is still retained.
 These structure functions are proportional to the cross-section
 for the transverse real photon to interact with different
 polarisation states of the virtual photon: transverse (T),
 longitudinal (L), transverse--longitudinal interference (A)
 and interference between the two transverse polarisations (B).
 They are connected to the unintegrated forms of \stt, \slt, \ttl and
 \ttt respectively.
 The structure function $\Fttqed\equiv 2x\FtTqed+\FtLqed$
 is a combination of these structure functions.
 Using this relation and the limit $\epsilon(y)= \rho(y) = 1$,
 Eq.~(\ref{eqn:crossaz}) reduces to
%
\begin{eqnarray}
  \frac{\der^4\sigma_{\e\gamma\rightarrow\e\ff}}
       {\der x\,\der \qsq\, \der \zzp\, \der \az/2\pi} &=&
  \frac{2\pi\alpha^2}{xQ^4}
  \left[1+(1-y)^2\right]\cdot
  \nonumber \\
  &&
  \left[
  \Fttqed-\FtAqed\cos\az+\smfrac12\FtBqed\cos2\az
  \right].
 \label{eqn:crossazred}
\end{eqnarray}
%
 In this equation, \zzp and \az always refer to the produced fermion.
 However, to achieve a structure function \FtAqed,
 which does not vanish when integrated over \zzp, the angle \az
 is defined slightly differently, as the azimuth of whichever produced
 particle (fermion or anti-fermion) has the smaller
 value of \zzp, or \cts, as shown in Figure~\ref{fig:chap2_03}(b).
 This definition leaves all the structure functions
 unchanged except that \FtAqed now is symmetric in \zzp, thereby allowing
 for an integration over the full kinematically allowed range in \zzp,
 namely $(1\!-\!\beta)/2$ to $(1\!+\!\beta)/2$.
 The integration with respect to \zzp leads to
%
\begin{eqnarray}
  \frac{\der^3\sigma_{\e\gamma\rightarrow\e\ff}}
       {\der x\,\der \qsq\, \der \az/2\pi} &=&
  \frac{2\pi\alpha^2}{xQ^4}
  \left[1+(1-y)^2\right]\cdot
  \nonumber \\
  &&
  \left[
  \ftqed-\faqed\cos\az+\smfrac12\fbqed\cos2\az
  \right]\, .
 \label{eqn:crossazlim}
\end{eqnarray}
%
 This formula is used in the experimental determinations of
 \faqed and \fbqed.
 The full set of functions can be found in Ref.~\cite{SEY-9801}, here
 only the functions used for the determination of
 \faqed and \fbqed are listed:
%
\begin{eqnarray}
 \ftqed(x,\beta) &=&
 \frac{\efv\alpha}{\pi} x\left\{
 \left[x^2+\left(1-x\right)^2\right]\ln\left(\frac{1+\beta}{1-\beta}\right)
 \right.
 \nonumber\\&&
 \left.
 -\beta
 +8\beta x\left(1-x\right)
 -\beta\left(1-\beta^2\right)\left(1-x\right)^2
 +\left(1-\beta^2\right)
 \right.
 \nonumber\\&&
 \left.
 \left(1-x\right)
 \left[\frac{1}{2}\left(1-x\right)\left(1+\beta^2\right)-2x\right]
 \ln\left(\frac{1+\beta}{1-\beta}\right)
 \right\}\,,
 \nonumber\\
 \faqed(x,\beta) &=&
 \frac{4\efv\alpha}{\pi} x\sqrt{x\left(1-x\right)} \left(1-2x\right)
 \left\{
 \beta\left[1+\left(1-\beta^2\right)\frac{1-x}{1-2x}\right]
 \right.
 \nonumber\\&&
 \left.
 +\frac{3x-2}{1-2x}\sqrt{1-\beta^2}\arccos\left(\sqrt{1-\beta^2}\right)
 \right\},
 \nonumber\\
 \fbqed(x,\beta) &=&
 \frac{4\efv\alpha}{\pi}x^2\left(1-x\right)
 \left\{\beta\left[1-\left(1-\beta^2\right)\frac{1-x}{2x}\right]
 +\frac{1}{2}\left(1-\beta^2\right)
 \right.\nonumber\\ & & \left.
 \left[\frac{1-2x}{x}-\frac{1-x}{2x}\left(1-\beta^2\right)\right]
 \ln\left(\frac{1+\beta}{1-\beta}\right)\right\}\,.
 \label{eqn:mike1}
\end{eqnarray}
%
 Here \ef is the charge (in units of the electron charge) of the produced
 fermion.
 The structure functions \faqed and \fbqed are new.
 In contrast, the structure function \ftqed can be obtained
 from Eq.~(\ref{eqn:strucnull}), together with the cross-sections
 listed in Ref.~\cite{BUD-7501}, taking the appropriate limit.
 The corrections compared to the leading logarithmic approximation are of
 order \omfwq for \ftqed and \fbqed. For \faqed they are already of
 order \omfw.
 The structure functions in the leading logarithmic approximation can
 be obtained from Eqs.~(\ref{eqn:mike1}) in the limit $\beta\rightarrow 1$.
 They are listed, for example, in Ref.~\cite{AUR-9601},
 and have the following form
%
%
\begin{eqnarray}
 \ftqed(x,\beta=1) &=&
 \frac{\efv\alpha}{\pi}x
 \biggl\{\left[x^2+\left(1-x\right)^2\right]\ln\frac{\wsq}{\mf^2}-1+8x(1-x)
 \biggr\}\,,
 \nonumber\\
 \faqed(x,\beta=1) &=&
 \frac{4\efv\alpha}{\pi}
 \biggl\{x\left(1-2x\right)\sqrt{x\left(1-x\right)}\biggr\}\,,
 \nonumber\\
 \fbqed(x,\beta=1) &=&
 \frac{4\efv\alpha}{\pi}\biggl\{x^2\left(1-x\right)\biggr\}\,.
 \label{eqn:QPM}
\end{eqnarray}
%
 So far, the structure functions \faqed and \fbqed have only been
 measured for the \mumu final state using the \qsq range from
 1.5$-$30~\gevsq, as discussed in Section~\ref{sec:qedres}.
 The inclusion of the mass dependent terms significantly changes
 the structure functions in the present experimentally accessible
 range in \qsq.
 The numerical effect is most prominent at low values of \qsq
 and gets less important as \qsq increases, as demonstrated
 for the case of \mumu production.
 In Figure~\ref{fig:chap3_05} for $\qsq=1$~\gevsq, the mass corrections
 are extremely important, especially at large values of $x$, while
 in Figure~\ref{fig:chap3_06} for $\qsq=100$~\gevsq, they are small.
 \par
%
%
\begin{figure}
\begin{center}
{\includegraphics[width=1.0\linewidth]{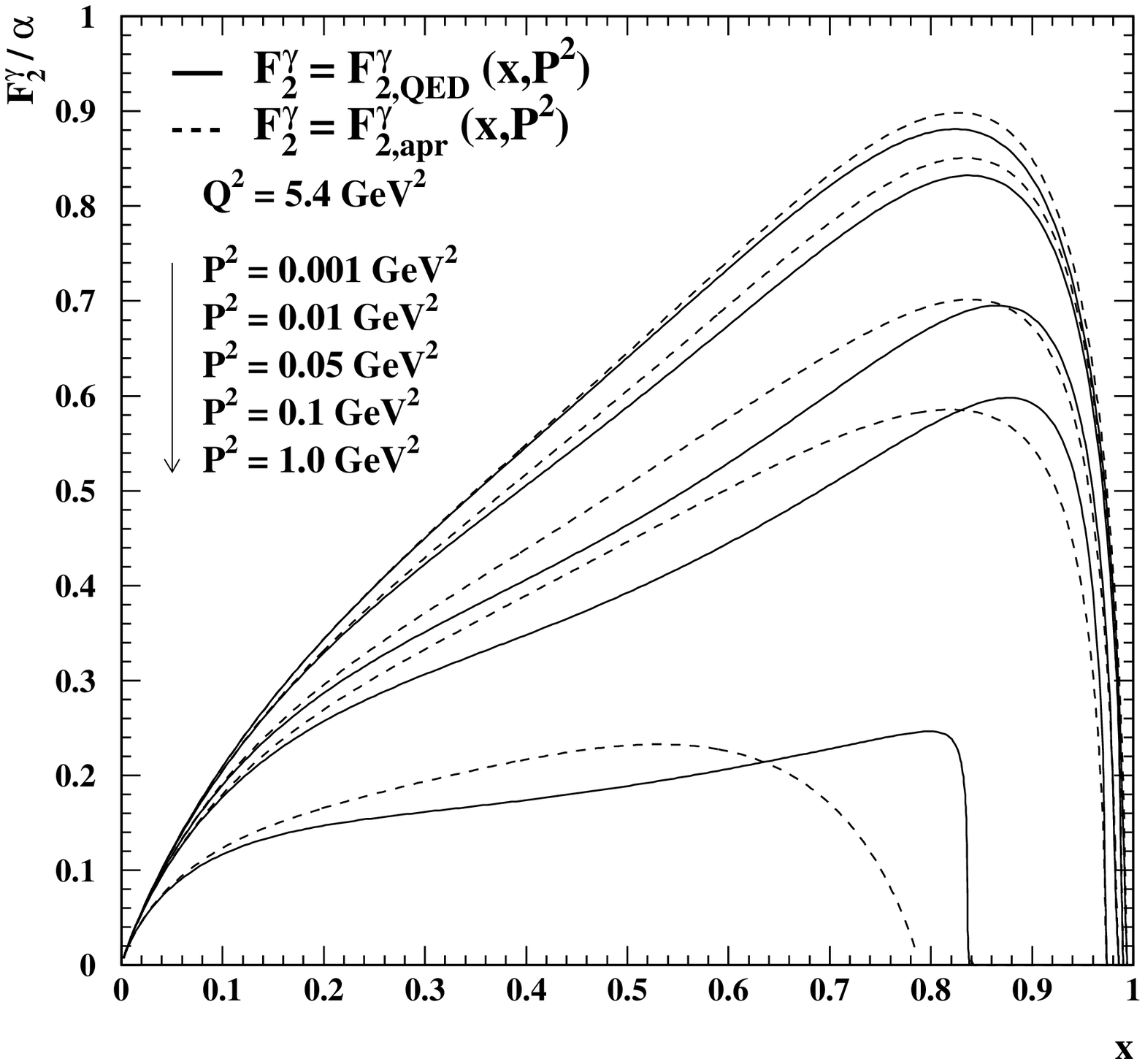}}
\caption[
         The \psq dependence of the structure function \ftqed.
        ]
        {
         The \psq dependence of the structure function \ftqed.
         The structure function \ftqed (full) and the approximation
         \ftapr (dash) are shown for $\qsq=5.4$~\gevsq, and for
         various \psq values, 0.001, 0.01, 0.05, 0.1 and 1.0~\gevsq.
        }\label{fig:chap3_07}
\end{center}
\end{figure}

%
 The second measurement of QED structure functions performed by the
 experiments is the measurement of \ftqed for $\qsq\gg\psq$, but
 keeping the full dependence on the small but finite virtuality
 of the quasi-real photon \psq.
 The structure function \ftqed for quasi-real photons in the limit
 $\qsq \gg \psq$ can be obtained from Eq.~(\ref{eqn:struc}), together with
 the cross-sections listed in Ref.~\cite{BUD-7501}.
 The resulting formula is very long and will not be listed here.
 The result is shown in Figure~\ref{fig:chap3_07},
 together with a compact approximation
%
\begin{eqnarray}
 \ftapr(x,\psq) &=&
 \frac{\efv\alpha}{\pi}x\biggl\{
 \left[x^2+\left(1-x\right)^2\right]
 \ln\frac{\wsq}{\mf^2+x\left(1-x\right)\psq}
 \nonumber \biggr.\\ & & \biggl.
 \phantom{\frac{\efv\alpha}{\pi}x\biggl\{}
 -1+8x(1-x)-\frac{x\left(1-x\right)\psq}{\mf^2+x\left(1-x\right)\psq}
 \biggr\}\, ,
 \label{eqn:f2approx}
\end{eqnarray}
%
 obtained in the limit $\mf^2\ll\qsq,\wsq$,
 which is rather accurate for small values of \psq.
 However, for $\psq>0.01$~\gevsq the approximation starts to deviate
 significantly from the exact formula and should not be used anymore.
 The structure function \ftqed is strongly suppressed as a function of \psq
 for increasing \psq, for example, for $x=0.5$ and $\qsq=5.4$~\gevsq
 the ratio of \ftqed for $\psq=0$ and $\psq=0.05$~\gevsq is 1.4.
 This suppression is clearly observed in the data, as discussed in
 Section~\ref{sec:qedres}.
 \par
%
%
 The QED structure functions defined above can only be used
 for the analysis of leptonic final states. For hadronic final states
 the leading order QED diagrams are not sufficient and QCD corrections
 are important.
 Therefore, the cross-sections and consequently also
 the structure functions cannot be calculated and
 parametrisations are used instead.
 This is the subject of the next section.
%
%
\subsection{Hadronic structure function \ft}
\label{sec:QCD}
 After the first suggestions that the structure functions of the
 photon may be obtained from deep inelastic electron-photon
 scattering at \epem colliders in Refs.~\cite{BRO-7102,WAL-7201},
 much theoretical work has been devoted to the investigation of the
 hadronic structure function \ft.
 The striking difference between the photon structure function \ft and
 the structure function of a hadron, for example, the proton structure
 function \ftp, is due to the point-like coupling of the photon to quarks,
 as shown in Figure~\ref{fig:chap1_02}(b).
 This point-like coupling leads to the fact that \ft rises
 towards large values of $x$, whereas the structure function of a
 hadron decreases.
 Furthermore, due to the point-like coupling,
 the logarithmic evolution of the photon structure function
 \ft with \qsq has a positive slope for all values of $x$, or in
 other words, the photon structure function \ft exhibits positive scaling
 violations for all values of $x$, even without accounting for QCD
 corrections.
 This is in contrast to the scaling violations observed for the
 proton structure function \ftp, which exhibits positive scaling
 violations at small values of $x$, and negative scaling violations
 at large values of $x$, caused by pair production of quarks from
 gluons and by gluon radiation respectively.
 In the case of the photon, the 'loss' of quarks at large values of $x$
 due to gluon radiation is overcompensated by the 'creation' of quarks
 at large values of $x$ due to the point-like coupling of the photon
 to quarks.
 \par
\begin{figure}[tbp]
\begin{center}
{\includegraphics[width=1.0\linewidth]{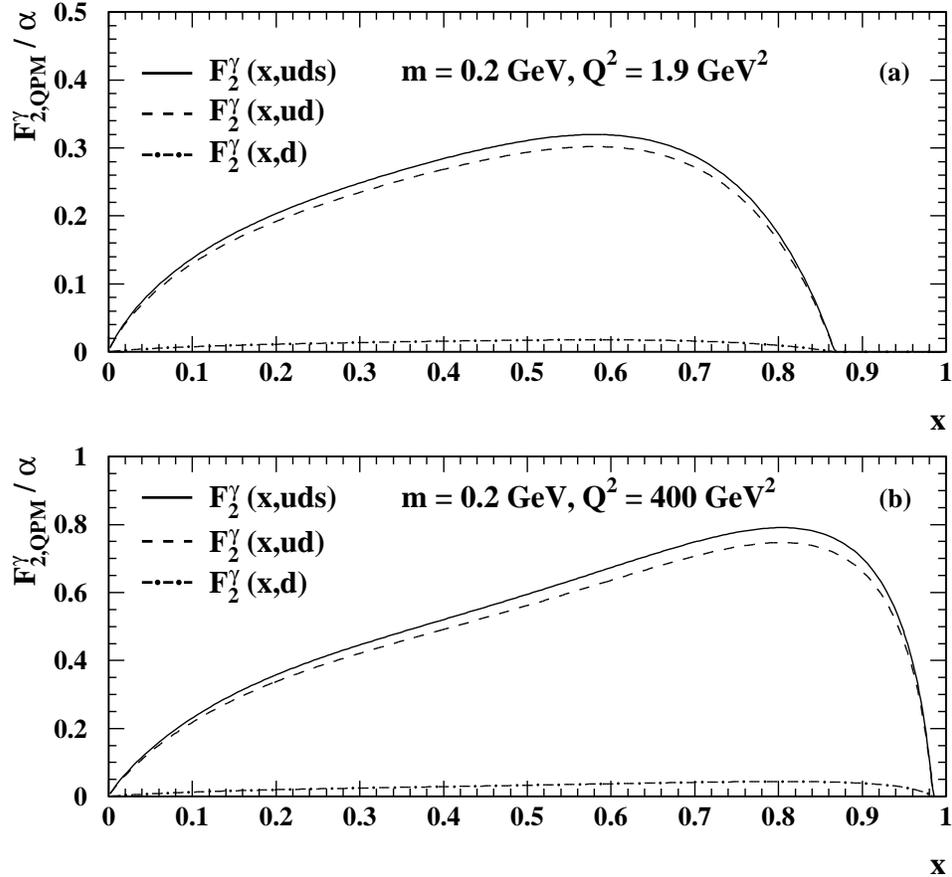}}
\caption[
         The QPM prediction for the structure function \ft for light quarks.
        ]
        {
         The QPM prediction for the structure function \ft for light quarks.
         Shown are the predictions of Eq.~(\ref{eqn:ZER}) adding up
         the contributions for light quarks using a mass of
         $m_{q_k}=m=0.2$~\gev for all quark species $k=u,d,s$.
         The different curves correspond to \ft from the down quarks alone
         (dot-dash), \ft from the down and up quarks (dash) and also adding
         strange quarks (full).
        }\label{fig:chap3_08}
\end{center}
\end{figure}

 The quark parton model, QPM, already predicts a logarithmic evolution
 of the photon structure function \ft with \qsq.
 This was first realised in Refs.~\cite{WAL-7301,KIN-7301} based on
 the calculation of the \qsq dependence of the so-called box diagram, for
 the reaction $\gsg\rightarrow\qqbar$, shown in Figure~\ref{fig:chap2_01}.
 The QPM result for quarks of mass $m_{q_k}$ is:
%
\begin{eqnarray}
 \ftqpm(x,\qsq) &=& \nc \sum_{k=1}^{\nf}\frac{\eqkv\alpha}{\pi}x
 \biggl\{\left[x^2+\left(1-x\right)^2\right]\ln\frac{\wsq}{m^2_{q_k}}
 \biggr.\nonumber\\&&
 \biggl.
 \phantom{\nc \sum_{i=1}^{\nf}\frac{\eqv\alpha}{\pi}x\biggl\{}
 -1+8x(1-x)\biggr\}\, ,
 \label{eqn:ZER}
\end{eqnarray}
%
 where \nc is the number of colours and the sum runs over all
 active flavours $i=1,\ldots,\nf$.
 The QPM formula is equivalent to the leading logarithmic approximation
 of \ftqed given in Eq.~(\ref{eqn:QPM}).
 This result, shown in Figure~\ref{fig:chap3_08} for three light quark
 species, is referred to as the calculation of \ft based on
 the Born term, the box diagram \ft, the QPM
 result for \ft, or as the QED structure function \ft.
 In Figure~\ref{fig:chap3_08} the contributions from the different quark
 species are added up for the smallest and
 largest value of \qsq for which measurements of \ft at LEP exist.
 In this \qsq range the photon structure function rises by about
 a factor of two at large values of $x$.
 Due to the dependence on the quark charge, the photon structure function
 \ft for light quarks is dominated by the contribution from up quarks.
 \par
%
%
 The pioneering investigation of the photon structure function in the
 framework of QCD was performed by Witten in Ref.~\cite{WIT-7701},
 using the technique of operator product expansion.
 The calculation showed that by including the leading logarithmic QCD
 corrections in the limit of large values of \qsq, the behaviour of
 \ft is logarithmic and similar to the QPM prediction.
 Schematically the result reads:
%
\begin{equation}
 \ftasy(x,\qsq)=\aem\frac{\tilde{a}(x)}{\al}=
             \aem \left[a(x)\ln\frac{\qsq}{\lamsq}\right]\, .
 \label{eqn:ASYLO}
\end{equation}
%
 The term $\ln(\wsq/m^2_{q_k})$ of Eq.~(\ref{eqn:ZER}) is replaced by
 $\ln(\qsq/\lamsq)$, which means the mass is replaced by the QCD
 scale \lam, and \wsq is replaced by \qsq, which in the leading
 logarithmic approximation is equivalent, because \wsq
 and \qsq are related by a term which depends only on $x$, as can be seen
 from Eq.~(\ref{eqn:wsq}) for $\psq=0$.
 However, the $x$ dependence of \ft, as predicted by the QPM, which
 treats the quarks as free, is altered by including the QCD corrections.
 The result from Witten is called the leading order asymptotic solution
 for the photon structure function \ft, since it is a calculation of \ft
 using the leading order logarithmic terms, but summing all orders in the
 strong coupling constant \al, and for the limit of asymptotically large
 values of \qsq.
 The photon structure function \ft in the leading order asymptotic solution
 is inversely proportional to \al, and the \qsq evolution, as well as the
 normalisation, are predicted by perturbative QCD at large values of \qsq.
 Therefore, there was hope that the measurement of the
 photon structure function would lead to a precise measurement of \al.
 However, the asymptotic calculation simplifies the full equations by
 retaining only the asymptotic terms, which means the terms which
 dominate for $\qsq\rightarrow\infty$.
 The non-asymptotic terms are connected to the
 contribution from the hadron-like part
 of the structure function, shown in Figure~\ref{fig:chap1_02}(c).
%
\begin{figure}
\begin{center}
{\includegraphics[width=1.0\linewidth]{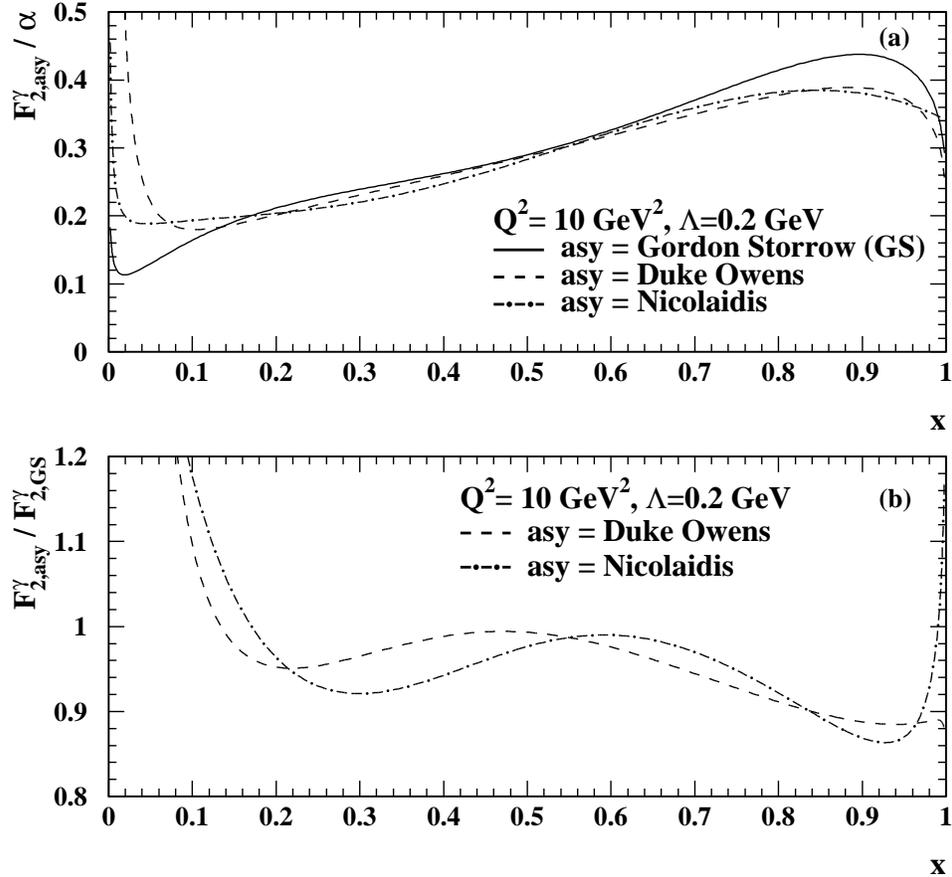}}
\caption[
         Comparison of the parametrisations of the $x$ dependence of the
         leading order asymptotic solution of \ft.
        ]
        {
         Comparison of the parametrisations of the $x$ dependence of the
         leading order asymptotic solution of \ft.
         The parametrisations are compared for $\qsq=10$~\gevsq,
         $\lam = 0.2$~\gev, and for three quark flavours.
         Shown are in
         (a) the structure function \ft obtained from
         the Gordon Storrow parametrisation (full), the Duke and Owens
         parametrisation (dash), and the Nicolaidis parametrisation
         (dot-dash).
         In (b) the differences are explored by dividing the
         older parametrisations by the Gordon Storrow parametrisation.
         The result for the Duke and Owens parametrisation is
         shown as a dashed line and the result for
         the Nicolaidis parametrisation as a dot-dashed line.
        }\label{fig:chap3_09}
\end{center}
\end{figure}
%
 The asymptotic solution is well behaved for $x\rightarrow 1$ and
 removes the divergence of the QPM result for vanishing quark
 masses, but in the limit $x\rightarrow 0$ it diverges like
 $x^{-0.5964}$, as was already realised in Ref.~\cite{WIT-7701}.
 The asymptotic solution has also been re-derived in a diagrammatic
 approach in Refs.~\cite{LLE-7801,FRA-7901,FRA-7902}, and in
 Ref.~\cite{DEV-7901}, by using the Altarelli-Parisi
 splitting technique from Refs.~\cite{ALT-7701,ALT-7801}.
 \par
 No closed analytic form of the $x$ dependence of the asymptotic solution
 can be obtained, since the asymptotic solution is given in moment space
 using Mellin moments.
 Consequently only parametrisations of the $x$ dependence of the
 parton distribution functions of the photon, based on the
 findings of the asymptotic solution of \ft, have been derived.
 The first parametrisation, given in Ref.~\cite{NIC-8001}, has been
 obtained by factoring out the singular behaviour at $x\rightarrow 0$
 and expanding the remaining $x$ dependence by Jacobi polynomials.
 Another parametrisation has been obtained in Ref.~\cite{DUK-8201}.
 The most recent available parametrisation has been derived in
 Ref.~\cite{GOR-9201} based on the technique of solving the
 evolution equations directly in $x$ space.
 In Ref.~\cite{GOR-9201}, it is compared to the two parametrisations
 discussed above and it is found to be the most accurate
 parametrisation of the asymptotic solution.
 The predictions of the parametrisations of the asymptotic solution
 are compared in Figure~\ref{fig:chap3_09} for $\qsq=10$~\gevsq,
 $\lam = 0.2$~\gev, and assuming three quark flavours.
 The three parametrisations are rather close to each other in the range
 $0.2<x<0.8$, where they agree to better than 10$\%$, but at
 larger and smaller values of $x$ the differences are much larger.
 \par
 The asymptotic solution, Eq.~(\ref{eqn:ASYLO}), factorises the $x$
 and \qsq dependence of \ft, which is not the case when solving the
 evolution equations as discussed in Appendix~\ref{sec:PDFTH}.
 Figure~\ref{fig:chap3_10} shows the difference between the asymptotic
 solution and the result from the GRV parametrisation of the photon
 structure function \ft from Refs.~\cite{GLU-9201,GLU-9202}.
 The GRV parametrisation is obtained by solving the
 full evolution equations.
 In this figure the logarithmic \qsq behaviour is factored out and
 the asymptotic solution is compared to the leading order GRV
 parametrisation of \ft for several values of \qsq.
 The asymptotic solution is consistently lower than the GRV
 parametrisation in the range $0.2<x<0.8$, and for all values of \qsq.
 However the agreement improves with increasing
 $x$ and \qsq. For example at $\qsq= 100$~\gevsq the agreement is
 better than 20$\%$, for the whole range $0.2<x<0.8$.
 \par
%
\begin{figure}
\begin{center}
{\includegraphics[width=1.0\linewidth]{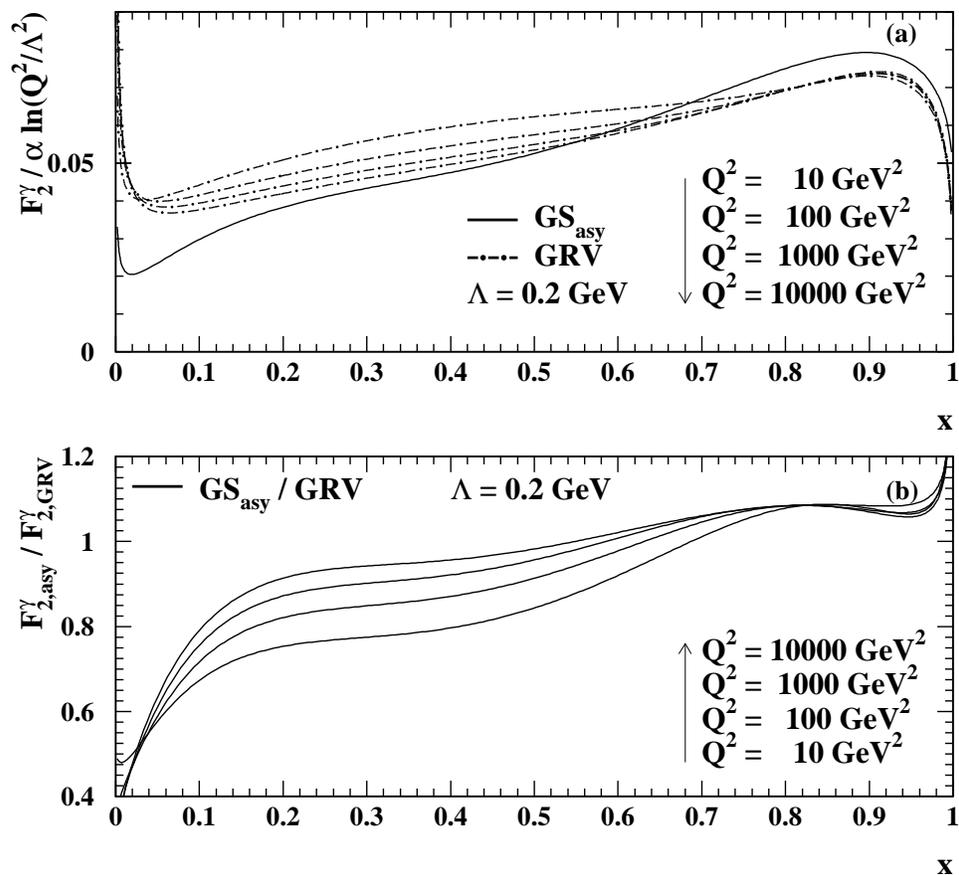}}
\caption[
         Comparison of the asymptotic solution and the leading order GRV
         parametrisation of \ft.
        ]
        {
         Comparison of the asymptotic solution and the leading order GRV
         parametrisation of \ft.
         The logarithmic \qsq dependence is factored out for
         $\lam = 0.2$~\gev, and three quark flavours are assumed.
         Shown are in
         (a) the asymptotic solution of \ft using
         the Gordon Storrow parametrisation (full),
         and the result of the GRV parametrisation  (dot-dash)
         obtained by solving the full evolution equations.
         The  GRV parametrisation is shown
         for several values of \qsq, 10, 100, 1000 and 10000~\gevsq.
         In (b) the differences are explored by dividing the
         Gordon Storrow parametrisation by the GRV result.
        }\label{fig:chap3_10}
\end{center}
\end{figure}
%
 The asymptotic solution has been extended to next-to-leading
 order in QCD in Ref.~\cite{BBU-7901}, leading to
%
\begin{equation}
 \ftasy(x,\qsq) = \aem\left[ a(x)\ln\frac{\qsq}{\lamsq}
                      + b(x)\right]\, .
 \label{eqn:ASYHO}
\end{equation}
%
 It was found in Ref.~\cite{DUK-8001} that the next-to-leading order
 corrections to the asymptotic solution are large at large $x$, and that
 the structure function \ft is negative for $x$ smaller than about
 0.2.
 In addition the divergence at low values of $x$ gets more and more
 severe in higher orders in QCD, and also extends to larger
 values in $x$, as discussed in Refs.~\cite{ROS-8301,ROS-8401}.
 The divergence at small $x$ of the perturbative, but asymptotic, result,
 which is cancelled by including the non-asymptotic contribution to the
 photon structure function, has attracted an extensive theoretical debate.
 For the real photon, the hadron-like part of the photon structure
 function \ft cannot be calculated in perturbative QCD, and only its
 \qsq evolution is predicted, as in the case of the proton structure function.
 Given this, the predictive power of QCD for the calculation of
 the photon structure function is reduced, and the scope for the
 determining \al from the photon structure function \ft is obscured.
 \par
 Several strategies have been taken to deal with this problem.
 It is clear from the singularities of the asymptotic, point-like,
 contribution that describing \ft as a simple superposition of the asymptotic
 solution and a regular hadron-like contribution, as derived, for example,
 based on VMD arguments, cannot solve the problem, because a hadron-like
 part, which is chosen to be regular, will never remove the singularity.
 Therefore, either a singular part has to be added by hand, to
 remove the singularities of the asymptotic solution, or the singularities
 have to be dealt with by including the non-asymptotic contribution as a
 supplement to the point-like part of the photon structure function \ft.
 The various approaches attempted along these lines will be discussed
 briefly.
 \par
 The first approach to deal with the singularities
 was suggested and outlined in Refs.~\cite{ANT-8301}.
 This method tries to retain as much as possible of the predictive power
 of the point-like contribution to the structure function, and the
 possibility to extract \al from the photon structure function \ft.
 The solution chosen to remove the divergent behaviour
 consists of a reformulation of the structure function by isolating
 the singular structure of the asymptotic, point-like part at low
 values of $x$, based on the analysis of the singular structure
 of \ft in moment space.
 Then, an ad-hoc term is introduced\footnote
 {The regularisation term of Ref.~\cite{ANT-8301}
 is based on the photon-parton splitting functions of Ref.~\cite{BBU-7901}.
 Unfortunately the photon-gluon splitting function in
 next-to-leading order erroneously contained a factor $\delta(1-x)$, which
 was removed later in Refs.~\cite{FON-9201,GLU-9201}.
 As discussed in Ref.~\cite{GLU-9201},
 this weakens the next-to-leading order singularity at low values of $x$,
 and therefore, it will also affect the exact form of the proposed
 regularisation term of Ref.~\cite{ANT-8301}.
 },
 which removes the singularity and regularises \ft, but depends
 on an additional parameter, which has to be obtained by experiments,
 for example, by performing a fit to the low-$x$ behaviour of \ft.
 Several data analyses using this approach have been performed,
 as summarised in Ref.~\cite{BER-8701}.
 \par
 A second way to separate the perturbative and the non-perturbative
 part of the photon structure function, known as the FKP approach,
 was developed in Refs.~\cite{FIE-8601,FIE-8701,KAP-8901}.
 Here, the separation into the perturbative and the non-perturbative
 parts of \ft is done on the basis of the transverse momentum
 squared \ptq of the quarks in the splitting $\gamma\rightarrow\qqbar$,
 motivated by the experimental observation that for transverse momenta
 above a certain minimum value, the data can be described by a
 purely perturbative ansatz.
 The minimum transverse momentum squared \ptnq was found to be of the
 order of $1-2$~\gevsq. Given this large scale, no significant sensitivity
 of the point-like part of the photon structure function to \al remains.
 It has been argued in Ref.~\cite{FRA-8701} that these values are
 too high, and that still some sensitivity to \al is left, even when
 using the FKP ansatz.
 The FKP approach has several weaknesses, which are discussed, for example,
 in Refs.~\cite{SCH-9501,DRE-9501}.
 The main shortcomings are that terms are included which
 formally are of higher order, and that the parametrisation
 is based only on 'valence quark' contributions,
 which means that \ft vanishes in the limit $x\rightarrow 0$, whereas
 the 'sea quarks' result in a rising \ft at small values of $x$.
 This ansatz is therefore currently not widely used.
 \par
 The last approach discussed here is outlined in
 Refs.~\cite{GLU-8301,GLU-8401}, and is driven by the observation that
 by using the full evolution equations, the solution to \ft is regular
 both in leading and in next-to-leading order for all values of $x$.
 The method is analogous to the proton case and the starting point is
 the definition of input parton distribution functions for the photon
 at a virtuality scale \qnsq.
 \par
 The relation between the quark parton distribution functions
 $q_k^{\gamma}$ and the structure function \ft in leading order
 is given by the following relation
%
\begin{eqnarray}
 \label{eqn:F2def}
 \ftxq &=& x \sum_{k=1}^{\nf} \eqkt
           \left[q_k^{\gamma}(x,\qsq)+\bar{q}_k^{\gamma}(x,\qsq)\right],
\end{eqnarray}
%
 The flavour singlet quark part $\Sigma^{\gamma}(x,\qsq)$ and the flavour
 non-singlet part $q^{\gamma}_{\rm NS}(x,\qsq)$ of the photon structure
 function are defined by
%
\begin{eqnarray}
 \ftxq &=& x \left[q^{\gamma}_{\rm NS}(x,\qsq) +
              \avet \Sigma^{\gamma}(x,\qsq)\right],
\end{eqnarray}
%
 such that
%
\begin{eqnarray}
 \Sigma^{\gamma}(x,\qsq) &=& \sum_{k=1}^{\nf}
 \left[q_k^{\gamma}(x,\qsq) + \bar{q}_k^{\gamma}(x,\qsq)\right],\nonumber\\
 q^{\gamma}_{\rm NS}(x,\qsq) &=& \sum_{k=1}^{\nf}
 \left[ \eqkt - \avet\right]
 \left[q_k^{\gamma}(x,\qsq) + \bar{q}_k^{\gamma}(x,\qsq)\right],
 \label{eqn:Sigdef}
\end{eqnarray}
%
 where $\avet=1/\nf\sum_{k=1}^{\nf} \eqkt$ is
 the average charge squared of the quarks.
 \par
 The  input distribution functions are evolved in \qsq using
 the QCD evolution equations.
 With this, the $x$ dependence at an input scale \qnsq has to be obtained
 either from theoretical considerations, which are usually based on VMD
 arguments if \qnsq is chosen as a low scale, or
 fixed by a measurement of the structure function \ft.
 This approach gives up the predictive power of QCD for the
 normalisation of the photon structure function and retains only,
 as in the proton case, the \al sensitivity of QCD to the \qsq evolution.
 Because the evolution with \qsq is only logarithmic, the length
 of the lever arm in \qsq is very important, and consequently
 the sensitivity to \al crucially depends on the range of \qsq
 where measurements of \ft can be obtained.
 \par
 There are several groups using this approach. They differ however in the
 choice of \qnsq, the factorisation scheme, and the assumptions concerning
 the input parton distribution functions at the starting scales.
 The mathematical framework is outlined in Appendix~\ref{sec:PDFTH},
 following the discussion given in Ref.~\cite{VOG-9701}, and the
 available parton distribution functions are reviewed in
 Section~\ref{sec:PDF}.
 Using this framework the predictions of perturbative QCD on the evolution
 of \ft can be experimentally tested by first fixing the non-perturbative
 input by measuring \ft at some value of \qsq and then exploring
 the evolution of \ft for fixed values of $x$ as function of \qsq.
 Given the large lever arm in \qsq from 1~\gevsq to about 1000~\gevsq
 when exploiting the full statistics from LEP at all \epem centre-of-mass
 energies, there is some sensitivity left for measuring \al from the
 photon structure function, especially at large values of $x$, as
 discussed in detail in Ref.~\cite{VOG-9401,AUR-9601}.
 This completes the discussion of the quasi-real photons, and virtual
 photons are discussed in the following.
 \par
%
%
 For virtual photons the point-like contribution to
 the photon structure function \ft has been derived in the limit
 $\qsq\gg\psq\gg\lamsq$ in leading order in Ref.~\cite{UEM-8101},
 and in next-to-leading order in Ref.~\cite{UEM-8201}.
 The solution is positive and finite for all values of $x$.
 It was expected that the contribution from the
 hadron-like component is negligible in this limit.
 However, a recent investigation discussed in Section~\ref{sec:PDF}
 showed that this is only true at large values of $x$ and \psq.
%
\begin{figure}[tbp]
\begin{center}
{\includegraphics[width=1.0\linewidth]{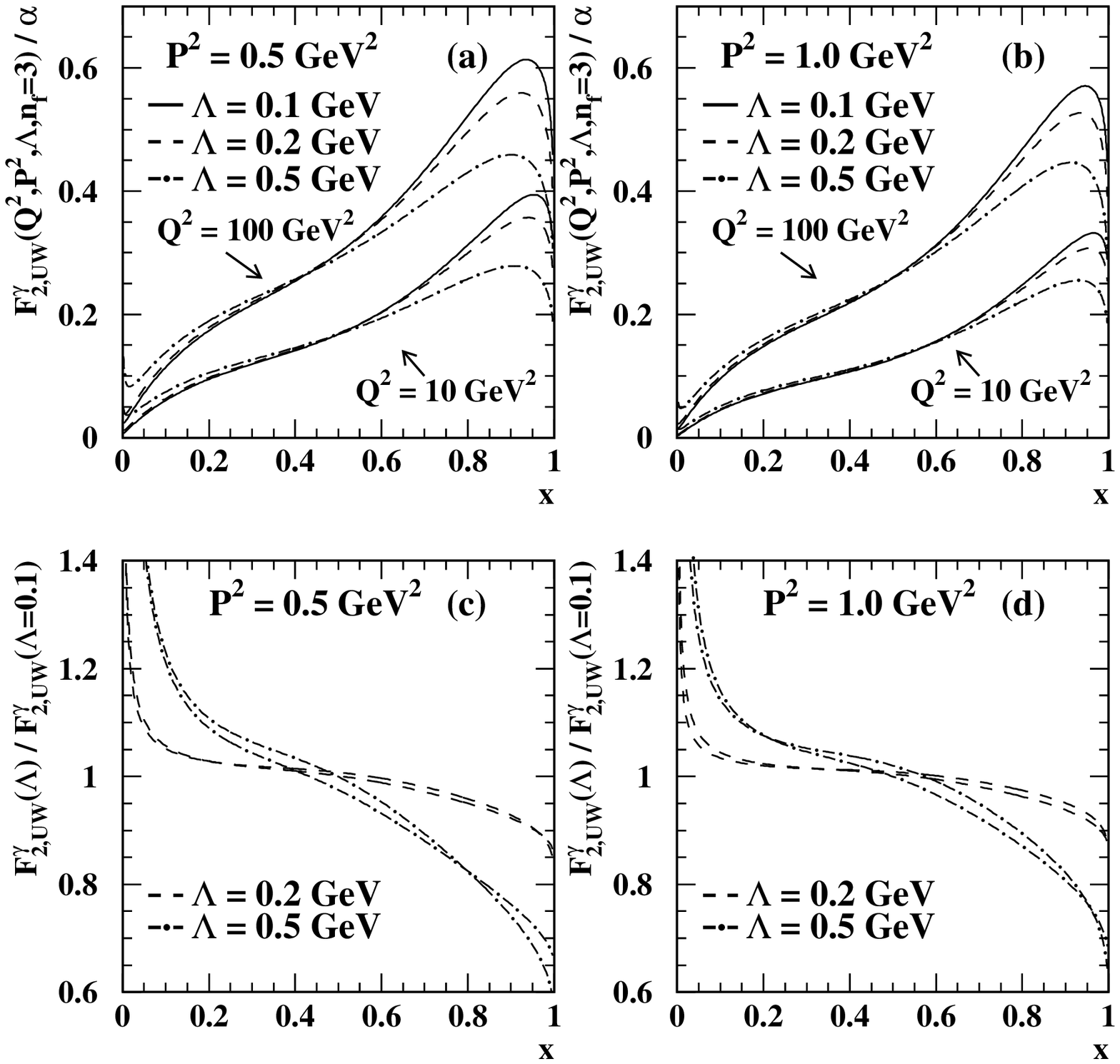}}
\caption[
         The structure function \ft in the limit $\qsq\gg\psq\gg\lamsq$.
        ]
        {
         The structure function \ft in the limit $\qsq\gg\psq\gg\lamsq$.
         Shown is the leading order structure function \ft in the limit
         studied by Uematsu and Walsh (UW) for three flavours
         and for several values of
         \lam, 0.1 (full), 0.2 (dash) and 0.5 (dot-dash) and
         using different values for \qsq and \psq.
         In all figures the predictions for $\qsq = 10$ and
         $100$~\gevsq are shown.
         In (a) and (c) $\psq = 0.5$~\gevsq is used, and
         (b) and (d) are for $\psq = 1$~\gevsq.
         The structure function \ft is shown in (a,b), and the figures
         (c,d) explore the dependence on \lam by showing
         the structure function \ft for
         $\lam = 0.2$ and 0.5~\gev divided by the prediction for
         $\lam = 0.1$~\gev.
        }\label{fig:chap3_11}
\end{center}
\end{figure}
%
 The  leading order result of the purely perturbative calculation
 from Ref.~\cite{UEM-8101} is shown for three light quarks
 in Figure~\ref{fig:chap3_11}, using two values of \qsq, 10 and
 $100$~\gevsq, and for two values of \psq, 0.5 and $1.0$~\gevsq, which
 are accessible within the LEP2 programme.
 In addition the dependence on the QCD scale \lam is shown, which,
 although not unambiguously defined in leading order, already gives
 an indication of the sensitivity to \al.
 The sensitivity to \lam does not change very much within the chosen
 range of \qsq and \psq, but there is a strong dependence on $x$.
 The most promising region is at large values of $x$, where the
 remaining contributions from the hadron-like part of the
 photon structure function \ft are very small.
 In this region \ft varies by about 10-20$\%$ if \lam is changed
 from 0.1 to 0.5~\gev.
 This means that a 5$\%$ measurement of \ft would be desirable in this
 region to constrain \lam, which is very challenging given the
 small cross section and the difficult experimental conditions.
 \par
\begin{figure}[tbp]\unitlength 1pt
\begin{center}
\begin{picture}(360,150)(0,0)
%
%
 \Line(40,110)(090,110)
 \Line(90,110)(150,140)
 \Photon(90,110)(120,080){4}{8.5}   
 \Line(120,080)(150,080)
 \Line(120,080)(120,050)
 \Line(120,050)(150,050)
 \Photon(90,20)(120,50){4}{7.5}   
 \Line(40,20)(090,20)
 \Line(90,20)(150,20)
 \Text(040,100)[lb]{\mbox{\Large \textbf{e}}}
 \Text(040,005)[lb]{\mbox{\Large \textbf{e}}}
 \Text(150,120)[lb]{\mbox{\Large \textbf{e}}}
 \Text(150,005)[lb]{\mbox{\Large \textbf{e}}}
 \Text(160,072)[lb]{\mbox{\Large \textbf{Q}}}
 \Text(160,042)[lb]{\mbox{\Large \textbf{Q}}}
 \Text(080,080)[lb]{\mbox{\boldmath\Large $\gamma^\star$}}
 \Text(080,030)[lb]{\mbox{\boldmath\Large $\gamma$}}
 \Text(040,050)[lb]{\mbox{\large \textbf{(a)}}}
%
%
 \Line(200,110)(250,110)
 \Line(250,110)(310,140)
 \Photon(250,110)(280,080){4}{8.5}   
 \Line(280,080)(310,080)
 \Line(280,080)(280,050)
 \Line(280,050)(310,050)
 \Photon(250,20)(265,30){4}{4.5}   
 \GOval(265,30)(04,05)(0){0.9}
 \Line(269,30)(280,33)
 \Line(269,30)(280,27)
 \Gluon(268,33)(280,50){4}{3.5}
 \Line(200,20)(250,20)
 \Line(250,20)(310,20)
 \Text(200,100)[lb]{\mbox{\Large \textbf{e}}}
 \Text(200,005)[lb]{\mbox{\Large \textbf{e}}}
 \Text(310,120)[lb]{\mbox{\Large \textbf{e}}}
 \Text(310,005)[lb]{\mbox{\Large \textbf{e}}}
 \Text(320,072)[lb]{\mbox{\Large \textbf{Q}}}
 \Text(320,042)[lb]{\mbox{\Large \textbf{Q}}}
 \Text(240,080)[lb]{\mbox{\boldmath\Large $\gamma^\star$}}
 \Text(240,030)[lb]{\mbox{\boldmath\Large $\gamma$}}
 \Text(200,050)[lb]{\mbox{\large \textbf{(b)}}}
\end{picture}
\caption[
         The  leading order contributions to \fth.
        ]
        {
         The  leading order contributions to \fth.
         Shown are examples of leading order diagrams contributing to
         (a) the point-like, and (b) the hadron-like part of the heavy quark
         structure function \fth, with $Q=c, b, t$.
        }\label{fig:chap3_12}
\end{center}
\end{figure}
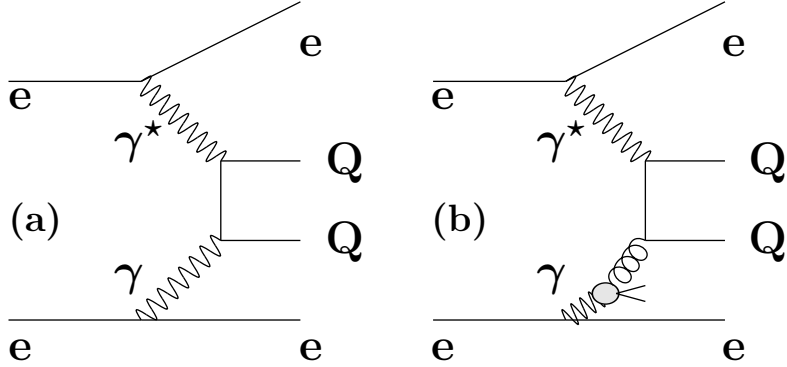

 The above discussion applies to the light quarks $u,d,s$.
 Due to the large scale established by their masses, the contribution of
 heavy quarks to the photon structure functions can be treated
 differently.
 At present collider energies, only the contribution of charm quarks
 to the structure function \ft is important.
 The contributions of the bottom quark and the even heavier top quark
 can, however, be calculated similarly to those of the charm quarks.
 Like the structure function \ft for light quarks, the structure
 function for heavy quarks \fth, $h=c,b,t$, receives contributions
 from the point-like and the hadron-like component of the photon.
 The leading order diagrams are shown in Figure~\ref{fig:chap3_12}.
%
%
 For invariant masses near the production threshold $W=2\mh$, the most
 accurate treatment of the point-like contribution of heavy quarks
 to the structure function is given by the prediction of the lowest
 order Bethe-Heitler formula.
 Due to the large mass scale QCD effects are small and
 this QED result is in general sufficient.
 The structure function can be obtained from Eq.~(\ref{eqn:struc}),
 together with the cross-sections listed in Ref.~\cite{BUD-7501}.
 The resulting formula is very long and the approximation
 made in Ref.~\cite{HIL-7901}, which is valid for $2x\psq/\qsq\gg1$,
 is sufficiently accurate in most cases and is used,
 for example, when constructing parton distribution functions.
 This approximation for virtual photons $\psq>0$ is given by
%
\begin{eqnarray}
 \fth &=& \nc
 \frac{\eqhv\alpha}{\pi}x
 \biggl\{
 \left[x^2+\left(1-x\right)^2\right]\ln\frac{1+\beta\gamma}{1-\beta\gamma}
 \biggr.\nonumber \\&& \biggl.
 -\beta+6\beta x\left(1-x\right)
 \biggr.\nonumber \\&& \biggl.
 +\left[
 2x\left(1-x\right)-\frac{1-\gamma^2}{1-\beta^2}
 -\left(1-\beta^2\right)\left(1-x\right)^2
 \right]
 \frac{\beta\gamma\left(1-\beta^2\right)}{1-\beta^2\gamma^2}
 \biggr.\nonumber \\&& \biggl.
 +\left(1-\beta^2\right)\left(1-x\right)
 \left[\frac{1}{2}\left(1-x\right)\left(1+\beta^2\right)-2x\right]
 \ln\left(\frac{1+\beta\gamma}{1-\beta\gamma}\right)
 \biggr\}\nonumber \\
 \mbox{with:}&& \nonumber \\
 \gamma &=&  \sqrt{1-\frac{4x^2\psq}{\qsq}},\quad\quad
             \mbox{and}\quad \beta=\sqrt{1-\frac{4\mh^2}{W^2}}\,.
\label{eqn:BH}
\end{eqnarray}
%
 For real photons, $\psq=0$ and $\gamma=1$, this reduces to
 \ftqed as given in Eq.~(\ref{eqn:mike1}).
 For real photons the next-to-leading order predictions
 have also been calculated in Ref.~\cite{LAE-9401}.
 \par
%
\begin{figure}[tbp]
\begin{center}
{\includegraphics[width=1.0\linewidth]{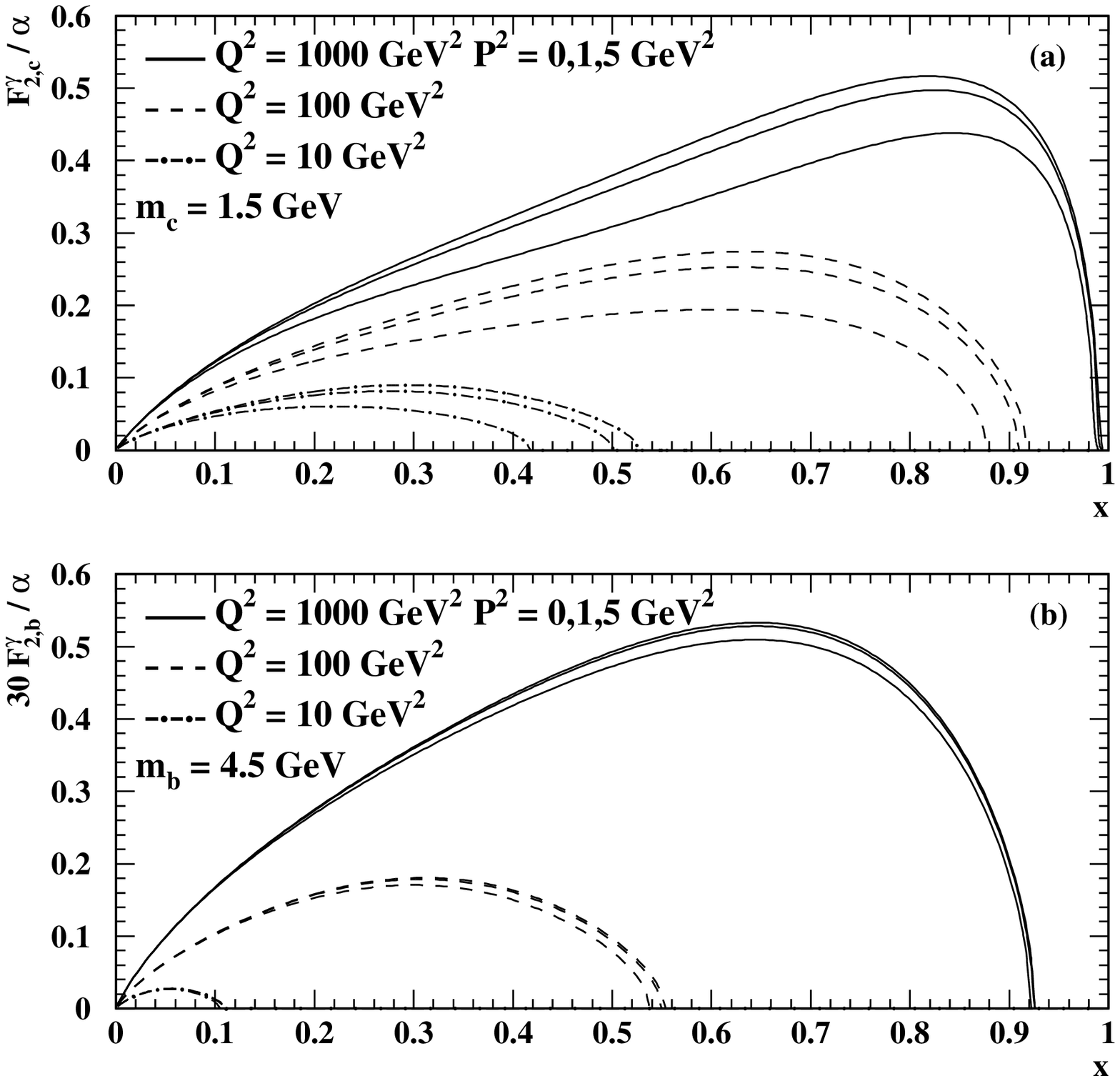}}
\caption[
         The point-like heavy quark contribution to \ft for various
         values of \qsq.
        ]
        {
         The point-like heavy quark contribution to \ft for various
         values of \qsq.
         Shown are in
         (a) the contribution of charm quarks, \ftc,
         and in
         (b) the contribution of bottom quarks, \ftb,
         to the photon structure function.
         Both contributions are calculated for three values of \qsq,
         10~\gevsq (dot-dash), 100~\gevsq (dash), and $1000$~\gevsq (full).
         At each \qsq, three curves are shown, which correspond to
         $\psq = 0,1$ and 5~\gevsq, where the suppression of \ft gets
         stronger for increasing \psq.
         To put both contributions on the same scale the bottom part
         has been multiplied by 30 as indicated in the figure.
        }\label{fig:chap3_13}
\end{center}
\end{figure}
%
 For the hadron-like contribution the photon-quark coupling
 must be replaced by the gluon-quark coupling,
 $\eqhv\alpha\rightarrow\eqht\al/6$, and the Bethe-Heitler formula has
 to be integrated over the allowed range in fractional momentum of the gluon.
 The hadron-like contribution, discussed in Section~\ref{sec:PDF},
 is only important at small values of $x$.
 The dominant point-like contribution to the structure function \ft
 for charm and bottom quarks, using $\mc=1.5$~\gev and $\mb=4.5$~\gev,
 is shown in Figure~\ref{fig:chap3_13} for three values of \qsq,
 10, 100 and $1000$~\gevsq, and three values of \psq, 0, 1 and $5$~\gevsq.
 Several observations can be made.
 The structure functions rises with \qsq and also, due to
 Eq.~(\ref{eqn:x}), the large $x$ part is more and more populated.
 Due to their small charge and large mass, the contribution from
 bottom quarks is very small.
 The suppression with \psq is stronger for the charm quarks since they
 are lighter than the bottom quarks.
 \par
 At large \qsq and large invariant masses, $W\gg2\mh$, the mass of the
 heavy quarks can be neglected in the evolution of \ft, provided that
 the usual continuity relations are respected and the appropriate number
 of flavours are taken into account in \lam.
 This concludes the discussion on the hadronic structure function
 \ft, and the remaining part of this section is devoted to the electron
 structure function and to radiative corrections to the deep inelastic
 scattering process.
 \par
 Recently, as described, for example, in
 Refs.~\cite{DRE-9401,SLO-9601,SLO-9801,SLO-9901} it has been proposed
 not to measure the photon structure function, but to
 measure the electron structure function instead.
 In measuring the electron structure function the situation is
 similar to the measurement of the proton structure function in the
 sense that the energy of the incoming particle, the electron in this
 case, is known.
 Therefore there is probably no need for an unfolding of $x$,
 explained in Section~\ref{sec:tools}, which is needed
 for the measurement of the photon structure function.
 This, on first sight, is an appealing feature since it promises
 greater precision in the measurement of the electron structure
 function than in the measurement of the photon structure function.
 But, as already discussed in Ref.~\cite{FOR-9501}, the advantage of
 greater measurement precision is negated by uncertainties which arise
 in interpreting the results in terms of the photon structure,
 because the differences in the predictions of the photon structure
 functions are integrated out.
 The region of low values of
 $x_{\rm e}=xz$ receives contributions from the regions of
 large momentum fraction $x$ and low scaled photon energy $z$,
 and small momentum fraction $x$ and large scaled photon energy $z$.
 Due to this, largely different photon structure functions lead to very
 similar electron structure functions, as was demonstrated in
 Ref.~\cite{FOR-9501}.
 Given this, further pursuit of this method does not seem very
 promising, since it does not give more insight into the structure
 of the photon.
 \par
 The last topic discussed in this section is the size of
 radiative corrections.
 Radiative corrections to the process
 $\ee \rightarrow \ee \gsg \rightarrow \ee X$ have been calculated
 for a (pseudo) scalar particle $X$ in
 Refs.~\cite{DEF-8101,DEF-8102,NEE-8401,LAN-8701}
 and for the \mumu final state in Refs.~\cite{LAN-8701,BDK-8502,BDK-8602}.
 It has been found that they are very small, on the per cent level, for
 the case where both photons have small virtualities and the scattered
 electrons are not observed.
 Consequently the equivalent photon approximation has only small
 QED corrections.
 For the case of deep inelastic electron-photon scattering a detailed
 analysis has been presented in Refs.~\cite{LAE-9603,LAE-9701}.
 The theoretical calculation is performed in the leading logarithmic
 approximation which means that the corrections
 are dominated by radiation from the deeply inelastically
 scattered electron. Only photon exchange is taken into account,
 since $Z$ boson exchange can be safely neglected at presently accessible
 values of \qsq.
 The calculation is analogous to the experimental
 determination of the kinematical variables.
 The momentum transfer squared \qsq is determined from the scattered electron,
 whereas $x$ is based on mixed variables, which means \qsq is obtained from
 the scattered electron and \wsq is taken from the hadronic
 variables.
 The radiative corrections are dominated by initial state radiation,
 whereas final state radiation and the Compton process are found
 to contribute very little.
 Final state radiation is usually not resolved experimentally
 due to the limited granularity of the electromagnetic calorimeters used.
 The Compton process contributes less than 0.5$\%$ to the cross-section
 for the range $3.2\cdot 10^{-4}<x<1$ and $3.2<\qsq<10^4$~\gevsq.
 The contribution of initial state radiation is usually
 negative and for a given \qsq its absolute value is largest at the
 smallest accessible $x$ and decreases with increasing $x$.
 For most of the phase space covered by the presently available
 experimental data the radiative corrections amount to less
 than 5$\%$.
 \par
 Due to the capabilities of the Monte Carlo programs used in the
 experimental analyses of photon structure functions discussed
 in Section~\ref{sec:gener}, the radiative corrections are usually
 neglected in the determination of \ft.
 They are, however, accounted for when measuring \ftqed.
%
%
\subsection{Vector meson dominance and the hadron-like part of \ft}
\label{sec:VMD}
 In this section the parametrisations of the hadron-like part \fthad
 of the photon structure function, which are constructed based on VMD
 arguments, are briefly reviewed.
 Only the main arguments needed to construct \fthad are given, for
 details the reader is referred to the original publications.
 There have been several attempts to derive the hadron-like part of the
 photon structure function \fthad based on VMD arguments motivated by
 the fact that the photon can fluctuate into a $\rho$ meson.
 No precise data on the structure function of the $\rho$ meson
 exist, and the structure function of the $\rho$
 is approximated by the structure function of the pion, \ftpion.
 In the first attempts to measure the photon structure function
 \ft, it was approximated by the sum of the point-like and the
 hadron-like part, where \fthad was constructed as a function
 of $x$ alone, and its \qsq evolution was ignored.
 In the context of the evolution of the parton distribution functions,
 the \qsq dependence given by perturbative QCD is also taken into
 account, and only the $x$ dependence at the scale \qnsq is obtained
 from VMD arguments.
 These two issues will be discussed in the following.
 \par
 The most widely used approximation for an \qsq-independent
 hadron-like component of the photon structure function \ft was
 obtained in Ref.~\cite{PET-8001,PET-8301}.
 The quark distribution functions of the $\rho$ meson are taken to be
 $xq_i^\rho (x) = 1/2 (1-x)$ and the photon is modelled as an
 incoherent sum of $\rho$ and $\omega$, leading to
%
\begin{equation}
 \label{eqn:VMDpet}
 \fthad = \frac{8}{9}\frac{4\pi\aem}{\frhoq}xq_i^\rho(x) =
          \aem \left[0.2 (1-x) \right] \, ,
\end{equation}
%
 where \frhoq is the $\rho$ decay constant with $4\pi/\frhoq = 1/2.2$,
 as taken from Ref.~\cite{BER-8701}.
 This approximation was used in several measurements of the photon
 structure function \ft given in
 Refs.~\cite{JAD-8401,PLU-8401,PLU-8701,TAS-8601,TOP-9402}.
 A similar parametrisation has been proposed by Duke and Owens
 in Ref.~\cite{DUK-8001}. This  parametrisation, which is assumed to be
 valid at $\qsq=3$~\gevsq, is given by
%
\begin{equation}
 \label{eqn:VMDowe}
 \fthad=\frac{4\pi\aem}{\frhoq}
        \left[ 0.417 \sqrt{x} (1-x) + 0.133 (1-x)^5\right] \, .
\end{equation}
%
 \par
\begin{figure}[tbp]
\begin{center}
{\includegraphics[width=1.0\linewidth]{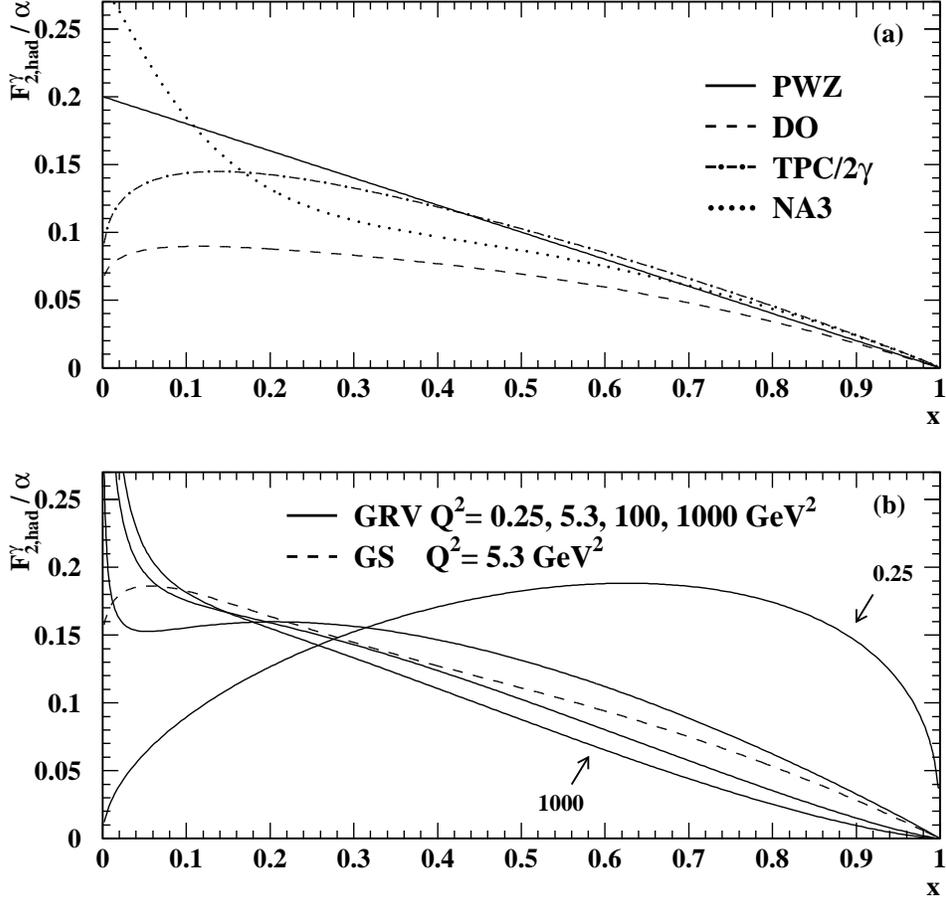}}
\caption[
         Comparison of parametrisations of the hadron-like contribution
         to the photon structure function \ft.
        ]
        {
         Comparison of parametrisations of the hadron-like contribution
         to the photon structure function \ft.
         Shown are in (a) the theoretically motivated parametrisations obtained
         by Peterson Walsh and Zerwas, (PWZ, full), and Duke and Owens,
         (DO, dash), together with the experimentally determined
         parametrisations from the TPC/2$\gamma$,
         (TPC/2$\gamma$, dot-dash), and the NA3, (NA3, dot), experiments.
         The evolution of the hadron-like part of \fthad is shown in (b),
         for the leading order parametrisations from Gl{\"u}ck, Reya and Vogt,
         (GRV, full), for several values of \qsq.
         In addition shown is the hadron-like input distribution from
         Gordon and Storrow, (GS, dash), which is valid for
         $\qsq=5.3$~\gevsq.
        }\label{fig:chap3_14}
\end{center}
\end{figure}

 Parametrisations of \fthad have been obtained experimentally from
 a measurement of the photon structure function \ft by the
 TPC/2$\gamma$ experiment and from measurements of the
 pion structure function \ftpion, for example, by the NA3 experiment.
 The parametrisation obtained in Ref.~\cite{TPC-8701} by the
 TPC/2$\gamma$ experiment, is based on a measurement of \ft in
 the range $0.3<\qsq<1.6$~\gevsq, with an average value of
 $\qzm=0.7$~\gevsq. The fit to the data yields
%
\begin{eqnarray}
 \label{eqn:VMDTPCfit}
 \fthad &=&\aem\left[ (0.22\pm 0.01) x^{0.31\pm 0.02} (1-x)^{0.95} +
           \right.\nonumber\\&&\left.\quad\quad\quad\quad\quad
           (0.06\pm 0.01) (1-x)^{2.5\pm 1.1}\right].
\end{eqnarray}
%
 The pion structure function \ftpion has been measured from the Drell-Yan
 process by the NA3 experiment for an average invariant mass squared
 of the \mumu system of 25~\gevsq, as detailed in Ref.~\cite{NA3-8301}.
 The NA3 data have been refitted by the TPC/2$\gamma$ experiment and
 the best fit to the data, as listed in Ref.~\cite{TPC-8701},
 is given by
%
\begin{equation}
 \label{eqn:NA3fit}
      \ftpion=\aem\left[ 0.22 x^{0.41} (1-x)^{0.95} +
                         0.26 (1-x)^{8.4}\right]\, ,
\end{equation}
%
 where the first part describes the contribution from valence quarks
 and the second part is the result for the sea quark contribution.
 In Figure~\ref{fig:chap3_14}(a), the theoretically motivated
 parametrisations, Eqs.~(\ref{eqn:VMDpet}) and~(\ref{eqn:VMDowe}),
 are shown, together with the experimentally determined parametrisations,
 Eqs.~(\ref{eqn:VMDTPCfit}) and~(\ref{eqn:NA3fit}).
%
%
 In the region of large values of $x$ the various parametrisations are
 rather similar.
 In contrast, for small values of $x$, where there was no precise
 data, the different parametrisations show a large spread.
 However, the \qsq dependence has not been taken into account
 in these parametrisations and the
 parametrisations are determined for different values of \qsq.
 \par
%
\begin{figure}[tbp]
\begin{center}
{\includegraphics[width=1.0\linewidth]{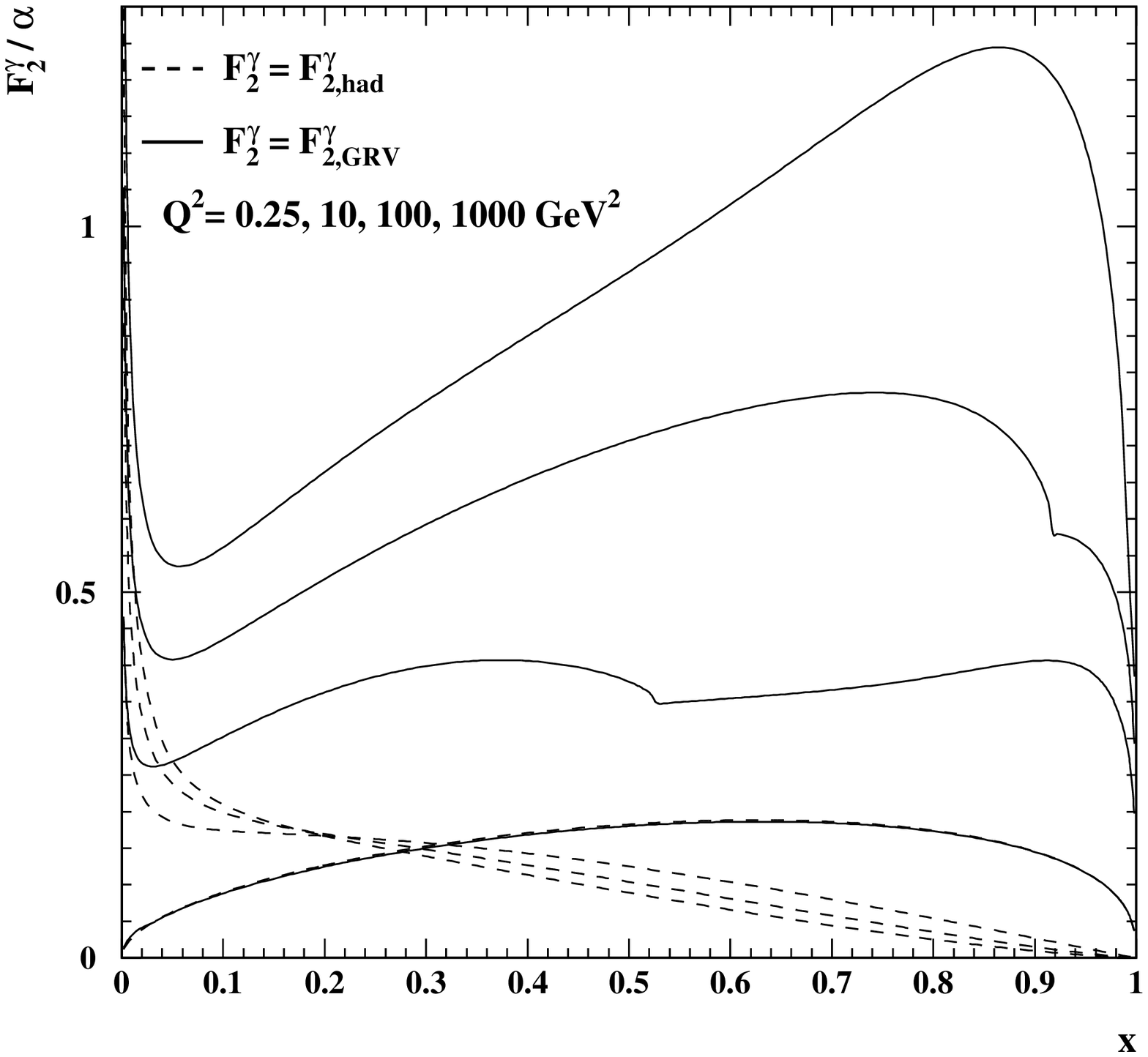}}
\caption[
         The \qsq dependence of the photon structure function \ft
         in comparison to the hadron-like contribution.
        ]
        {
         The \qsq dependence of the photon structure function \ft
         in comparison to the hadron-like contribution.
         The GRV parametrisation of the structure function \ft
         in leading order is compared to the hadron-like part of \ft
         taken as predicted by the evolved
         hadron-like input distribution function of the GRV
         parametrisation of the photon structure function \ft.
         Both functions are shown for four active flavours.
        }\label{fig:chap3_15}
\end{center}
\end{figure}
%
 The inclusion of the \qsq dependence of \fthad has been performed
 by several groups when constructing the parton distribution functions
 as discussed in Section~\ref{sec:PDF}.
 As examples, the leading order parton distribution functions of Gordon
 and Storrow, taken from Ref.~\cite{GOR-9201}, and Gl{\"u}ck, Reya and
 Vogt  taken from Refs.~\cite{GLU-9201,GLU-9202}, are discussed, which use
 VMD motivated input distribution functions based on measurements of \ftpion.
 In deriving the input distribution functions several assumptions
 are made.
%
\begin{Enumerate}
 \item The photon is assumed to behave like a $\rho$ meson,
       which means that \fthad can be expressed as
       $$ \fthad=\kappa\frac{4\pi\aem}{\frhoq}
          \sum_{k=1}^{\nf} \eqkt x q^\rho_k(x)\, ,$$
       where \frhoq has been defined above and $\kappa$ is a
       proportionality factor to take into account higher
       mass mesons using an incoherent sum.
 \item The structure function of the $\rho$ meson is assumed to be
       the same as the structure function of the $\pi^0$, which is
       expressed as half the sum of the $\pi^+$ and $\pi^-$ structure
       functions.
 \item The constituent quarks of the pions have a valence, $v$, and
       a sea, $\xi$, contribution, and the other quarks have only
       a sea contribution.
       For example, in the $\pi^+$ the up quark has valence and sea
       contributions, whereas the $\bar{d}$ has only a sea contribution.
       In addition, the valence quark distributions are assumed to be
       the same for all quark species, as are the sea quark contributions.
\end{Enumerate}
%
 Then by using Eq.~(\ref{eqn:F2def}), for example, for three light
 quark species, $\fthad = \aem \, 2.0 / 2.2 \,(5v-9\xi)/9$ is
 obtained.
 In the case of the leading order GRV parton distribution functions of
 the photon the valence and the sea parts are expressed at
 $\qnsq=0.25$~\gevsq by the published
 parton distribution functions of the pion, as given by Ref.~\cite{GLU-9204}.
 In the case of the Gordon Storrow parton distribution functions
 the VMD contribution is derived using basically the same assumptions.
 The parametrisation at the input scale $\qnsq=5.3$~\gevsq, and
 for three light quark species, taken from Ref.~\cite{GOR-9201},
 is given by
%
\begin{eqnarray}
 \label{eqn:gordon storr}
 \fthad &=& \aem \left[1.3360 \sqrt{x} (1-x)+ 0.641 (1-x)^5 +
            \frac{6}{27}0.0742 \sqrt{x} (1-x)^5
            \right]\!.
\end{eqnarray}
%
 In Figure~\ref{fig:chap3_14}(b) the two parametrisations are
 compared for three flavours, and in addition the \qsq evolution of
 the GRV prediction is studied.
 The parametrisation from GRV is shown at the scale where the
 evolution starts, $\qnsq=0.25$~\gevsq, at the
 scale where the parametrisation from GS is derived, $\qsq=5.3$~\gevsq,
 and for two large scales $\qsq=100$ and 1000~\gevsq.
 The evolution slowly reduces \fthad at large values of $x$ with
 increasing \qsq, and also creates a steep rise of  \fthad
 at low values of $x$, as in the case of the proton structure
 function \ftp.
 At $\qsq=5.3$~\gevsq the two parametrisations are similar for
 $x>0.2$, but at smaller values of $x$ the GRV parametrisation
 has already evolved a steep rise, which is purely driven by
 the evolution equations and not based on data.
 This rise cannot be obtained in the case of the GS parametrisation,
 because this parametrisation is obtained from a fit to data for
 $\qsq>5.3$~\gevsq, which do not cover the region of small $x$.
 \par
 The importance of the hadron-like contribution to the structure function
 \ft decreases for increasing \qsq, as can be seen from
 Figure~\ref{fig:chap3_15}, where the hadron-like contribution is shown
 together with the full structure function \ft as predicted by the
 leading order GRV parametrisation, both using $\nf=4$, for increasing
 values of \qsq.
 At the input scale the two functions coincide by construction.
 However, as \qsq increases there is a strong rise of \ft
 and a slow decrease of \fthad at large values of $x$.
%
%
\subsection{Alternative predictions for \ft}
\label{sec:alter}
 There have been several attempts to construct the photon structure
 function \ft differently from the leading twist procedure to derive
 \ft from the evolution equations, as described in
 Appendix~\ref{sec:PDFTH}.
 These attempts, which include power corrections, will be summarised
 briefly below.
 \par
 The model for \ft from Ref.~\cite{BAD-9901} is an extension of
 the model constructed for the proton case in
 Refs.~\cite{BAD-9201,BAD-8901}.
 It describes \ft as a superposition of a hadron-like part based on
 a VMD estimate and a point-like part given by the perturbative QCD
 solution of \ft, suppressed however by $1/\qnsq$ at low values of
 \qsq.
%
\begin{eqnarray}
 \ft(\wsq,\qsq)
 &=& \fthad(\wsq,\qsq) +  \ftpl(\wsq,\qsq)\nonumber\\
 &=& \pz\frac{3\qsq}{4\pi^2\aem^2}\sum_{\rho,\omega,\phi}
     \frac{\Mv^3\,\Gamma_{\epem}^{V} \,\siggv(\wsq)}{(\qsq+\Mv^2)^2}
     \nonumber\\
 &&  +\frac{\qsq}{\qsq+\qnsq}\cdot
     \ft\biggl(\frac{\qsq+\qnsq}{\qsq+\wsq},\qsq+\qnsq\biggr)
\label{eqn:badel}
\end{eqnarray}
%
 Here \Mv is the mass and $\Gamma_{\epem}^{V}$ the leptonic width
 of the vector meson $V$, and $\qnsq=1.2$~\gevsq, as in
 Ref.~\cite{BAD-9201}.
 The total cross-sections \siggv are represented by the sum of pomeron
 and reggeon contributions with parameters given in Ref.~\cite{DON-9201}.
 For moderate values of \qsq the structure function \ft is given by
 this ad-hoc superposition and in the limit of high \qsq the
 perturbative QCD solution of \ft is recovered, but with
 $1/\qsq$ corrections from the hadron-like part.
 The model has been shown to describe the results of the measured
 \siggsg cross-sections from Ref.~\cite{TPC-9003} for the ranges
  $0.2<\qsq< 7$~\gevsq and $2<W<10$~\gev.
 \par
 The model for \ft from Ref.~\cite{ABR-9801} relies on the Gribov
 factorisation described in Ref.~\cite{GRI-6201}.
 This factorisation is based on the assumption that at high energies the
 total cross-section of two interacting particles can be described by a
 universal pomeron exchange.
 In the model for \ft it is assumed that this factorisation also holds
 for virtual photon exchange at low values of $x$,
 as explained in Ref.~\cite{LEV-9701}.
 Using this, the Gribov factorisation
 relates the ratio of the photon-proton and proton-proton cross-sections
 to  the ratio of the photon and proton structure functions
%
\begin{equation}
 \ftxq = \ftpxq\frac{\siggp(\wsq)}{\sigpp(\wsq)}.
\label{eqn:gribov}
\end{equation}
%
 In this framework a prediction for the photon structure function
 at low values of $x$ can be obtained from the measurement of the
 proton structure function \ftp at low values of $x$.
 This extends the knowledge of  \ft to lower values of $x$
 because the results on \ftp reach down to $x\approx10^{-4}$, whereas
 the data on \ft probe only the photon structure down to
 $x\approx 10^{-3}$.
 However, this information can never replace a real measurement
 of \ft.
 The parton distribution functions are constructed using a
 phenomenological ansatz similar to the LAC case described
 in Section~\ref{sec:PDF} for four massless quark flavours.
 All quark distribution functions have the same functional form and
 the strange and charm quarks are suppressed with respect
 to the up and down quarks simply by constant factors.
 The parametrisation of \ft is obtained for $\qnsq=4$~\gevsq from
 a fit to the data of the photon structure function \ft from
 Refs.~\cite{AMY-9002,AMY-9501,AMY-9701,DEL-9601,JAD-8401,OPALPR185,%
 PLU-8701,TAS-8601,TOP-9402,TPC-8701}
 and the proton structure function data for $x<0.01$.
 Unfortunately the starting scale of the evolution is too
 high so that no valid comparisons with the low \qsq measurements
 of \ft can be made.
 \par
 The model for \ft from Ref.~\cite{DON-9802} is based on the
 the assumption that for $\wsq\gg\qsq$ the cross-section and,
 by using Eq.~(\ref{eqn:strucnull}), also the structure function \ft can
 be described mainly by pomeron exchange.
 The published data from Refs.~\cite{L3C-9803,OPALPR207,OPALPR213},
 and preliminary results from Refs.~\cite{FIN-9701},
 are used in the comparison which is performed for $\wsq > 225$~\gevsq.
 It is found that for the region where data exist the contribution
 from the pomeron exchange is insufficient to describe the data, and
 that the contribution from the hadron-like part of the photon
 structure is important.
 The hadron-like component is modelled by the valence-like
 pion structure function but, even including this component,
 the prediction is significantly below the  data for $\qsq>2$~\gevsq.
 \par
 The models discussed above will not be considered further.
 In contrast, in this review all comparisons of data and theory
 will be based on the asymptotic solution of \ft and on the
 parametrisations of \ft reviewed in the next section.
%
%

%
%
\section{Parton distribution functions}
\label{sec:PDF}
 There exist several parton distribution functions for real, and also
 for virtual photons, in leading and next-to-leading order, which are
 based on the full evolution equations discussed in Appendix~\ref{sec:PDFTH}.
 They are constructed very similarly to the parton distribution functions
 of the proton.
 The various parton distribution functions for the photon
 differ in the assumptions made about the starting scale \qnsq,
 the input distributions assumed at this scale, and also in
 the amount of data used in fitting their parameters.
 The distributions basically fall into three classes depending
 on the theoretical concepts used.
 The first class, consisting of the DG, LAC and WHIT
 parton distribution functions\footnote{The parton distribution
 functions are usually abbreviated with the first letters of the names
 of the corresponding authors, which will be mentioned below.},
 are purely phenomenological fits to the data, starting from an
 $x$-dependent ansatz for the parton distribution functions.
 The second class of parametrisations
 base their input distribution functions on theoretical prejudice and obtain
 them from the measured pion structure function, using VMD arguments and
 the additive quark model, as done in the case of GRV, GRSc and AFG, or
 on VMD plus the quark parton model result mentioned above, as done
 in the GS parametrisation.
 The third class consists of the SaS distributions which use
 ideas of the two classes above, and in addition relate the
 input distribution functions to the measured photon-proton
 cross-section.
 The main features of the different sets are described
 below, concentrating on the predictions for \ft derived from
 the parametrisations.
 The individual parton distribution functions, for example, the gluon
 distribution functions are not addressed, only their impact on
 \ft is discussed.
 For more details the reader is referred to the original publications.
%
\begin{Enumerate}
 \item \underline{\emph{DG}}~\cite{DRE-8503}:
 The first parton distribution functions were obtained by Drees and
 Grassie.
 This approach uses the evolution equations in leading order with
 $\lam = 0.4$~\gev.
 The $x$-dependent ansatz for the input distributions at
 $\qnsq = 1$~\gevsq is parametrised by 13 parameters and fitted to
 the only data available at that time, the preliminary PLUTO data at
 $\qsq = 5.3$~\gevsq from Ref.~\cite{BER-8301}.
 Due to the limited amount of data available, further assumptions
 had to be made.
 The quark distribution functions for quarks carrying the same charge
 are assumed to be equal, $q_d^{\gamma}=q_s^{\gamma}$ and
 $q_u^{\gamma}=q_c^{\gamma}$,
 and the gluon distribution function is generated purely dynamically,
 which means the gluon input distribution function is set to zero.
 Three independent sets are constructed for $\nf=3,4,5$, which means
 that they are not necessarily smooth at the flavour thresholds.
 The charm and bottom quarks are treated as massless and enter only
 via the number of flavours used in the evolution equations.
 They are included for $\qsq>20$ and 200~\gevsq respectively.
 The parametrisations clearly suffer from limited experimental input
 and they are not widely used today for measurements of \ft.
%
%
 \item \underline{\emph{LAC}}~\cite{ABR-9102}:
 The parametrisations from Levy, Abramowicz and Charchula use
 essentially the same procedure as the ones from Drees and Grassie,
 but are based on much more data, and therefore no assumptions on the
 relative sizes of the quark input distribution functions are made.
 An $x$-dependent ansatz, similar to the DG ansatz, using 12 parameters
 is evolved using the leading order evolution equations for four
 massless quarks, where \lam is fixed to 0.2~\gev.
 The charm quark contributes only for $W>2\mc$, otherwise the charm
 quark is treated as massless. No parton distribution function for
 bottom quarks is available.
 Three sets are constructed which differ from each other
 in the starting scale \qnsq and in
 the assumptions made concerning the gluon distribution.
 The sets LAC1 and LAC2 start from $\qnsq = 4$~\gevsq, whereas LAC3 uses
 $\qnsq = 1$~\gevsq.
 In addition, the sets LAC1 and LAC2 differ in the
 parametrisation of the gluon distribution.
 In the set LAC1 the gluon distribution is assumed to be
 $xg(x)\sim x^b(1-x)^c$, where $b$ and $c$ are fitted to the data,
 while the set LAC2 fixed $b=0$.
 The data used in the fits are from Refs.~\cite{AMY-9002,CEL-9001,%
 JAD-8401,PLU-8401,PLU-8403,PLU-8404,PLU-8701,TAS-8601,TPC-8501,%
 TPC-8702,TPC-8701}.
 The structure function \ft obtained from the LAC parametrisations is
 shown in Figure~\ref{fig:chap4_01} for two typical values of \qsq
 where data are available from the LEP experiments, $\qsq=5$ and
 135~\gevsq.
 The sets LAC1 and LAC2 are almost identical for $x>0.2$ for both values
 of \qsq, and although the gluon distribution function of the set
 LAC3 is very different from the ones used in the sets LAC1 and LAC2,
 as can be seen from Ref.~\cite{ABR-9102},
 the structure function \ft differs by less than 15$\%$ for $x>0.2$.
 For $x<0.2$ and at low values of \qsq however the differences in the
 predictions are larger than the experimental errors.
%
\begin{figure}[tbp]
\begin{center}
{\includegraphics[width=1.0\linewidth]{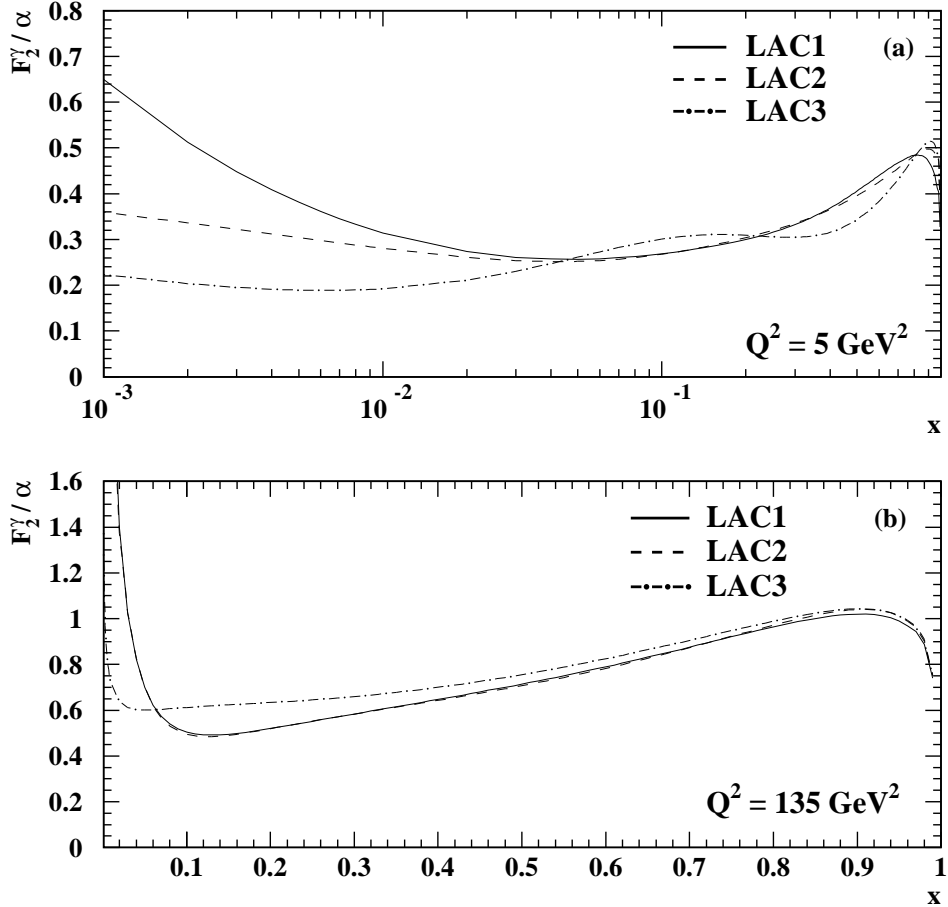}}
\caption[
         The structure function \ft from the LAC parton distribution
         functions.
        ]
        {
         The structure function \ft from the LAC parton distribution
         functions.
         Shown is the predicted  structure function \ft for the three
         sets LAC1-3 for two values of \qsq.
         In (a) the prediction is shown on a logarithmic scale in $x$ and
         for $\qsq=5$~\gevsq, whereas in (b) a linear scale is used
         for $\qsq=135$~\gevsq.
        }\label{fig:chap4_01}
\end{center}
\end{figure}
%
%
 \item \underline{\emph{WHIT}}~\cite{HAG-9501}:
 The parametrisations of parton distribution functions of the photon
 from Watanabe, Hagiwara, Izubuchi and Tanaka
 use a leading order approach, with three light flavours and a
 starting scale of $\qnsq=4$~\gevsq.
 The charm contribution, with $\mc=1.5$~\gev, is added
 according to the Bethe-Heitler formula in the region
 $4<\qsq<100$~\gevsq, while for higher values, $\qsq>100$~\gevsq,
 the massive evolution equations from Ref.~\cite{GLU-8201} are used.
 No parton distribution function for bottom quarks is available.
 The distributions of the light quarks are separated into
 distributions for valence quarks and distributions for sea quarks,
 which are linear combinations of the flavour singlet and non-singlet
 contributions to \ft, introduced in Eq.~(\ref{eqn:Sigdef}).
 The valence quark distributions
 describe the quarks which directly stem from the photon and they are
 parametrised as functions of $x$ at \qnsq.
 The sea quark distributions, account for the quarks produced in the
 process $\gamma^\star g\rightarrow\qqbar$, and at $\qsq = \qnsq$ they
 are approximated by the Bethe-Heitler formula using 0.5~\gev
 for the mass of the three light quark species.
 The QCD scale is taken to be $\lam  = 0.4$~\gev.
 The gluon distribution function is parametrised as
 $xg(x)= a(c+1)(1-x)^c$ and six sets with  $a=0.5,1$ and $c=3,9,15$ are
 constructed, all being consistent with the data of the structure
 function \ft used in the fits.
 The data used are published data from
 Refs.~\cite{AMY-9002,JAD-8401,OPALPR092,PLU-8401,PLU-8701,TAS-8601,TPC-8701}
 and preliminary data from Refs.~\cite{TPC-8802,TOP-9406,VEN-9301}.
 They are subject to an additional requirement of
 $\xmin>\qsq/(\qsq+W^{\rm{max}}_{\rm{vis}})$, which is introduced
 to remove the part of the data that was taken at the upper acceptance
 boundary in \wsq, which means at low values of $x$.
 \par
%
\begin{figure}[tbp]
\begin{center}
{\includegraphics[width=1.0\linewidth]{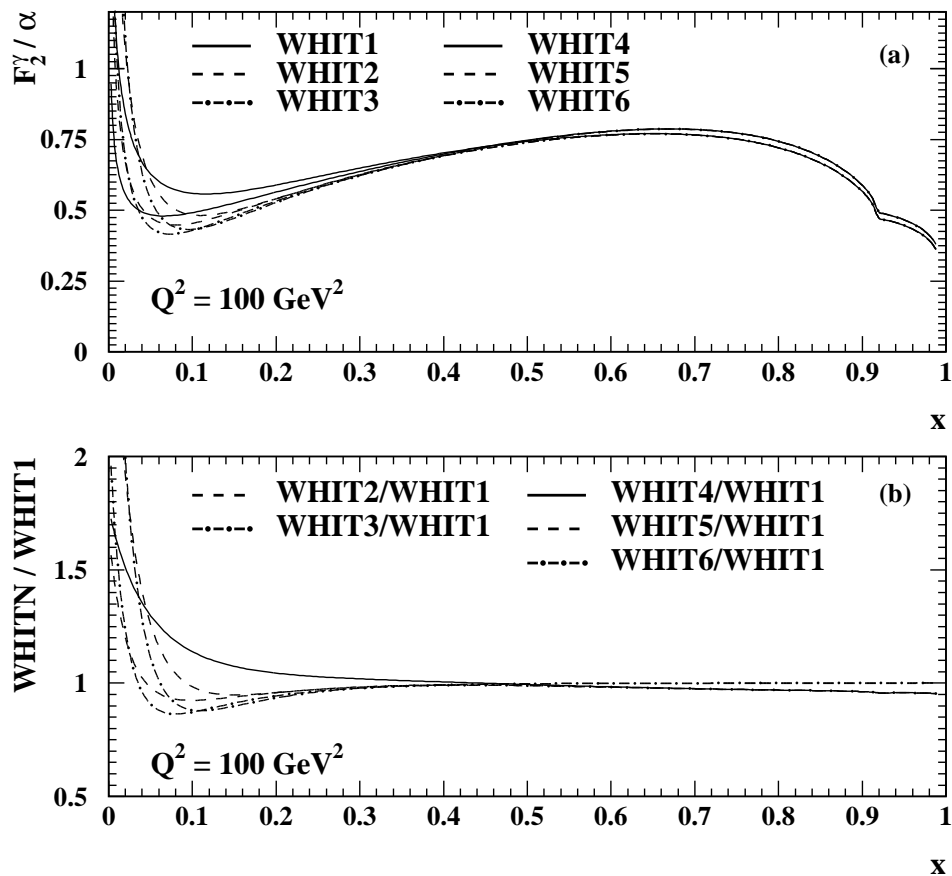}}
\caption[
         The structure function \ft from the WHIT parton distribution
         functions.
        ]
        {
         The structure function \ft from the WHIT parton distribution
         functions.
         The structure function \ft is shown in (a) for the individual
         sets WHIT1-6 and in (b) the sets WHIT2-6 are all divided by
         the set WHIT1.
         The individual sets in (a) fall into two groups containing three
         curves each, which coincide at large values of $x$.
         The sets WHIT1-3 predict a higher structure function at
         large values of $x$ than the sets WHIT4-6 and at low
         values of $x$ the sets WHIT4-6 start rising earlier
         for decreasing values of $x$ than the sets WHIT1-3.
        }\label{fig:chap4_02}
\end{center}
\end{figure}
%
 The different predictions for \ft of the various sets are shown in
 Figure~\ref{fig:chap4_02} for $\qsq=100$~\gevsq.
 Figure~\ref{fig:chap4_02}(a) shows the individual sets and in
 Figure~\ref{fig:chap4_02}(b) they are all normalised to the set WHIT1.
 The kink in the distributions in Figure~\ref{fig:chap4_02}(a) at
 $x\approx 0.9$ is typical for all 4 flavour parametrisations
 of \ft using massive charm quarks, and is due to the charm quark mass
 threshold.
 Only for $x$ values to the left of the threshold is charm production
 possible, and the threshold varies with \qsq, as can be seen from
 Eq.~(\ref{eqn:x}) and Figure~\ref{fig:chap3_13}.
 The sets fall into two groups depending in the value of the parameter
 $a$, with WHIT1-3 having $a=0.5$, and WHIT4-6 using $a=1$.
 The larger value of $a$ makes the sets WHIT4-6 start rising
 earlier for decreasing values of $x$.
 The two groups agree with each other to better than 5$\%$ for $x>0.3$,
 and for small $x$, where the gluon part becomes important, they differ by
 more than a factor of two.
 For most of the data on \ft the difference between the individual sets
 is much smaller than the experimental accuracy.
 But at small values of $x$ the data are precise enough
 to disentangle the very different predictions of the various sets.
%
%
 \item \underline{\emph{GRV}}~\cite{GLU-9201,GLU-9202}:
 The parton distribution functions from Gl{\"u}ck, Reya and Vogt
 are constructed using
 basically the same strategy which is also successfully used for
 the description of the proton and pion structure functions.
 The parton distribution functions are available in leading order
 and next-to-leading order.
 They are evolved from $\qnsq = 0.25$~\gevsq in leading order and from
 $\qnsq = 0.30$~\gevsq in next-to-leading order.
 The starting distribution is a hadron-like
 contribution based on VMD arguments, by using
 the parton distribution functions of the pion from Ref.~\cite{GLU-9204},
 the similarity of the $\rho$ and $\pi$ mesons, and a proportionality
 factor, $\kappa$, to account for the sum of $\rho,\omega$ and $\phi$
 mesons, as explained in detail in Section~\ref{sec:VMD}.
 The functional form of the starting distribution is
 $\qg=\qbg=\Gg=\kappa\frac{4\pi\alpha}{\frhoq}\ffp(x,\qnsq)$,
 where $x\ffp(x,\qnsq)\sim x^b(1-x)^c$ with $b>0$.
 The parameter $1/\frhoq =2.2$ is taken from Ref.~\cite{BER-8701},
 leaving $\kappa$ as the only free parameter, which is obtained
 from a fit to the data in the region $0.71<\qsq<100$~\gevsq,
 for $W>2$~\gev, to avoid resonance production.
 The point-like contribution is chosen to
 vanish at $\qsq=\qnsq$ and for $\qsq>\qnsq$, it is generated
 dynamically using the full evolution equations, as is also done
 for the evolution of the hadron-like component.
 The full evolution equations for massless quarks,
 with $\lam = 0.2$~\gev, are used in the \disg factorisation scheme,
 while removing all spurious higher order terms.
 The charm and bottom quarks are included via the Bethe-Heitler
 formula for $\mc=1.5$~\gev and $\mb=4.5$~\gev, and at high values
 of $W$ they are treated as massless quarks in the evolution.
 The data used in the fits are published data from
 Refs.~\cite{AMY-9002,JAD-8401,PLU-8404,PLU-8401,PLU-8403,PLU-8701,%
 TAS-8601,TPC-8501,TPC-8702,TPC-8701}
 and preliminary data from Refs.~\cite{CEL-9001}, all subject to the
 additional requirement $W>2$~\gev mentioned above.
 The leading order and next-to-leading order predictions are shown in
 Figure~\ref{fig:chap4_03} for several values of \qsq.
 The values chosen are: a very low scale, the lowest \qsq value where
 a measurement of \ft from LEP is available, and two typical values
 of \qsq for structure function analyses at LEP,
 $\qsq = 0.8, 1.9, 15$ and 100~\gevsq.
 The behaviour of the leading order and next-to-leading order
 predictions are rather different at very low and at high values of $x$.
 In the central part $0.1<x<0.9$, and for $\qsq = 1.9$~\gevsq they differ
 by no more than 20$\%$.
 Because none of the predictions is consistently higher in this region,
 and since the experiments integrate over rather large ranges in $x$
 when measuring the photon structure function \ft,
 it will be very hard to disentangle the two in this region
 in the near future.
 At lower values of $x$ however the data start to be precise enough.
%
\begin{figure}[tbp]
\begin{center}
{\includegraphics[width=1.0\linewidth]{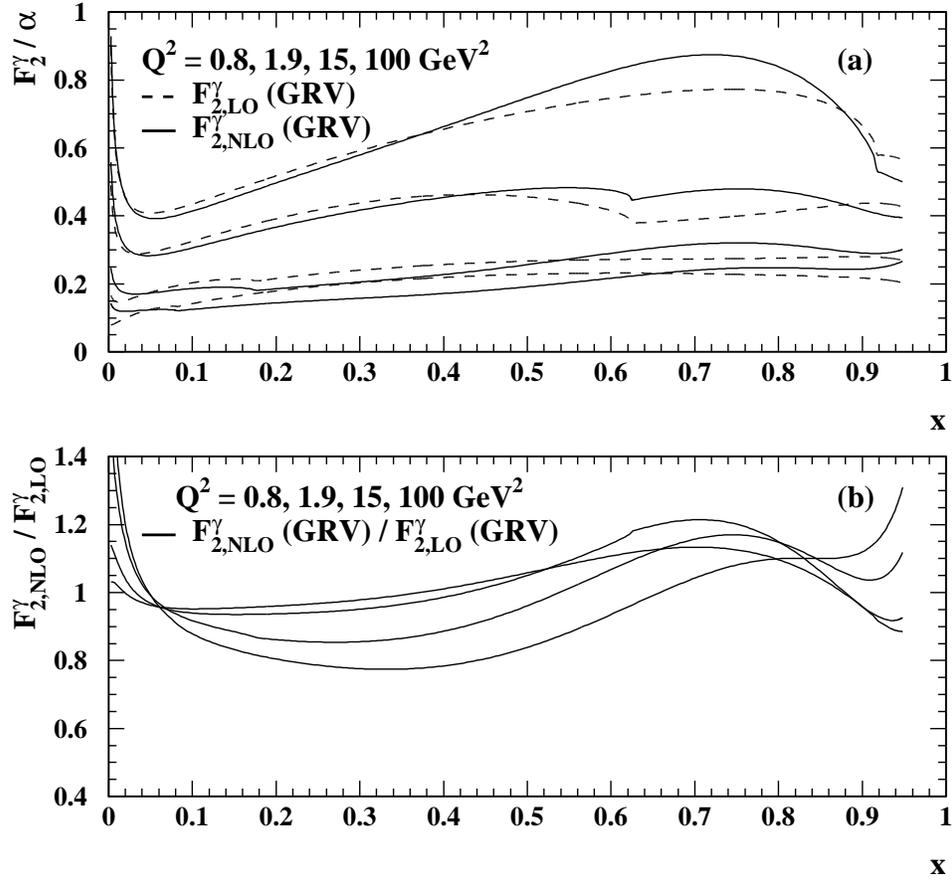}}
\caption[
         Comparison of the GRV leading order and next-to-leading order
         parametrisations of the photon structure function \ft.
        ]
        {
         Comparison of the GRV leading order and next-to-leading order
         parametrisations of the photon structure function \ft.
         In (a) the photon structure function \ft is shown in
         leading order (dash) and next-to-leading order (full), for
         four active flavours and for several values of \qsq,
         0.8, 1.9, 15 and 100~\gevsq, and in (b) the ratio of
         the next-to-leading order and the leading order
         parametrisations is explored for the same values of \qsq.
        }\label{fig:chap4_03}
\end{center}
\end{figure}
%
 \item \underline{\emph{AFG}}~\cite{AUR-9402}:
 The strategy used in constructing these parametrisations
 by Aurenche, Fontannaz and Guillet
 is very similar to the one used for the GRV parametrisations.
 The starting scale for the evolution is very low, $\qnsq=0.5$~\gevsq.
 This value is obtained from the requirement that the point-like
 contribution to the photon structure function vanishes at
 $\qsq=\qnsq$.
 Consequently, the input is taken as purely hadron-like, based
 on VMD arguments, where a coherent sum of low mass vector
 mesons $\rho,\omega$ and $\phi$ is used.
 The AFG distributions are obtained in the \msbl
 factorisation scheme. Therefore the input distributions contain an
 additional technical input, as shown in Eq.~(\ref{eqn:c2pr}), which was
 derived from a study of the factorisation scheme dependence
 and the momentum integration of the box diagram.
 With this choice of the factorisation scheme and the technical input,
 the parton distribution functions are universal and process independent.
 In contrast, the \disg scheme introduces a process dependence, because
 the $C_{2,\gamma}$ as given by Eq.~(\ref{eqn:c2ph}), which is absorbed
 into the quark distribution functions when using the \disg scheme,
 contains process dependent terms, as explained in Ref.~\cite{AUR-9402}.
 The evolution is performed in the massless scheme for three flavours
 for $\qsq<\mc^2=2$~\gevsq and for four flavours for $\qsq>\mc^2$,
 always using $\lam = 0.2$~\gev.
 No parton distribution function for bottom quarks is available.
 An additional scale factor, $K$, is provided to adjust the VMD
 contribution. In the standard set this parameter is fixed to
 $K=1$.
 Otherwise $K$ is obtained from a fit to published data taken from
 Refs.~\cite{AMY-9002,JAD-8401,PLU-8401,PLU-8701,TAS-8601}.
 In Figure~\ref{fig:chap4_04} the higher order prediction of \ft from
 AFG is compared to the GRV prediction for three values of \qsq,
 2, 15 and 100~\gevsq.
 At low \qsq there are large differences between the two predictions
 which tend to get smaller as \qsq increases.
%
\begin{figure}[tbp]
\begin{center}
{\includegraphics[width=1.0\linewidth]{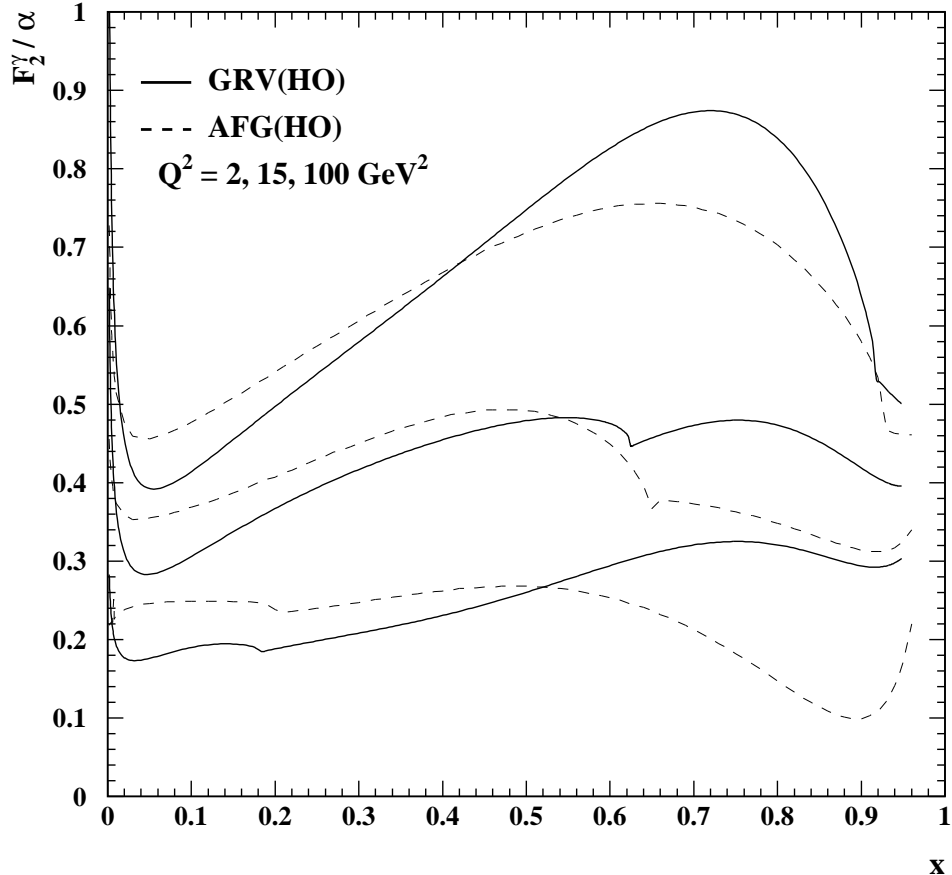}}
\caption[
        Comparison of the higher order structure function \ft from
        AFG and GRV.
        ]
        {
        Comparison of the higher order structure function \ft from
        AFG and GRV.
        The predicted higher order structure function \ft from
        AFG (dash) is compared to the prediction from GRV (full)
        for the three values of \qsq, 2, 15 and 100~\gevsq.
        }\label{fig:chap4_04}
\end{center}
\end{figure}
%
 \item \underline{\emph{GS}}~\cite{GOR-9201,GOR-9701}:
 The parton distribution functions from Gordon and Storrow
 are available in leading and in next-to-leading order.
 They were first constructed in Ref.~\cite{GOR-9201}, starting the
 evolution at $\qnsq = 5.3$~\gevsq, and later updated in
 Ref.~\cite{GOR-9701} by including  more data,
 and reducing the starting scale to $\qnsq = 3.0$~\gevsq.
 Since the data on the photon structure function \ft only indirectly
 constrain the gluon distribution of the photon, a first attempt was made
 to fit jet production data from TOPAZ~\cite{TOP-9301,TOP-9402},
 and AMY~\cite{AMY-9401},
 which show some sensitivity to the gluon distribution via the
 contributions of resolved photon processes to the jet production.
 However, the data are not precise enough to considerably constrain
 the  gluon distribution function.
 Due to the large starting scale, the input distributions cannot be
 based only on VMD arguments. The authors choose a VMD input similar to
 the one used in the GRV ansatz, but supplement it with an ansatz of
 the point-like component, based on the lowest order Bethe-Heitler
 formula for three light quarks. The quark masses are constrained
 to fulfill $0.25<m_u=m_d<0.4$~\gev and $0.35<m_s<0.55$~\gev and are
 fitted to the data, resulting in masses of 0.29~\gev for up and down quarks
 and 0.41~\gev for strange quarks, as explained in Ref.~\cite{GOR-9201}.
 As both contributions, the hadron-like and the point-like,
 vanish as $x\rightarrow 1$, the GS quark distribution
 functions are greatly suppressed at high values of $x$ compared to,
 for example, the quark distribution functions from GRV.
 The contribution from charm quarks is added via the Bethe-Heitler
 formula with a charm quark mass of $\mc=1.5$~\gev.
 The evolution is performed for three light flavours using
 $\lam = 0.2$~\gev and this is supplemented with the Bethe-Heitler charm
 contribution up to $\qsq=50$~\gevsq.
 At $\qsq=50$~\gevsq this result is matched to a four flavour evolution
 ansatz, which was started at $\qsq=10$~\gevsq, in such a way that \ft is
 continuous. For $\qsq>50$~\gevsq a four flavour massless approach is chosen,
 which is known to overestimate the charm contribution.
 To remove the negative structure function \ft obtained at large
 $x$ when working in next-to-leading order in the \msbl scheme, the
 quark distributions are supplemented by a technical
 input, defined in Eq.~(\ref{eqn:c2ph}), which removes the divergence.
 The leading order and  next-to-leading order parton distribution functions
 are connected to each other by the requirements that they are
 identical at \qnsq, and that the gluon distribution is the
 same for the leading order and the next-to-leading order parametrisations.
 The data used in the fits are published data from
 Refs.~\cite{AMY-9002,AMY-9501,CEL-8302,JAD-8401,JAD-8301,OPALPR092,%
 PLU-8401,PLU-8701,%
 TOP-9301,
 TAS-8601,%
 AMY-9401,
 TPC-8501,TPC-8701,TOP-9402}
 and preliminary data from Refs.~\cite{CEL-9001,TPC-8802,VEN-9302}.
%
\begin{figure}[tbp]
\begin{center}
{\includegraphics[width=1.0\linewidth]{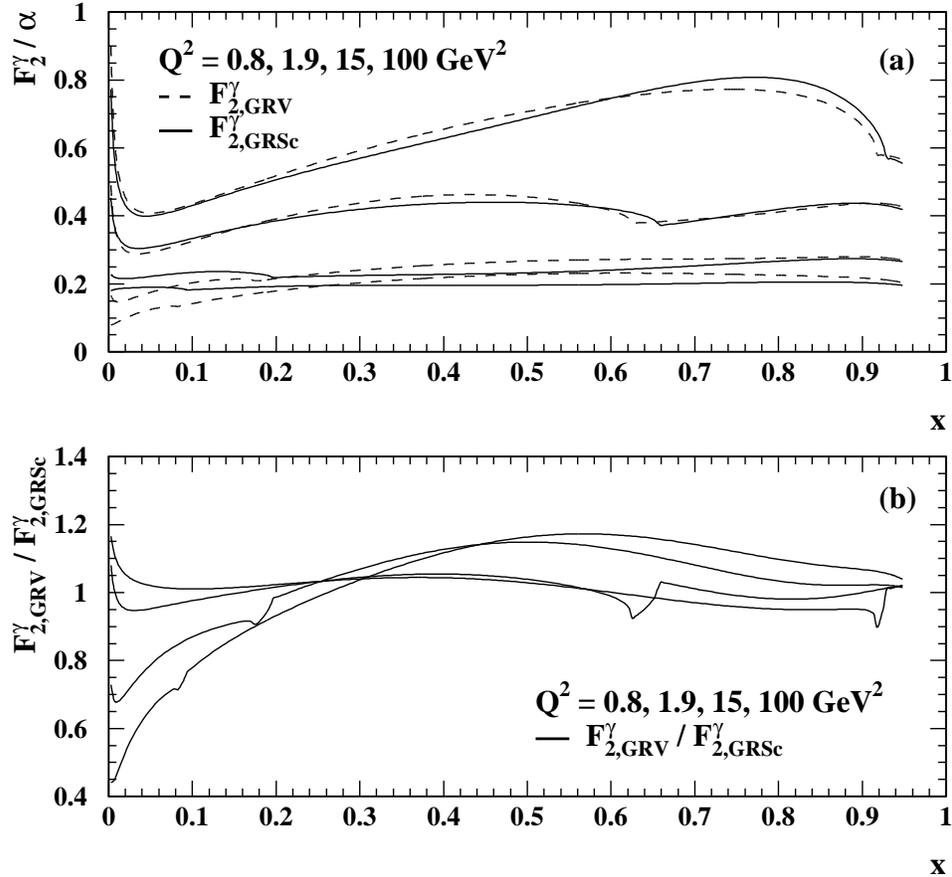}}
\caption[
         Comparison of the leading order GRV and GRSc parametrisations of \ft.
        ]
        {
         Comparison of the leading order GRV and GRSc parametrisations of \ft.
         In (a) the photon structure function \ft is shown in
         leading order for the GRV (dash) and the GRSc (full)
         parametrisations, for four active flavours and for several
         values of \qsq, 0.8, 1.9, 15 and 100~\gevsq, and in (b) the
         ratio of the GRSc and the GRV parametrisations is explored
         for the same values of \qsq.
        }\label{fig:chap4_05}
\end{center}
\end{figure}
%
 \item \underline{\emph{GRSc}}~\cite{GLU-9902}:
 The parton distribution functions from Gl{\"u}ck, Reya and Schienbein
 are constructed using basically the same strategy as the GRV
 parametrisations. However, in addition to the use of the new pion
 input from Ref.~\cite{GLU-9901}, some conceptual changes have been made.
 As a result, as in the case of AFG, no parameters have to be
 obtained from a fit to the \ft data.
 Like the AFG parametrisations, the GRSc parametrisations use a coherent
 sum of vector mesons, and therefore there is no free parameter $\kappa$,
 which was used in the case of the GRV parametrisations when
 using an incoherent sum.
 The treatment of \al has been changed from the approximate
 next-to-leading order formula to an exact solution of the
 renormalisation group equation for \al in next-to-leading order,
 using $\lam = 0.204/0.299$~\gev in leading/next-to-leading order.
 The contribution of charm quarks is taken from the leading order
 Bethe-Heitler formula, both for the leading order and the higher
 order parametrisations, and the mass of the charm quarks was changed
 from 1.5 to 1.4~\gev.
 The numerical differences between \ft as predicted by the
 leading order GRV and GRSc parametrisations is shown in
 Figure~\ref{fig:chap4_05} for several values of \qsq.
 At low values of \qsq the two parametrisations are very different
 especially at low values of $x$.
 For increasing \qsq they get closer, and for $\qsq>15$~\gevsq the
 differences are smaller than 10$\%$.
%
%
 \item \underline{\emph{SaS}}~\cite{SCH-9301,SCH-9302,SCH-9501}:
 Two sets are constructed by Schuler and Sj{\"o}strand
 in leading order, SaS1 using $\qnsq=0.36$~\gevsq as starting scale and
 SaS2 for $\qnsq=4$~\gevsq.
 Both sets use $\lam = 0.2$~\gev, the massless evolution equations for
 light quarks, and the Bethe-Heitler formula, Eq.~(\ref{eqn:mike1}),
 for contributions of charm and bottom quarks with masses $\mc=1.3$~\gev
 and $\mb=4.6$~\gev.
 The leading order parton distribution functions
 are derived both in the \msbl and the \disg scheme\footnote
 {The scheme dependence is introduced by hand into the
  leading order parton distribution functions, by including
  the universal part of
  $C_{2,\gamma}(x)=3\big\{\left[x^2+(1-x^2)\right]\ln(1/x)-1+6x(1-x)\big\}$,
  which formally is of
  next-to-leading order. This choice is motivated by the fact that
  although formally $C_{2,\gamma}(x)$ is of higher order, numerically it
  is important.
 }.
 \par
 The parameters are fitted to data for $\qsq>\qnsq$ and the dependence
 on the photon virtuality \psq is kept to allow for an extension to
 virtual photons, discussed below.
 The motivation for the choice of the two sets SaS1 and SaS2
 is an investigation of the correlation between the size of
 the hadron-like input function and the starting scale \qnsq.
 Consequently, the main difference between the two sets is that the set SaS2
 contains a larger VMD contribution compared to the set SaS1, which is
 needed to still fit the data, while starting at a much larger scale
 \qnsq.
 For the set SaS1 the normalisation of the VMD contribution, as well
 as the starting scale is determined from the analysis of $\gamma$p
 scattering data, only the shape of the VMD distribution is fitted
 to the data of the photon structure function.
 In contrast, for the set SaS2, the starting scale is fixed
 to $\qnsq=4$~\gevsq, the functional form of the distribution
 functions is changed, and an additional proportionality factor $K$
 is introduced for the VMD contribution, and fitted to the data,
 resulting in $K=2.422$. This factor corresponds to
 an inclusion of higher mass vector mesons to compensate for the
 fact that no point-like contribution is allowed to evolve from
 $0.36<\qsq<4$~\gevsq.
 \par
 The subdivision into point-like and hadron-like parton distribution
 functions is made explicit in the SaS distribution functions, allowing
 for an independent treatment of the two, for example,
 in the simulation of the properties of hadronic final states,
 originating from the hadron-like and the point-like part of the photon
 structure function.
 The point-like part is further factorised into a term which describes
 the probability of the photon to split into a \qqbar state at a
 perturbatively large scale, and a so-called \emph{state\/} distribution
 which describes the parton distribution functions within
 this \qqbar state. This subdivision is made to facilitate the
 proper use of the parton distribution functions in Monte Carlo
 programs, when using the parton shower concept.
 The data used in the fits are published data from
 Refs.~\cite{AMY-9002,JAD-8401,OPALPR092,PLU-8404,PLU-8701,TAS-8601,TOP-9402}
 and preliminary data from Refs.~\cite{CEL-9001}.
%
\begin{figure}[tbp]
\begin{center}
{\includegraphics[width=0.98\linewidth]{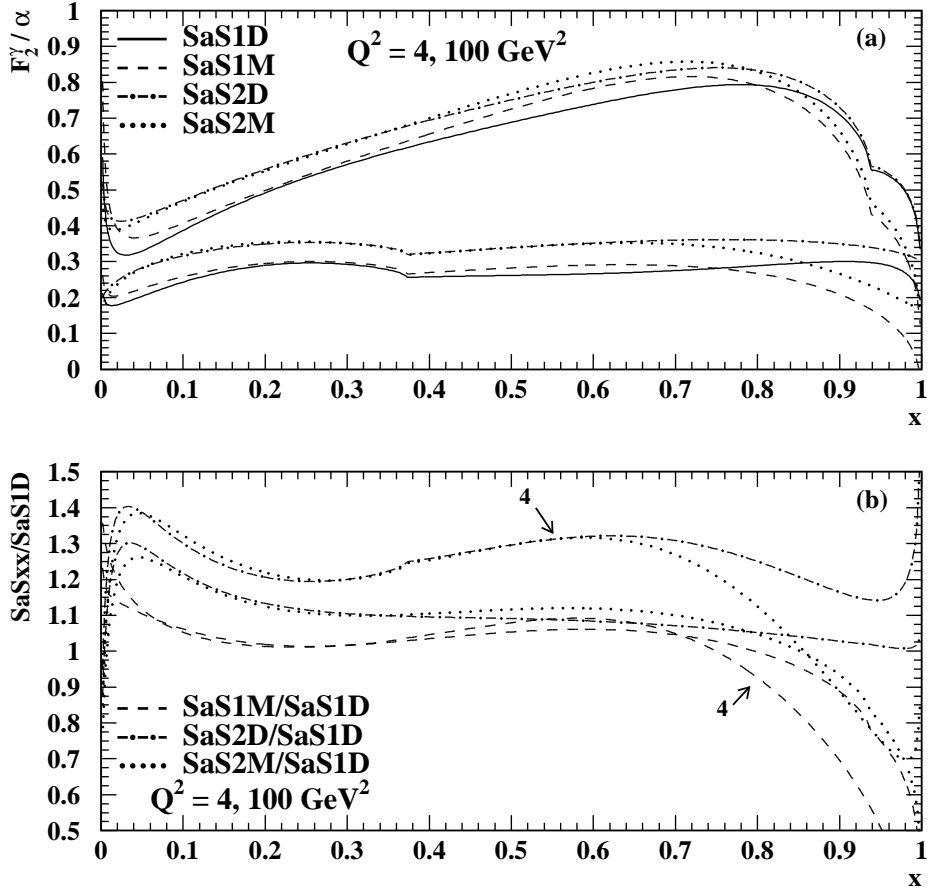}}
\caption[
         The structure function \ft from the SaS parton distribution
         functions for several values of \qsq.
        ]
        {
         The structure function \ft from the SaS parton distribution
         functions for several values of \qsq.
         The values chosen are the starting scale of the evolution
         of the sets SaS2, $\qsq=4$~\gevsq and $\qsq=100$~\gevsq.
         The structure function \ft is shown in (a) for the individual
         sets 1D, 1M, 2D and 2M
         and in (b) the sets 1M, 2D and 2M are divided by the
         prediction of the set SaS1D.
        }\label{fig:chap4_06}
\end{center}
\end{figure}
%
 Figure~\ref{fig:chap4_06}(a) shows the \ft prediction of the
 individual sets for $\qsq=4$ and $100$~\gevsq, and in
 Figure~\ref{fig:chap4_06}(b) they are normalised to the set SaS1D.
 Some general trends can be seen from the figure.
 The SaS2 sets predict a larger hadron-like part and therefore
 they are larger at small values of $x$, than the SaS1 sets.
 This difference decreases with increasing \qsq, as can be seen
 from the ratios displayed in Figure~\ref{fig:chap4_06}(b).
 At large values of $x$ the hadron-like part is small and the
 difference mainly comes from the different
 treatment of $C_{2,\gamma}(x)$, which makes the sets 1D and 2D
 agree with each other for $100$~\gevsq and also the sets 1M and 2M.
 At large values of $x$ the structure function \ft is more strongly
 suppressed when using the \msbl scheme, which makes the sets
 1M and 2M vanish faster as $x$ approaches unity.
 \end{Enumerate}
%
 The main features of the parton distribution functions for real
 photons described above, are listed in Table~\ref{tab:chap4_01}.
 \par
%
\renewcommand{\arraystretch}{1.0}
\begin{table}[tbp]
\caption[
         The parton distribution functions for real photons.
        ]
        {
         The parton distribution functions for real photons.
         The table contains a compilation of the most recent versions
         of the available parton distribution functions for real photons.
         The abbreviations \emph{evol\/} means that the charm contribution
         is included as massless or massive quark in the evolution
         equations, whereas \emph{BH\/} denotes the inclusion of massive charm
         quarks via the Bethe-Heitler formula.\\
        }\label{tab:chap4_01}
\begin{center}
\begin{tabular}{cccccccc} \hline
 authors & set & $\qnsq\,\left[\gevsq\right]$ & scheme &
 $\lam\,\left[\gev\right]$ & gluon & charm & Ref. \\ \hline
  DG &   & 1.0  &   & 0.400 &           & evol &\protect\cite{DRE-8503}\\
                                                                        \hline
 LAC & 1 & 4.0  &   & 0.200 &           & evol     &\protect\cite{ABR-9102}\\
     & 2 &  "   &   &  "    & b=0       &  "       & \\
     & 3 & 1.0  &   &  "    &           &  "       & \\ \hline
 WHIT& 1 & 4.0  &   & 0.400 &a=0.5,c=3  & BH, evol & \protect\cite{HAG-9501}\\
     & 2 &  "   &   &  "    &a=0.5,c=9  &  "       & \\
     & 3 &  "   &   &  "    &a=0.5,c=15 &  "       & \\
     & 4 &  "   &   &  "    &a=1.0,c=3  &  "       & \\
     & 5 &  "   &   &  "    &a=1.0,c=9  &  "       & \\
     & 6 &  "   &   &  "    &a=1.0,c=15 &  "       & \\
                                                                        \hline
                                                                        \hline
 GRV & LO & 0.25 &         & 0.200 &  & BH, evol & \protect\cite{GLU-9202}\\
     & HO & 0.30 & \disg   &  "    &  &  "       & \\ \hline
 AFG &    & 0.5  & \msbl   & 0.200 &  & BH, evol & \protect\cite{AUR-9402}\\
                                                                        \hline
  GS & LO & 3.0  &         & 0.200 &  & BH, evol & \protect\cite{GOR-9701}\\
     & HO &  "   & \msbl   &  "    &  &  "       & \\     \hline
 GRSc& LO & 0.5  &         & 0.204 &  & BH, evol & \protect\cite{GLU-9902}\\
     & HO &  "   & \disg   & 0.299 &  &  "       &                   \\ \hline
 SaS & 1D & 0.36 & '\disg' & 0.200 &  & BH, evol & \protect\cite{SCH-9501}\\
     & 2D & 4.0  & '\disg' &  "    &  &  "       &\\
     & 1M & 0.36 & '\msbl' &  "    &  &  "       &\\
     & 2M & 4.0  & '\msbl' &  "    &  &  "       &\\                 \hline\\
\end{tabular}
\end{center}\end{table}
%
 In general the parametrisations of the parton distribution functions
 do not differ much in the quark distribution functions at medium
 values of $x$, because they are constrained by the \ft data used
 in the fits.
 In contrast, in the region of high and low values of $x$ the available
 parton distribution functions are not well constrained.
 The highest value of $x$ reached in the measurements of \ft is restricted
 by the minimum value of invariant mass required to be well
 above the region of resonance production.
 Therefore, for example, the very different quark distribution functions
 of GS and GRV are still consistent with the data on \ft.
 For low values of $x$ measurements became available only recently, and
 they are not yet incorporated in the presently available parton
 distribution functions.
 Consequently, there was considerable freedom in the gluon distribution
 function, which is important at low-$x$. This freedom has been exploited
 in the differences of the various sets constructed by several groups.
 This results in very different gluon distribution functions but also,
 driven by the gluons, in different quark distribution functions at low
 values of $x$.
 The new data on \ft now start to constrain the parton distribution
 functions also at low values of $x$, and, although in leading order
 the photon only couples to quarks, the measurements of \ft give an
 indirect constraint on the gluon distribution function as well.
 \par
 Also promising is the inclusion of jet production data, either from the
 reaction $\gamma\gamma\rightarrow$~jets or $\gamma{\rm p}\rightarrow$~jets,
 which are studied at LEP and HERA.
 This data are directly sensitive to the gluon distribution function
 at low values of $x$, for example, via the boson-gluon fusion diagram
 or gluon-gluon scattering.
 In addition, jet production can be used to explore the region of high
 values of $x$ and to further constrain the quark distribution functions
 in this region.
 Because the main subject here is the photon structure function \ft, and
 the HERA results have not yet been incorporated in the construction of
 parton distribution functions, this interesting topic is not discussed
 in further detail here.
 However,
 the results from the HERA experiment on jet production cross-sections,
 and what can be learned about the parton distribution function of
 the photon will be briefly discussed in Section~\ref{sec:ggreal}.
 \par
\begin{figure}[tbp]
\begin{center}
{\includegraphics[width=1.1\linewidth]{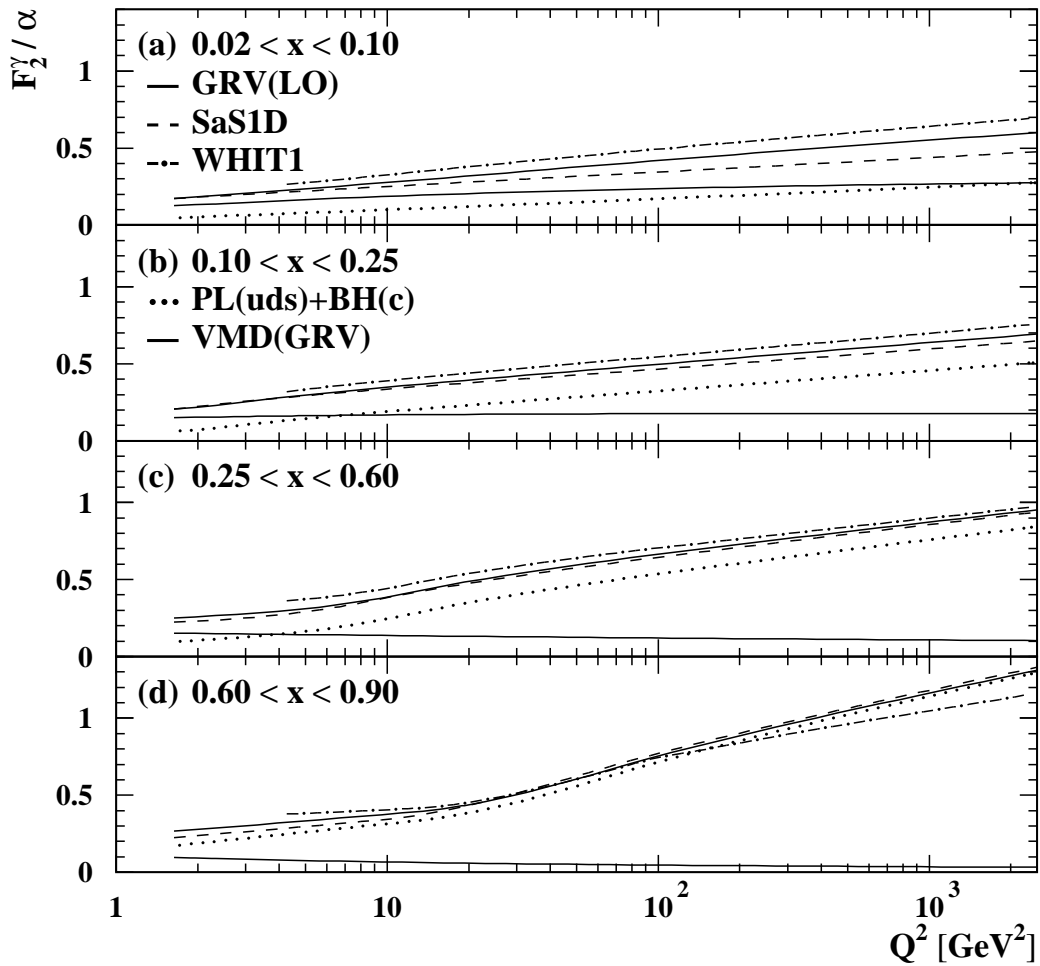}}
\caption[
         Predictions for the \qsq evolution of the photon structure
         function \ft for various $x$ ranges.
        ]
        {
         Predictions for the \qsq evolution of the photon structure
         function \ft for various $x$ ranges.
         The predictions of the GRV, SaS and WHIT1 parametrisations are
         compared to the evolution of the purely point-like part for
         three light flavours for $\lam_3 = 0.232$~\gev,
         combined with \ftc as predicted by
         the Bethe-Heitler formula, denoted with PL(uds)+BH(c), and
         to the hadron-like part of the GRV parametrisation,
         labelled VMD(GRV).
        }\label{fig:chap4_07}
\end{center}
\end{figure}

 As there is the freedom to choose both, the input distribution functions,
 and the value of \lam, differently for the
 leading order and next-to-leading order parton distribution functions,
 the predicted structure functions \ft in leading and next-to-leading
 order are similar.
 However, to perform a meaningful investigation
 of the sensitivity of the photon structure function \ft to the running
 coupling constant \al, by studying the \qsq evolution of \ft,
 it is mandatory to use the next-to-leading order
 approach, in order for the scale \lam to be fixed.
 For illustration, the predicted \qsq evolution of \ft for four
 active flavours, and for various leading order parametrisations of \ft
 is shown in Figure~\ref{fig:chap4_07}.
 The predictions of the GRV, SaS and WHIT1 parametrisations are compared
 to the evolution of the purely point-like part for three light flavours
 for $\lam_3 = 0.232$~\gev, combined with \ftc as predicted by the
 Bethe-Heitler formula, denoted with PL(uds)+BH(c), and to the
 hadron-like part of the GRV parametrisation, labelled VMD(GRV).
 The bins in $x$ used, correspond to the experimental analyses of
 Ref.~\cite{OPALPR207}.
 The hadron-like part of \ft dominates at low values of $x$
 for all values of \qsq, but for $x>0.1$ and $\qsq >10$~\gevsq
 the point-like part is more important, and for  $x>0.6$ the
 hadron-like part is negligible for $\qsq >10$~\gevsq.
 As the importance of the hadron-like part of \ft decreases for
 increasing $x$, the predictions of the GRV, SaS and WHIT1
 parametrisations get closer to each other.
 All parametrisations predict a strongly increasing slope for increasing
 $x$, driven by the point-like contribution, with WHIT1 showing the
 flattest behaviour.
 \par
%
%
 Several parton distribution functions for transverse virtual photons
 have been constructed. They can be applied to any process
 which is dominated by the contribution of transverse virtual photons.
 There is no unique prescription on how to extend the parton
 distribution functions for $\psq>0$, and different approaches have been
 performed.
 All parton distribution functions are constructed such that they reproduce
 the correct limits for small and large values of \psq.
 For $\psq=0$ the parton distribution functions for real photons are
 recovered, and the limit $\lamsq\ll\psq\ll\qsq$ is given by
 the perturbative QCD results of Refs.~\cite{UEM-8101,UEM-8201}.
 There exist one simple approach by Drees and Godbole, which is independent
 of the specific choice of parton distribution functions used for
 $\psq=0$, and therefore can be applied on top of any of the existing
 parton distribution functions listed above.
 In addition three parton distribution functions for real photons,
 GRV, GRSc and SaS, have been extended to also incorporate the region of
 $\psq>0$. The extension of the GRV parton distribution functions is
 called GRS.
%
\begin{Enumerate}
 \item \underline{\emph{DG}}~\cite{DRE-9401}:
 In the simple model by Drees and Godbole the parton distribution functions
 for virtual photons are obtained by simple, \psq-dependent, multiplicative
 factors from the parton distribution functions for real photons,
 where by construction the gluon distribution function is more suppressed
 than the quark distribution functions, as suggested in
 Ref.~\cite{BZS-9301}.
%
 \begin{eqnarray}
 q_k^{\gamma}(x,\qsq,\psq) &=&
 q_k^{\gamma}(x,\qsq) \cdot
 \Bigl[
 \frac{\ln\frac{\qsq+\psq}{\psq+\pcsq}}{\ln\frac{\qsq+\pcsq}{\pcsq}}
 \Bigr]
 \equiv
 q_k^{\gamma}(x,\qsq) \cdot L
 , \nonumber\\
 g^{\gamma}(x,\qsq,\psq)   &=&   g^{\gamma}(x,\qsq) \cdot L^2,
 \label{eqn:DGsup}
 \end{eqnarray}
%
 where \pcsq should be chosen to be a typical hadronic scale, which
 means, to be in the range $\lamsq \le \pcsq \le 1$~\gevsq.
 With this, the parton distribution functions are globally suppressed,
 which means the $x$ dependence of the parton distribution functions for real
 photons is not altered for $\psq>0$.
%
%
 \item \underline{\emph{GRS}}~\cite{GLU-9501}:
 In the parton distribution functions from Gl{\"u}ck, Reya and Stratmann
 a boundary condition, similar to the one used for real photons, is
 applied at $\qsq=\max(\psq,\qnsq)$.
 This condition allows to smoothly interpolate between $\psq=0$ and
 $\psq\gg\lamsq$, using a frozen non-perturbative input for
 $0<\psq<\qnsq$, where \qnsq is the starting scale of the evolution
 for the parton distribution functions of the real photon.
 Then the parton distribution functions are obtained by solving
 the leading order or next-to-leading order inhomogeneous evolution
 equations, which are chosen to be the same as for the real photon.
 For the inclusion of heavy quarks the extension of the Bethe-Heitler
 formula to the region $\psq>0$, Eq.~(\ref{eqn:BH}) have been taken.
 The analysis of the sensitivity to the non-perturbative input
 distribution functions shows large non-perturbative contributions at
 small values of $x$ up to $\psq=10$~\gevsq, which will be discussed
 below.
%
\begin{figure}[tbp]
\begin{center}
{\includegraphics[width=0.98\linewidth]{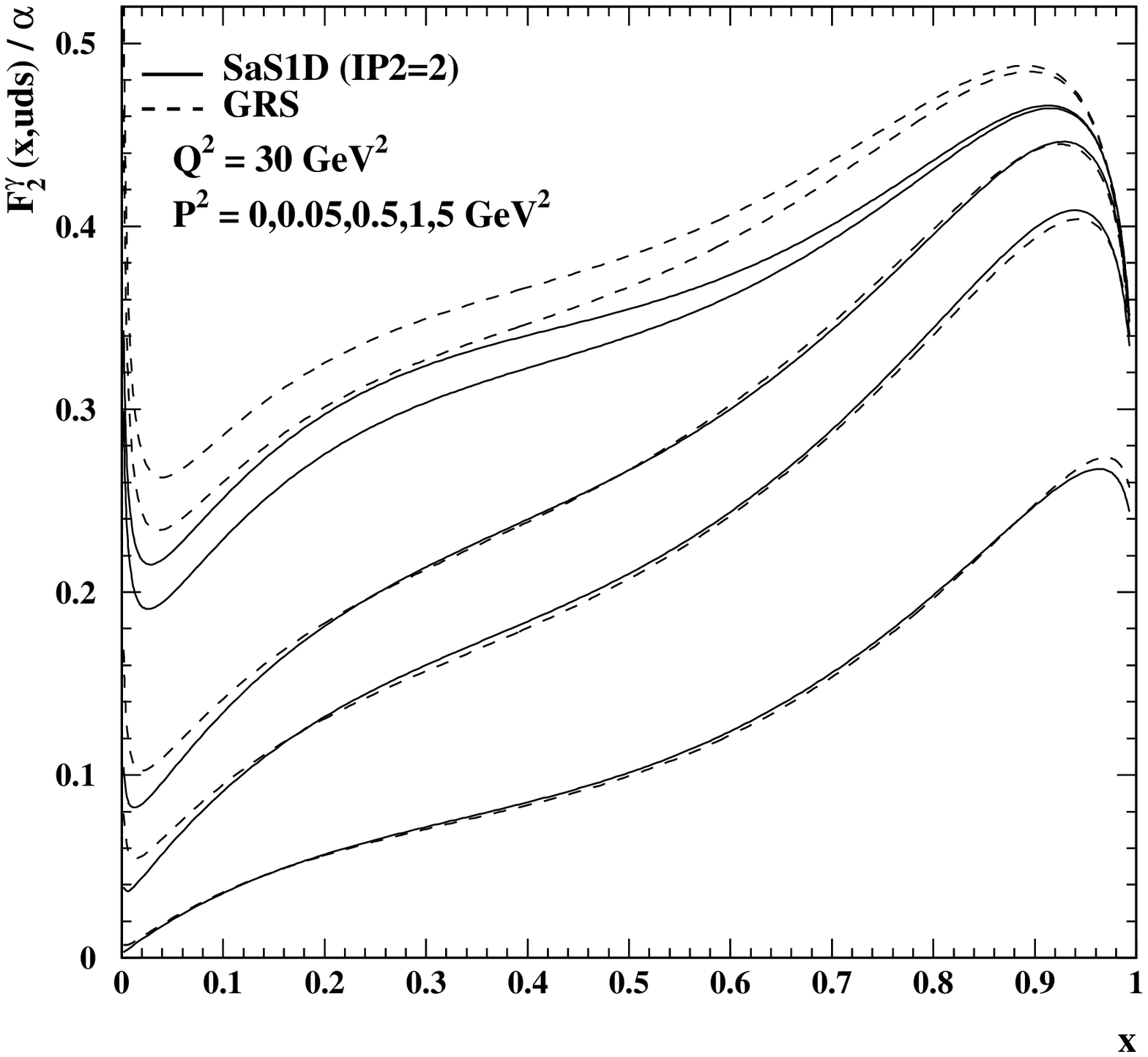}}
\caption[
         Comparison of leading order parametrisations of the photon
         structure function \ft for virtual photons.
        ]
        {
         Comparison of leading order parametrisations of the photon
         structure function \ft for virtual photons.
         The photon structure function \ft is shown for the SaS1D (full)
         and the GRS (dash) prediction for $\qsq=30$~\gevsq and
         for several values of \psq,  0, 0.05, 0.5, 1 and $5$~\gevsq,
         always for three light flavours, $\nf=3$.
         The predictions decrease with increasing \psq.
         For the SaS1D prediction the choice of \psq suppression is made
         such to have the result which is closest to the GRS prediction,
         resulting in $\mbox{IP2}=2$.
        }\label{fig:chap4_08}
\end{center}
\end{figure}
%
 \item \underline{\emph{GRSc}}~\cite{GLU-9901}:
 The parton distribution functions from Gl{\"u}ck, Reya and Schienbein
 for the real photon are extended for virtual photons, $\psq>0$, based
 on the assumption that for virtual photons the photon virtuality
 should entirely be taken care of
 by the flux factors, which are valid for $\qsq\gg\psq$.
 As a consequence of this all partonic subprocess
 cross-sections are calculated as if $\psq=0$.
 Therefore, in the GRSc approach the process $\ggs\rightarrow\qqbar$
 is used to evaluate $C_\gamma(x)$, instead of the process
 $\ggss\rightarrow\qqbar$ which was used in the case of GRS.
 Also the charm contribution for $\psq>0$ is based on the
 Bethe-Heitler formula for $\psq=0$, Eq.~(\ref{eqn:mike1}).
%
\begin{figure}[tbp]
\begin{center}
{\includegraphics[width=1.00\linewidth]{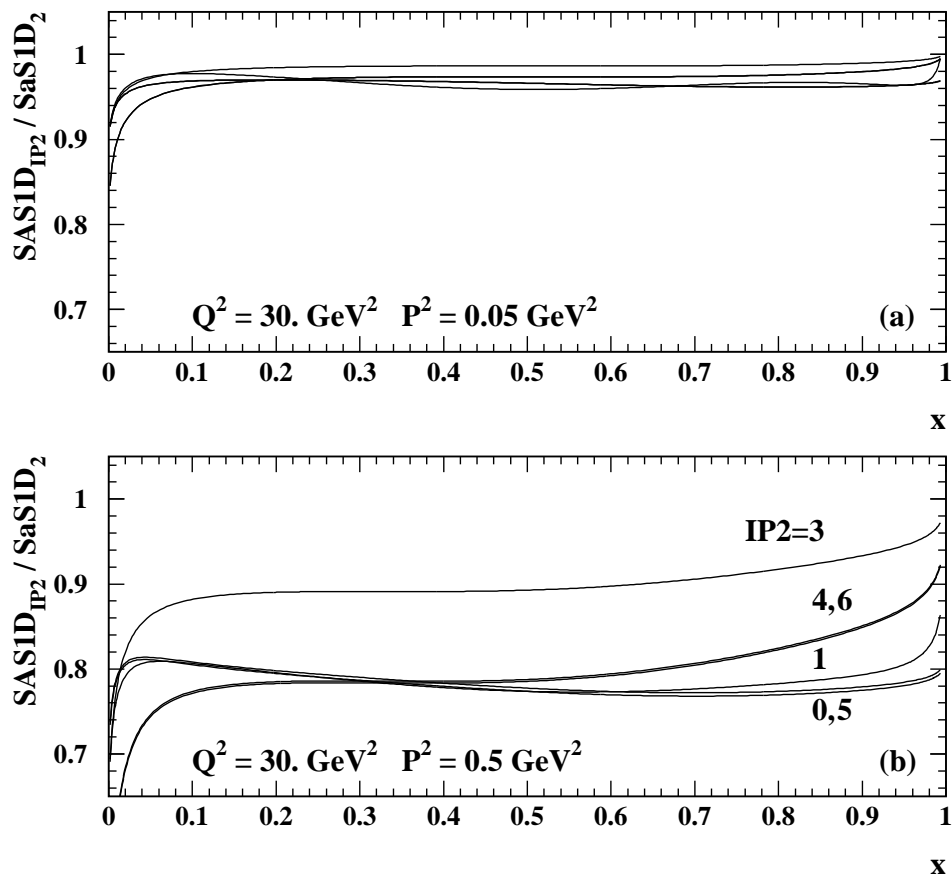}}
\caption[
         The variation of the \psq suppression of \ft in the SaS1D
         parametrisations.
        ]
        {
         The variation of the \psq suppression of \ft in the SaS1D
         parametrisations.
         The different choices of the \psq suppression of \ft are shown
         by varying the parameter IP2 for $\qsq=30$~\gevsq
         and for two values of \psq.
         The curves shown in (a) and (b) are for $\psq=0.05$
         and $0.5$~\gevsq respectively. The predictions for the
         different choices of IP2 are all divided by the result
         obtained for $\mbox{IP2}=2$.
        }\label{fig:chap4_09}
\end{center}
\end{figure}
%
 \item \underline{\emph{SaS}}~\cite{SCH-9501,SCH-9601}:
 The starting point for the Schuler and Sj{\"o}strand parton
 distribution functions is a representation of the parton distribution
 functions of the virtual photon as a dispersion integral in the mass
 of the \qqbar fluctuations, as discussed in Ref.~\cite{BJO-8901}.
 The \qqbar fluctuations are separated into a discrete sum of vector
 meson states, and a high mass continuous perturbative spectrum
 from the point-like contribution.
 Both terms are suppressed by different \psq-dependent terms.
 Various sets of boundary conditions and different evolution equations
 for virtual photons, which differ by terms of the order of \psq/\qsq,
 are proposed, accessible in the parametrisations via the parameter IP2.
 The heavy quarks are included as in the GRS case.
 \end{Enumerate}
%
 In Figure~\ref{fig:chap4_08} the GRS predictions are compared
 to the SaS predictions using the set SaS1D.
 The structure function is shown for three light flavours,
 for $\qsq = 30$~\gevsq and exploring the \psq suppression for
 $\psq=0, 0.05, 0.5, 1$ and 5~\gevsq.
 The parametrisations show some differences for quasi-real photons,
 where the GRS prediction is higher at all values of $x$.
 For $x>0.2$ the GRS prediction is about 8$\%$ higher and for smaller
 values of $x$ the rise is much faster than in the case of the SaS1D
 prediction.
 As soon as $\psq> 0.5$~\gevsq they perfectly agree with
 each other for $x>0.1$.
 \par
 The theoretical uncertainty on how the structure function \ft
 is suppressed for increasing \psq is explored in the SaS
 distribution functions.
 In Figure~\ref{fig:chap4_09} the various choices are compared to the
 choice which is closest to the GRS prediction.
 The larger the value of \psq the more the various choices differ,
 as can be seen from Figure~\ref{fig:chap4_09}(a), where at
 $\psq = 0.05$~\gevsq the predictions are close together,
 whereas at $\psq = 0.5$~\gevsq, Figure~\ref{fig:chap4_09}(b),
 sizeable differences are seen.
 Taking the variations as an estimate of the theoretical uncertainty, it
 amounts to about 20$\%$ at $\psq = 0.5$~\gevsq.
 \par
%
\begin{figure}[tbp]
\begin{center}
{\includegraphics[width=1.0\linewidth]{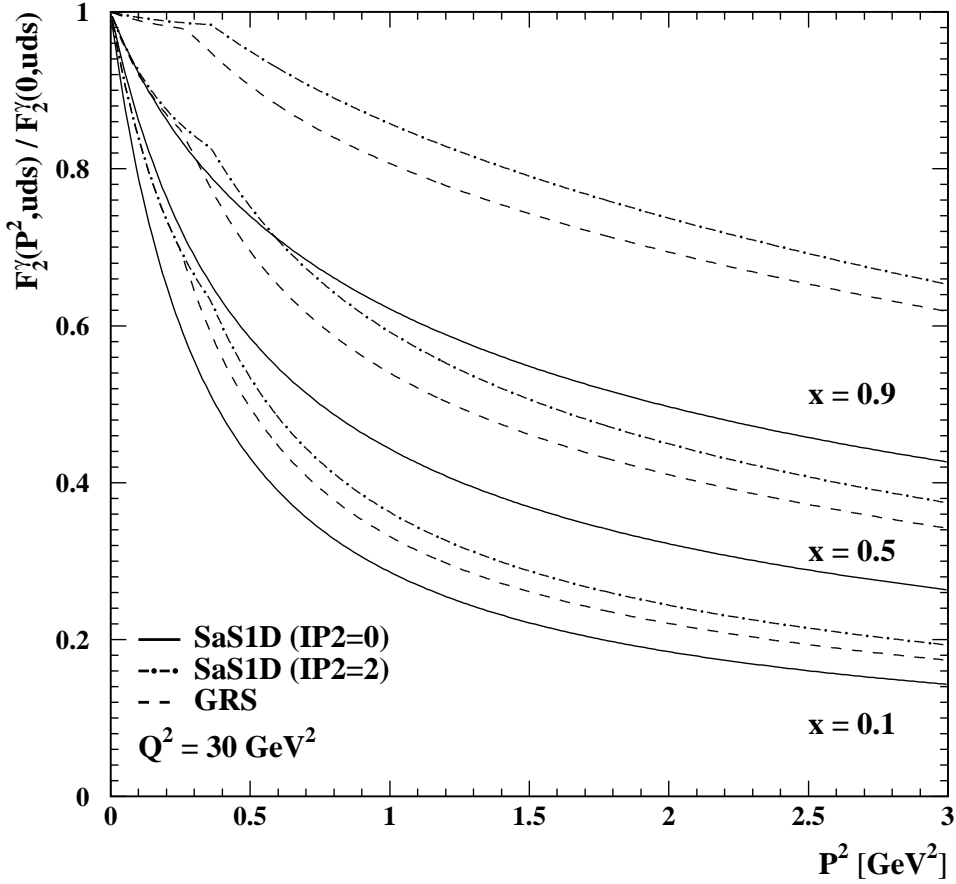}}
\caption[
         The predicted \psq suppression of the photon structure function
         \ft(uds).
        ]
        {
         The predicted \psq suppression of the photon structure function
         \ft(uds).
         Shown are the prediction of \ft from the GRS parton distribution
         functions (dash), together with the prediction from the SaS1D
         parton distribution functions for two modes of the \psq
         suppression. The two modes chosen are the recommended
         suppression ($\mbox{IP2}=0$, full) and the one which is most
         similar to the GRS suppression ($\mbox{IP2}=2$, dot-dash).
         The values of \ft are normalised to the prediction for
         real photons, $\psq=0$.
         The curves are calculated for $\qsq = 30$~\gevsq, for three
         values of $x$, 0.1, 0.5 and 0.9, and for three flavours.
        }\label{fig:chap4_10}
\end{center}
\end{figure}
%
 Although the absolute predictions for \ft from the GRS and
 SaS1D parametrisations agree quite well for $\psq>0.5$~\gevsq,
 they differ in the relative suppression as a function of \psq,
 shown in Figure~\ref{fig:chap4_10} for $\qsq = 30$~\gevsq, and
 for several values of $x$.
 The GRS predictions are compared to the ones from SaS1D using
 $\mbox{IP2}=0$ and 2.
 The suppression decreases with increasing $x$ and the suppression
 as predicted by SaS, using the recommended scheme, $\mbox{IP2}=0$,
 is always stronger than the one predicted by GRS.
 The kinks in the distributions at $\psq=0.36$ and $0.25$~\gevsq
 for the SaS and GRS predictions are due to the boundary conditions
 applied.
 \par
%
\begin{figure}[tbp]
\begin{center}
{\includegraphics[width=1.0\linewidth]{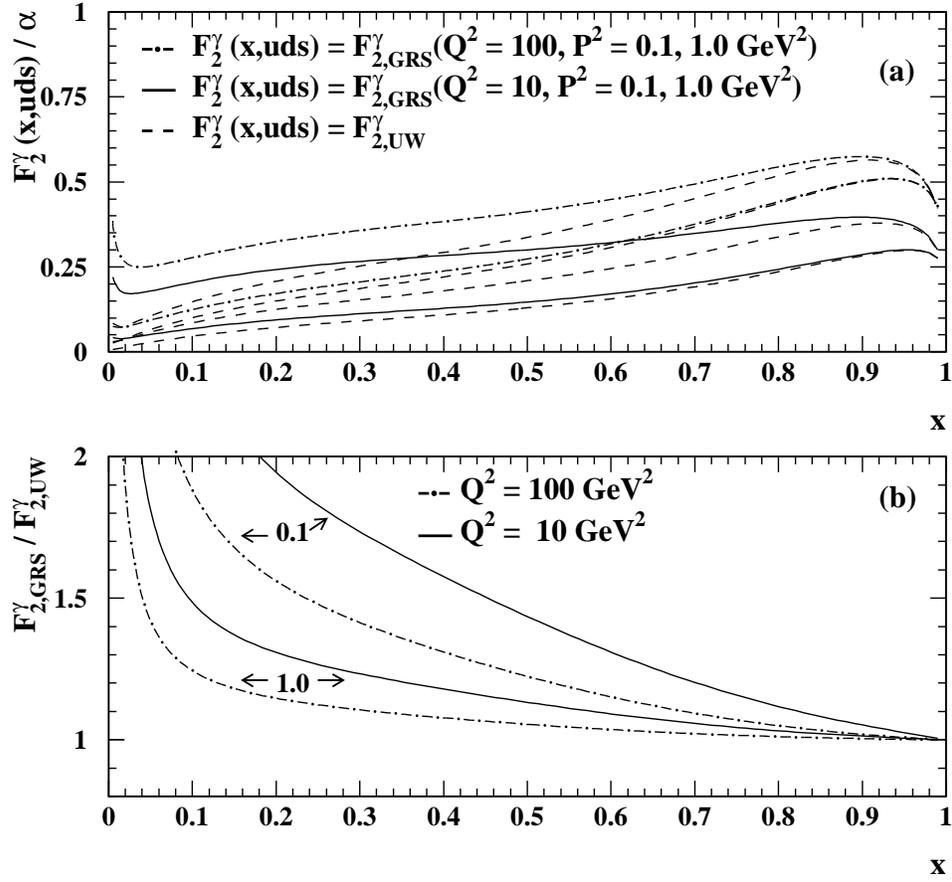}}
\caption[
         Comparison of the photon structure function \ft for virtual
         photons with the purely perturbative point-like part.
        ]
        {
         Comparison of the photon structure function \ft for virtual
         photons with the purely perturbative point-like part.
         The predictions of the structure function \ft are shown
         for three active
         flavours, for two values of \qsq, 10 and $100$~\gevsq,
         and for two values of \psq, 0.5 and $1.0$~\gevsq, and
         for $\lam_3 = 0.232$~\gev.
         The prediction from the GRS parametrisations is compared to
         the purely perturbative point-like part which is equivalent to
         the prediction from Uematsu and Walsh (UW).
         In (a) the individual predictions are shown and in
         (b) the GRS predictions are divided by the contribution of
         the purely point-like part.
        }\label{fig:chap4_11}
\end{center}
\end{figure}
%
 Based on the GRS parametrisations the sensitivity to the
 non-perturbative input distribution functions can be studied.
 In Figure~\ref{fig:chap4_11} the full solution for three light quark
 species is compared to the purely point-like part which is equivalent
 to the prediction from Uematsu and Walsh from Ref.~\cite{UEM-8101}.
 The comparison is done for two values of \qsq, 10 and $100$~\gevsq,
 and for two values of \psq, 0.1 and $1.0$~\gevsq, and
 all predictions are for $\lam_3 = 0.232$~\gev.
 The GRS parametrisations predict a significant hadron-like contribution
 at small values of $x$ for all values of \psq.
 The importance decreases for increasing \qsq and even stronger for
 increasing \psq, as can be seen from Figure~\ref{fig:chap4_11}(b),
 where the ratio of the GRS prediction and the purely point-like part
 is shown.
 At large values of $x$ the hadron-like contribution is less important.
 For example, the hadron-like contribution amounts to less than 10$\%$
 for $x>0.3$ for $\qsq = 100$~\gevsq and $\psq = 1.0$~\gevsq.
 This is a region which is still accessible within the LEP2 programme,
 however only with very limited statistics.
 \par
%
\begin{figure}[tbp]
\begin{center}
{\includegraphics[width=1.0\linewidth]{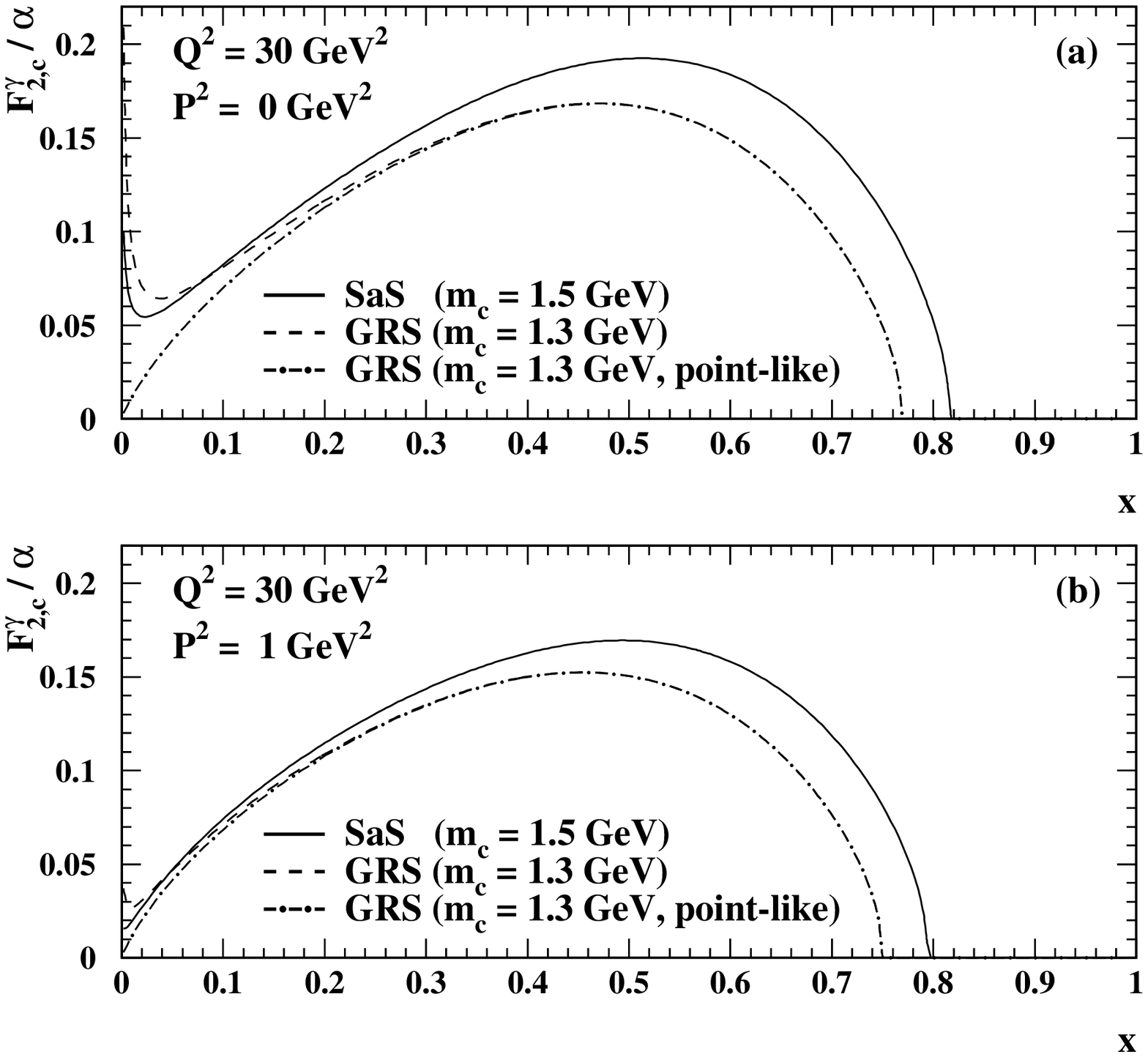}}
\caption[
         The point-like and hadron-like contributions to \ftc.
        ]
        {
         The point-like and hadron-like contributions to \ftc.
         The predictions of the SaS1D (full) and the GRS
         (dash) parametrisations are shown for $\qsq=30$~\gevsq and
         for two values of \psq.
         Figure (a) is for $\psq=0$ and (b) uses $\psq=1$~\gevsq.
         For the GRS prediction in addition the contribution from the
          point-like process alone (dot-dash) is shown.
        }\label{fig:chap4_12}
\end{center}
\end{figure}
%
 The last issue discussed in the comparison of the SaS1D and
 the GRS parametrisations is the contribution to \ft from the
 point-like and hadron-like production of charm quark pairs.
 The two predictions are shown in Figure~\ref{fig:chap4_12} for
 $\qsq=30$~\gevsq and for two values of \psq, 0 and 1~\gevsq.
 The mass of the charm quark is $\mc=1.5$~\gev for the GRS
 parametrisation, whereas SaS uses $\mc=1.3$~\gev.
 For the GRS prediction in addition the contribution from the
 point-like part alone is shown.
 The point-like contribution is found to dominate for $x>0.1$, whereas
 at smaller values of $x$ the hadron-like component gives a significant
 contribution and dominates as $x$ approaches zero.
 The difference between the two predictions for the point-like  part
 is entirely due to the different choice for the mass, which means
 when changing the mass in GRS to $\mc=1.3$~\gev they are identical.
 However, GRS and SaS differ in the contribution from the hadron-like
 part, with GRS predicting a faster rise for small values of $x$.
 This difference is due to the different gluon distribution functions.
 For increasing \psq the hadron-like part gets less important.
 \par
 Experimentally, the measurement of the heavy quark contributions
 to \ft is very difficult, mainly because of the low statistics
 available.
 Firstly, the heavy quark production is suppressed by the large quark
 masses and secondly, to establish a heavy quark contribution, the quark
 flavour has to be identified, which can only be done with small
 efficiencies.
 With the available data the measurement of \ftb is hopeless, because,
 due to the large bottom mass and the small electric charge
 the number of events is too small.
 However, the measurement of the charm contribution to \ft is likely
 to be performed soon for the first time, because experimentally
 about 30 events with positively identified charm quarks are
 available, as has been reported in Ref.~\cite{PAT-9901}.
%
%

%
%
\section{Tools to extract the structure functions}
\label{sec:tools}
 The general experimental procedure to measure structure functions is the
 following. The data are divided into ranges in \qsq and the structure
 functions are obtained as functions of $x$ from the
 distributions of measured values of $x$, usually denoted by \xvis.
 If the energies of both incoming particles are known, like in the case
 of deep inelastic charged lepton-nucleon scattering as, for example,
 in electron-proton scattering at HERA, the values of \qsq and $x$ can
 be obtained from measuring the energy and angle of the scattered electron.
 Consequently, in regions of acceptable resolution in \qsq and $x$ as
 measured from the scattered electron, the proton structure function
 \ftp can be derived without the measurement of the hadronic final state.
 In addition the known energy of the proton can be used to replace
 some less well measured quantities and to obtain \qsq and $x$
 from the hadronic final state.
 \par
 For deep inelastic electron-photon scattering the energy of the incoming
 quasi-real photon is not known. It could only be obtained from the
 measurement of the energy of the corresponding electron.
 For most of the phase space of quasi-real photons the corresponding
 electrons are not observed in the detectors and no
 measurement of the photon energy can be performed.
 Only in the situation of the exchange of two highly virtual photons
 both electrons are observed in the detectors and the invariant mass of
 the photon-photon system, as well as $x$, can be obtained from the two
 scattered electrons.
 Consequently, for the measurement of the structure functions of the
 quasi-real photon, $x$ has to be derived using Eq.~(\ref{eqn:x})
 from measuring the invariant mass of the final state $X$.
 \par
 In the case of lepton pair production it is required that
 both leptons are measured in the tracking devices of the detectors,
 and an accurate measurement of $W$ can be performed.
 In contrast, the hadronic final state is usually only partly observed
 in the detectors and the measurement of $W$ is much less precise.
 Due to this, a good description of the observed hadronic final state
 by the Monte Carlo models is much more important for
 the measurement of the photon structure function, than
 for the measurement of the proton structure function at HERA.
 The value of \xvis is obtained from the measurement of the visible hadronic
 mass \Wvis, together with the well measured \qsq, and therefore the
 uncertainty of \xvis is completely dominated by the uncertainty of \Wvis.
 The uncertainty of \Wvis receives two contributions.
 Firstly \Wvis is affected by the uncertainty of the measurement of the
 seen hadrons, which are observed by tracking devices
 and electromagnetic as well as hadronic calorimeters, and secondly
 the measurement of \Wvis suffers from the fact that some of the hadrons
 are scattered outside of the acceptance of the detectors.
 To account for these deficiencies,
 in most of the analyses the structure function \ft is obtained
 from an unfolding procedure which relies on the correlation between
 the measured \xvis and the underlying value of $x$, as predicted by
 the Monte Carlo programs.
 Therefore the Monte Carlo programs and the unfolding programs are
 the most important tools used in the measurement of photon structure
 functions. They are discussed in Section~\ref{sec:gener} and
 Section~\ref{sec:unfol} respectively.
%
%
\subsection{Event generators}
\label{sec:gener}
 Event generators are extensively used in the determination of photon
 structure functions. In this section the most commonly used programs
 are discussed.
 Only the main features of the Monte Carlo programs relevant for
 deep inelastic electron-photon scattering are described,
 details can be found in the individual program manuals, and
 a general overview is given in Ref.~\cite{LON-9601}.
 \par
 There exist two groups of programs relevant for the measurement
 of photon structure functions.
 The first group deals with low multiplicity final states, like
 resonances, charm quark pairs, or lepton pairs.
 The programs used for the measurement of the QED structure functions
 are BDK, GALUGA and Vermaseren, where by now the Vermaseren program
 can be regarded as the standard Monte Carlo for lepton pair production.
%
\begin{Enumerate}
 \item \underline{\emph{GALUGA}}:~\cite{SCH-9801}
 The GALUGA  Monte Carlo is a more recent program, which contains
 an implementation of the full cross-section
 formula from Ref.~\cite{BUD-7501}.
 It is generally not used as an event generator, but as a useful
 tool to investigate the importance of the individual terms
 to the differential cross-section, as listed in Eq.~(\ref{eqn:true}).
 \item \underline{\emph{Vermaseren}}:~\cite{SMI-7701,VER-7901,VER-8301,%
 BHA-7701}
 In most applications the Vermaseren program is based only on the
 cross-section for the multipheripheral diagram, shown in
 Figure~\ref{fig:chap2_02}(a).
 The full dependence on the mass of the muon and on \psq is kept.
 The program generally is used to generate
 large size event samples, which are compared to the data.
 Sometimes this program is also abbreviated with JAMVG by using the
 initials of the author.
 \item \underline{\emph{BDK}}:~\cite{BDK-8501,BDK-8601,BDK-8602,BDK-8603}
 The BDK program is similar to the Vermaseren program, and in
 addition QED radiative corrections to the process are contained.
 This program is mainly used for the determination of the radiative
 corrections to be applied to the data which, after correction,
 are compared to the predictions of the Vermaseren Monte Carlo.
\end{Enumerate}
%
 The QED predictions of the three programs are very similar and they
 nicely agree with the data, as discussed in Section~\ref{sec:qedres}.
 \par
 The second group of programs is used for the
 determination of the hadronic structure function \ft.
 The situation for the multi-particle hadronic final state is more
 complex than for the case of the leptonic final state,
 as it involves QCD.
 Because the multi-particle hadronic final state cannot be predicted by
 perturbative QCD, there is some freedom on how to model it, and
 the available programs follow different philosophies
 to predict the properties of the multi-particle hadronic final state.
 The programs can be further subdivided into two classes.
 The first class consists of the special purpose Monte Carlo
 programs TWOGAM~\cite{LON-9601}\footnote{No detailed description of
 this  Monte Carlo program is publically available.}
 and TWOGEN~\cite{BUI-9401},
 which contain only electron-photon scattering
 reactions, and are therefore very hard to test thoroughly,
 except by using the electron-photon scattering data themselves.
 This is dangerous, as the measurement of the hadronic structure function
 and the modelling of the hadronic final state are intimately related.
 For this reason, and also because the programs do not contain
 parton showers, their importance is gradually decreasing.
 The second class consists of the general purpose Monte Carlo programs
 HERWIG~\cite{MAR-8801,KNO-8801,CAT-9101,ABB-9001,SEY-9201,MAR-9201,MAR-9601},
 PHOJET~\cite{ENG-9501,ENG-9601}
 and PYTHIA~\cite{SJO-9401}.
 These general purpose Monte Carlo programs are also successfully
 used to describe electron-proton and proton-proton interactions.
 Even more important for electron-photon scattering is the fact that
 some of the parameters are constrained by electron-proton
 and proton-proton scattering data, therefore leaving less freedom
 for adjustments to the electron-photon scattering data.
 \par
 The general procedure for the event generation by Monte Carlo methods
 splits the reaction into different phases, and in each of these phases,
 specific choices are made.
 For deep inelastic electron-photon scattering first a photon is
 radiated from one of the electrons using an approximation for the
 photon flux, discussed in Section~\ref{sec:EPA}.
 Then a parton inside the photon is selected according to the
 prediction of one the various parton distribution functions,
 discussed in Section~\ref{sec:PDF}.
 The selected parton takes part in the hard sub-process, which is
 generated using fixed order matrix elements.
 A configuration for the photon remnant is chosen.
 The emission of further partons is generated from the initial partons,
 using a prescription of the backward evolution of the initial
 state parton shower, and from
 the  outgoing partons, modelled by the final state parton shower,
 which is identical to the parton shower used in the \epem annihilation
 events.
 Finally, all partons are converted into hadrons by means of some
 hadronisation model, and these hadrons are allowed to decay,
 using decay tables.
 \par
 The special purpose Monte Carlo programs used are:
%
\begin{Enumerate}
 \item \underline{\emph{TWOGAM}}:~\cite{LON-9601}
 The special purpose Monte Carlo program TWOGAM was developed within
 the DELPHI collaboration. The events are separated into three event classes,
 point-like events, hadron-like events and the so-called resolved photon
 component.
 The simulation of point-like events is based on a full implementation
 of Eq.~(\ref{eqn:true}) using the QED cross-sections, with free values
 chosen for the light quark masses.
 The hadron-like events are generated according to some VMD prescription.
 The resolved photon component is added for the scattering of two real,
 or virtual, photons with transverse polarisation in the following way.
 The probability to find a parton in a photon is given by a set of parton
 distribution functions for real photons, suppressed by a factor which depends
 on the  virtuality of the photon. The generated partons then undergo a hard
 $2\rightarrow 2$ scattering process.
 No parton showers are included, and the hadronisation is based on the
 Lund string model.
 By using this concept also the virtual photon is allowed to fluctuate
 into a hadronic state.
 \item \underline{\emph{TWOGEN}}:~\cite{BUI-9401}
 The special purpose Monte Carlo program TWOGEN was developed within
 the OPAL collaboration.
 The version used for structure function analyses is called F2GEN.
 This program is in principle based on Eq.~(\ref{eqn:true}), but neglects
 all but the term proportional to \stt.
 Then the differential cross-section is
 expressed as a product of the transverse-transverse luminosity function
 for real and virtual photons, and the cross-section \stt.
 The cross-section \stt is implemented only for real photons, and
 as given in Eq.~(\ref{eqn:struc}), it is proportional to \ft.
 The program generates only the multipheripheral diagram.
 The angular distribution of the quark anti-quark final state
 is chosen to be like in the case of leptons for point-like events and
 according to a limited transverse momentum model, called peripheral,
 for hadron-like events.
 A mixture of the point-like and peripheral events can be generated
 based on a hit and miss method. This combination is called perimiss.
 No parton showers are included, and the hadronisation is based on the
 Lund string model.
\end{Enumerate}
%
  The general purpose Monte Carlo programs used are:
%
\begin{Enumerate}
 \item \underline{\emph{HERWIG}}
 :~\cite{MAR-8801,KNO-8801,CAT-9101,ABB-9001,SEY-9201,MAR-9201,MAR-9601}
 The general purpose Monte Carlo program HERWIG has been extended
 to electron-photon processes in the LEP2 workshop,
 Ref.~\cite{LON-9601}.
 The first available version was HERWIG5.8d. The next version used
 in experimental analyses is HERWIG5.9, which got improved as detailed
 in Section~\ref{sec:qcdreshad}.
 The improvements are called HERWIG5.9+\kt and HERWIG5.9+\kt(dyn),
 reflecting the changes applied to the intrinsic transverse momentum
 of the quarks in the photon, \kt, either with fixed or dynamically
 (dyn) adjusted upper limit.
 In the HERWIG model the photon flux is based on the EPA,
 Eq.~(\ref{eqn:epa}), and the hard interaction is simulated as
 $\rm e q \rightarrow \rm e q$ scattering, where the incoming quark
 is generated according to a set of parton distribution functions.
 The incoming quark is subject to an initial state parton shower
 which contains the $\gamma\rightarrow\qqbar$ vertex.
 The initial state parton shower is designed in such a way that the hardest
 emission is matched to the sum of the matrix elements for the higher order
 resolved processes, $g\rightarrow\qqbar$ and $q \rightarrow qg$ and
 the point-like $\gamma\rightarrow\qqbar$ process.
 The parton shower uses the
 transverse momentum as evolution parameter and obeys angular ordering.
 This procedure dynamically separates the events into point-like and
 hadron-like events, and this separation will be different from the
 choice made in the parton distribution functions.
 For hadron-like events the photon remnant gets a transverse momentum
 \kt with respect to the direction of the incoming photon discussed
 above, where originally the transverse momentum was generated from a
 gaussian distribution.
 The outgoing partons undergo final state parton showers as in the
 case of \epem annihilations.
 The hadronisation is based on the cluster model.
%
%
 \item \underline{\emph{PHOJET}}:~\cite{ENG-9501,ENG-9601}
 The general purpose Monte Carlo program PHOJET is based on the
 dual parton model from Ref.~\cite{CAP-9401}.
 It was designed for photon-photon collisions, where originally only
 real or quasi-real photons were considered.
 It has recently been extended to match the deep inelastic electron-photon
 scattering case, if one of the photons is highly virtual.
 It can also be used for the scattering of two highly virtual photons.
 Both photons are allowed to fluctuate into a hadronic state before they
 interact.
 For the case of deep inelastic scattering the program is not based on
 the DIS formula, but rather the \gsg cross-section is calculated
 from the $\gamma\gamma$ cross-section by extrapolating in \qsq on the
 basis of the Generalised Vector Dominance model using
 Ref.~\cite{GIN-8201}.
 The events are generated from soft and hard partonic processes,
 where a cut-off on the transverse momentum of the scattered partons
 in the photon-photon centre-of-mass system is used to separate the
 two classes of events.
 The present value of this cut-off is $2.5$~\gev, which means
 that for $W<5$~\gev only soft processes are generated.
 This results in a strange behaviour of the $W$ distribution for
 $W<5$~\gev, which has to be treated with special care.
 The sum of the processes is matched to the deep inelastic scattering
 cross-section, or in other words to \ft. However, in the present version
 this matching is imperfect, which results in the fact that the actual
 distribution in $x$ generated is not the same as one would expect from
 the input photon structure function \ft used in the simulation.
 This makes it difficult to use PHOJET for a direct unfolding procedure,
 but rather it should only be used to determine the transformation
 matrix relating the generated value of $x$ to the observed value \xvis.
 Initial state parton showers are simulated with a backward evolution
 algorithm using the transverse momentum as evolution scale.
 Final state parton showers are generated with the Lund scheme.
 Both satisfy angular ordering implied by coherence effects.
 The hadronisation is based on the Lund string model.
%
%
 \item \underline{\emph{PYTHIA}}:~\cite{SJO-9401}
 In the general purpose Monte Carlo program PYTHIA the process
 is implemented rather similar than in the HERWIG program.
 PYTHIA\footnote{A new version of the PYTHIA program exists,
 PYTHIA6.0. The description given here is still based on the
 capabilities of the version PYTHIA5.7, because this is
 the version which was used in the experimental analyses.
 }
 includes the reaction as deep inelastic electron-quark
 scattering, where the quarks are generated according to
 parton distribution functions of the quasi-real photon.
 The flux of the quasi-real photon has to be externally provided,
 and the corresponding electron is only modelled in the collinear
 approximation.
 The program relies on the leading order matrix element
 for the $\rm e q \rightarrow \rm e q$ scattering process.
 Higher order QCD processes are subsequently generated via parton
 showers, without a matching prescription to the exact matrix elements.
 Initial and final state parton showers are implemented,
 using the parton virtuality as the evolution parameter.
 The separation into point-like and hadron-like
 events is taken from the parton distribution functions, if available.
 In this way, a consistent subdivision into point-like and hadron-like
 events can be achieved in the event generation and the
 parton distribution functions, by using the SaS parton distribution
 functions together with PYTHIA.
 For hadron-like events the photon remnant gets a \kt
 generated from a gaussian distribution, and for
 point-like events the  transverse momentum follows a powerlike
 behaviour, $\der\ktsq/\ktsq$, with $\ktsqm=\qsq$ as the upper limit.
 The hadronisation is based on the Lund string model.
\end{Enumerate}
%
 Due to the different choices made in the various
 steps of the event generation the predictions of the
 Monte Carlo programs differ significantly.
 The quality of the description of the data by the various programs
 is an active field of research. The results of the investigations
 are discussed in Section~\ref{sec:qcdreshad}.
%
%
\subsection{Unfolding methods}
\label{sec:unfol}
 The determination of the structure function \ftxq involves the
 measurement of $x$ and \qsq.
 For the hadronic final state the resolution in \wsq, and therefore
 the resolution in $x$, is not very good, due to mismeasurements of the
 hadrons and losses of particles outside of the acceptance of the detectors.
 Therefore unfolding programs are used to relate
 the observed hadronic final state to the underlying value of $x$.
 The unfolding problem as well as the programs used for the unfolding
 are described below.
 \par
 The principle problem which is solved by the unfolding is the following.
 The distribution \gdet
 of a quantity $u$ (e.g. \xvis) directly measured by the detector is
 related to the distribution \fpar of a partonic variable $\omega$
 (e.g. $x$) by an integral equation which expresses the convolution
 of the true distribution with all effects that occur
 between the creation of the hard process and the measurement
%
 \begin{equation}
 \gdet(u) = \int A(u,\omega)\,\fpar(\omega)\,\der\omega + B(u)\,\, ,
 \label{eqn:unf}
 \end{equation}
%
 where $B(u)$ represents an additional contribution from
 background events.
 The task of the unfolding procedure is to obtain the underlying
 distribution \fpar, from the measured distribution \gdet using the
 transformation $A$ and the background contribution, usually
 obtained from Monte Carlo simulations.
 This is done either by discretising and inverting the equation, or
 by using Bayes' theorem.
 The relevant programs used, are based on different statistical
 methods and have slightly different capabilities.
 They are discussed below.
%
\begin{Enumerate}
 \item \underline{\emph{RUN}}:~\cite{BLO-8401,BLO-9601,BLO-9901}
 The RUN program by Blobel is used since long in structure function analyses.
 It is based on a regularised unfolding technique and allows for
 an unfolding in one dimension.
 The integral equation is transformed into a matrix equation,
 and solved numerically, leading to the histogram \fpar($\omega$).
 This simple method can produce spurious oscillating components in the
 result due to limited detector resolution and statistical fluctuations.
 Therefore the method is improved by a regularisation procedure which
 reduces these oscillations.
 The regularisation is implemented in the program using the assumption
 that the resulting underlying distribution has minimum curvature.
 Technically, the unfolding program RUN works as follows.
 A set of Monte Carlo events is used as an input to the unfolding program.
 These events are based on an input \ftxq and
 implicitly carry the information about the response function $A(\xvis,x)$.
 A continuous weight function \fmult is defined which depends
 only on $x$. This function is used to calculate an individual
 weight factor for each Monte Carlo event.
 The weight function is obtained by a fit of the \xvis distribution
 of the Monte Carlo sample to the measured \xvis distribution of the data,
 such that the reweighted Monte Carlo events
 describe as well as possible the \xvis distribution of the data.
 After the unfolding both distributions agree with each other
 on a statistical basis.
 The unfolded \ftxq from the data is then obtained by multiplying the input
 \ftxq of the Monte Carlo with the weight function \fmult.
%
 \item \underline{\emph{GURU}}:~\cite{KAR-9601}
 The GURU program by H\"ocker and Kartvelishvili is a more recent,
 slightly different, implementation of an regularised unfolding
 technique based on the method of Single Vector Decomposition, SVD.
 In this method the matrix $A$ is decomposed into the product
 $A=USV^T$, where $U$ and $V$ are orthogonal matrices and $S$
 is a diagonal matrix with non-negative diagonal elements,
 the so-called singular values.
 The regularisation procedure of the GURU program is very similar to
 the one used in the RUN program.
 The problem is regularised by adding a regularisation term
 proportional to the regularisation parameter $\tau$.
 In contrast to the automated procedure to determine the value of
 $\tau$ implemented in the RUN program, in the GURU program the value
 of $\tau$ has to be
 adjusted to the problem under study, by determining the number of
 terms of the decomposition which are statistically significant,
 as explained in detail in Ref.~\cite{KAR-9601}.
 However, there is one practical advantage of the GURU program,
 it allows for an unfolding in several dimensions.
 This is a very interesting feature, as two-dimensional unfolding is
 a promising candidate to improve on the error of \ft stemming from
 the dependence of the unfolded result of \ft on the underlying
 Monte Carlo program used to simulate the hadronic final state.
%
 \item \underline{\emph{BAYES}}:~\cite{AGO-9401}
 The BAYES program by D'Agostini is based on Bayes' theorem.
 This method is completely different from the two above, because the
 matrix inversion is avoided by using Bayes' theorem.
 Starting point is the existence of a number of independent causes
 $C_i,i=1,2\ldots n_c$ which can produce one effect, $E$, for
 example, an observed event.
 Then if one knows the initial probability of the cause, $P(C_i)$,
 and the conditional probability, $P(E\vert C_i)$
 of the cause $C_i$ to produce the effect $E$, then Bayes' theorem
 can be formulated as
%
 \begin{equation}
 P(C_i \vert E) = \frac{P(E \vert C_i)P(C_i)}
                  {\sum_{k=1}^{n_c}P(E \vert C_k) P(C_k)}.
 \label{eqn:bayes}
 \end{equation}
%
 This formula can be used for multidimensional unfolding.
 In the one dimensional case the following identifications can be made;
 $P(C_i)=\fpar$, $P(E \vert C_i) = A$ and the distribution of the
 effects $E$ is equivalent to \gdet.
 The best results are obtained if one uses some a priori knowledge
 on $P(C_i)$, then after some iterations the final result is obtained.
 A careful study of the possible bias due to the choice of the initial
 distribution has to be performed.
\end{Enumerate}
%
 In general, when using the one-dimensional unfolding, there is not
 much difference in the results obtained with the various methods.
 Clearly, for all programs the dependence on the transformation
 between the generated variables and the measured ones as given by
 $P(E \vert C_i)$, or $A$ has to be carefully investigated by using the
 predictions of several Monte Carlo models.
 \par
 Traditionally the unfolding was performed only in the variable $x$.
 Motivated by the limited quality of the description of the
 observed hadronic final state by the Monte Carlo programs, discussed
 in Section~\ref{sec:qcdreshad}, there have been investigations
 to study the unfolding in two dimensions to accommodate this shortcoming,
 as discussed, for example, in Ref.~\cite{CAR-9701}.
 The main idea is the following. In the one-dimensional unfolding
 using the variable $x$, the result is independent of
 the actual shape of the input distribution
 function $\fpar(x)$ used in the unfolding and depends only on the
 transformation $A(\xvis,x)$, which partly depends on the Monte Carlo model
 used, but also to a large extend on the detector capabilities which
 are independent of the chosen model\footnote{This is of course only
 true if the same detector parts are populated with particles by the
 different Monte Carlo models.}.
 By using a second variable, $v$, the same argument applies
 to this variable.
 Now the result is independent of the joint input distribution
 function $\fpar(x,v)$ of $x$ and $v$ and only the transformation
 $A(\xvis,\vvis,x,v)$ matters, which now also depends on the
 transformation of $v$.
 Because only the transformation of $v$ but not its actual
 distribution affects the unfolding result, a part of the dependence
 on the Monte Carlo model is removed.
 There are some indications, for example, shown in Ref.~\cite{CAR-9701},
 that the unfolding in two dimensions may reduce the systematical error of
 the structure function results, and this technique has been used in
 recent structure function analyses, as explained in
 Section~\ref{sec:qcdresf2}.
 \par
 Although some improvements of the unfolding procedure has been
 achieved, the main emphasis should be on the understanding of the
 reasons for the shortcomings of the Monte Carlo programs and on
 the improvement of their description of the data.
 This work has been started
 by the ALEPH, L3 and OPAL collaborations and the LEP Working Group for
 Two-Photon Physics, reported in Ref.~\cite{FIN-9901}, but meanwhile,
 better tools to cope with the situation are certainly useful.
%
%

%
%
\section{Measurements of the QED structure of the photon}
\label{sec:qedres}
 The QED structure of the photon has been investigated for
 all leptonic final states \epem, \mupmum and \tauptaum.
 Most results are obtained for \mupmum final states for various
 reasons.
 For \epem final states more Feynman diagrams contribute,
 which makes the analysis in terms of the photon structure more
 difficult.
 The \tauptaum final states are rare, as the $\tau$ is heavy, and
 also \tauptaum final states are more difficult to identify,
 because only the decay products of the $\tau$ can be observed.
 The hadronic decays of the $\tau$ suffer from large background from
 \qqbar production and the muonic decays of the $\tau$ from
 the \mupmum production process.
 The most promising channel is the one where one $\tau$ decays to a muon
 and the other to an electron, but these are very rare.
 In contrast,
 the \mupmum  final states has a clear signature, a large cross section
 and is almost background free, which makes it ideal for the
 measurements of the QED structure of the photon.
 \par
 Several measurements of QED structure functions have been
 performed by various experiments.
 Prior to LEP, mainly the structure function \ftqed was measured.
 The LEP experiments refined the analysis of the \mumu final
 state, to derive more information on the QED structure of the photon.
 The \mumu final state is such a clean environment that it allows for much
 more subtle measurements to be performed, than in the
 case of hadronic final states.
 Examples are, the measurement of the dependence
 of \ftqed on the small, but finite, virtuality of the quasi-real
 photon, \psq, which is often referred to as the target mass effect,
 and the measurement of the structure functions \faqed and \fbqed,
 which are deduced from the distribution of the azimuthal angle \az,
 as outlined in Section~\ref{sec:QED}.
 The interest in the investigation of QED structure functions is
 threefold.
 Firstly the investigations serve as tests of QED to order
 $\mathcal{O}(\aem^4)$, secondly, and also very important,
 the investigations are used to refine the experimentalists tools in a real
 but clean situation to investigate the possibilities of extracting
 similar information from the much more complex hadronic final state, and
 thirdly, the measurement of the QED structure of the photon can give
 some information on the hadronic structure of the photon as well,
 because at large values of $x$ the quark parton model, which is nothing
 but QED, gives a fair approximation of the hadronic structure of the photon.
 \par
 The various results are discussed below, starting with the measurements
 of \ftqed, followed by the measurements of the structure
 functions \faqed and \fbqed of quasi-real photons.
 The final topic discussed is the investigation of the structure of
 highly virtual photons by a measurement of the differential
 cross-section for the exchange of two highly virtual photons, which
 only recently has been performed quantitatively for the first time.
 \par
 The structure function \ftqed has been measured for average
 virtualities in the range $0.45<\qzm<130$~\gevsq.
 The results from the
 CELLO~\cite{CEL-8301},
 DELPHI~\cite{DEL-9601},
 L3~\cite{L3C-9801},
 OPAL~\cite{OPALPR271},
 PLUTO~\cite{PLU-8501}
 and
 TPC/2$\gamma$~\cite{TPC-8401} experiments can be found in
 Tables~\ref{tab:chap12_01}$-$\ref{tab:chap12_07}.
 In addition there exist preliminary results from the
 ALEPH and DELPHI experiment presented in Refs.~\cite{BRE-9701}
 and~\cite{ZIN-9901}.
 The ALEPH results are preliminary since more than two years and therefore
 they are not considered here.
 The DELPHI results are listed in Table~\ref{tab:chap12_08}.
 The result at $\qzm=12.5$~\gevsq\ is going to replace the published
 measurement at $\qsq=12$~\gevsq, which will still be used here.
 \par
%
\begin{figure}[tbp]
\begin{center}
{\includegraphics[width=1.0\linewidth]{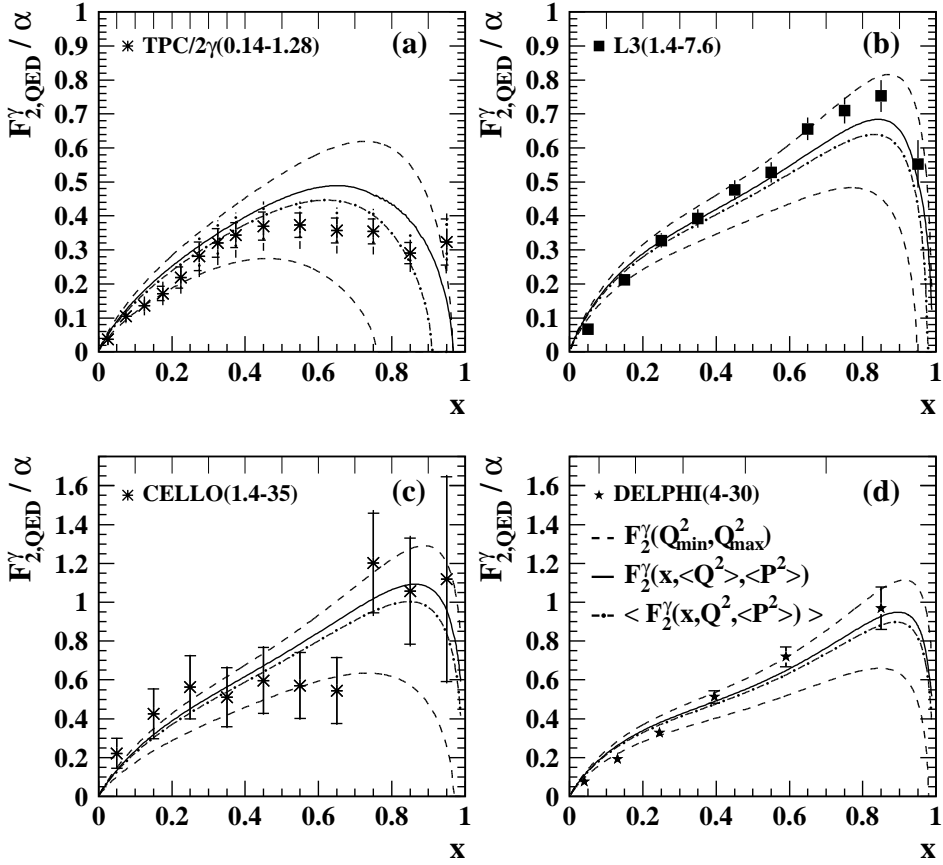}}
\caption[
         The measured average structure function  $\langle\ftqed\rangle$
         from the CELLO, DELPHI, L3 and the TPC/2$\gamma$ experiments,
         compared to QED predictions.
        ]
        {
         The measured average structure function $\langle\ftqed\rangle$
         from the CELLO, DELPHI, L3 and the TPC/2$\gamma$ experiments,
         compared to QED predictions.
         The points represent the data with their statistical
         (inner error bars) and total errors (outer error bars).
         The tic marks at the top of the figures indicate the bin
         boundaries.
         The results of the four experiments are compared to four
         different QED predictions, namely
         the structure function \ftqed at the lower and upper limit of
         the \qsq range studied (dash), and the two quantities
         $\langle\ftqed(x,\qsq,\pzm)\rangle$ (dot-dash)
         and $\ftqed(x,\qzm,\pzm)$ (full) explained in the text.
        }\label{fig:chap6_01}
\end{center}
\end{figure}
%
 Special care has to be taken when comparing the experimental results
 to the QED predictions, because slightly different quantities are
 derived by the experiments.
 Some of the experiments express their result
 as an average structure function, $\langle\ftqed(x,\qsq,\pzm)\rangle$,
 measured within their experimental \qsq acceptance, whereas the other
 experiments unfold their result as a structure function for an average
 \qsq value, $\ftqed(x,\qzm,\pzm)$.
 The second choice is much more appropriate for comparisons to theory,
 because in this case all experimental dependence is removed, whereas in
 the first case the measured average structure function still depends on the
 experimental acceptance, which can only approximately be modelled by
 theory.
 Fortunately, for not too large bins in \qsq, and assuming a constant
 experimental acceptance as a function of \qsq,
 the two quantities $\langle\ftqed(x,\qsq,\pzm)\rangle$ and
 $\ftqed(x,\qzm,\pzm)$ are very similar, as can be seen from
 Figure~\ref{fig:chap6_01}, where the experimental results on the
 average structure function $\langle\ftqed\rangle$ from
 the CELLO, DELPHI, L3 and TPC/2$\gamma$ experiments
 are shown together with several QED predictions.
 The measurements are compared to \ftqed at the lower and upper limit of
 the \qsq range studied, and to the two quantities
 $\langle\ftqed(x,\qsq,\pzm)\rangle$ and $\ftqed(x,\qzm,\pzm)$,
 using the values of \pzm listed below.
 Here the average structure function $\langle\ftqed(x,\qsq,\pzm)\rangle$ is
 calculated as the average of \ftqed within the \qsq range used by the
 experiments, but without
 taking into account the \qsq dependence of the cross section.
 The \qsq range is divided into 100 bins on a linear scale in \qsq
 and for each point in $x$ the average is calculated from all non-zero
 values of \ftqed.
 The difference between $\langle\ftqed(x,\qsq,\pzm)\rangle$ and
 $\ftqed(x,\qzm,\pzm)$ is small compared to the experimental errors of
 the CELLO, DELPHI and TPC/2$\gamma$ measurements. However, for the
 measurement of L3, the size of the difference is comparable to the
 experimental uncertainty, especially at large values of $x$.
 \par
%
\begin{figure}[tbp]
\begin{center}
{\includegraphics[width=0.96\linewidth]{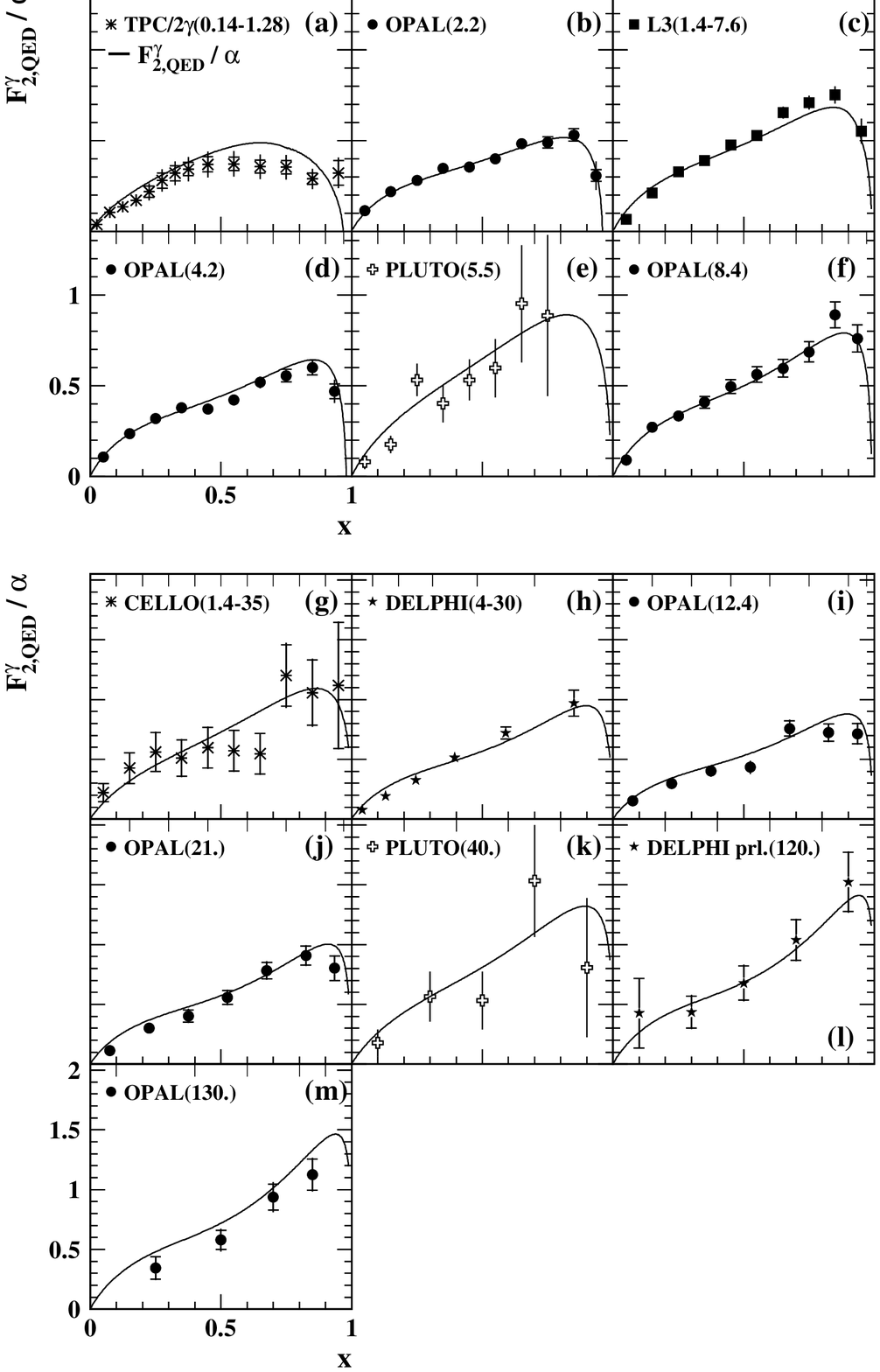}}
\caption[
         The world summary of \ftqed measurements.
        ]
        {
         The world summary of \ftqed measurements.
         The data are compared to \ftqed(x,\qzm,\pzm), or
         $\langle\ftqed(x,\qsq,\pzm)\rangle$, with numbers as given in
         the text.
         The points represent the data with their statistical
         (inner error bars) and total errors (outer error bars).
         The quoted errors for (h) are statistical only.
         The tic marks at the top of the figures indicate the bin
         boundaries.
        }\label{fig:chap6_02}
 \end{center}
 \end{figure}
%
 Figure~\ref{fig:chap6_02} shows the world summary of the \ftqed
 measurements, where the experimental results are compared either
 to the predicted $\langle\ftqed(x,\qsq,\pzm)\rangle$ or to
 $\ftqed(x,\qzm,\pzm)$.
 For the measurements which quote an average \psq for their dataset,
 where \pzm is either obtained from the Monte Carlo prediction, or from a
 best fit of the QED prediction to the data,
 this value is chosen in the comparison.
 For the comparisons of the other results $\psq=0$ is used.
 The curves shown correspond to
 (a) $\langle\ftqed(x,0.14$-$1.28,0)\rangle$,
 (b) \ftqed(x,2.2,0.05),
 (c) $\langle\ftqed(x,1.4$-$7.6,0.033)\rangle$,
 (d) \ftqed(x,4.2,0.05),
 (e) \ftqed(x,5.5,0),
 (f) \ftqed(x,8.4,0.05),
 (g) $\langle\ftqed(x,1.4$-35$,0)\rangle$,
 (h) $\langle\ftqed(x,4$-$30,0.04)\rangle$,
 (i) \ftqed(x,12.4,0.05),
 (j) \ftqed(x,21,0.05),
 (k) \ftqed(x,40,0),
 (l) \ftqed(x,120,0.066),
 and
 (m) \ftqed(x,130,0.05), where all numbers are given in \gevsq.
 There is agreement between the data and the QED expectations
 to order $\mathcal{O}(\aem^4)$ for three orders of magnitude in
 \qsq.
 Some differences are seen for the  TPC/2$\gamma$ result, but at these
 low values of \qsq\ this could also be due to the simple averaging
 procedure used for the theoretical prediction.
 \par
%
\begin{figure}[tbp]
\begin{center}
{\includegraphics[width=1.05\linewidth]{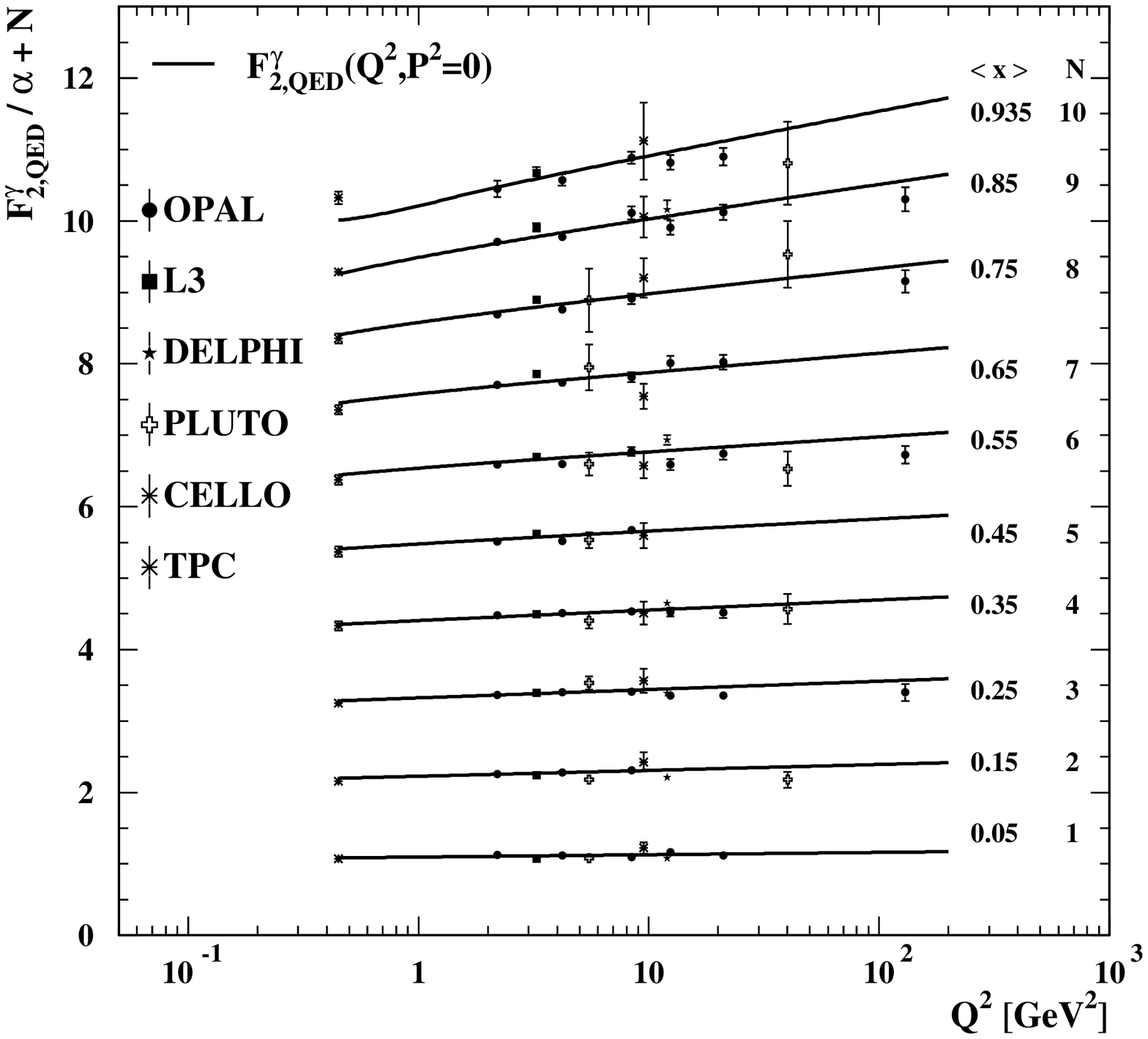}}
\caption[
         The measured \qsq evolution of \ftqed.
        ]
        {
         The measured \qsq evolution of \ftqed.
         The measurements of \ftqed as a function of \qsq for various
         $x$ ranges compared to QED.
         The points represent the data with their total errors.
         The data from Figure\protect~\ref{fig:chap6_02} are shown
         after correcting for the effect of non zero \psq in the data.
         The curves correspond to the QED prediction for $\psq=0$.
         The upper two PLUTO points at $\qsq=40$~\gevsq belong
         to $N=8$ and $N=10$.
        }\label{fig:chap6_03}
 \end{center}
 \end{figure}
%
%
\begin{figure}[tbp]
\begin{center}
{\includegraphics[width=1.0\linewidth]{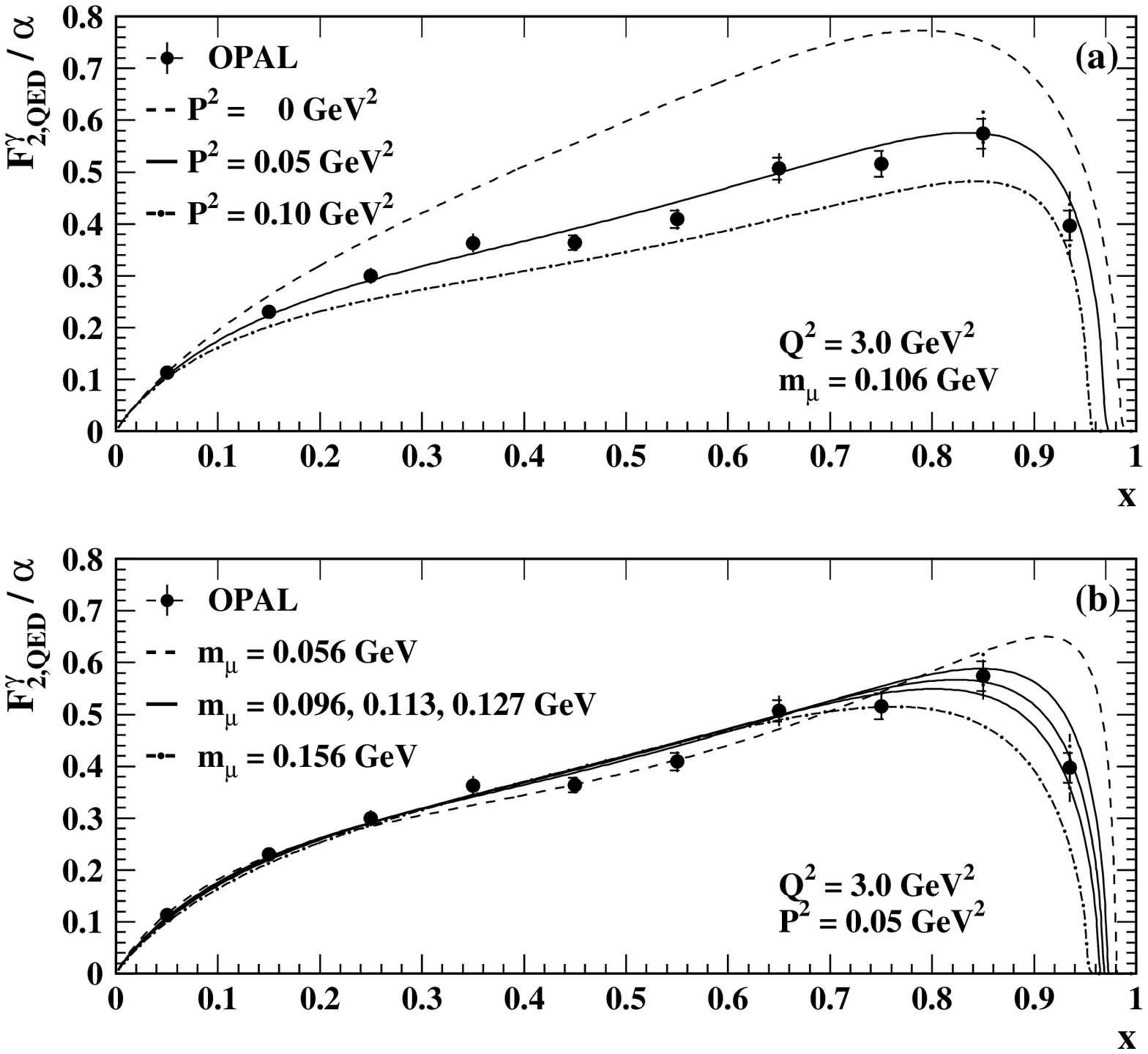}}
\caption[
         The dependence of \ftqed on \psq and on the mass of the muon.
        ]
        {
         The dependence of \ftqed on \psq and on the mass of the muon.
         The OPAL data for $\qzm=3$~\gevsq are compared to several
         QED predictions of \ftqed(x,\qzm,\pzm,$m_\mu$), where
         in (a) \pzm is varied for a fixed mass of the muon
         of $m_\mu=0.106$~\gev and in (b) the mass of the
         muon is varied for fixed $\pzm=0.05$~\gevsq.
         The variations shown in (a) are $\pzm=0$ (dash), 0.05 (full)
         and 0.1~\gevsq (dot-dash),
         and the chosen masses in (b) are $m_\mu=0.056$ (dash),
         $ \Y{0.113}{0.014}{0.017}$(full),
         and 0.156~\gev (dot-dash).
         The points represent the data with their statistical
         (inner error bars) and total errors (outer error bars).
         The tic marks at the top of the figures indicate the bin
         boundaries.
        }\label{fig:chap6_04}
\end{center}
\end{figure}
%
 Another way to compare data and theory is exploited in
 Figure~\ref{fig:chap6_03}, where the same data is displayed as a
 function of \qsq in bins of $x$ with bin sizes of 0.1 if possible,
 and with central values of $x$ as indicated in the figure.
 To separate the measurements from each other an integer value, N,
 counting the bin number is added to the measured \ftqed.
 To be able to compare all results to the same QED
 prediction all data which quote an average \psq for their measurement
 are corrected for this effect by multiplying the quoted result by the
 ratio of \ftqed calculated at $\psq=0$ and at $\psq=\pzm$.
 The measurements which were obtained for different bin sizes in $x$
 than the ones used in the figure are displayed at the closest
 central value.
 All curves represent the predicted average \ftqed in the $x$
 bin under study, for $\psq=0$.
 In general the agreement between the data and the predictions is acceptable
 and the prediction clearly follows the increasing slope for increasing
 $x$ observed in the data.
 \par
 The LEP data are precise enough that the effect of the small virtuality
 \psq of the quasi-real photon can be investigated in detail.
 As an illustration the comparison is made for the most precise
 data coming from the OPAL experiment in Figure~\ref{fig:chap6_04}.
 The data consist of the dataset at $\qzm=3.0$~\gevsq listed in
 Table~\ref{tab:chap12_05}.
 The dependence of \ftqed on \psq can be clearly established, and
 the experimental result shown in Figure~\ref{fig:chap6_04}(a)
 is consistent with the QED expectation for the average value of
 \psq predicted by the QED Monte Carlo program Vermaseren,
 $\pzm=0.05$~\gevsq.
 The dependence of \ftqed on the mass squared of the muon and on \psq is
 similar, as can be seen from Eq.~(\ref{eqn:BH}) and from its
 approximation Eq.~(\ref{eqn:f2approx}).
 Consequently, the data can also be used to measure the mass of the
 muon, by assuming the \pzm value predicted by QED.
 A precision of about 14$\%$ on the mass of the muon can be derived
 from Figure~\ref{fig:chap6_04}(b) using the following procedure.
 A fit to the data using the QED prediction for $\qsq=3.0$~\gevsq
 and $\psq=0.05$~\gevsq yields as a best fit result
 $m_\mu=0.113$~\gev, for a \chiq of 12.2 for nine degrees of freedom.
 The shape of the \chiq distribution is close to a parabola, and
 by varying the mass in each direction until the minimum
 \chiq increases by one unit, the error on $m_\mu$ is determined.
 The final result is $m_\mu=\Y{0.113}{0.014}{0.017}$~\gev.
 Although this is not a very precise measurement of the mass of the
 muon it can
 serve as an indication on the precision possible for the determination
 of \lam, if it only were for the point-like contribution to
 the hadronic structure function \ft.
 \par
 The structure functions \faqed and \fbqed are obtained from the
 measured shape of the distribution of the azimuthal angle \az,
 which can be written as
%
 \begin{equation}
 \frac{\der\,N}{\der\,\az} \sim (1 - A\cos\az + B\cos 2\az).
 \label{eqn:fit}
 \end{equation}
%
 For small values of $y$, the two parameters A and B can be
 identified with \faoft and 1/2\fboft, by comparing to
 Eq.~(\ref{eqn:crossazlim}), which is valid in the limit
 $\rho(y) = \epsilon(y) = 1$.
 The two parameters A and B are fitted to obtain the structure
 function ratios.
 By multiplying the measured structure function ratios with the
 measured \ftqed, the structure functions \faqed and \fbqed are obtained.
 The error of this measurement is completely dominated by the
 error on the fitted values of A and B, and the main contribution
 to this error is of statistical nature.
 The structure functions \faqed and \fbqed were measured by L3
 in Ref.~\cite{L3C-9801} and by OPAL in Ref.~\cite{OPALPR271}, and
 they are listed in Table~\ref{tab:chap12_09} and~\ref{tab:chap12_10}
 respectively.
 In addition preliminary results on \faoft and 1/2\fboft from ALEPH
 and DELPHI are available in Ref.~\cite{BRE-9701} and
 Ref.~\cite{ZIN-9901}.
 For the same reasons as mentioned above for \ft the ALEPH results
 are not considered here, the DELPHI results are listed in
 Table~\ref{tab:chap12_11}.
 \par
%
\begin{figure}[tbp]
\begin{center}
{\includegraphics[width=1.0\linewidth]{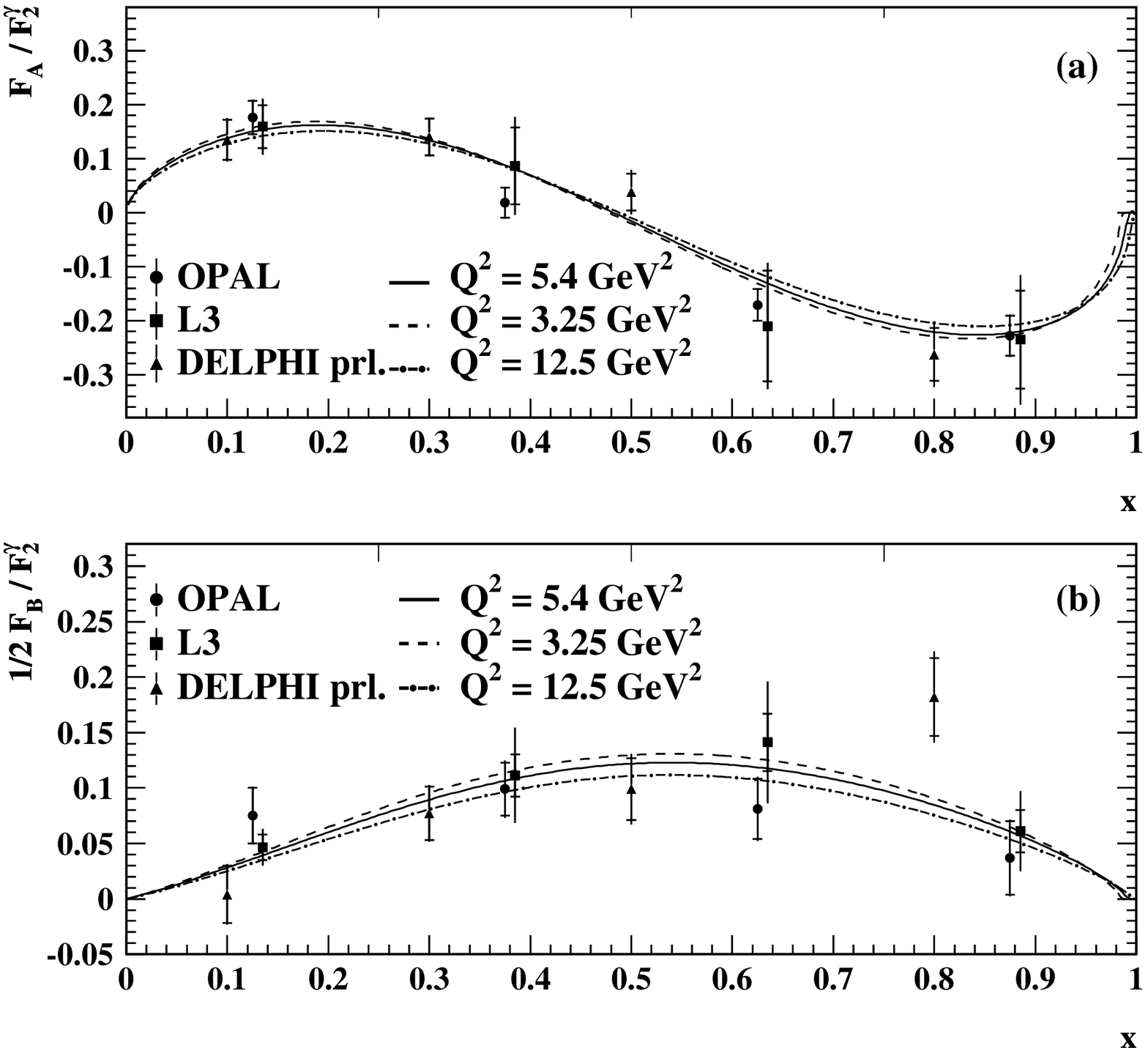}}
\caption[
         The measurements of \faoft and 1/2\fboft.
        ]
        {
         The measurements of \faoft and 1/2\fboft.
         In (a) the OPAL, the L3 and the preliminary DELPHI results
         are compared to the theoretical prediction of \faoft(x,\qzm) and
         in (b) to 1/2\fboft(x,\qzm), always
         using the structure functions given in Eq.~(\ref{eqn:mike1}).
         The points represent the data with their statistical
         (inner error bars) and total errors (outer error bars).
         The tic marks at the top of the figures indicate the bin
         boundaries of the OPAL and L3 analyses.
         The different curves correspond to the different values
         of \qzm, 3.25 (dash), 5.4 (full) and 12.5~\gevsq (dot-dash).
        }\label{fig:chap6_05}
\end{center}
\end{figure}
%
%
\begin{figure}[tbp]
\begin{center}
{\includegraphics[width=1.0\linewidth]{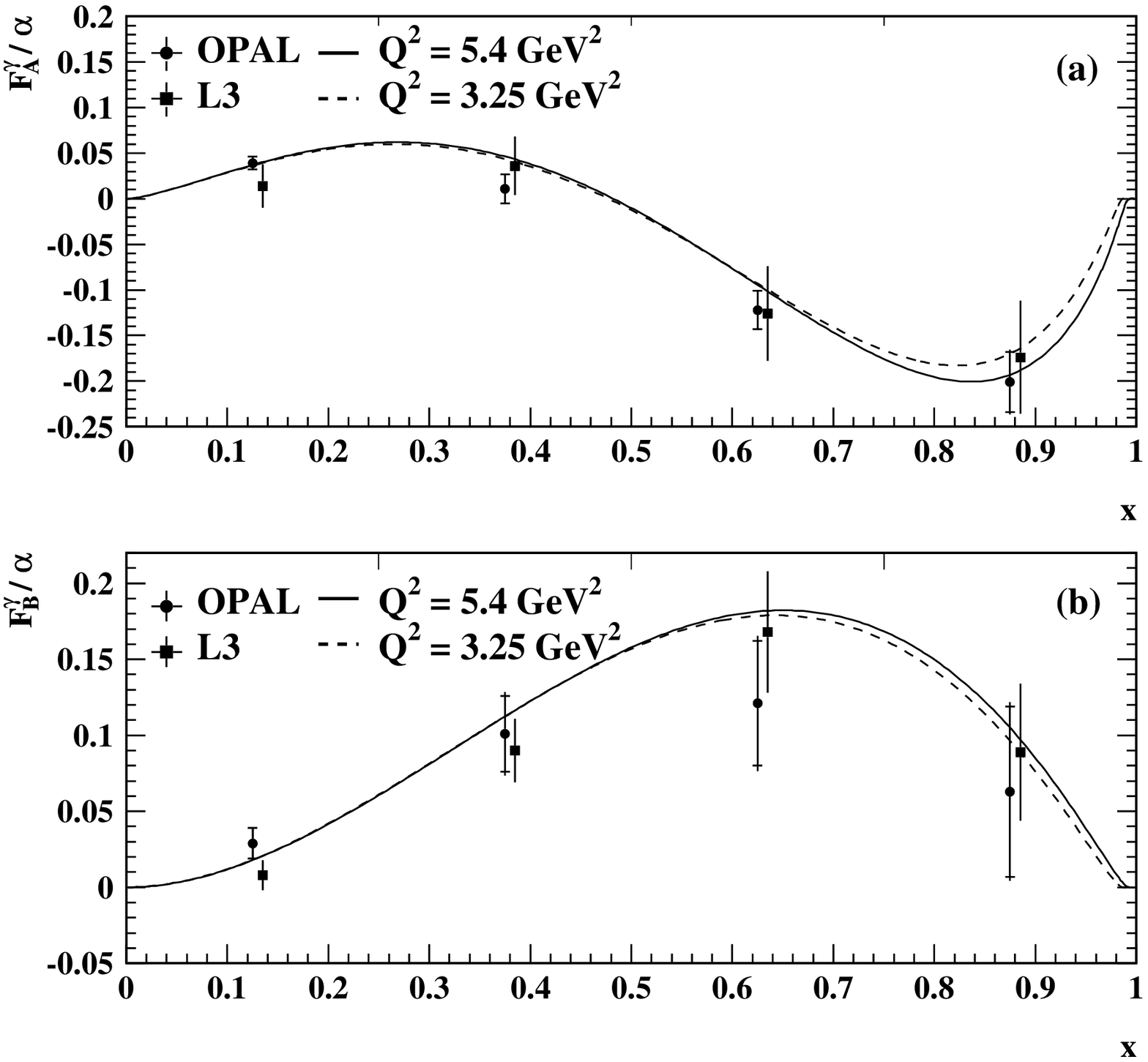}}
\caption[
         The measurements of \faqed and \fbqed.
        ]
        {
         The measurements of \faqed and \fbqed.
         In (a) the OPAL and the L3 results are compared
         to the theoretical prediction of \faqed(x,\qzm) and
         in (b) to \fbqed(x,\qzm), always
         using the structure functions given in Eq.~(\ref{eqn:mike1}).
         The points represent the data with their statistical
         (inner error bars) and total errors (outer error bars).
         The tic marks at the top of the figures indicate the bin
         boundaries of the analyses.
         The different curves correspond to the different values
         of \qzm, 3.25 (dash) and 5.4 (full).
        }\label{fig:chap6_06}
\end{center}
\end{figure}
%
 The measurements of \faoft and 1/2\fboft are compared in
 Figure~\ref{fig:chap6_05}. They all agree with each other and with
 the QED prediction from Ref.~\cite{SEY-9801}.
 The measurements of \faqed and \fbqed from the L3 and OPAL
 experiments are compared in Figure~\ref{fig:chap6_06}.
 The measurements from L3 and OPAL are performed in slightly different ways.
 The strength of the \az dependence varies with the scattering
 angle $\cts$ of the muons in the photon-photon centre-of-mass system.
%
\begin{figure}[tbp]
\begin{center}
{\includegraphics[width=1.05\linewidth]{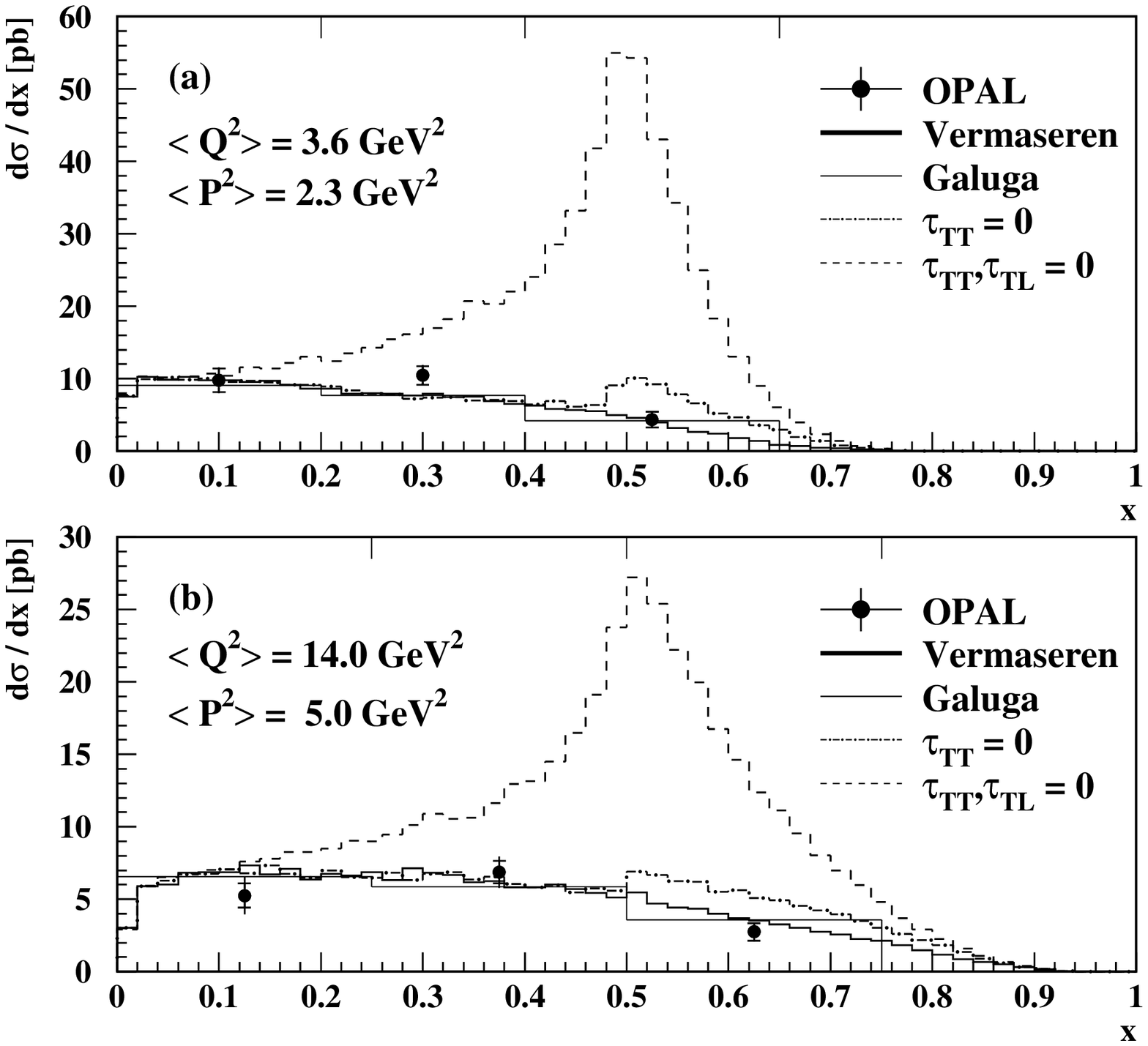}}
\caption[
         The measurement of the differential QED cross section \dsigdx
         for highly virtual photons.
        ]
        {
         The measurement of the differential QED cross section \dsigdx
         for highly virtual photons.
         The differential cross-sections \dsigdx, for the reaction
         $\ee\rightarrow\ee\gamma^\star\gamma^{\star}\rightarrow \ee \mumu$,
         unfolded from the data,
         (a) for $1.5< \qsq < 6$~\gevsq and $1.5< \psq < 6$~\gevsq
         and
         (b) for $5< \qsq < 30$~\gevsq and $1.5< \psq < 20$~\gevsq.
         The points represent the data with their statistical
         (inner error bars) and total errors (outer error bars).
         The tic marks at the top of the figures indicate the bin
         boundaries.
         The data are compared to various QED predictions explained
         in the text.
        }\label{fig:chap6_07}
\end{center}
\end{figure}
%
 Reducing the acceptance of \cts enhances the \az
 dependence but, to obtain a result for \faqed and \fbqed which is
 valid for the full range of \cts, the measurement has to be extrapolated
 using the predictions of QED.
 The measurements from L3 are obtained in the range
 $\vert\cts\vert<0.7$, and extrapolated to the full range in \cts, whereas
 the measurements presented by OPAL are valid for the full angular
 range $\vert\cts\vert<1$.
 There are two other differences in the analyses.
 The OPAL measurements uses the predictions including the mass corrections,
 Eq.~(\ref{eqn:mike1}), whereas the result from the L3 experiment is
 obtained based on the leading logarithmic approximation, Eq.~(\ref{eqn:QPM}).
 As the predictions for \faqed and \fbqed are only valid for $\psq=0$, the
 OPAL measurement of \ftqed is corrected for the effect of non-zero \psq in
 the data by multiplying the result of the unfolding for $\pzm=0.05$~\gevsq
 by the ratio of \ftqed for $\psq=0$ and \ftqed for $\pzm=0.05$~\gevsq,
 both as predicted by QED.
 In the case of L3 the measured ratios \faoft and 1/2\fboft are
 multiplied with the measured, and therefore \psq-dependent, \ftqed,
 without correcting for the effect of non-zero \psq in the data.
 Despite the differences in the analyses strategies, the measurements are
 consistent with each other, and they are nicely described by the QED
 prediction.
 \par
 With these measurements it can be experimentally established that
 both \faqed and \fbqed are different from zero.
 The shape of \fbqed cannot be accurately determined, however it is
 significantly different from a constant.
 The best fit to a constant value leads to $\fbqed/\aem=0.032$
 and 0.042 with \chidof of 8.9 and 3.1 for the L3 and OPAL
 results respectively.
 Because the precision of the measurements is limited mainly by the
 statistical error, and the luminosities used for the results by the
 experiments amount to about 100~\invpb, taken at LEP1 energies,
 a significant improvement is expected from exploiting the full
 expected statistics of 500~\invpb of the LEP2 programme.
 Several investigations to also measure \fa and \fb for
 hadronic final states are underway by the LEP experiments,
 but no results are available yet.
 This concludes the discussion on the QED structure of the quasi-real
 photon and the remaining part of this section deals with the structure
 of highly virtual photons.
 \par
 Following the discussion of Section~\ref{sec:cross} the experimentally
 extracted quantity is the differential QED cross section \dsigdx for
 highly virtual photons, given in Eq.~(\ref{eqn:true}).
 The measurement from OPAL is listed in Table~\ref{tab:chap12_12} and
 shown in Figure~\ref{fig:chap6_07} for two ranges in photon
 virtualities.
 In Figure~\ref{fig:chap6_07}(a) the ranges
 $1.5< \qsq < 6$~\gevsq and $1.5< \psq < 6$~\gevsq are used,
 and
 Figure~\ref{fig:chap6_07}(b) is for
 $5< \qsq < 30$~\gevsq and $1.5< \psq < 20$~\gevsq.
 The data are compared to various QED predictions.
 The full line denotes the differential cross-sections as predicted
 by the Vermaseren Monte Carlo using the same bins as for the data.
 The additional three histograms represent the cross-section
 calculations from the GALUGA Monte Carlo for three different
 scenarios: the full cross-section (full), the
 cross-section obtained for vanishing \ttt (dot-dash) and the
 cross-section obtained for vanishing \ttt and \ttl (dash),
 all as defined in Eq.~(\ref{eqn:truedb}).
 There is good agreement between the data and the QED predictions from
 the Vermaseren and the GALUGA Monte Carlo programs, provided all terms
 of the differential cross-section, Eq.~(\ref{eqn:truedb}), are used.
 However, if either \ttt or both \ttt and \ttl are neglected in the
 QED prediction as implemented in the GALUGA Monte Carlo, there is a
 clear disagreement between the data and the QED prediction.
 This measurement clearly establishes the contributions of
 the interference terms \ttl and \ttt, described in Section~\ref{sec:cross},
 to the cross-section.
 Both terms, \ttt and \ttl, are present in the data, mainly at $x>0.1$, and
 the corresponding contributions to the cross-section are negative.
 Especially, the contribution from \ttl is very large
 in the specific kinematical region of the OPAL analysis.
 \par
 Since the kinematically accessible range in terms of \qsq and \psq
 for the measurement of the leptonic and the hadronic structure
 of the photon is the same, and given the size of the interference
 terms in the leptonic case, special care has to be taken when the
 measurements on the  hadronic structure are interpreted in terms
 of structure functions of virtual photons.
%
%

%
%
\section{Measurements of the hadronic structure of the photon}
\label{sec:qcdres}
 One of the most powerful methods to investigate the hadronic structure
 of quasi-real photons is the measurement of photon structure functions
 in deep inelastic electron-photon scattering at \epem colliders.
 These measurements have by now a tradition of almost twenty years,
 since the first \ft was obtained in 1981 by the
 PLUTO experiment in Ref.~\cite{PLU-8102}.
 The main idea of this measurement is given by Eq.~(\ref{eqn:approx}), which
 means that by measuring the differential cross-section, and accounting
 for the kinematical factors, the photon structure function \ft is obtained.
 The photon structure function \ft in leading order is proportional
 to the quark content of the photon, Eq.~(\ref{eqn:F2def}), and
 therefore the measurement of \ft reveals the structure of the photon.
 The discussion of the experimental results is divided into
 three parts.
 The description of the experimentally observed distributions
 of the hadronic final state by the Monte Carlo models is reviewed
 first, followed by the discussion of the measurements of the
 hadronic structure function \ft, and the description of the
 measurements concerning the hadronic structure for the exchange of
 two virtual photons.
%
%
\subsection{Description of the hadronic final state}
\label{sec:qcdreshad}
 As has been explained in Section~\ref{sec:tools}
 the adequate description of the hadronic final state by the Monte Carlo
 models is very important for measurements of the photon structure.
 With the advent of the LEP2 workshop general purpose Monte Carlo
 programs, for the first time also containing the deep inelastic
 electron-photon scattering reaction, became available.
 The first serious attempt to confront these models with the
 experimental data has been performed by the OPAL experiment in
 Ref.~\cite{OPALPR185}.
 None of the Monte Carlo programs available at
 that time was able to satisfactorily reproduce the data distributions.
 Therefore, the full spread of the predictions was included in the
 systematic error of the \ft measurement, which consequently suffered
 from large systematic errors.
 This observation initiated an intensive work on the understanding
 of the shortcomings of the Monte Carlo models.
 The results of these studies and the attempts to improve on the Monte
 Carlo models are summarised in this section.
 \par
%
\begin{figure}[tbp]
\begin{center}
{\includegraphics[width=1.0\linewidth]{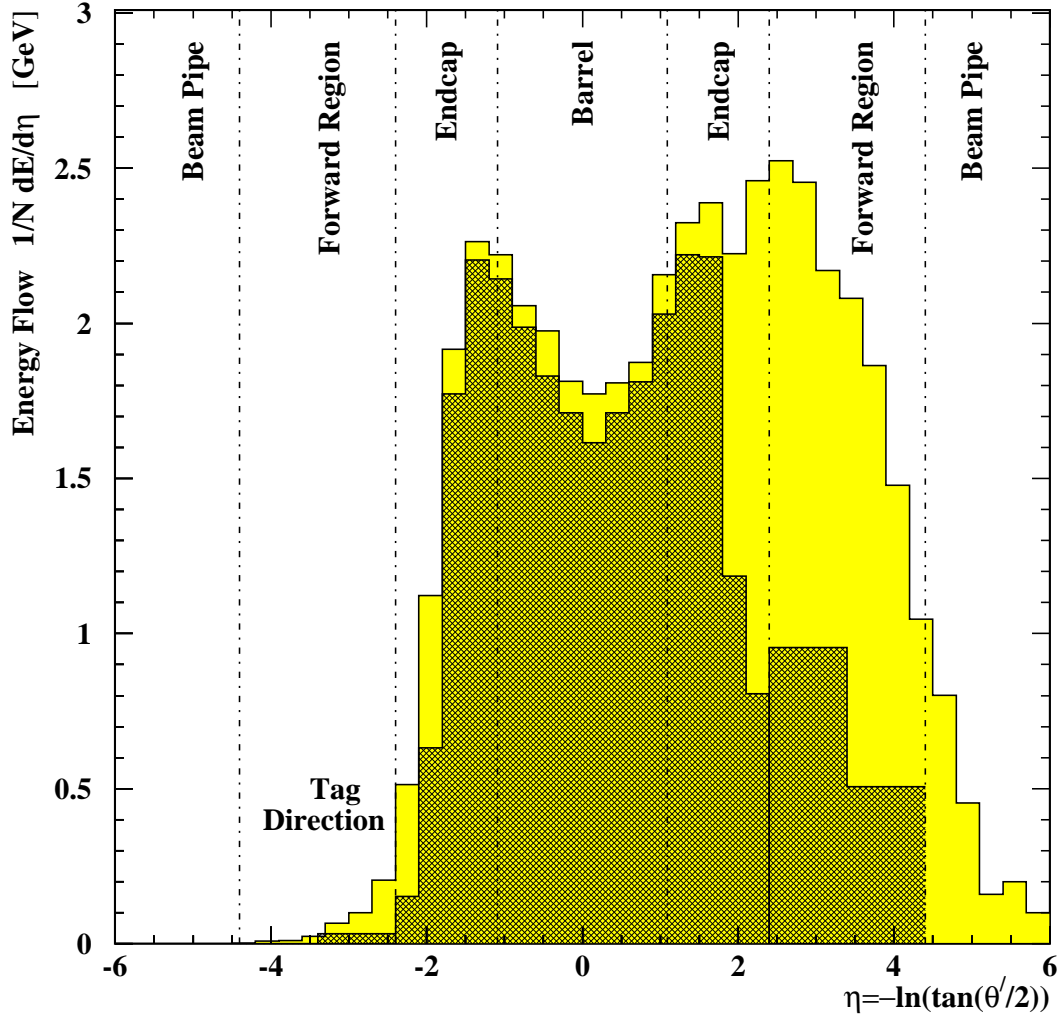}}
\caption[
         The general behaviour of the hadronic energy flow.
        ]
        {
         The general behaviour of the hadronic energy flow.
         The hadronic energy flow per event based on the HERWIG5.8d generator
         is shown
         as a function of the pseudorapidity $\eta$  for $\qzm = 13$~\gevsq.
         The observed electron is always at negative rapidities,
         $-3.5<\eta<-2.8$, and is not shown.
         The dark shaded histogram represents the energy reconstructed by the
         OPAL detector after the simulation of the detector response to the
         HERWIG5.8d events. The generated energy distribution for these
         events is represented by the lightly shaded histogram.
         The vertical lines show the acceptance regions of the
         OPAL detector components.
        }\label{fig:chap7_01}
\end{center}
\end{figure}
%
 The flow of hadronic energy as a function of the pseudorapidity,
 for an average event, \flow, is shown in
 Figure~\ref{fig:chap7_01}, taken from Ref.~\cite{OPALPR185}.
 The generated hadronic energy flow as predicted by the HERWIG5.8d
 Monte Carlo is compared to the visible flow of the hadronic energy
 as observed after simulating the response of the OPAL detector.
 The hadronic energy flow sums over all charged and neutral particles.
 The pseudorapidity is defined as $\eta=-\ln(\tan(\theta'/2))$,
 where $\theta'$ is the polar angle of the particle measured from the
 direction of the beam that has produced the quasi-real photon,
 so the observed electron is at $-3.5<\eta<-2.8$, but is not shown.
 This figure demonstrates that a significant fraction of the
 energy flow in events from the HERWIG5.8d generator goes into the
 forward region of the detector.
 About two thirds of the energy is deposited in the central
 region of the detector, and 30$\%$ goes into the forward region.
 As little as 5$\%$ of the total hadronic energy is lost
 in the beampipe.
 \par
%
\begin{figure}[tbp]
\begin{center}
{\includegraphics[width=0.8\linewidth]{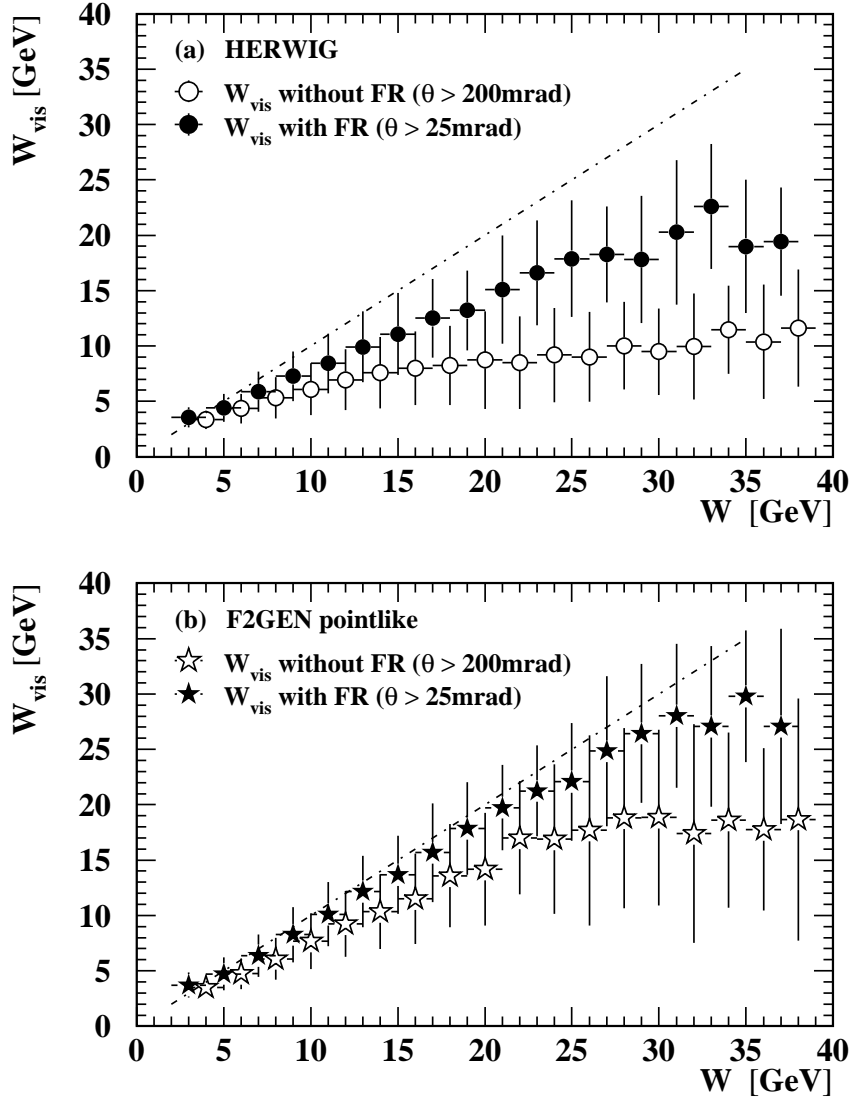}}
\caption[
         The correlation between the measured and generated hadronic
         invariant mass.
        ]
        {
         The correlation between the measured and generated hadronic
         invariant mass.
         The correlation between the generated hadronic invariant
         mass $W$ and the visible mass \Wvis with and without the hadronic
         energy sampled in the forward region (FR) of the OPAL detector.
         In (a) the correlation is shown for HERWIG5.8d and (b) for F2GEN;
         in each case for two cuts on the minimum polar angle of the
         acceptance region.
         The symbols represent the average \Wvis in each bin,
         and the vertical error bar its standard deviation.
         The dashed line corresponds to a perfect correlation $\Wvis = W$.
        }\label{fig:chap7_02}
\end{center}
\end{figure}
%
 The reconstructed energy flow is rather similar to the generated energy
 flow in the central region, but significantly different in the forward
 region.
 The small inefficiency in the central detector
 region is mostly due to the fact that some hadrons in this region
 carry low energy, and therefore fail quality cuts applied to the
 events in the experimental analyses.
 The forward regions of the LEP detectors are only equipped with
 electromagnetic calorimeters and the hadronic energy in the forward
 region can only be sampled by using these electromagnetic calorimeters.
 Consequently, for example in the case of the OPAL detector, only
 about 42$\%$ of the total hadronic energy in the forward region can be
 recovered, with an energy resolution of $\Delta E/E=30\%$ of the
 seen energy, as explained in Ref.~\cite{OPALPR185}.
 Given this, the detectors are precise enough to disentangle various
 predictions in the central part of the detector.
 However, they are not able to distinguish well between models which
 produce different energy flow distributions in the forward region.
 \par
 The different Monte Carlo models produce rather different hadronic
 energy flows also within the clear acceptance of the detectors, which
 leads to the fact that for a given value of $W$, the visible invariant
 mass \Wvis is rather different when using different Monte Carlo models.
 The correlation between the generated and visible invariant masses is
 shown in Figure~\ref{fig:chap7_02}, taken from Ref.~\cite{OPALPR185},
 for two Monte Carlo models described in Section~\ref{sec:gener}.
 The level of correlation achieved between \Wvis and $W$, strongly
 depends on the acceptance region for the hadronic final state and
 also on the model chosen.
 A Monte Carlo model like F2GEN predicts a much stronger correlation than,
 for example, the HERWIG5.8d Monte Carlo model. This strongly effects the
 acceptance of the events and therefore the \xvis distributions,
 especially at low values of \xvis and correspondingly low values of $x$.
 \par
%
\begin{figure}[tbp]
\begin{center}
{\includegraphics[width=1.0\linewidth]{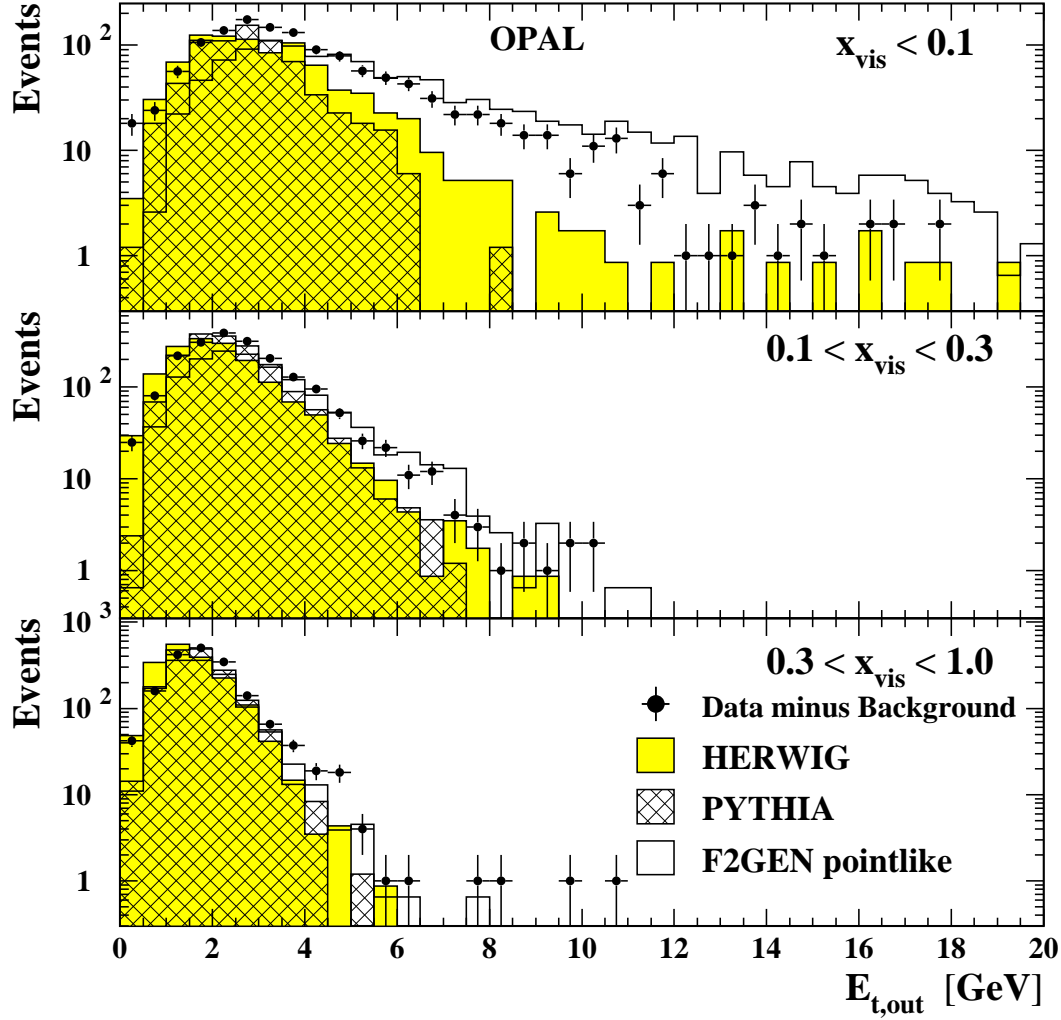}}
\caption[
         The measured \etout distribution for $\qzm = 13$~\gevsq in bins
         of $x$.
        ]
        {
         The measured \etout distribution for $\qzm = 13$~\gevsq in bins
         of $x$.
         The energy transverse to the plane defined by the beam axis
         and the momentum vector of the observed electron,
         \etout, is shown at the detector level
         for three ranges in \xvis.
        }\label{fig:chap7_03}
\end{center}
\end{figure}
\begin{figure}[tbp]
\begin{center}
{\includegraphics[width=0.96\linewidth]{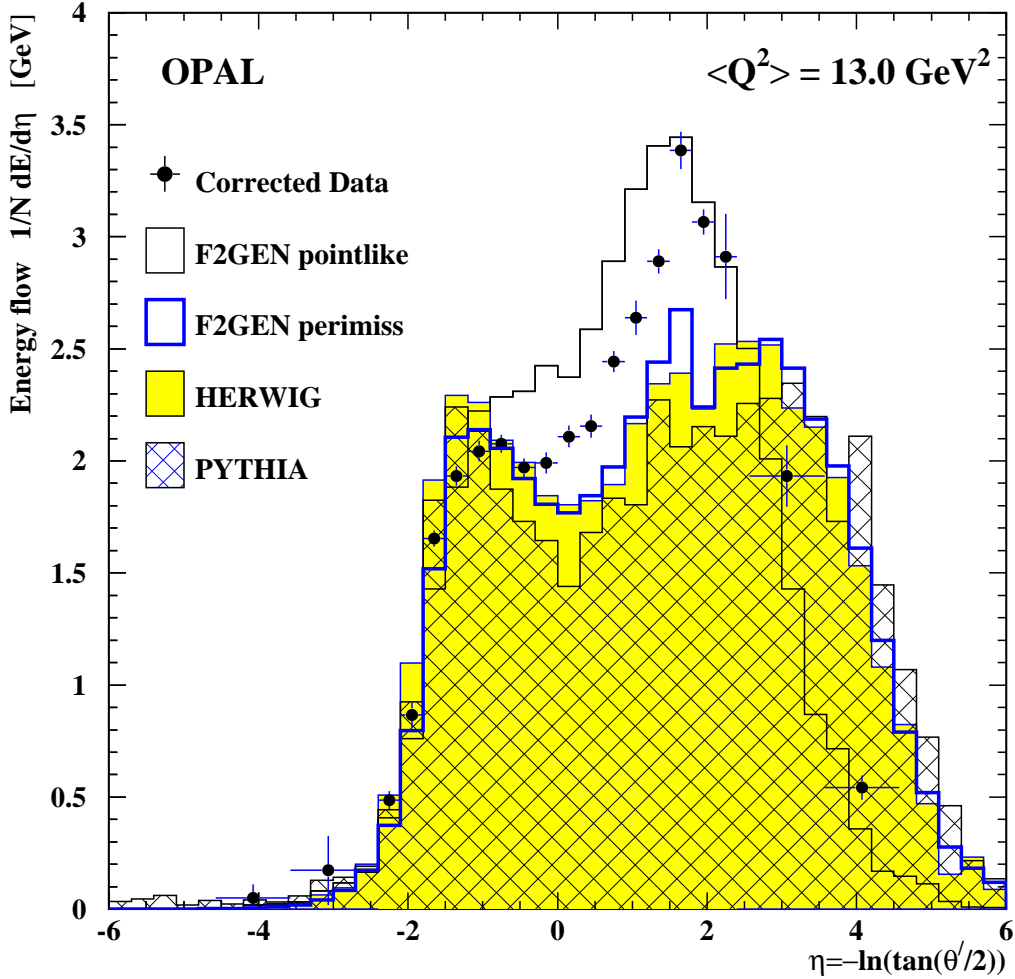}}
\caption[
         The corrected hadronic energy flow for $\qzm = 13$~\gevsq.
        ]
        {
         The corrected hadronic energy flow for $\qzm = 13$~\gevsq.
         The measured energy flow per event is
         corrected for the detector inefficiencies,
         as a function of pseudorapidity $\eta$, and compared to the
         generated energy flow of the HERWIG5.8d and PYTHIA Monte Carlo models
         and the energy flow of samples of point-like and perimiss
         events from the F2GEN model.
         The vertical error bars on the data points are the sum of the
         statistical and systematic errors, and the horizontal bars
         indicate the bin widths.
        }\label{fig:chap7_04}
\end{center}
\end{figure}
\begin{figure}[tbp]
\begin{center}
{\includegraphics[width=0.96\linewidth]{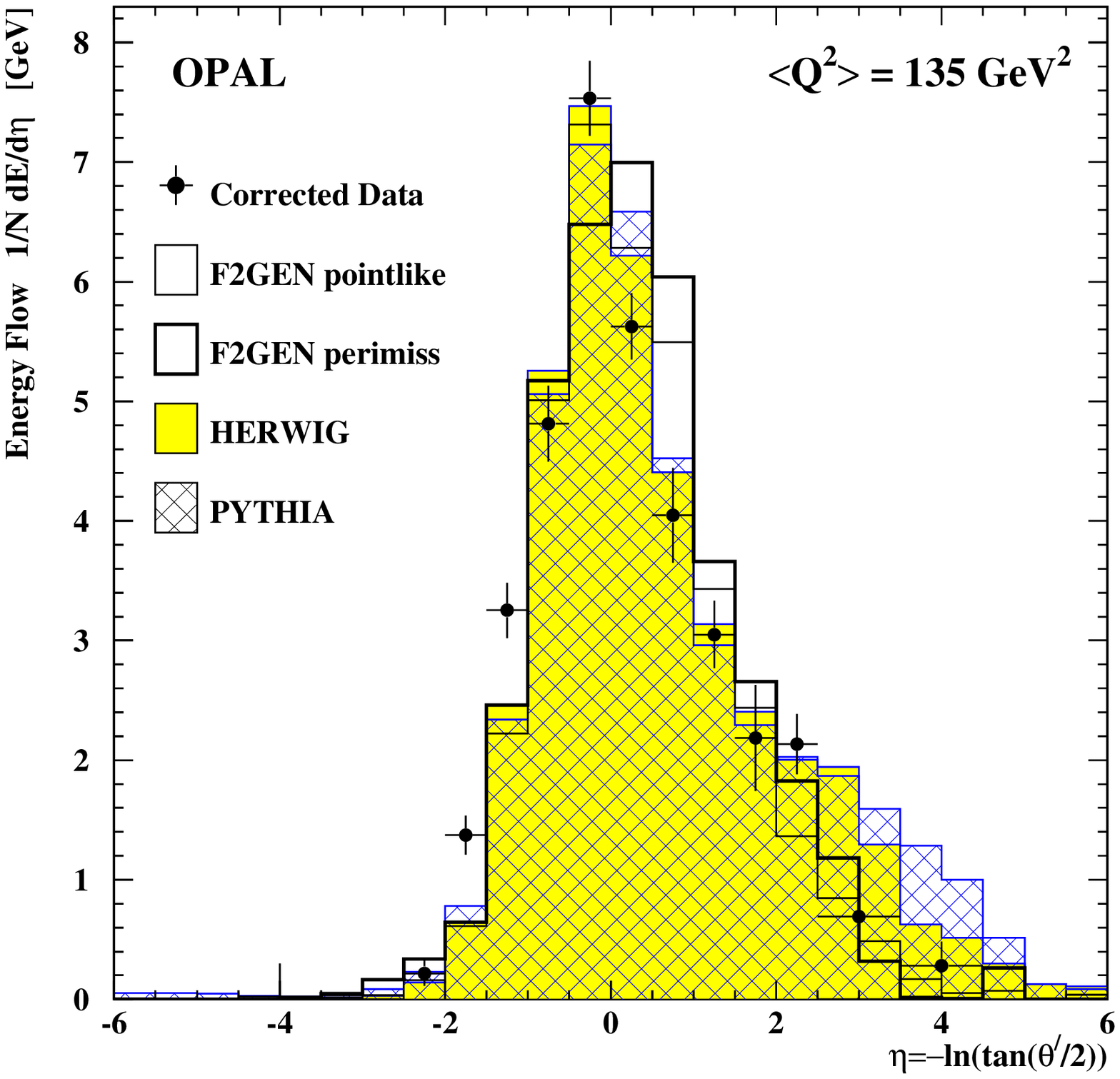}}
\caption[
         The corrected hadronic energy flow for $\qzm = 135$~\gevsq.
        ]
        {
         The corrected hadronic energy flow for $\qzm = 135$~\gevsq.
         Same as Figure\protect~\ref{fig:chap7_04} but for
         $\qzm = 135$~\gevsq.
        }\label{fig:chap7_05}
\end{center}
\end{figure}

%
 The differences of the predictions can most clearly be seen in
 variables like the energy transverse to the plane defined by the
 beam axis and the momentum vector of the observed electron, \etout,
 which is shown in Figure~\ref{fig:chap7_03},
 taken from Ref.~\cite{OPALPR185}, in bins of \xvis.
 The value of \etout is obtained by summing up the absolute values of
 the energy transverse to the tag plane for all objects.
 The F2GEN model predicts the hardest spectrum and lies above the data,
 whereas the HERWIG5.8d model lies below the data and the PYTHIA
 prediction does not even populate the tail of the \etout distribution.
 It is apparent from this figure that the largest differences occur
 at low values of \xvis.
 Taking the differences of the models into account in a bin-by-bin
 correction procedure the observed hadronic energy flow can be corrected
 to the \emph{hadron level\/}\footnote{The term \emph{hadron level\/}
 means that all cuts are applied to generated quantities and that the
 observable shown is calculated from generated stable particles,
 which are usually defined with lifetimes of more than 1~ns. In contrast,
 the \emph{detector level\/} distributions are obtained by applying
 cuts to the measured quantities and also calculating the observable
 under study from measured objects after applying quality cuts to observed
 tracks and calorimetric clusters.
 } and compared to various model predictions.
 Examples of this are shown in Figure~\ref{fig:chap7_04} for
 $\qzm = 13$~\gevsq, and in Figure~\ref{fig:chap7_05} for
 $\qzm = 135$~\gevsq, both taken from Ref.~\cite{OPALPR185}.
 The errors take into account the dependence of the correction
 on the Monte Carlo model chosen for correcting the data.
 A detailed discussion of the various models used can be found in
 Section~\ref{sec:gener}.
 The shape of the hadronic energy flow drastically changes from
 a two peak structure at low values of \qsq to a one peak structure
 for increasing \qsq, which also means increasing values of $x$.
 The differences between the models shrink considerably, and in addition
 the predictions lie much closer to the data.
 This shows firstly that the problem is located in the region of low
 values of $x$ and large values of $W$, and secondly that the data
 are certainly precise enough to further constrain the models.
 None of the models shown is able to describe the data at
 $\qzm = 13$~\gevsq, but the agreement improves for $\qzm = 135$~\gevsq.
 \par
%
%
 After these findings were reported, several methods were investigated
 to reduce the dependence of the measured \ft on the Monte Carlo models.
 A first attempt to improve on the HERWIG5.9 model was made in
 Refs.~\cite{LAU-9701,CAR-9701} by altering the distribution of the
 transverse momentum, \kt, of the quarks inside the
 photon from the program default.
 The default Gaussian behaviour was replaced by a power-law function of
 the form $\der\ktsq/(\ktsq+\kn^2)$ with
 $\kn=0.66$~GeV, motivated by the observation made in photoproduction
 studies at HERA that a better description of the data is achieved
 if the intrinsic transverse momentum distribution is changed to
 the power-law behaviour, as explained in Ref.~\cite{ZEU-9501}.
 The upper limit of \ktsq in HERWIG5.9+\kt was fixed at
 $\ktsqm=25$~GeV$^2$, which is almost the upper limit of \qsq for the
 OPAL analysis from Ref.~\cite{OPALPR185}.
 This led to some improvements in the description of the OPAL
 data by the HERWIG5.9+\kt Monte Carlo, as reported, for example,
 in Ref.~\cite{LAU-9701}.
 \par
%
%
 A second attempt to improve on the situation is based on a purely
 kinematic consideration already explained in Ref.~\cite{LON-9601}.
 The longitudinal momentum of the photon-photon system is unknown,
 but the transverse momentum is well constrained by measuring
 the transverse momentum of the scattered electron.
 In addition, when the $+z$ axis is chosen in the hemisphere of the
 observed electron, the unseen electron which radiated the quasi-real
 photon escapes with $\etwop-\vert\pztwop\vert\approx 0$
 along the beam line.
 Here \etwop and \pztwop denote the energy and
 longitudinal momentum of the unseen electron.
 This fact can be used to replace $\ehad+\pzhad$, the sum of the energy
 and longitudinal momentum of the total hadronic system, by quantities
 obtained from the scattered electron.
%
\begin{figure}[tbp]
\begin{center}
{\includegraphics[width=0.9\linewidth]{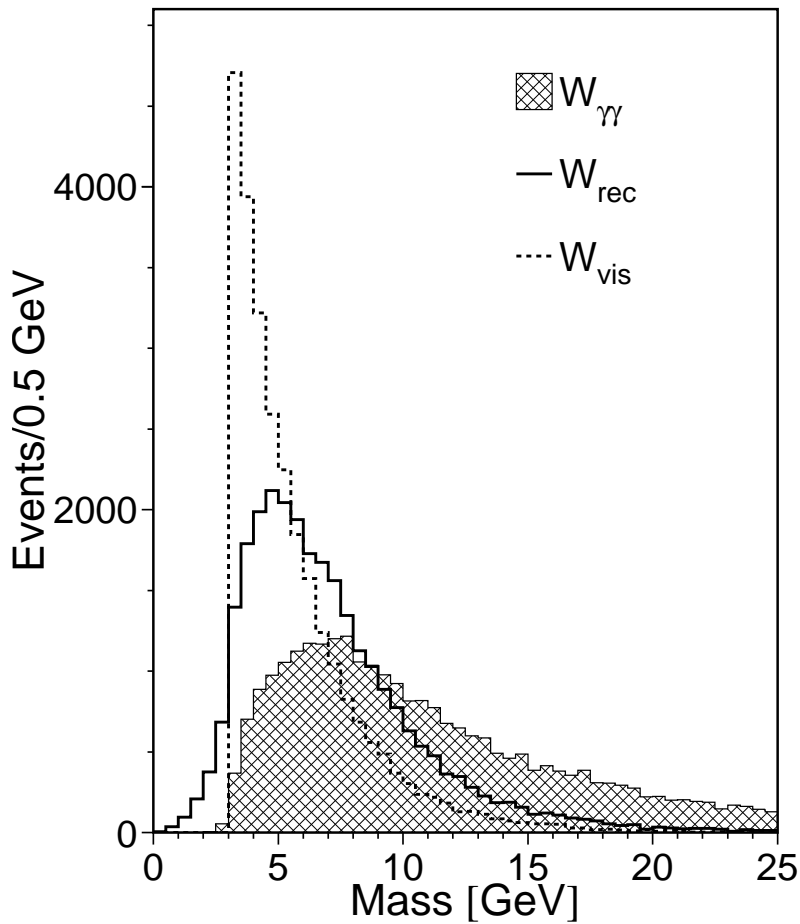}}
\caption[
         Comparison of different methods for the reconstruction of the
         invariant mass of the hadronic final state.
        ]
        {
         Comparison of different methods for the reconstruction of the
         invariant mass of the hadronic final state.
         Shown are the generated mass $W=W_{\gamma\gamma}$, the
         visible mass obtained from the observed hadrons in the L3
         detector, \Wvis and \Wrec, defined in Eq.~(\ref{eqn:wrec}).
         All distributions are for the PHOJET Monte Carlo
         model in the \qsq range $1.2-9$~\gevsq.
        }\label{fig:chap7_06}
\end{center}
\end{figure}
%
 If in addition the transverse momentum squared of the hadronic system,
 \pthadq, is replaced by that of the scattered electron, \ptonepq,
 a part of the uncertainty of the measurement of the hadronic final
 state can be eliminated.
 The value of $W$ reconstructed in this scheme is called
 \Wrec and has the following form
%
 \begin{eqnarray}
  \Wrec &=& \left(\ehad+\pzhad\right)\cdot
            \left(\ehad-\pzhad\right)-\pthadq \nonumber\\
        &=& \left[2E-\left(\eonep+\pzonep\right)\right]
            \left(\ehad-\pzhad\right)-\ptonepq\, .
 \label{eqn:wrec}
 \end{eqnarray}
%
 Because the quantities which are replaced depend on the properties of
 the hadronic final state, the improvement is expected to show
 some dependence on the Monte Carlo programs used.
 For example, for the PHOJET Monte Carlo,
 the improvement on the resolution in $W$ can be seen from
 Figure~\ref{fig:chap7_06}, taken from Ref.~\cite{L3C-9803}.
 The generated values of $W=W_{\gamma\gamma}$ are compared to \Wvis and
 \Wrec using the PHOJET Monte Carlo model in the \qsq range $1.2-9$~\gevsq.
 The improvement is expected to be largest for L3, because this detector,
 on top of the general problems discussed above, suffers from a dead
 region in acceptance, as can be seen from
 Figure~\ref{fig:chap7_07}, taken from Ref.~\cite{L3C-9803}.
 The distribution of \Wrec is closer to the $W$ distribution than the
 \Wvis distribution, but still the agreement with $W$ is not very good.
 \par
%
\begin{figure}[tbp]
\begin{center}
{\includegraphics[width=1.0\linewidth]{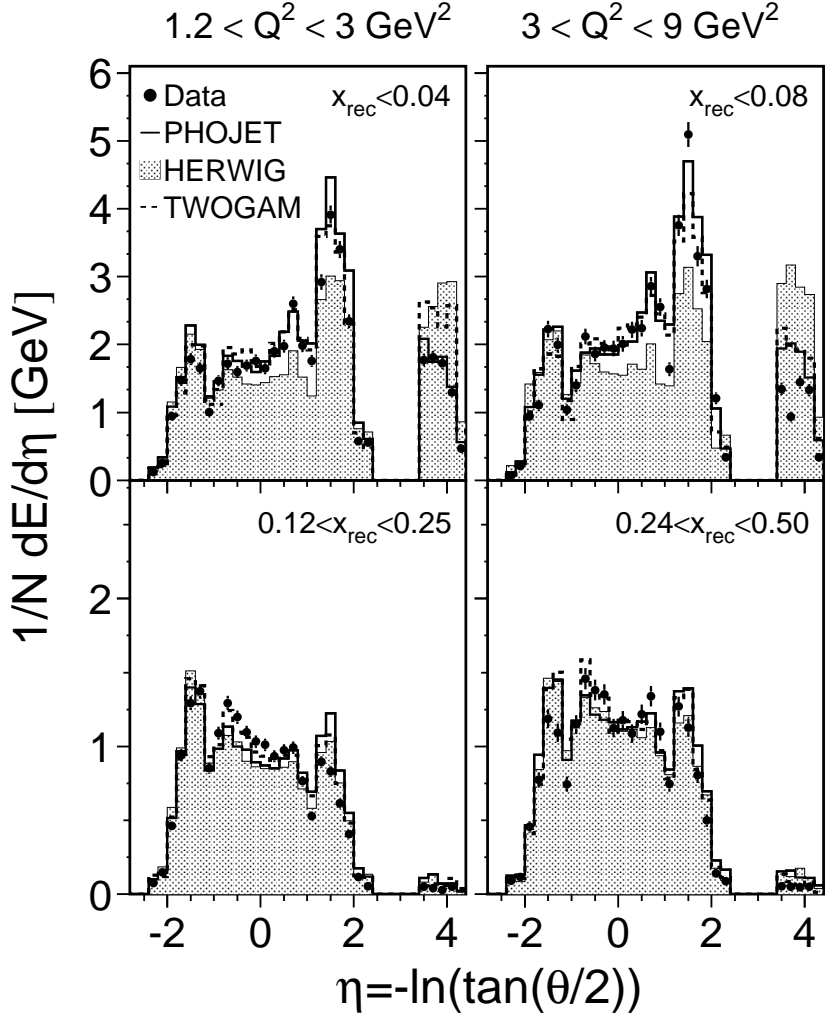}}
\caption[
         The measured hadronic energy flow from L3.
        ]
        {
         The measured hadronic energy flow from L3.
         The measured hadronic energy flow is compared to the Monte Carlo
         predictions in bins of \xrec, obtained from \Wrec and \qsq,
         and in bins of \qsq.
         The models used are HERWIG5.9+\kt, PHOJET and TWOGAM.
        }\label{fig:chap7_07}
\end{center}
\end{figure}
%
 Recently, in Ref.~\cite{L3C-9803}, the PHOJET Monte Carlo model has
 been used for the first time in a structure function analysis.
 Also for the first time in this analysis the TWOGAM Monte Carlo program
 was used outside the DELPHI collaboration.
 In Figure~\ref{fig:chap7_07}, taken from Ref.~\cite{L3C-9803},
 the prediction of the hadronic energy flow for these two models
 are compared to the L3 data for the \qsq range $1.2-9$~\gevsq.
 Again, these two models, although closer to the data than the
 HERWIG5.8d and PYTHIA predictions in the case of the OPAL analysis,
 do not accurately account for the features observed in the data
 distributions.
 In the case of L3 the HERWIG5.9+\kt model, which was tuned for the
 \qsq region of $6-30$~\gevsq, does not provide a satisfactory
 description of the data taken in the region $1.2<\qsq<9$~\gevsq,
 and therefore the L3 analysis of the photon structure function \ft
 is only based on the PHOJET and TWOGAM models.
 \par
 The above information is valuable in understanding the discrepancies,
 however, the investigations suffer from three main shortcomings.
 Firstly, always slightly different cuts are applied to the experimental
 data, and therefore, although the data are rather indicative, it is not
 clear, whether a consistent picture emerges from the results of the
 different experiments.
 Secondly, in the present form, the data cannot be directly compared to
 the generated quantities obtained without simulating the detector response.
 This is because either the data are not corrected for
 detector effects, as in the case of L3, Figure~\ref{fig:chap7_07},
 or they still depend on cuts applied to the data, which is the case
 for the OPAL distributions, Figures~\ref{fig:chap7_04}
 and~\ref{fig:chap7_05}, which are only obtained for events
 fulfilling the experimental cuts applied at the detector level.
 However, in order for the authors to improve on their models it is
 mandatory that they can compare to corrected distributions provided
 by the experiments, without the need of simulating the detector
 response.
 Thirdly, it is hard to get a reliable estimate of the systematic
 error of the experimental result within one experiment, because in this
 case it can only be obtained from varying the Monte Carlo predictions using
 models which do not accurately describe the data, certainly not a very
 reliable method.
 \par
 To overcome these shortcomings a combined effort by the ALEPH,
 L3 and OPAL collaborations and the LEP Working Group for Two-Photon
 Physics has been undertaken, and preliminary results of this work have been
 reported in Ref.~\cite{FIN-9901}.
 The data of the experiments have been analysed in two regions of
 \qsq, $1.2-6.3$~\gevsq and $6-30$~\gevsq, using identical cuts
 and also identical Monte Carlo events passed through the respective
 programs of the individual experiments to simulate the detector response.
 The data are corrected to the hadron level in a phase space
 region which is purely determined by cuts at the hadron level, and
 the systematic error is estimated by the spread of the corrected
 distributions of the three experiments.
 Then the data are compared to predictions from the HERWIG5.9+\kt and
 the PHOJET Monte Carlo models.
 Several distributions are studied, namely,
 the reconstructed invariant hadronic mass, defined
 within a restricted range in polar angles,
 the transverse energy out of the plane defined by the beam direction and the
 direction of the observed electron, the number of tracks,
 the transverse momenta of tracks with respect to the beam axis, and
 the hadronic energy flow as a function of the pseudorapidity.
 Preliminary results of this investigation have been reported
 in Ref.~\cite{FIN-9901}.
 It is found that for large regions in most of the distributions studied,
 the results of the different experiments are closer to each other
 than the sizeable differences which are observed between the data and the
 models.
 Since the data distributions are corrected to the hadron level,
 they can be directly compared to the predictions of the
 Monte Carlo models. Therefore the combined LEP data serve as an
 important input to improve on the Monte Carlo models.
 The investigation already led to an improved version of the
 HERWIG5.9+\kt program obtained by again altering the modelling of the
 intrinsic transverse momentum of the quarks within the photon.
 While the fixed limit of $\ktsqm=25$~GeV$^2$ was adequate for the
 region $6<\qsq<30$~\gevsq, for lower values of \qsq,
 it introduces too much transverse momentum.
 This has been overcome by dynamically adjusting the upper limit
 of \kt by the event kinematic on an event by event basis.
 In this scheme, called HERWIG5.9+\kt(dyn)
 the maximum transverse momentum is set to $\ktsqm\approx\qsq$.
 This change leads to an improved description of the data also for
 the region $1.2<\qsq<6.3$~\gevsq.
 \par
 Another way of reducing the model dependence of the measured \ft is
 to perform the unfolding in two dimensions, as described
 in Section~\ref{sec:unfol}.
 Recent preliminary results from the ALEPH and OPAL experiments,
 presented in Refs.~\cite{BOE-9901} and~\cite{CLA-9901} respectively,
 show that this indeed reduces the systematic uncertainty on the
 structure function measurements.
 \par
 From the discussion above it is clear that the error on the
 measurement of \ft will vary strongly with the selection of Monte Carlo
 models chosen to obtain the size of the systematic uncertainty.
 However, given the improved understanding of the shortcomings
 and the combined effort in improving on the Monte Carlo
 description of the data, it is likely that the error on \ft
 will shrink considerably in future measurements.
 This closes the discussion about the description of the hadronic
 final state by the Monte Carlo models, and the measurements
 of \ft will be discussed next.
%
%
\subsection{Hadronic structure function \ft}
\label{sec:qcdresf2}
 Many measurements of the hadronic structure function \ft have been
 performed at several \epem colliders.
 Because in some cases it is not easy to correctly derive the errors
 of several of the measurements, a detailed survey of
 the available results has been performed, the outcome of which is
 presented in Appendix~\ref{sec:tabqcd}.
 The measurements and what can be learned from them about the
 structure of the photon and on its description by perturbative QCD
 is discussed in the following.
 The interpretation of the data will only be based on published results,
 and on preliminary results from the LEP experiments, which are likely
 to be published soon.
 In contrast the preliminary results from
 Refs.~\cite{CEL-9001,TOP-9406,TPC-8802,VEN-9301,VEN-9302},
 which were used in the fit procedures of several of the
 parton distribution functions of the photon, but which never got
 published, are regarded as obsolete, and will not be considered here.
 In all summary figures presented below only those preliminary results
 from the LEP experiments are included which are based on data which
 have not yet been published.
 For the preliminary results which are meant to replace a previous
 measurement in the near future the previously published result will
 be shown until the new result is finalised.
 \par
%
\begin{figure}[tbp]
\begin{center}
{\includegraphics[width=1.0\linewidth]{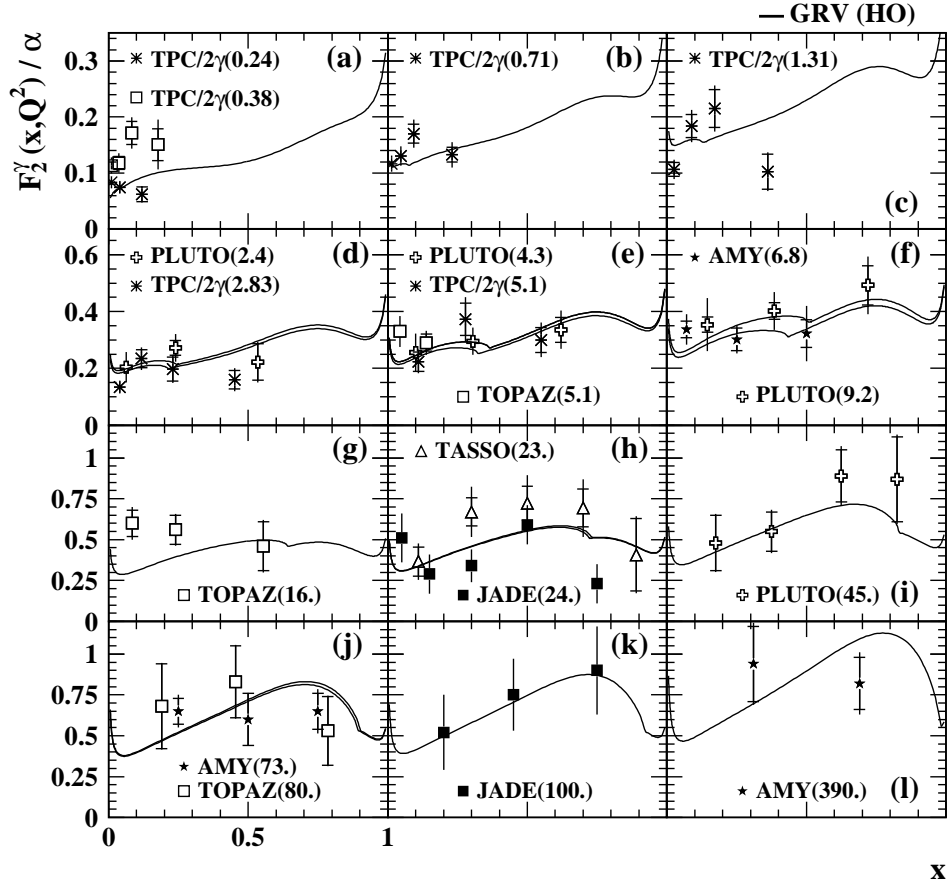}}
\caption[
         Measurements of the hadronic structure function \ft
         except from LEP.
        ]
        {
         Measurements of the hadronic structure function \ft
         except from LEP.
         The points represent the data with their statistical
         (inner error bars) and total errors (outer error bars),
         if available, otherwise only the full errors are shown.
         The measured data are shown in comparison to the prediction
         of \ft obtained from the GRV higher order parton distribution
         function, using the \qsq values given in brackets.
        }\label{fig:chap7_09}
\end{center}
\end{figure}
%
 The range in \qzm covered by the various experiments is
 $0.24<\qzm<400$~\gevsq, which is impressive given the small cross
 section of the process.
 The published results from the
 ALEPH~\cite{ALE-9901},
 AMY~\cite{AMY-9501,AMY-9701},
 DELPHI~\cite{DEL-9601},
 JADE~\cite{JAD-8401},
 L3~\cite{L3C-9803,L3C-9804},
 OPAL~\cite{OPALPR185,OPALPR207,OPALPR213},
 PLUTO~\cite{PLU-8401,PLU-8701},
 TASSO~\cite{TAS-8601},
 TPC/2$\gamma$~\cite{TPC-8701}
 and
 TOPAZ~\cite{TOP-9402}
 experiments can be found in
 Tables~\ref{tab:chap13_01}$-$~\ref{tab:chap13_10}.
 The additional preliminary results from the
 ALEPH~\cite{BOE-9901},
 L3~\cite{ERN-9901}
 and
 DELPHI~\cite{TIA-9701,TIA-9801}
 experiments are listed in Tables~\ref{tab:chap13_11}$-$\ref{tab:chap13_13}
 \par
 In the present investigations of the photon structure function \ft
 two distinct features of the photon structure are investigated.
 Firstly, the shape of \ft is measured as a function of $x$ at fixed \qsq.
 Particular emphasis is put on measuring the low-$x$ behaviour of
 \ft in comparison to \ftp as obtained at HERA.
 Secondly, the evolution of \ft with \qsq is investigated.
 As explained in Section~\ref{sec:QCD}, this evolution is predicted by QCD
 to be logarithmic. These two issues are discussed.
 \par
%
\begin{figure}[tbp]
\begin{center}
{\includegraphics[width=0.9\linewidth]{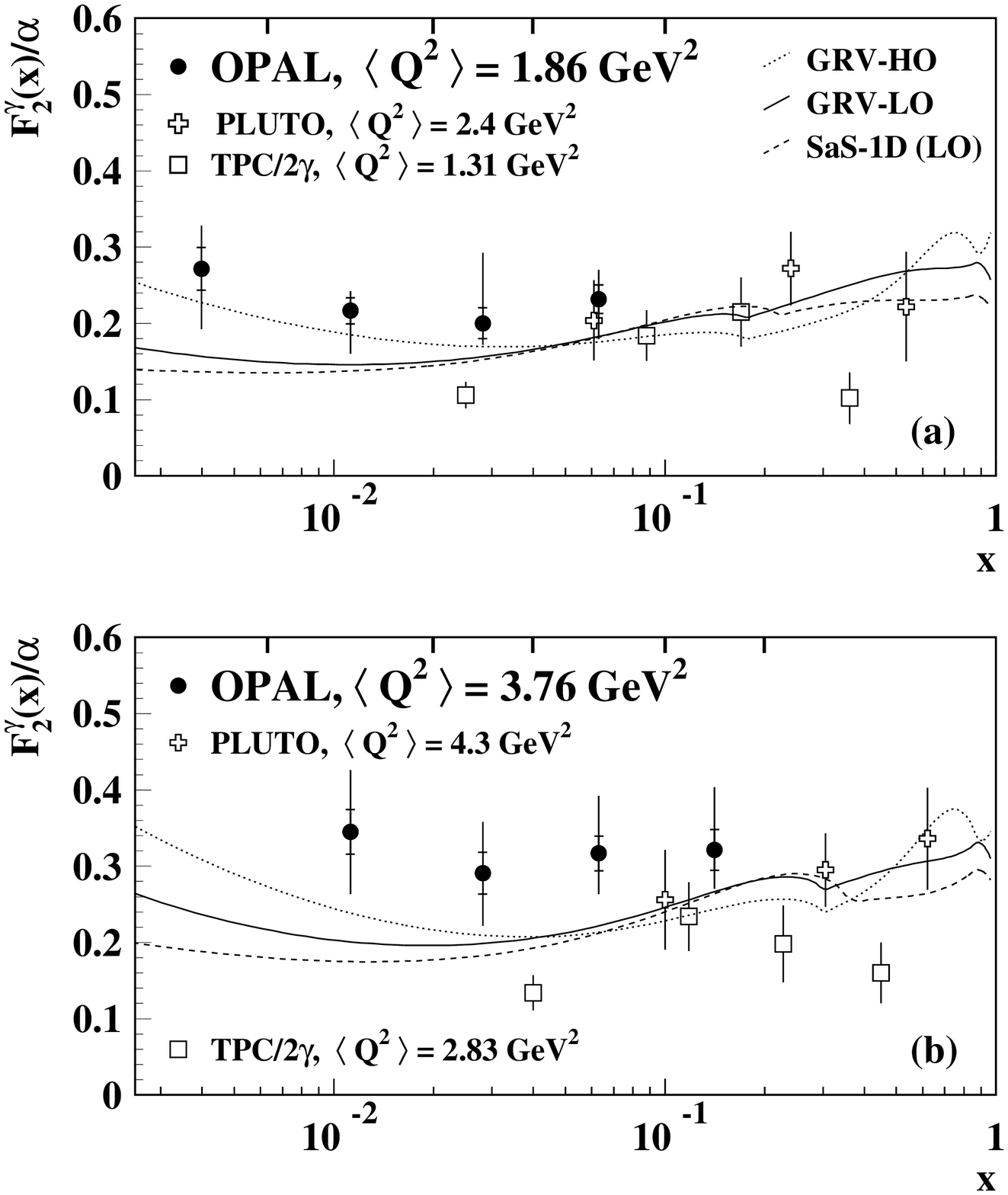}}
\caption[
         Comparison of measurements of \ft at low values of \qsq.
        ]
        {
         Comparison of measurements of \ft at low values of \qsq.
         The OPAL data at \qzm of 1.86 and 3.76~\gevsq are compared to
         previous results from the PLUTO and TPC/2$\gamma$ experiments.
         The points represent the OPAL data with their statistical
         (inner error bars) and total errors (outer error bars).
         For the previous data only the total errors are shown.
         The tic marks at the top of the figures indicate the bin
         boundaries of the OPAL analysis.
        }\label{fig:chap7_10}
\end{center}
\end{figure}
%
 The collection of measurements on \ft which have been performed
 at \epem centre-of-mass energies below the mass of the $Z$ boson
 is shown in Figure~\ref{fig:chap7_09}.
 Their precision is mainly limited
 by the statistical error and, due to the simple assumptions
 made on the hadronic final state, the systematic errors are small,
 but in light of the discussion above, they may be underestimated.
 The global behaviour of the data is roughly described, for example,
 by the \ft obtained from the GRV higher order parton distribution
 function.
 However, some of the data show quite unexpected features.
 For example, the structure function as obtained from the
 TPC/2$\gamma$ experiment shows an unexpected shape at low
 values of $x$, and also the results from TOPAZ rise very
 fast towards low values of $x$.
 In addition there is a clear disagreement between
 the TASSO and JADE data at $\qzm=23-24$~\gevsq.
 Certainly at this stage much more data were needed to clarify
 the situation.
 \par
%
\begin{figure}[tbp]
\begin{center}
{\includegraphics[width=0.9\linewidth]{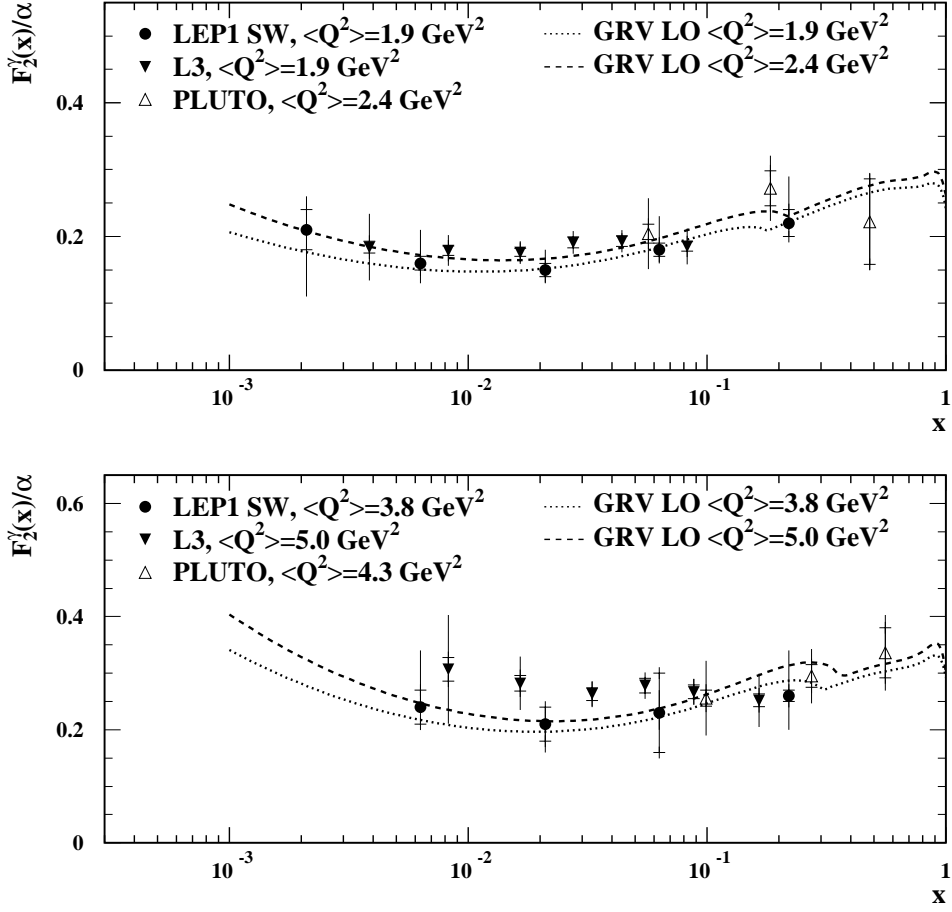}}
\caption[
         Preliminary update of \ft at low values of \qsq from OPAL.
        ]
        {
         Preliminary update of \ft at low values of \qsq from OPAL.
        The preliminary OPAL data at \qzm of 1.9 and 3.8~\gevsq are
        compared to the results from the PLUTO and L3 experiments.
        The points represent the data with their statistical
        (inner error bars) and total errors (outer error bars).
        For the PLUTO result only the total errors are shown.
        For the L3 result the errors are obtained as explained
        in Table~\protect\ref{tab:chap13_05}.
        }\label{fig:chap7_11}
\end{center}
\end{figure}
%
 The measurement of \ft has attracted much interest at LEP over
 the last years.
 The LEP Collaborations have measured the photon structure function
 \ft in the range $0.002<x$ \lsim 1 and $1.86<\qzm<400$ \gevsq.
 The first published result of the low-$x$ behaviour of \ft at low values
 of \qsq performed on a logarithmic scale in $x$ is shown in
 Figure~\ref{fig:chap7_10}.
 The data have been unfolded based on the HERWIG5.8d Monte Carlo model.
 Only a weak indication of a possible rise at low values of $x$ for
 $\qsq <4$~\gevsq is observed.
 More important, the data seem to be consistently higher than what is
 predicted by the GRV and SaS1D parametrisations.
 In addition, there emerges an inconsistency at $\qsq \approx 4$~\gevsq,
 between the OPAL and PLUTO data of Refs.~\cite{OPALPR213}
 and~\cite{PLU-8401} on one hand, and the TPC/2$\gamma$ data of
 Ref.~\cite{TPC-8701} on the other hand.
 \par
%
\begin{figure}[tbp]
\begin{center}
{\includegraphics[width=1.0\linewidth]{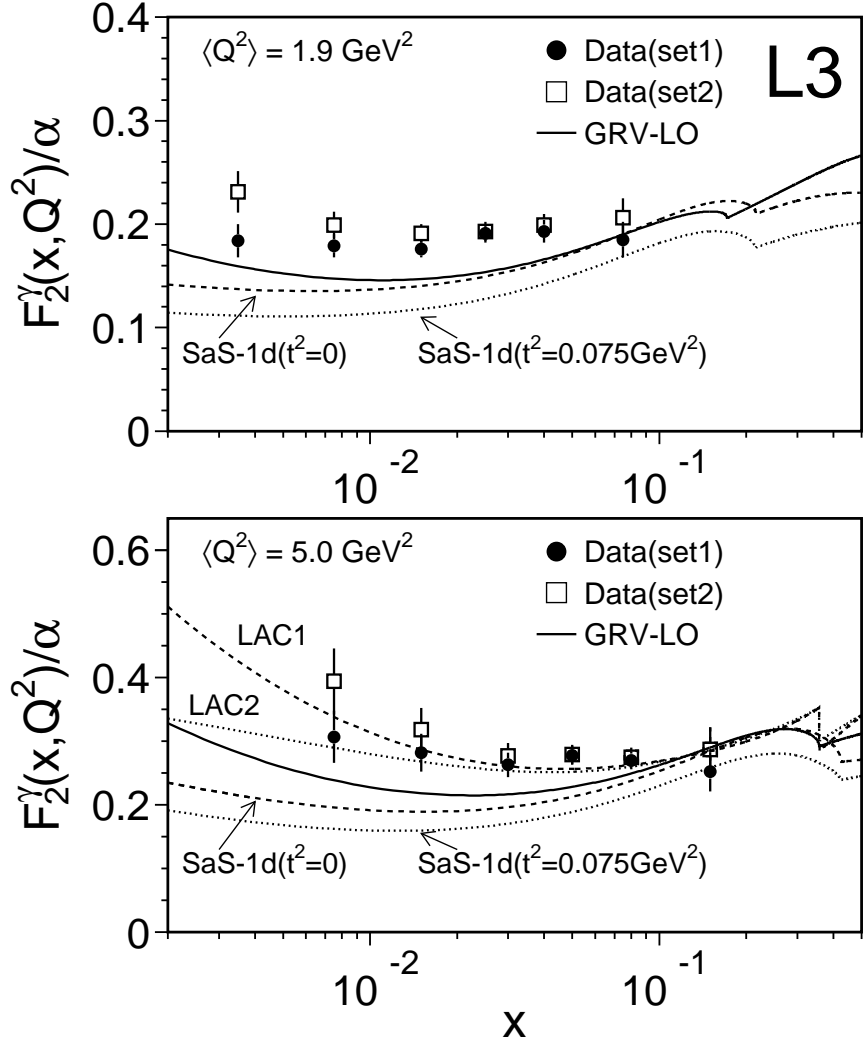}}
\caption[
         The measurements of \ft at low values of \qsq from L3.
        ]
        {
         The measurements of \ft at low values of \qsq from L3.
         The points represent the L3 data with their total experimental
         errors, but excluding the large error stemming from the choice
         of Monte Carlo model.
         The models chosen for the unfolding of \ft from the data
         are PHOJET (set1) and TWOGAM (set2).
        }\label{fig:chap7_12}
\end{center}
\end{figure}
%
 Recently a preliminary update of the OPAL measurement
 at low values of \qsq, shown in
 Figure~\ref{fig:chap7_11}, has been presented in Ref.~\cite{CLA-9901}.
 The new OPAL analysis is based on the same data as the published
 results, but uses the PHOJET and the
 improved HERWIG5.9+\kt Monte Carlo models explained above, which much
 better describe the experimentally observed distributions than
 the HERWIG5.8d model used for the results in Figure~\ref{fig:chap7_10}.
 In addition the method of two-dimensional unfolding based on the GURU
 program has been explored.
 With these improvements the systematic errors could be considerably
 reduced and now, this new measurement is consistent with the GRV leading
 order prediction.
 By repeating the analysis with the HERWIG5.8d Monte Carlo model,
 but using two-dimensional unfolding, the new analysis leads to results
 consistent with the published results from Ref.~\cite{OPALPR213},
 however with reduced errors.
 From this it can be concluded that firstly, the high values of the
 published results are due to the bad description of the data by the
 HERWIG5.8d model, and secondly that the two-dimensional unfolding
 reduces the error of the measurement, even when using an inaccurate
 model for the unfolding of \ft.
 \par
%
\begin{figure}[tbp]
\begin{center}
{\includegraphics[width=1.0\linewidth]{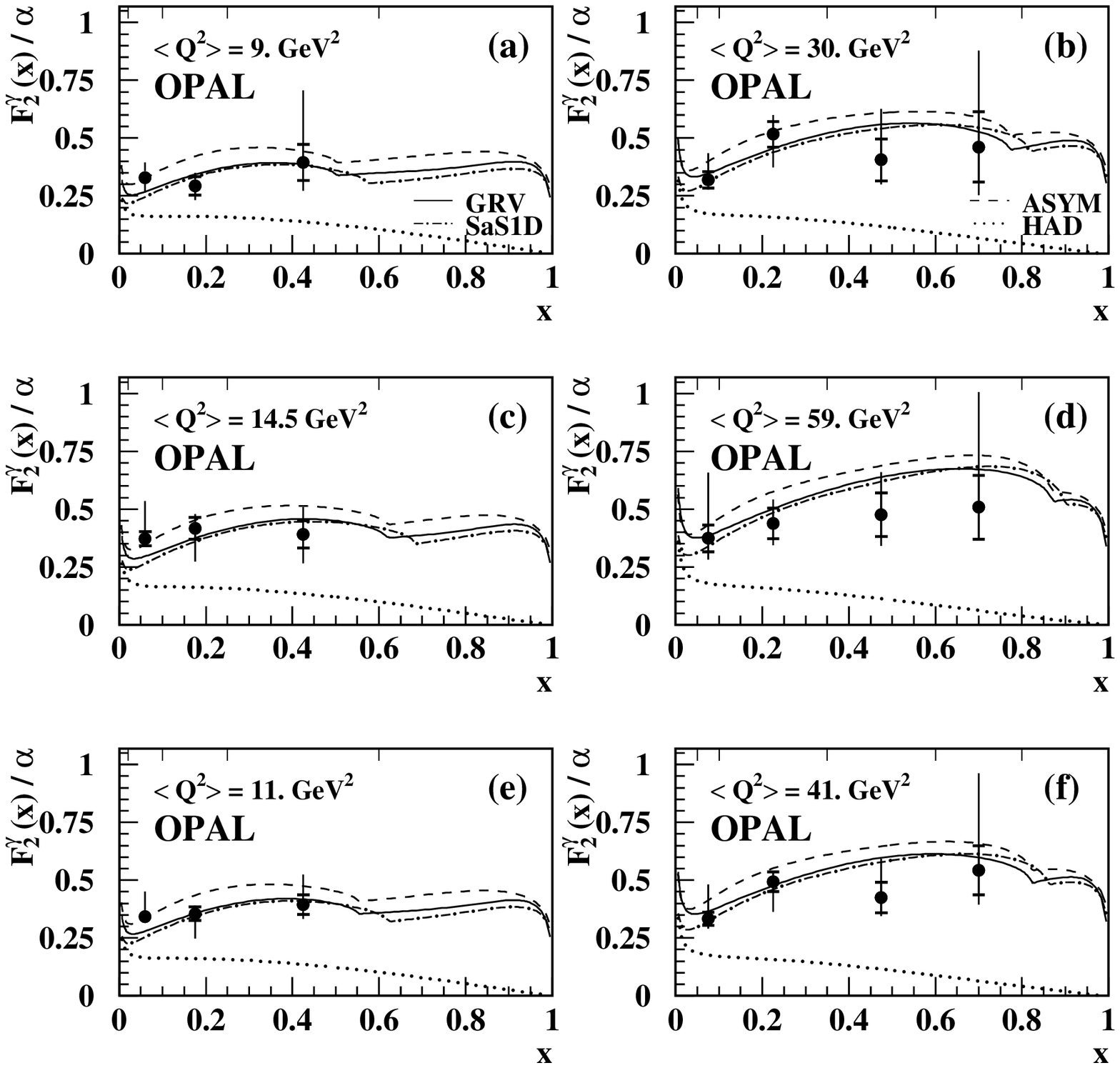}}
\caption[
         The first measurement of \ft for $\ssee>m_Z$.
        ]
        {
         The first measurement of \ft for $\ssee>m_Z$.
         The structure function \ft is measured for four active flavours
         in four bins in \qsq with mean values of
         (a) \qzm =  9~\gevsq, (b) \qzm = 30~\gevsq, (c) \qzm = 14.5~\gevsq,
         and (d) \qzm = 59~\gevsq.
         In (e) the measurement for the combined data sets of (a) and (c),
         and in (f) the measurement for the combined data sets of
         (b) and (d) is shown.
         The points represent the OPAL data with their statistical
         (inner error bars) and total errors (outer error bars).
         The tic marks at the top of the figures indicate the bin
         boundaries. The data are compared to several predictions
         described in the text.
        }\label{fig:chap7_13}
\end{center}
\end{figure}
%
 A similar analysis at low values of $x$ and \qsq, shown in
 Figure~\ref{fig:chap7_12},
 was performed by the L3 experiment in Ref.~\cite{L3C-9803}.
 Two results for \ft were presented which differ in the model chosen for
 the unfolding of \ft from the data.
 The published results are for PHOJET (set1) and TWOGAM (set2).
 From the measurement it is clear that although some improved description
 of the hadronic final state can be achieved by the PHOJET and TWOGAM
 models, the model uncertainty still is the dominant error source at
 low values of $x$.
 At $\qsq=5$~\gevsq the LAC1 and LAC2 predictions are consistent with
 the L3 result.
 However, the L3 result is consistently higher than the SaS1D and the
 leading order GRV parametrisations of \ft for both values of \qsq.
 Given the quoted errors of the L3 result the GRV and SaS
 parametrisations need to be revisited.
 In addition, as can be seen from Figure~\ref{fig:chap7_11}, the
 preliminary OPAL and the L3 measurements are consistent with each other.
 \par
 The measurements discussed above are based on the entire data of the
 LEP1 running period at \epem centre-of-mass energies around the mass
 of the $Z$ boson.
 The first published result based on data for $\ssee>m_Z$ is shown in
 Figure~\ref{fig:chap7_13}, taken from Ref.~\cite{OPALPR207}.
 Due to the higher energy of the beam electrons the \qsq acceptance
 of the detectors is changed, see Figure~\ref{fig:chap2_05} and
 Eq.~(\ref{eqn:q2}).
 As a rule of thumb the values of \qsq accepted at LEP2 energies is
 about a factor of four higher than those accepted at LEP1 energies,
 when using the same detector device to measure the scattered electron.
 The results in Figure~\ref{fig:chap7_13} cover the \qzm range from
 9 to 59~\gevsq.
 The measured \ft as a function of $x$ is almost flat within this region
 and the absolute normalisation of \ft is well described by
 the predictions of the leading order GRV (solid) and the
 SaS1D (dot-dash) parametrisations of \ft evaluated at the
 corresponding values of \qzm.
 \par
 Figure~\ref{fig:chap7_13} also shows an augmented asymptotic
 prediction for \ft (ASYM).
 The contribution to \ft from the three light flavours is approximated
 by the leading order asymptotic form from Ref.~\cite{WIT-7701},
 using the parametrisation given in Ref.~\cite{GOR-9201}.
 This has been augmented by adding a point-like charm contribution
 and a prediction for the hadron-like part of \ft for
 $\lam_3 = 0.232$~\gev.
 The point-like charm contribution has been evaluated from the leading
 order Bethe-Heitler formula, Eq.~(\ref{eqn:BH}) for $\psq = 0$ and
 $\mc=1.5$~\gev.
 The estimate of the hadron-like part of \ft is given by the
 hadron-like part of the GRV leading order parametrisation of \ft
 for four active flavours, and evolved to the corresponding values of \qzm.
 It is known that the asymptotic solution has deficits in the region
 of low-$x$, because of the divergences in the solution
 which do not occur in the solution of the full evolution
 equations, as explained in Section~\ref{sec:QCD}.
 However, the asymptotic solution has the appealing feature that
 it is calculable in QCD, even at higher order and
 for medium $x$ and with increasing \qsq it should be more reliable.
 In addition, at high values of $x$ and \qsq the hadron-like
 contribution is expected to be small.
 In the region of medium values of $x$ studied in Figure~\ref{fig:chap7_13}
 this asymptotic prediction in general lies higher than the GRV and SaS
 predictions but it is still consistent with the data.
 The importance of the hadron-like contribution to \ft (HAD), which is
 shown separately at the bottom of the figure, decreases
 with increasing $x$ and \qsq, and it accounts for only 15$\%$ of
 \ft at $\qsq = 59$~\gevsq and $x = 0.5$.
 The asymptotic solution increases with decreasing \lam.
 For $\qsq = 59$~\gevsq and $x = 0.5$ the change in \ft is $+24\%$
 and $-16\%$ if \lam is changed from $\lam_3 = 0.232$~\gev
 to 0.1~\gev and 0.4~\gev respectively.
 \par
%
\begin{figure}[tbp]
\begin{center}
{\includegraphics[width=1.0\linewidth]{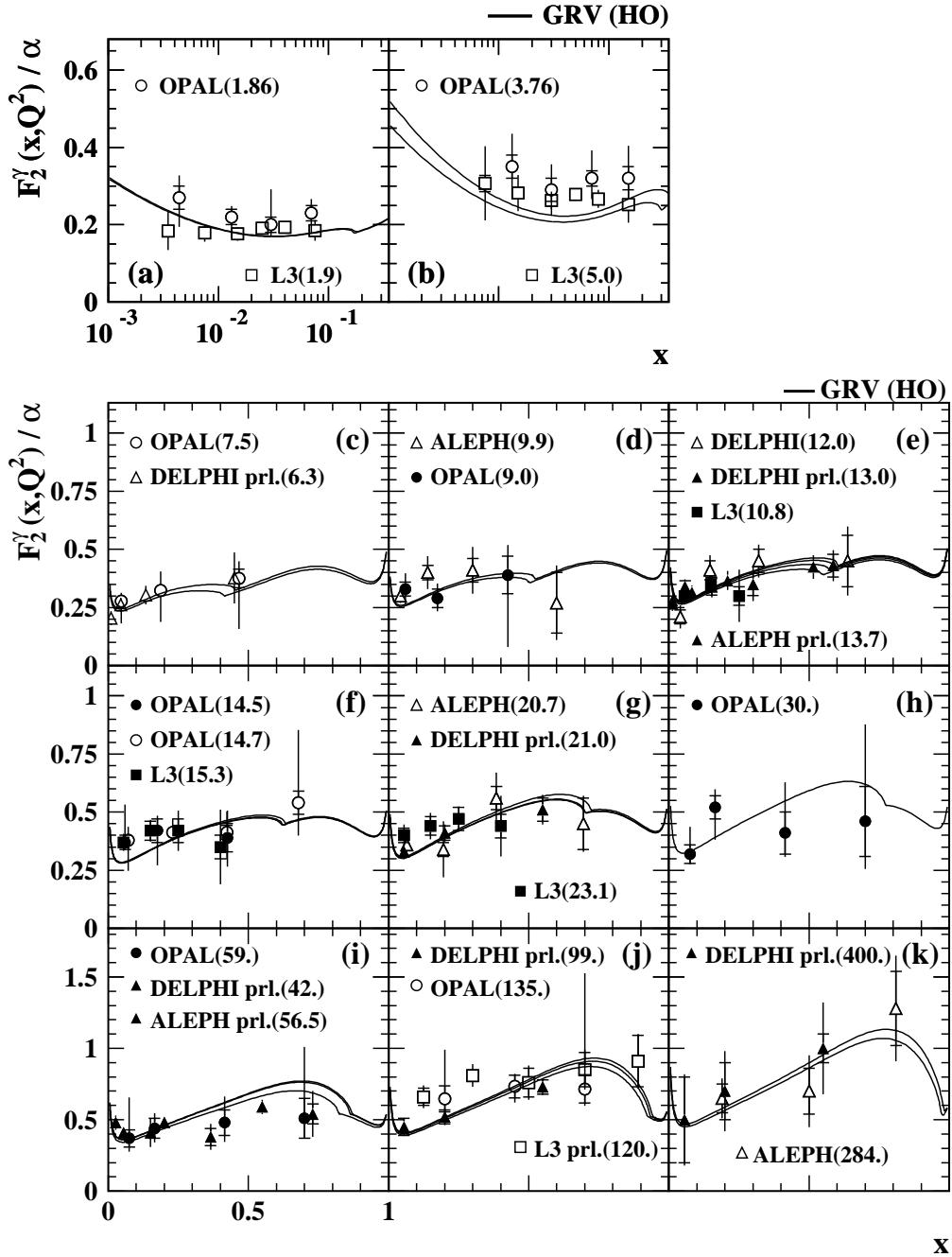}}
\caption[
         Measurements of the hadronic structure function \ft from LEP.
        ]
        {
         Measurements of the hadronic structure function \ft from LEP.
         Same as Figure\protect~\ref{fig:chap7_09} but showing
         the results from the LEP experiments.
        }\label{fig:chap7_14}
\end{center}
\end{figure}
%
 By now, many more measurements for $\ssee>m_Z$ using much higher
 data luminosities have been performed by the LEP experiments.
 All LEP measurements are displayed in Figure~\ref{fig:chap7_14}.
 The measurements obtained at LEP1 energies are shown with open
 symbols, whereas those obtained at LEP2 energies are shown with
 closed symbols.
 The varying energies of the beam electrons give the opportunity
 to compare data at similar \qsq but using different detector devices
 to measure the scattered electron.
 The results are consistent with each other, which gives confidence
 that the systematic errors are well enough controlled.
 \par
%
\begin{figure}[tbp]
\begin{center}
{\includegraphics[width=1.0\linewidth]{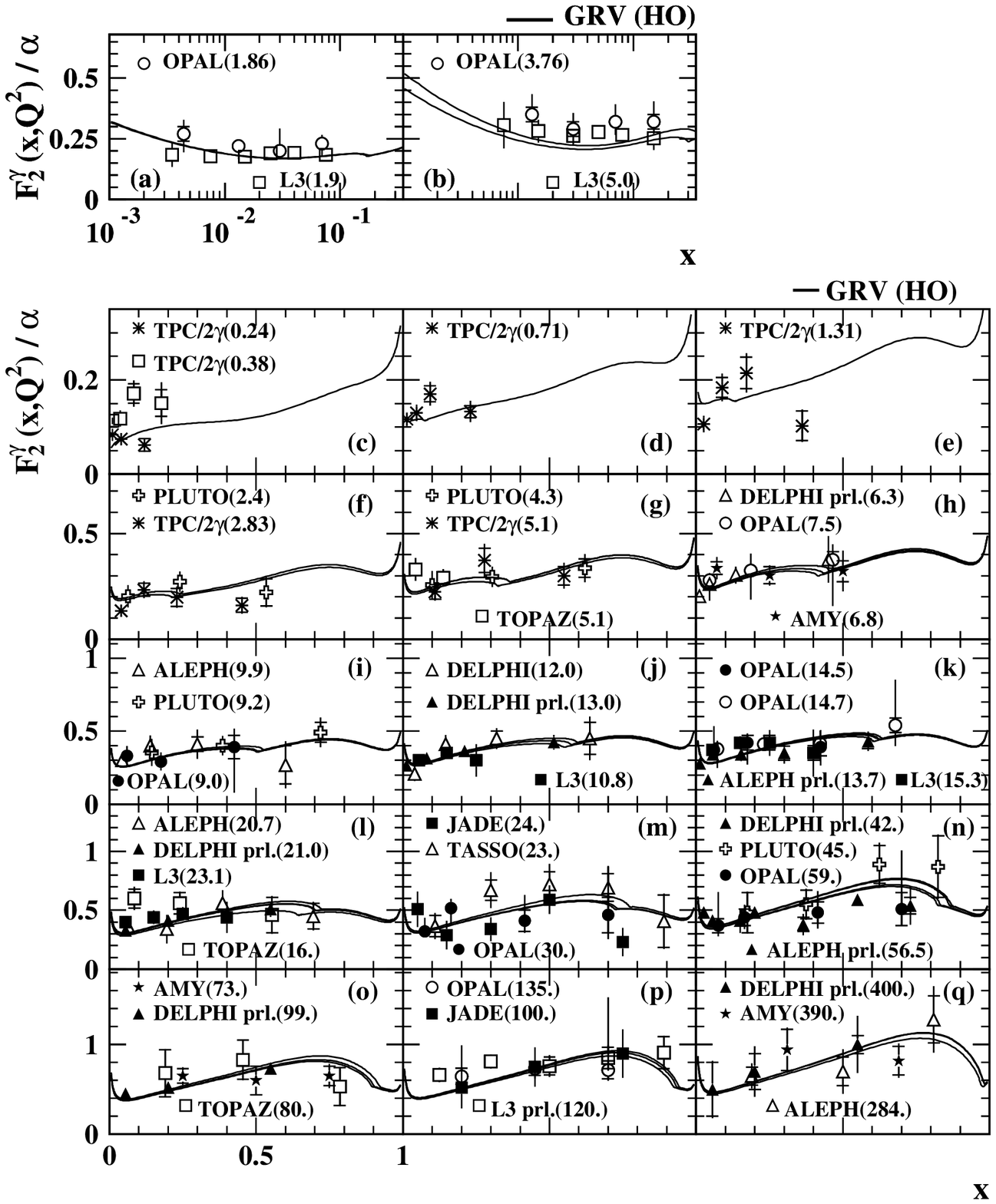}}
\caption[
         Summary of measurement of the hadronic structure function \ft.
        ]
        {
         Summary of measurement of the hadronic structure function \ft.
         Same as Figure\protect~\ref{fig:chap7_09} but showing
         all available results on \ft.
        }\label{fig:chap7_15}
\end{center}
\end{figure}
%
 A summary of all measurements of \ft is shown in Figure~\ref{fig:chap7_15}.
 The comparison to the GRV higher order prediction of \ft shows an
 overall agreement, but also some regions where the prediction does not
 so well coincide with the data.
 This large amount of data, which partly is rather precise, gives the
 possibility to study the consistency of the predictions with the
 data.
 The quality of the agreement is evaluated by a simple \chiq method
 based on
%
 \begin{equation}
  \chiq = \sum_{i=1}^{\mathrm{dof}} \left(
                \frac{\fti-\langle\ft(x,\qzm,0)\rangle}{\sigma_i}
                \right)^2\, ,
 \label{eqn:chiq}
 \end{equation}
%
 where the sum runs over all experiments, all \qzm values, and all bins
 in $x$.
 The term \fti denotes the measured value of \ft in the $i^{th}$ bin and
 $\sigma_i$ is its total error. The theoretical expectation is
 approximated by the average \ft in that bin in $x$ at $\qsq=\qzm$ and
 at $\psq=0$, abbreviated with $\langle\ft(x,\qzm,0)\rangle$.
 If, as in the case of the OPAL, the experiments quote asymmetric errors,
 this is taken into account by chosing the appropriate error depending
 on whether the prediction lies above or below the measured value.
 The procedure is not very precise, as it does not take into account the
 correlation of the errors between the data points and the experiments.
 However, because there are common sources of errors, it is most likely
 that by using this method the experimental error is overestimated.
 Given this, the predictions which are not compatible with the data are
 probably even worse approximations of the data when the comparison is
 done more precisely.
 A more accurate analysis would require to study in detail the
 correlation between the results within one experiment, but even more
 difficult, the correlation between the results from different experiments.
 This is a major task which is beyond the scope of the
 comparison presented here.
 \par
 The predictions used in the comparison are the WHIT
 parametrisations, which are the most recent parametrisations based
 on purely phenomenological fits to the data, and the GRV, GRSc and SaS
 predictions, which use some theoretical prejudice to construct
 \ft as detailed in Section~\ref{sec:PDF}.
 The theoretical expectation is approximated by
 $\langle\ft(x,\qzm,0)\rangle$.
 If $\ft(\langle x\rangle,\qzm,0)$, the structure function at the average
 value of $x$, is taken instead, the results only slightly differ,
 which means that the comparison is not very sensitive to the
 shape of \ft within the bins chosen.
 The results of this comparison are listed in
 Tables~\ref{tab:chap7_01} and~\ref{tab:chap7_02}.
 \par
%
\renewcommand{\arraystretch}{1.1}
\begin{table}[tbp]
\caption[
         Comparison of the GRV, GRSc and SaS1 predictions with
         measurements of \ft.
        ]
        {
         Comparison of the GRV, GRSc and SaS1 predictions with
         measurements of \ft.
         The values are calculated using all data shown in
         Figure\protect~\ref{fig:chap7_15}, apart from
         TPC/2$\gamma$ at $\qzm = 0.24$~\gevsq  which has a \qzm below the
         lowest value for which a parametrisation of \ft exists.
         Listed are the number of data points (dof) as well
         as the values of \chidof as calculated from
         Eq.\protect~(\ref{eqn:chiq}) for the individual experiments and
         for all data points (all).
         The minimum \qsq for the GRSc parametrisation is larger than
         the value for TPC/2$\gamma$ at $\qzm = 0.38$~\gevsq, therefore
         for GRSc the number of points is only 15 and 161 for
         TPC/2$\gamma$ and
         for all data, compared to the total number of 19 and 165.
         For experiments which have measured \ft for \qzm values
         below and above 4~\gevsq,
         in addition the \chidof values for the comparison based only on
         data for $\qzm>4$~\gevsq are shown in a second row.\\
        }\label{tab:chap7_01}
\begin{center}
\begin{tabular}{ccccccc}\hline
 \multicolumn{2}{c}{}&
 \multicolumn{2}{c}{GRV}   & GRSc & \multicolumn{2}{c}{SaS} \\\cline{3-7}
 \multicolumn{2}{c}{}&LO&HO&   LO &1D & 1M \\\hline
 Exp.            &   dof & \multicolumn{5}{c}{\chidof} \\\hline
 AMY             &       8& 0.75& 1.03& 0.71& 0.99& 0.86\\\hline
 JADE            &       8& 1.01& 1.16& 1.03& 1.09& 1.00\\\hline
 PLUTO           &      13& 0.50& 0.46& 0.53& 0.60& 0.51\\
 $\qzm>4$~\gevsq &      10& 0.55& 0.38& 0.60& 0.71& 0.61\\\hline
 TASSO           &       5& 0.97& 0.77& 1.03& 1.03& 0.85\\\hline
 TPC/2$\gamma$   &   19/15& 4.52& 4.11& 7.34& 2.00& 3.98\\
 $\qzm>4$~\gevsq &       3& 0.54& 0.67& 0.92& 0.40& 0.58\\\hline
 TOPAZ           &       8& 1.89& 2.15& 1.67& 2.40& 1.89\\\hline
 ALEPH           &      20& 0.96& 1.41& 0.90& 1.74& 1.01\\\hline
 DELPHI          &      24& 0.69& 1.47& 0.99& 1.12& 0.74\\\hline
 L3              &      28& 2.40& 2.10& 1.82& 3.93& 2.36\\
 $\qzm>4$~\gevsq &      22& 1.91& 2.21& 1.20& 3.43& 1.95\\\hline
 OPAL            &      32& 0.84& 0.80& 0.41& 1.22& 0.85\\
 $\qzm>4$~\gevsq &      24& 0.45& 0.55& 0.37& 0.79& 0.50\\\hline
\hline
 all             & 165/161& 1.55& 1.64& 1.58& 1.81& 1.50\\
 $\qzm>4$~\gevsq &     132& 0.98& 1.29& 0.90& 1.56& 1.02\\
\hline\\
\end{tabular}
\end{center}\end{table}
%
%
\renewcommand{\arraystretch}{1.1}
\begin{table}[tbp]
\caption[
         Comparison of the SaS2 and WHIT predictions with measurements of \ft.
        ]
        {
         Comparison of the SaS2 and WHIT predictions with measurements of \ft.
         Same as Table~\ref{tab:chap7_01}, but including only data
         with $\qzm>\qnsq=4$~\gevsq.\\
        }\label{tab:chap7_02}
\begin{center}
\begin{tabular}{cccccccccc}\hline
 \multicolumn{2}{c}{}    &
 \multicolumn{2}{c}{SaS} & \multicolumn{6}{c}{WHIT} \\\cline{3-10}
 \multicolumn{2}{c}{}&2D & 2M &1&2&3&4&5&6\\\hline
 Exp.  &dof &\multicolumn{8}{c}{\chidof} \\\hline
 AMY           &  8& 1.01& 1.09& 0.74& 0.67&  0.71&  0.76&  0.63&  0.58\\\hline
 JADE          &  8& 1.52& 1.51& 1.25& 1.09&  1.05&  1.27&  1.11&  1.08\\\hline
 PLUTO         & 10& 0.58& 0.60& 0.62& 0.57&  0.56&  0.81&  0.65&  0.63\\\hline
 TASSO         &  5& 0.52& 0.45& 0.64& 0.66&  0.66&  0.83&  0.80&  0.74\\\hline
 TPC/2$\gamma$ &  3& 2.38& 2.50& 1.83& 1.53&  1.26&  3.02&  2.23&  1.55\\\hline
 TOPAZ         &  8& 1.29& 1.34& 1.10& 1.05&  1.19&  0.63&  0.89&  1.07\\\hline
 ALEPH         & 20& 1.09& 1.18& 1.76& 8.39& 14.43& 14.04& 52.58& 81.73\\\hline
 DELPHI        & 24& 0.95& 0.88& 1.72& 7.90& 13.76& 13.69& 46.39& 73.32\\\hline
 L3            & 22& 1.22& 1.37& 0.86& 1.26&  1.88&  3.78&  9.61& 11.40\\\hline
 OPAL          & 24& 0.45& 0.48& 0.40& 0.45&  0.53&  1.14&  1.99&  2.18\\\hline
\hline
 all           &132& 0.97& 1.01& 1.10& 3.27&  5.37&  5.78& 18.66& 28.29\\\hline
\\
\end{tabular}
\end{center}\end{table}
%
 None of the parametrisations has difficulties to describe the
 AMY, JADE, PLUTO and TASSO data, and they all disfavour
 the TPC/2$\gamma$ results, which show an unexpected shape as function
 of $x$.
 The WHIT parametrisations predict a faster rise at low-$x$
 than the GRV, GRSc and the SaS parametrisations.
 Therefore, the agreement with the TOPAZ data is satisfactory for
 the WHIT parametrisations, whereas the GRV, GRSc and the SaS1
 parametrisations yield values of \chidof of around 2,
 and the SaS2 parametrisations lie somewhere between these extremes.
 For the same reason the WHIT parametrisations fail to describe the
 ALEPH and DELPHI data which tend to be low at low values of $x$,
 thereby leading to large \chidof
 for the WHIT parametrisations, especially for the sets WHIT4-6
 which use $a=1$, as explained in Section~\ref{sec:PDF}.
 The only acceptable agreement is achieved by using the set WHIT1.
 The OPAL results tend to be high at low values of $x$ and also
 they have larger errors, therefore only the extreme cases
 WHIT5-6 lead to unacceptable values of \chidof.
 The L3 experiment quotes the smallest uncertainties on their
 results which tend to be high at low values of \qsq.
 Consequently, none of the parametrisations which are valid
 below $\qsq=4$~\gevsq is able to describe the L3 data and all lead
 to large values of \chidof.
 For $\qsq>4$~\gevsq the agreement improves but the values of \chidof
 are still too large, except for GRSc.
 For the parametrisations valid for $\qsq>4$~\gevsq the best agreement
 with the L3 data is obtained for WHIT1.
 This comparison shows that already at the present level of
 accuracy the measurements of \ft are precise enough to constrain
 the parametrisations and to discard those which predict
 a fast rise at low-$x$ driven by large gluon distribution functions.
 \par
 In conclusion, for the parametrisations valid for $\qsq>4$~\gevsq
 satisfactory agreement is found with the SaS2 and the WHIT1
 parametrisations, except for the measurements of TPC/2$\gamma$.
 For the parametrisations evolved from lower scales, agreement is found
 for $\qsq>4$~\gevsq with the exception of the L3 and TOPAZ data,
 and at lower values of \qsq they are not able to account
 for the TPC/2$\gamma$ and L3 results.
 \par
%
\begin{figure}[tbp]
\begin{center}
{\includegraphics[width=1.0\linewidth]{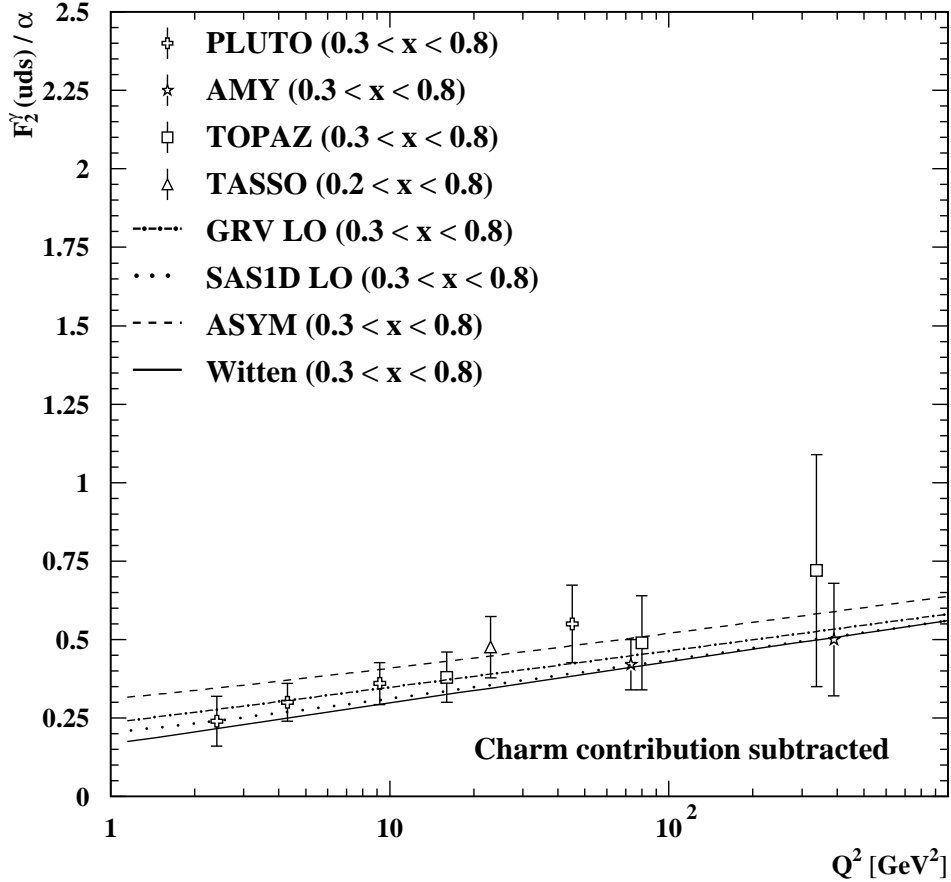}}
\caption[
         The measured \qsq evolution of \ft for three active flavours.
        ]
        {
         The measured \qsq evolution of \ft for three active flavours.
         The data with their full errors are compared to
         leading order predictions of \ft for the region $0.3<x<0.8$.
         Shown are the GRV and SaS1D parametrisations of \ft,
         the augmented asymptotic \ft (ASYM), and the purely asymptotic
         prediction (Witten), described in the text. The asymptotic
         predictions are evaluated for $\lam_3 = 0.232$~\gev.
        }\label{fig:chap7_16}
\end{center}
\end{figure}
%
 The second topic which is extensively studied using the large lever
 arm in \qsq, is the evolution of \ft with \qsq.
 The first measurements of this type were performed for \ft for three light
 flavours and the contribution to \ft from charm quarks
 was subtracted from the data based on the QPM prediction.
 This was motivated by the fact that at low values of \qsq the
 charm contribution is small and that the main aim of the analyses was
 to compare to the perturbative predictions for light quarks based
 on the asymptotic solution.
 At the time of most of the measurements no parton distribution
 functions of the photon were available.
 A summary of the published measurements of the \qsq evolution of \ftqd
 is given in Table~\ref{tab:chap13_14} and shown in
 Figure~\ref{fig:chap7_16}, where the point for TASSO has been obtained
 from combining the three middle bins listed in Table~\ref{tab:chap13_08}.
 The data are nicely described by the predictions from the SaS1D and
 GRV parametrisations of \ft.
 The purely asymptotic prediction from Ref.~\cite{WIT-7701}, using the
 parametrisation given in Ref.~\cite{GOR-9201} for $\lam_3 = 0.232$~\gev
 (Witten), predicts a slightly lower \ft than is seen in the data, whereas
 the augmented asymptotic solution (ASYM) is somewhat high compared to the
 data, but both are still consistent with the measured \ft.
 \par
%
\begin{figure}[tbp]
\begin{center}
{\includegraphics[width=0.95\linewidth]{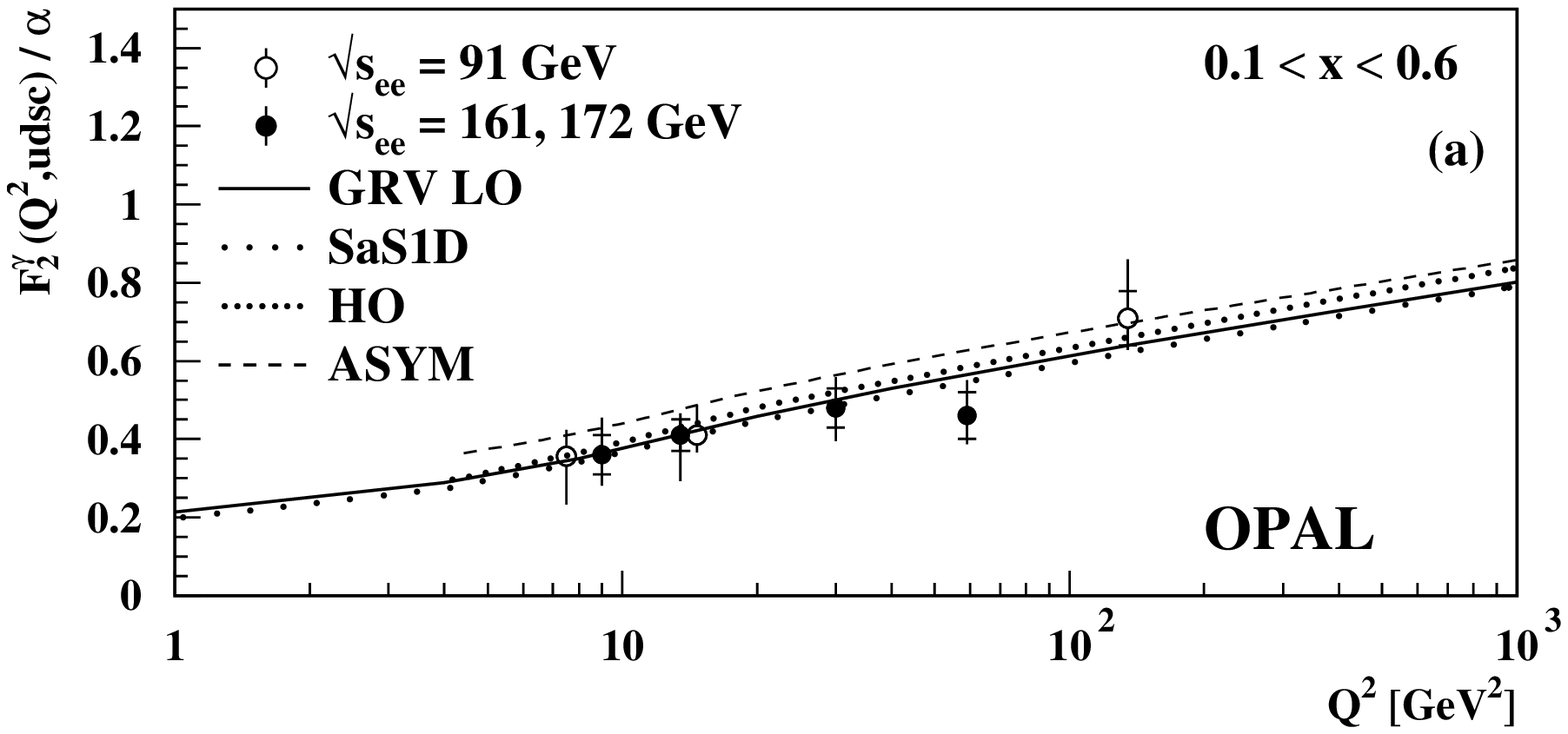}}
{\includegraphics[width=0.95\linewidth]{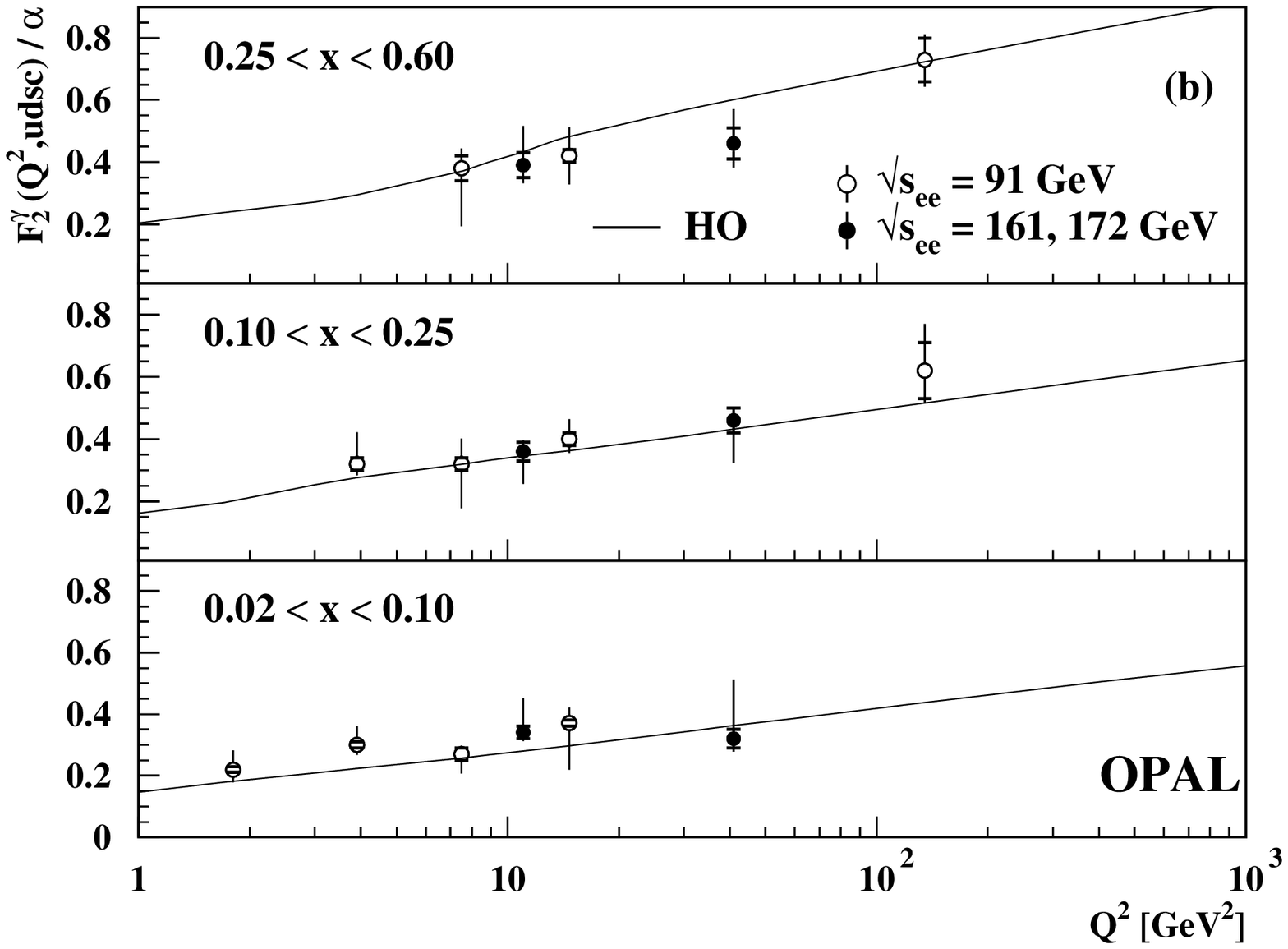}}
\caption[
         The measured \qsq evolution of \ft from OPAL.
        ]
        {
         The measured \qsq evolution of \ft from OPAL.
         The measurement of \ft is shown for four active flavours as
         a function of \qsq, in (a) for the range $0.1<x<0.6$, and in
         (b) subdivided into $0.02<x<0.10$, $0.10<x<0.25$ and $0.25<x<0.60$.
         In addition shown in (a) are the \ft of the GRV
         leading order (LO) and the SaS1D parametrisations,
         the \ft of the augmented asymptotic prediction
         (ASYM) and the result of a higher order calculation (HO),
         where the last two predictions are only shown for $\qsq >4$~\gevsq.
         In (b) the data are only compared to the higher order prediction.
         The points represent the OPAL data with their statistical
         (inner error bars) and total errors (outer error bars).
         In some of the cases the statistical errors are not visible
         because they are smaller than the size of the symbols.
        }\label{fig:chap7_17}
\end{center}
\end{figure}
%
 At higher values of \qsq the charm quark contribution to \ft gets
 larger and today also parametrisations of \ft for four active flavours
 are available.
 Consequently, more recent analyses of the evolution of \ft with
 \qsq are based on  measurements of \ft for four active flavours.
 In addition, due to the larger statistics available, the experiments
 start to look into the evolution using several ranges in $x$
 for the same value of \qzm.
 The first LEP measurement of this type is shown in
 Figure~\ref{fig:chap7_17}, taken from Ref.~\cite{OPALPR185}.
 In Figure~\ref{fig:chap7_17}(a) the result for the range $0.1<x<0.6$
 is compared to several parametrisations of \ft.
 Shown are the leading order
 (LO) predictions of the GRV and the SaS1D parametrisations, both
 including the contribution to \ft from massive charm quarks, and
 a higher order (HO) calculation provided by E.~Laenen, based on the
 GRV higher order parametrisation for three light quarks, complemented
 by the contribution of charm quarks to \ft based on the higher
 order calculation using massive charm quarks of Ref.~\cite{LAE-9401}.
 The differences between the three predictions are small compared to the
 experimental errors, and all predictions nicely agree with the data.
 In addition, the data are compared to the augmented asymptotic
 prediction as detailed above. This approximation lies higher than the
 data at low \qsq and approaches the data at the highest \qsq reached.
 \par
 The evolution of \ft with \qsq is measured by fitting a linear
 function of the form $a + b\,\ln (\qsq/\gevsq)$ to the data for the
 region $0.1<x<0.6$.
 Here $a$ and $b$ are parameters which are taken to be independent
 of $x$ within the bin in $x$ chosen.
 The fit to the OPAL data in the \qsq range of 7.5$-$135~\gevsq yields
%
 \begin{displaymath}
 \ftqn=(0.16 \pm 0.05^{+0.17}_{-0.16}) +
         (0.10 \pm 0.02^{+0.05}_{-0.02})\,\ln\,(\qsq)\, ,
 \end{displaymath}
%
 where \qsq is in \gevsq, with $\chidof = 0.77$ for the central value,
 as quoted in Ref.~\cite{OPALPR207}.
 The slope \slop is significantly different from zero but not
 precisely measured yet.
 \par
%
\begin{figure}[tbp]
\begin{center}
{\includegraphics[width=1.0\linewidth]{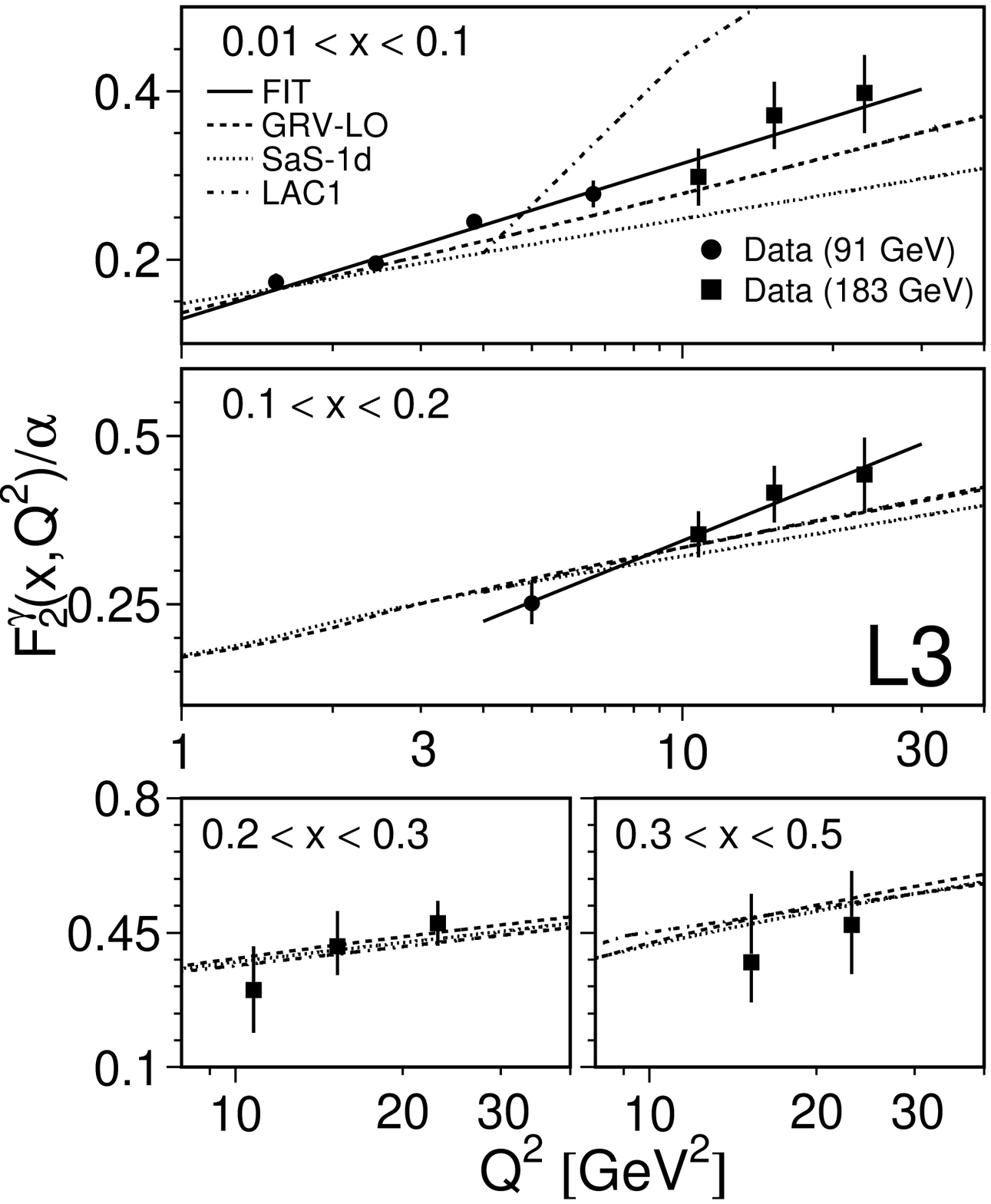}}
\caption[
         The measured \qsq evolution of \ft from L3.
        ]
        {
         The measured \qsq evolution of \ft from L3.
         The measurement of \ft is shown for four active flavours as
         a function of \qsq, for four ranges in $x$, $0.01<x<0.1$,
         $0.1<x<0.2$, $0.2<x<0.3$ and $0.3<x<0.5$.
         The data are compared to \ft from the leading order GRV,
         SaS1D, and LAC1 parametrisations.
         In addition shown is a fit to the data explained in the text.
         The points represent the L3 data with their total errors.
        }\label{fig:chap7_18}
\end{center}
\end{figure}
%
 The photon structure function \ft is expected to increase with \qsq
 for all values of $x$, but the size of the scaling violation is
 expected to depend on $x$, as shown in Figure~\ref{fig:chap4_07}.
 To examine whether the data exhibit the predicted variation in
 \slop, the \qsq range 1.86$-$135~\gevsq is analysed
 using common $x$ ranges.
 Figure~\ref{fig:chap7_17}(b) shows the measurement in comparison
 to the higher order calculation explained above.
 The points of inflection of \ft for \qsq below 15~\gevsq are caused
 by the charm threshold.
 The data show a slightly steeper rise with \qsq for increasing values
 of $x$, which is reproduced by the prediction of the higher order
 parametrisation of \ft.
 However, to experimentally observe the variation of \slop with $x$ the
 inclusion of more data and a reduction of the systematic error are needed.
 \par
%
\begin{figure}[tbp]
\begin{center}
{\includegraphics[width=1.0\linewidth]{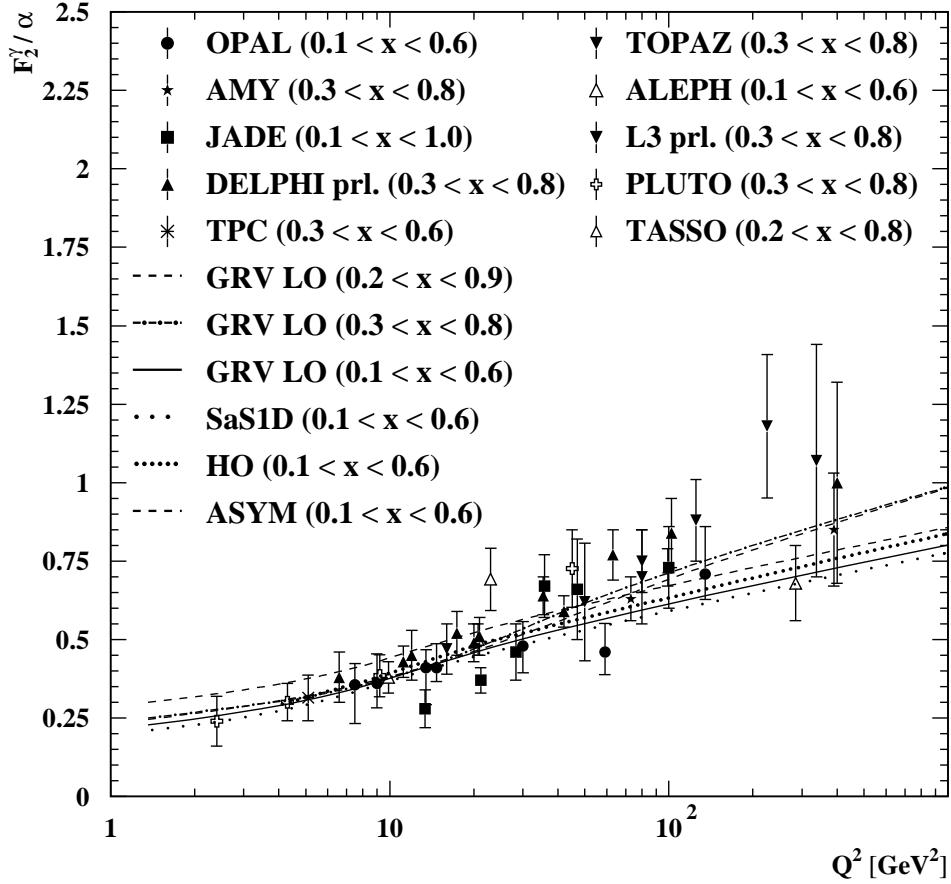}}
\caption[
         The measured \qsq evolution of \ft at medium $x$.
        ]
        {
         The measured \qsq evolution of \ft at medium $x$.
         The data with their full errors are compared to the
         predictions based on the leading order GRV and SaS1D
         parametrisations and to a higher order prediction (HO),
         as well as to an augmented asymptotic \ft (ASYM),
         both described in the text, all for the range $0.1<x<0.6$.
         In addition shown are the leading order GRV predictions for
         two other ranges in $x$, $0.2<x<0.9$ and $0.3<x<0.8$.
        }\label{fig:chap7_19}
\end{center}
\end{figure}
%
 A similar analysis performed by the L3 experiment is shown
 in Figure~\ref{fig:chap7_18}, taken from Ref.~\cite{L3C-9804}.
 Unfortunately the $x$ ranges are slightly different and the data
 cannot easily be combined.
 For $x>0.2$ the L3 data are described by all shown leading order
 parametrisations of \ft, from LAC1, SaS1D and GRV.
 For smaller values of $x$ some differences are seen.
 In the range $0.1<x<0.2$ the data show a steeper behaviour than
 what is predicted by the three parametrisations of \ft, and
 for $0.01<x<0.1$ they are higher than the SaS1D and GRV predictions,
 but show a similar slope, whereas the \ft based on the LAC1
 parametrisation predicts a much too fast rise with \qsq.
 \par
 The L3 data were fitted, as explained for the OPAL result above, in
 two regions of $x$, $0.01<x<0.1$ and $0.1<x<0.2$, using
 the \qsq range of 1.2$-$30~\gevsq. The results for the two regions are
%
 \begin{eqnarray*}
 \ftqn&=&(0.13\pm 0.01\pm 0.02)+(0.080\pm 0.009\pm 0.009)\ln\,(\qsq/\gevsq),\\
 \ftqn&=&(0.04\pm 0.08\pm 0.08)+(\phantom{0}0.13 \pm\phantom{0} 0.03
         \pm\phantom{0} 0.03)\ln\,(\qsq/\gevsq),
 \end{eqnarray*}
%
 with $\chidof$ of 0.69 and 0.13 for the central values.
 The results obtained by the L3 experiment are consistent with the OPAL
 result, which is valid for $0.1<x<0.6$.
 \par
%
\begin{figure}[tbp]
\begin{center}
{\includegraphics[width=1.0\linewidth]{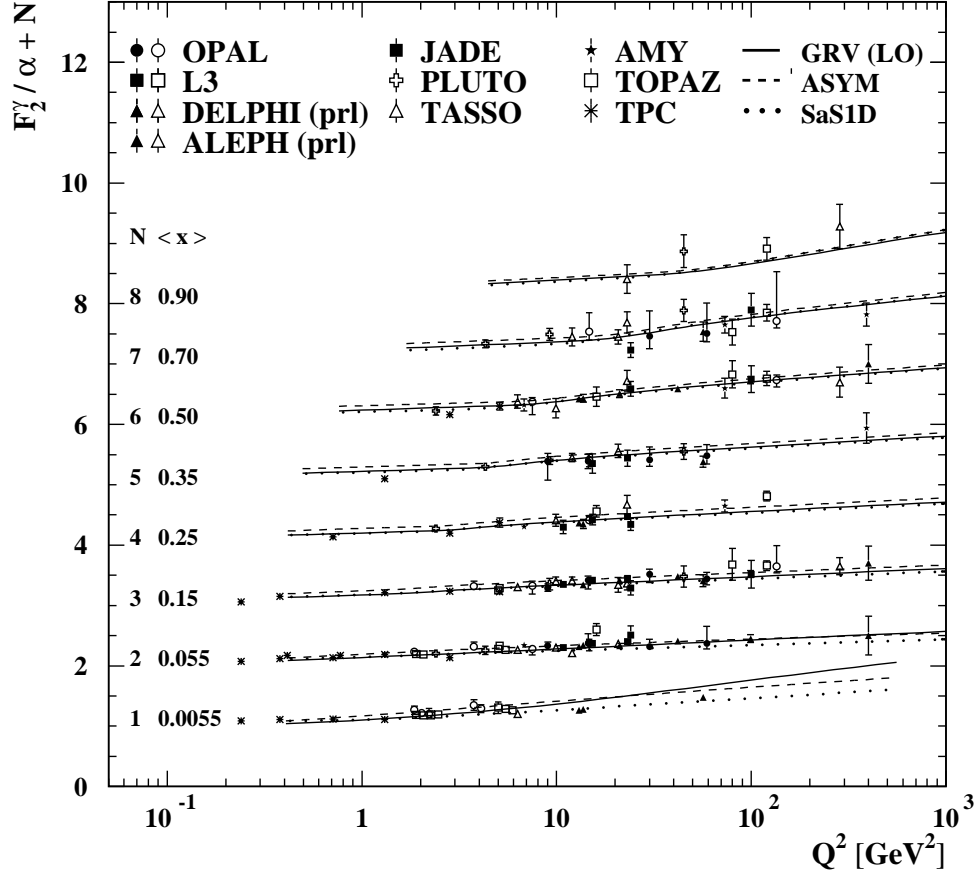}}
\caption[
         Summary of the measurements of the \qsq evolution of \ft.
        ]
        {
         Summary of the measurements of the \qsq evolution of \ft.
         The data with their full errors are compared to the
         predictions based on the leading order GRV and SaS1D
         parametrisations and to an augmented asymptotic \ft,
         described in the text.
         The bins used have boundaries of 0.001, 0.01, 0.1, 0.2, 0.3,
         0.4, 0.6, 0.8, 0.99, with central values as shown in the figure.
         To separate the measurements from each other an integer value, N,
         counting the bin number is added to the measured \ft.
        }\label{fig:chap7_20}
\end{center}
\end{figure}
%
 A collection of all available measurements of the evolution of
 \ft for four active flavours and at medium values of $x$ is listed
 in Table~\ref{tab:chap13_15} and shown in Figure~\ref{fig:chap7_19}.
 For the PLUTO result the average \ftc in the range $0.3<x<0.8$ for
 the \qzm values of the analyses has been added to the published
 three flavour result. The charm contribution has been obtained from
 Eq.~(\ref{eqn:BH}), for $\psq = 0$ and $\mc=1.5$~\gev.
 The only significant contribution from charm quarks is for
 $\qzm = 45$~\gevsq, where \ft increases from 0.55 to 0.73.
 As above for the three flavour result, the point for TASSO has
 been obtained from combining the three middle bins listed in
 Table~\ref{tab:chap13_08}.
 \par
%
\begin{figure}[tbp]
\begin{center}
{\includegraphics[width=1.0\linewidth]{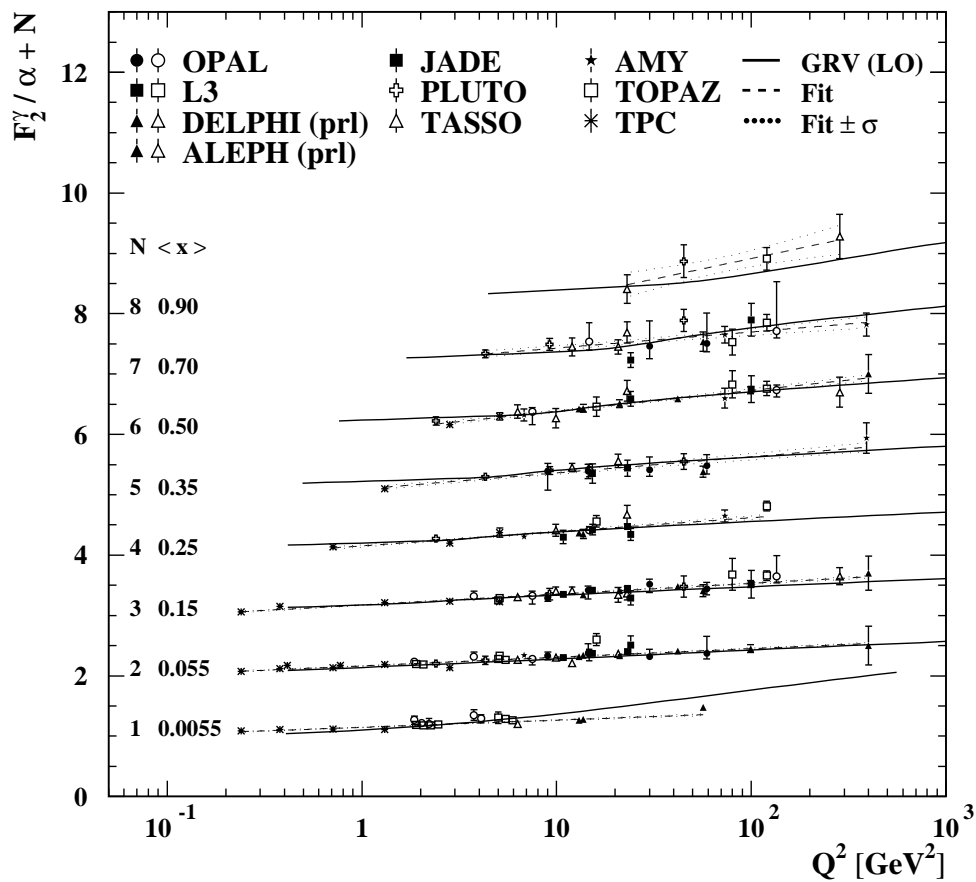}}
\caption[
         Measurements of the \qsq evolution of \ft compared to a linear fit.
        ]
        {
         Measurements of the \qsq evolution of \ft compared to a linear fit.
         The same data as in Figure\protect~\ref{fig:chap7_20}
         are compared to the results of linear fits using
         the function $a + b\,\ln (\qsq/\lamsq)$, with
         $\lam = 0.2$~\gev (dash) together with the errors of the fit (dot).
         In addition, the predictions of the leading order
         GRV parametrisations of \ft are shown as full lines.
        }\label{fig:chap7_21}
\end{center}
\end{figure}
%
 Unfortunately the different experiments quote their results for different
 ranges in $x$ which makes the comparison more difficult because
 the predictions for the various ranges in $x$ start to be significantly
 different for $\qsq>100$~\gevsq, as can be seen from the GRV predictions
 for three different ranges of $x$ shown in Figure~\ref{fig:chap7_19}.
 The measurements are consistent with each other and a clear rise of
 \ft with \qsq is observed.
 Again, this rise can be described reasonably well by the leading
 order augmented asymptotic prediction for $\lam_3 = 0.232$~\gev.
 \par
 In Figure~\ref{fig:chap7_19} only the medium $x$ region is studied.
 The large amount of data shown in Figure~\ref{fig:chap7_15}
 enables to investigate the variation of the scaling
 violation as a function of $x$ in more detail.
 For this purpose the data from Figure~\ref{fig:chap7_15} are displayed
 differently in Figure~\ref{fig:chap7_20}.
 The data are shown as a function of \qsq, divided in bins of $x$,
 with bin boundaries of 0.001, 0.01, 0.1, 0.2, 0.3,
 0.4, 0.6, 0.8, 0.99 and central values as shown in the figure.
 Each individual measurement is attributed to the bin with the closest
 central value in $x$ used.
 To separate the measurements from each other an integer value, N,
 counting the bin number is added to the measured \ft.
 The theoretical predictions are taken as the average \ft in the bin.
 The \qsq ranges used for the predictions are the maximum ranges
 possible for $1<W<250$~\gev and $\qnsq<\qsq<1000$~\gevsq, where
 \qnsq is the starting scale of the evolution for the respective
 parametrisation of \ft.
 The general trend of the data is followed by the predictions
 of the augmented asymptotic solution, and
 the GRV and SaS1D leading order parametrisations of \ft,
 however, differences are seen in specific ranges in $x$ which were
 discussed above in connection with Figure~\ref{fig:chap7_15}.
 \par
%
\renewcommand{\arraystretch}{1.1}
\begin{table}[tbp]
\caption[
         Fit results of the \qsq evolution of \ft.
        ]
        {
         Fit results of the \qsq evolution of \ft.
         Listed are the results for the parameters $a$ and $b$ of the
         fit using the function $a + b\,\ln (\qsq/\lamsq)$, with
         $\lam=0.2$~\gev. The errors are the one $\sigma$
         parameter errors as defined by the MINUIT program.
         The number of degrees of freedom is denoted with dof, and the
         correlation of the two parameters with cor.\\
        }\label{tab:chap7_03}
\begin{center}
\begin{tabular}{cccccc}\hline
        $x$     & $a\pm\sigma_a$   & $b\pm\sigma_b$    & dof&\chidof&cor\\
\hline
  $0.001- 0.01$ & $-0.02 \pm 0.01$ & $0.052 \pm 0.004$ & 18 & 1.69 &-0.95\\
  $0.01 - 0.1 $ & $-0.04 \pm 0.01$ & $0.062 \pm 0.003$ & 36 & 1.36 &-0.94\\
  $0.1  - 0.2 $ & $-0.08 \pm 0.02$ & $0.078 \pm 0.004$ & 33 & 0.51 &-0.92\\
  $0.2  - 0.3 $ & $-0.18 \pm 0.04$ & $0.102 \pm 0.009$ & 15 & 1.12 &-0.95\\
  $0.3  - 0.4 $ & $-0.28 \pm 0.08$ & $0.12  \pm 0.02 $ & 13 & 0.65 &-0.97\\
  $0.4  - 0.6 $ & $-0.44 \pm 0.08$ & $0.15  \pm 0.01 $ & 20 & 0.28 &-0.98\\
  $0.6  - 0.8 $ & $-0.21 \pm 0.17$ & $0.12  \pm 0.03 $ & 15 & 0.87 &-0.98\\
  $0.8  - 0.98$ & $-1.4  \pm 1.1 $ & $0.30  \pm 0.15 $ &  2 & 0.35 &-0.99\\
\hline\\
\end{tabular}
\end{center}\end{table}
%
 To quantify the increasing slope as function of \qsq for increasing values
 of $x$, the data are fitted, in bins of $x$ by a linear function of the form
 $a + b\,\ln (\qsq/\lamsq)$, with $\lam = 0.2$~\gev.
 The results of the fit are displayed in Figure~\ref{fig:chap7_21}
 and listed in Table~\ref{tab:chap7_03}.
 Because some of the data contain asymmetric errors, the central values
 and errors of the fit parameters are not obtained from analytically solving
 the problem, but rather the MINUIT program from Ref.~\cite{JAM-9501}
 has been used to perform the fit.
 The fitted values for the parameters $a$ and $b$, as well as their errors
 are given. The errors are calculated as the one $\sigma$ parameter errors
 defined by the MINUIT program, as explained in Ref.~\cite{JAM-9501}.
 They reflect the change of a given parameter, when the \chiq is changed
 from \chiqm to $\chiqm + 1$.
 The parameters $a$ and $b$ are almost 100$\%$ anticorrelated.
 The errors of the fitted functions are indicated in
 Figure~\ref{fig:chap7_21} using the full error matrix.
 For comparison the GRV leading order predictions are shown as well.
 Although the prediction, for example, from the GRV leading order
 parametrisation is not compatible with such a linear
 approximation, the data, at the present level of accuracy,
 can be fitted with linear functions with acceptable values of \chidof.
 In some cases \chidof is much smaller than unity
 indicating that using the full error is overestimating the errors.
 Consequently, the data are more precise and a combined fit with a careful
 estimation of the correlation of the errors should be performed soon.
 The results of the fits listed in Table~\ref{tab:chap7_03}, show
 a significant increase in slope for increasing $x$ in accordance
 with the theoretical expectation.
 \par
%
\begin{figure}[tbp]
\begin{center}
{\includegraphics[width=1.0\linewidth]{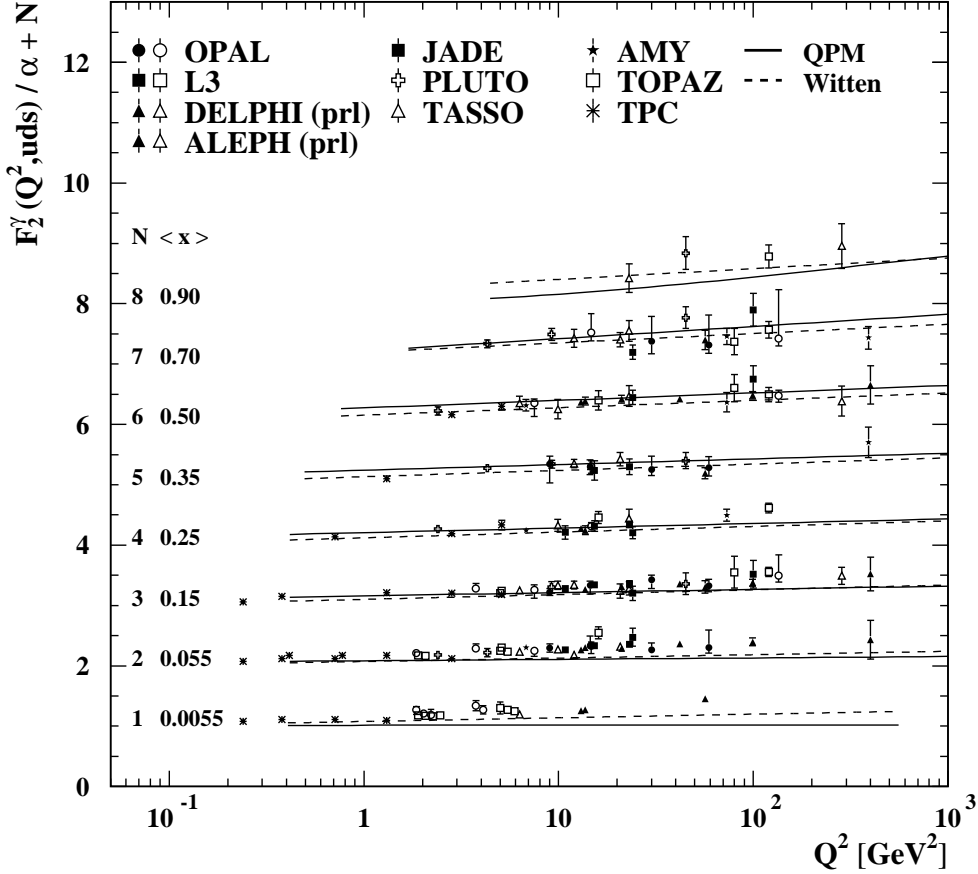}}
\caption[
         The \qsq evolution of \ft for three active
         flavours compared to the asymptotic solution and the QPM prediction.
        ]
        {
         The \qsq evolution of \ft for three active
         flavours compared to the asymptotic solution and the QPM prediction.
         The same data as in Figure\protect~\ref{fig:chap7_20} are used and
         the charm contribution is subtracted as explained in the text.
         The three flavour results are compared to the prediction of the
         asymptotic solution with $\lam_3 = 0.232$~\gev (Witten, full)
         and to the QPM prediction assuming $m_{q_k}=0.2$~\gev (QPM, dash).
        }\label{fig:chap7_22}
\end{center}
\end{figure}
%
 To compare the data more directly to the asymptotic solution of \ft,
 without the complication due to the heavy quark contribution, the
 charm quark contribution is subtracted from the measurements based
 on the point-like QPM prediction, Eq.~(\ref{eqn:BH}),
 for $\psq = 0$ and for a charm quark mass of $\mc=1.5$~\gev.
 In Figure~\ref{fig:chap7_22} the three flavour result is compared
 to the leading order asymptotic prediction from Ref.~\cite{WIT-7701},
 using the parametrisation given in Ref.~\cite{GOR-9201}, for
 $\lam_3 = 0.232$~\gev, and to the QPM prediction for the three
 light quarks assuming $m_{q_k}=0.2$~\gev.
 The values of \chidof as calculated from Eq.~(\ref{eqn:chiq})
 are given in in Table~\ref{tab:chap7_04}.
 \par
%
\renewcommand{\arraystretch}{1.1}
\begin{table}[tbp]
\caption[
         Comparison of the \qsq evolution of \ft to the asymptotic
         solution and to the QPM prediction.
        ]
        {
         Comparison of the \qsq evolution of \ft to the asymptotic
         solution and to the QPM prediction.
         Listed are the ranges in $x$ used, the number of data points
         in each range, dof, and the \chidof values for the predictions
         of \ft from the asymptotic solution (Witten), for
         $\lam_3 = 0.232$~\gev and the quark parton model (QPM),
         for $m_{q_k}=0.2$~\gev.\\
        }\label{tab:chap7_04}
\begin{center}
\begin{tabular}{cccc}\cline{3-4}
                &     &  Witten &      QPM \\
\hline
        $x$     & dof & \chidof & \chidof \\
\hline
  $0.001- 0.01$ &  20 &    14.7 &   56.9   \\
  $0.01 - 0.1 $ &  38 &    16.2 &   18.8   \\
  $0.1  - 0.2 $ &  35 &    3.76 &   2.85   \\
  $0.2  - 0.3 $ &  17 &    2.84 &   1.69   \\
  $0.3  - 0.4 $ &  15 &    1.01 &   1.86   \\
  $0.4  - 0.6 $ &  22 &    1.02 &   1.27   \\
  $0.6  - 0.8 $ &  17 &    0.84 &   0.99   \\
  $0.8  - 0.98$ &   4 &    0.76 &   1.94   \\
\hline
\end{tabular}
\end{center}\end{table}
%
 At low values of $x$ both predictions undershoot the data, and the
 agreement improves with increasing values of $x$.
 For $x>0.3$ the asymptotic prediction gives a slightly better
 description than the QPM prediction resulting in \chidof values
 around 1.
 It is a very interesting observation that the perturbative prediction
 is able to describe the behaviour for large $x$ for a reasonable
 value of the only free parameter $\lam_3$.
 \par
 To make a more quantitative statement on the description
 of the measured \ft by the perturbative prediction, an
 $x$-dependent parametrisation of the next-to-leading order asymptotic
 prediction must be available.
 Then, the data should be compared to the next-to-leading order
 asymptotic prediction to fix the QCD scale \lam, with a proper
 definition of the region of validity of this approximation to
 avoid the singularities.
 In addition, then the charm subtraction could also be based on the
 next-to-leading order calculation from Ref.~\cite{LAE-9401}.
 Finally the contribution of the hadron-like component of the charm
 production should be investigated, especially at low values of $x$.
%
%
\subsection{Hadronic structure of virtual photons}
\label{sec:qcdresvirt}
 The structure functions of virtual photons can be determined
 in the region $\qsq\gg\psq\gg\lamsq$ by measuring the cross-sections
 for events where both electrons are observed.
 In this region Eq.~(\ref{eqn:truedb}) can be used to define an
 effective structure function \feff as explained in
 Section~\ref{sec:cross}.
 Due to the \psq suppression of the cross-section these measurements
 suffer from low statistics.
 \par
%
\begin{figure}[tbp]
\begin{center}
{\includegraphics[width=0.7\linewidth]{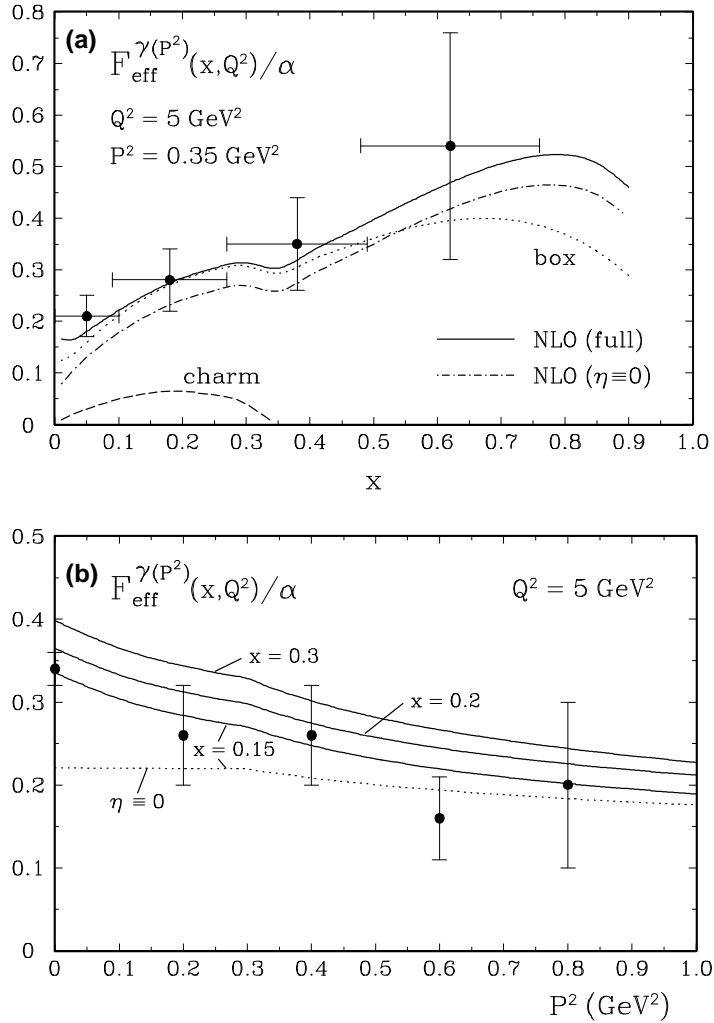}}
\caption[
         The effective photon structure function \feff from PLUTO.
        ]
        {
         The effective photon structure function \feff from PLUTO.
         In (a) \feff as a function of
         $x$ is compared to the next-to-leading order (NLO) result
         of the GRS parametrisation of the parton distribution
         functions of virtual photons with, and without, including
         a hadron-like component at the starting scale
         of the evolution, denoted with NLO (full) and NLO ($\eta=0$)
         respectively.
         In (b) the \psq evolution is shown in comparison to the full
         next-to-leading order result for three values of $x$,
         0.3, 0.2 and 0.15, and
         in comparison to the prediction for $\eta=0$ at $x=0.15$.
        }\label{fig:chap7_23}
\end{center}
\end{figure}
%
 The first measurement of this type has been performed by the
 PLUTO experiment in Ref.~\cite{PLU-8405},
 for $\pzm=0.35$~\gevsq and $\qzm=5$~\gevsq.
 In Figure~\ref{fig:chap7_23}, taken from Ref.~\cite{GLU-9501},
 the PLUTO measurement is compared to the theoretical
 predictions from the GRS parametrisations.
 In Figure~\ref{fig:chap7_23}(a) the
 structure function \feff as a function of $x$ is compared
 to three theoretical predictions in next-to-leading order.
 The best description of the data is obtained using the next-to-leading
 order result including a non-perturbative input at the starting scale
 of the evolution.
 If the hadron-like input is neglected, which corresponds to $\eta=0$,
 the prediction is consistently lower than the data, but still
 consistent with it, within the experimental errors.
 Also the prediction obtained by calculating \feff solely from the
 box diagram is still consistent with the data, although it is the lowest
 at high values of $x$.
 In Figure~\ref{fig:chap7_23}(b) the structure function \feff is shown as
 a function of \psq, including the result of \ft for the quasi-real photon.
 The measurement suggests a slow decrease with increasing \psq, but
 it is also consistent with a constant behaviour.
 The full next-to-leading order prediction is shown for three values
 of $x$ and, for $x=0.15$, in addition the purely perturbative solution
 is shown.
 Also for the \psq evolution  the full next-to-leading order prediction
 gives the best description of the data and the purely perturbative
 prediction is at the low end.
 \par
%
\begin{figure}[tbp]
\begin{center}
{\includegraphics[width=0.7\linewidth]{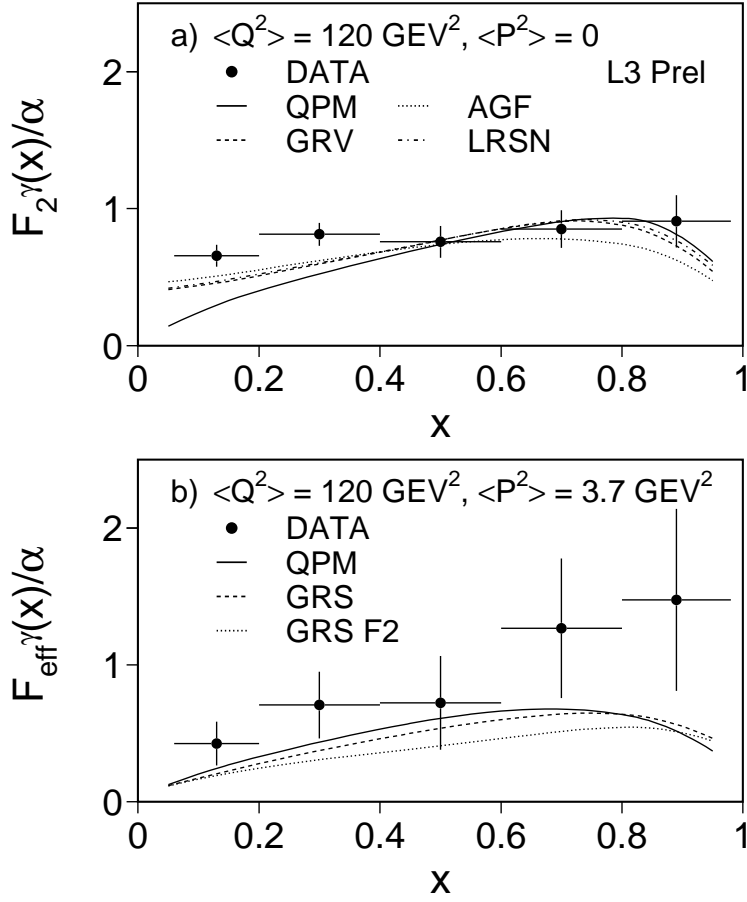}}
\caption[
         The effective photon structure function \feff from L3.
        ]
        {
         The effective photon structure function \feff from L3.
         In (a) the preliminary measurement of
         the structure function \ft for quasi-real photons is
         compared to the quark parton model prediction (QPM), to \ft as
         predicted by the GRV and AFG parametrisation of the parton
         distribution functions of real photons, and to a higher order
         calculation based on Ref.\protect~\cite{LAE-9401},
         denoted with LRSN.
         In (b) the effective structure function \feff is compared to the QPM
         prediction, to \ft predicted from the GRS parametrisation of
         the parton distribution functions of virtual photons (GRS F2),
         and to the full GRS prediction obtained from the GRS \ft together
         with the contribution of \fl as given by the  quark parton model.
        }\label{fig:chap7_24}
\end{center}
\end{figure}
%
 Recently preliminary results of a  similar measurement, using the full
 data taken at LEP1 energies, has been presented by the L3 experiment in
 Ref.~\cite{ERN-9901}.
 The average virtualities for the L3 result are $\qzm = 120$~\gevsq
 and $\pzm = 3.7$~\gevsq.
 In Figure~\ref{fig:chap7_24}, taken from Ref.~\cite{ERN-9901}, the
 measurement of \ft for quasi-real photons and the effective structure
 function, both as  functions of $x$, are compared to several
 theoretical predictions.
 As in the case of PLUTO the QPM result is too low compared to the
 data. Taking only \ft as calculated from the GRS parametrisation of the
 parton distribution functions of the photon, labelled as
 GRS F2, gets closer to the data, and the best description is found
 if the contribution of \fl, based on the prediction from the QPM,
 is added to this, denoted with GRS.
 The data show a faster rise with $x$ than any of the predictions,
 however with large errors for increasing $x$, which are mainly due
 to the low statistics available.
 \par
%
\begin{figure}[tbp]
\begin{center}
{\includegraphics[width=0.7\linewidth]{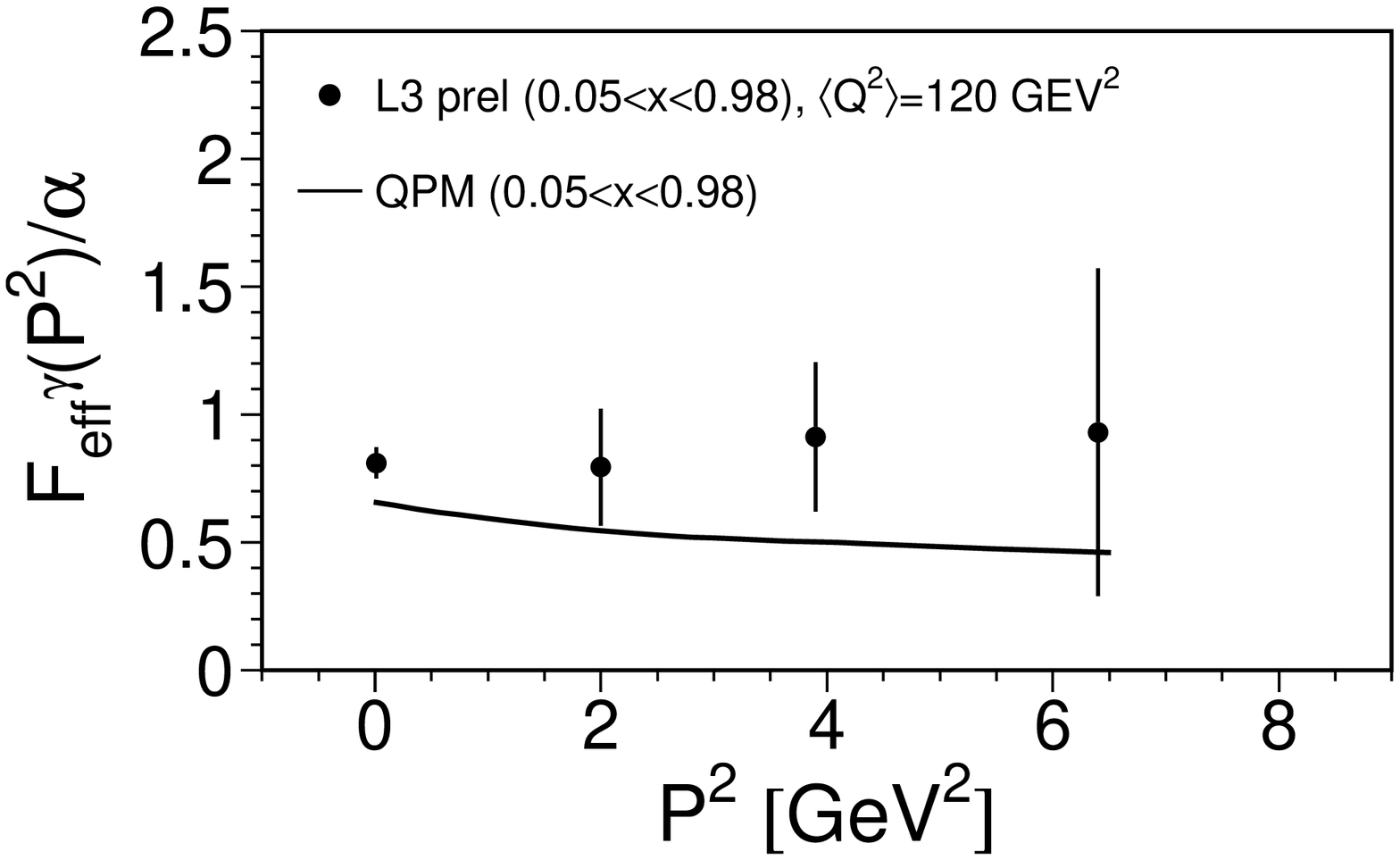}}
{\includegraphics[width=0.7\linewidth]{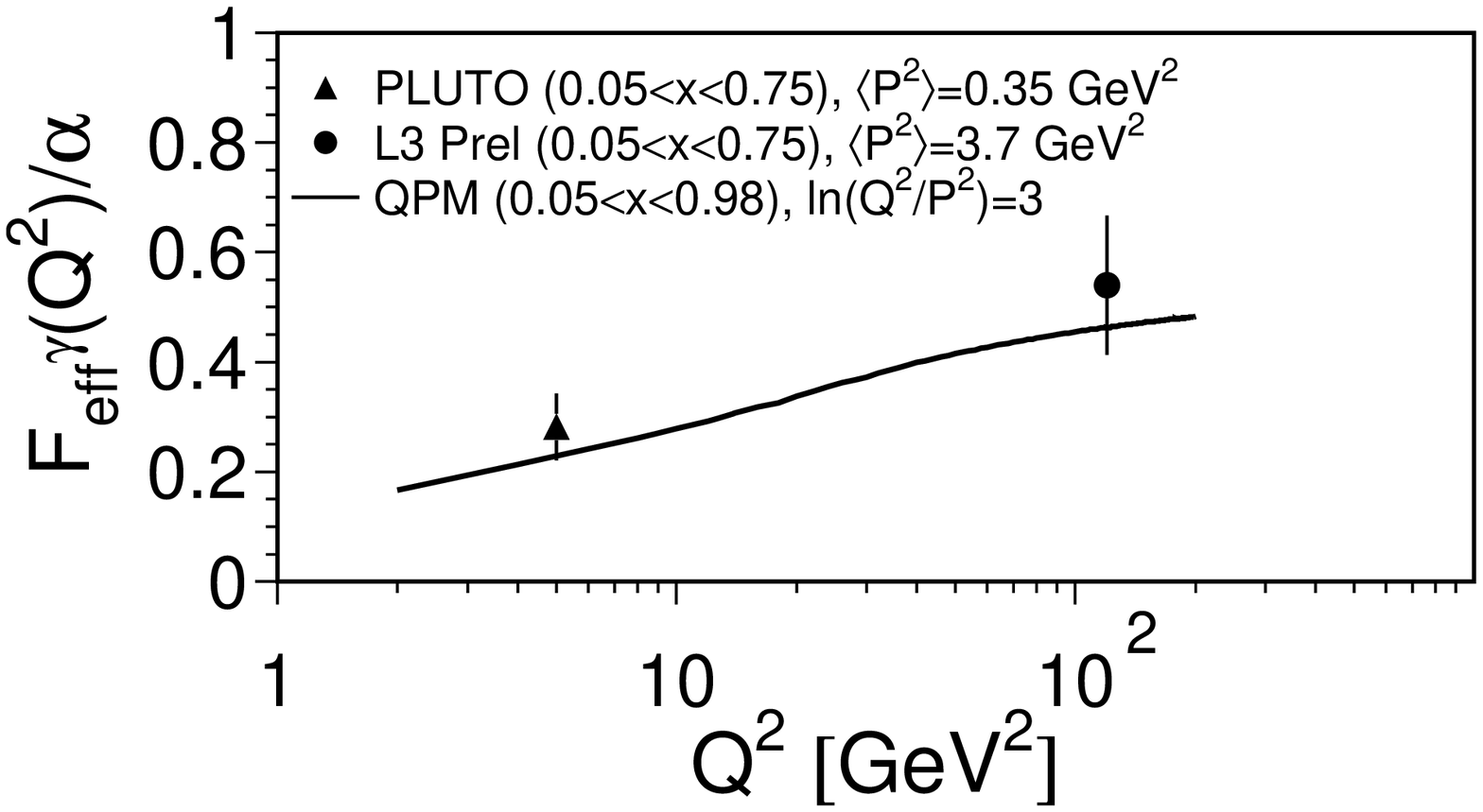}}
\caption[
         The virtuality dependence of \feff.
        ]
        {
         The virtuality dependence of \feff.
         In (a) the \psq dependence of the L3 result of \feff is compared
         to the quark parton model result (QPM), and in (b) the
         \qsq dependence of the results from PLUTO and L3
         are shown in comparison to the QPM prediction, for the average
         value of $\ln(\qzm/\pzm)$, which is around three.
        }\label{fig:chap7_25}
\end{center}
\end{figure}
%
 The \psq evolution of the L3 result of \feff is shown in
 Figure~\ref{fig:chap7_25}(a).
 The QPM prediction is consistent in shape with the data, but the
 predicted \feff is too low.
 However, the main difference comes from \ft for $\psq=0$, which
 is not described by the quark parton model for $x<0.4$.
 This is expected, because in this region the hadron-like part is
 predicted to be largest, as can be seen from
 Figure~\ref{fig:chap7_24}(a).
 But in this region the data are even higher than the predictions
 of all parametrisations of \ft which do contain a hadron-like
 contribution.
 The measurement for $\psq>0$ cannot rule out the quark parton model
 prediction, although it is consistently higher and does not favour
 the QPM prediction.
 The ratio of \qzm/\pzm is similar for the PLUTO and the L3 measurements,
 leading to values for $\ln(\qzm/\pzm)$ of 2.6 and 3.5 respectively.
 This enables to compare the \qsq evolution of the two measurements,
 as shown in Figure~\ref{fig:chap7_25}(b), for $0.05<x<0.75$.
 The evolution is consistent with the expectation of the quark parton
 model for $\ln(\qsq/\psq)=3$, and using the range $0.05<x<0.98$.
 \par
 In summary a consistent picture is found for the effective
 structure function \feff of the virtual photon between the PLUTO and
 preliminary L3 data and the general features of both
 measurements are described by the next-to-leading order predictions.
 However, the data do not constrain the predictions strongly and
 for detailed comparisons to be made the full statistics of the LEP2
 programme has to be explored.
 \par
 If both photons have similar virtualities the photon
 structure function picture can no longer be applied and the data are
 interpreted in terms of the differential cross-section.
 Due to the large virtualities the cross-section is small and
 large integrated luminosities are needed to precisely measure it.
 The main interest is the investigation of the hadronic structure
 of the interaction of two virtual photons.
 However,
 the interest in performing these measurements increased considerably
 in the last years, because calculations in the framework of the
 leading order BFKL evolution equation, which sums $\ln(1/x)$
 contributions, predicted a large cross-section for this
 kinematical range, see Refs.~\cite{BAR-9601,BAR-9701,BRO-9701,BRO-9702}.
 The predicted cross-section is so large that already measurements
 with low statistics are able to decide whether the BFKL picture is
 in agreement with the experimental observations.
 Recently theoretical progress has been made in
 Refs.~\cite{FAD-9801,CAM-9801} to also include next-to-leading order
 pieces in the calculations in the BFKL picture, as explained
 in Ref.~\cite{BRO-9901}.
 Large negative corrections to the leading order results were found,
 for example, shown in Refs.~\cite{BON-9801}. Consequently, there is
 some doubt about the perturbative stability of the BFKL calculation.
 The theoretical development is underway and this should be kept in mind
 in all comparisons to the BFKL predictions.
 The most suitable region for the comparison is
 $\wsq\gg\qsq\approx\psq\gg\lamsq$, which ensures
 similar photon virtualities and large values of 1/$x$.
 This is needed to ensure only little evolution in \qsq and large
 contributions from $\ln(1/x)$ terms.
 These requirements strongly reduce the available statistics,
 therefore compromises have been made in these comparisons.
 The variable of interest is
%
 \begin{equation}
  Y = \ln\frac{2p\cdot q}{\sqrt{\qsq\psq}}
    = \ln\frac{\wsq+\psq+\qsq}{\sqrt{\qsq\psq}}
    \approx \ln\frac{\wsq}{\sqrt{\qsq\psq}}
 \label{eqn:Y}
 \end{equation}
%
 where the approximation is only valid for $\wsq\gg\qsq,\psq$.
 However, experimentally this inequality is not very strong.
 \par
%
\begin{figure}[tbp]
\begin{center}
{\includegraphics[width=0.9\linewidth]{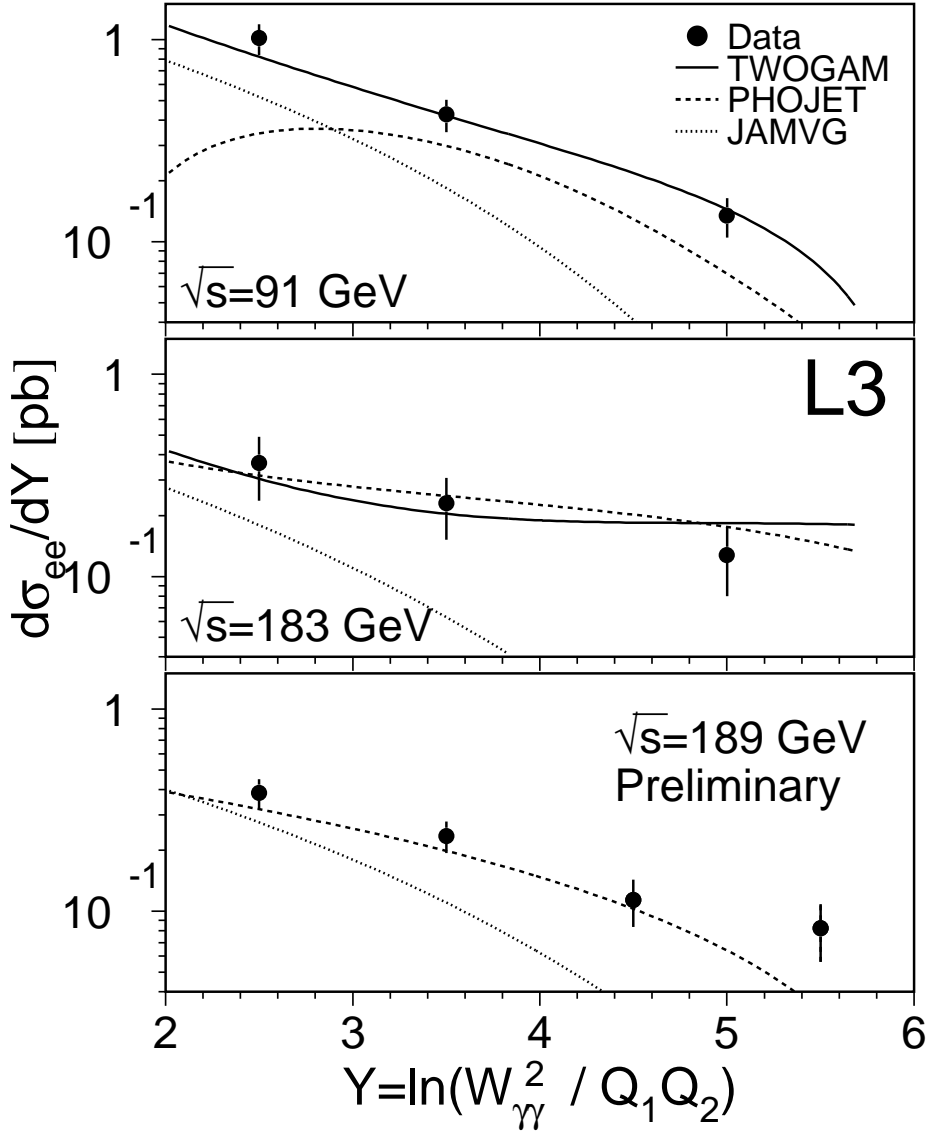}}
\caption[
         The hadronic cross-section for the exchange of two virtual
         photons from L3.
        ]
        {
         The hadronic cross-section for the exchange of two virtual
         photons from L3.
         The data of the L3 experiment are compared to the predictions
         of the PHOJET and TWOGAM Monte Carlo models and to an
         estimation in the framework of the quark parton model obtained
         from the Vermaseren Monte Carlo (JAMVG).
        }\label{fig:chap7_26}
\end{center}
\end{figure}
%
 The first measurement in this kinematical region has been performed by
 the L3 experiment in Ref.~\cite{L3C-9903} using data at LEP1 energies,
 $\ssee=91$~\gev,
 and at $\ssee=183$~\gev, and in addition preliminary results were reported
 recently in Ref.~\cite{ACH-9901} for data taken at $\ssee=189$~\gev.
 Electrons are selected for energies above 30~\gev at $\ssee=91$~\gev
 and for energies above 40~\gev at higher centre-of-mass energies.
 The average photon virtualities \qzm,\pzm are 3.5, 14 and 14.5~\gevsq
 respectively. The $W$ ranges used are $2-30 / 5-70 / 5-75$~\gev for
 the three centre-of-mass energies, which means the lowest values of
 $x$ probed are about $2-3\cdot 10^{-3}$, and the inequality
 $\wsq\gg\qsq,\psq$ reads $4\gg3.5$ at LEP1 and $25\gg14$ at LEP2
 energies, when using the minimum value of $W$ and the average
 virtualities.
 The differential cross-section as a function of
 $Y=\ln(\wsq/\sqrt{\qsq\psq})$ is shown in Figure~\ref{fig:chap7_26},
 taken from Ref.~\cite{ACH-9901}.
 The data are described by the TWOGAM Monte Carlo for
 $\ssee=91$~\gev and $\ssee=183$~\gev.
 The PHOJET model gives an adequate description at $\ssee=183$~\gev and
 $\ssee=189$~\gev, whereas it fails to describe the data $\ssee=91$~\gev,
 probably due to the low cut in $W$ applied for this data.
 For low values of $W$ the PHOJET Monte Carlo is known to
 be not very reliable, as explained in Section~\ref{sec:gener}.
 A prediction in the framework of the quark parton model obtained from
 the Vermaseren Monte Carlo is found to be too low at all energies.
 The presently predicted cross-sections by the BFKL calculation (not shown)
 are much higher than the measurements and are ruled out by the data.
 \par
%
\begin{figure}[tbp]
\begin{center}
{\includegraphics[width=1.0\linewidth]{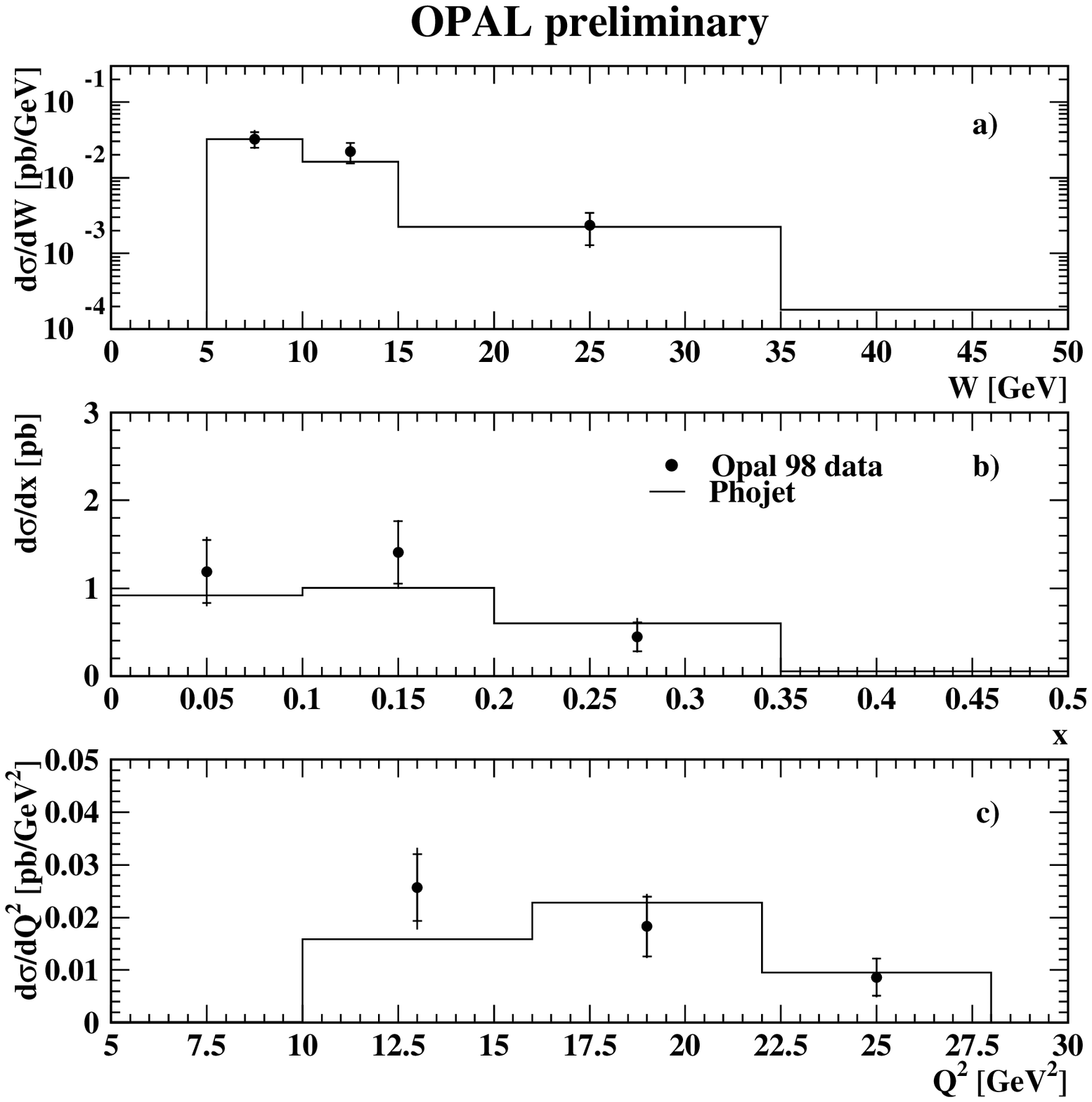}}
\caption[
         The hadronic ross-section for the exchange of two virtual
         photons from OPAL.
        ]
        {
         The hadronic cross-section for the exchange of two virtual
         photons from OPAL.
         The data of the OPAL experiment are compared to the prediction
         of the PHOJET Monte Carlo model for data taken at $\ssee=189$~\gev.
         In (a)$-$(c) the corrected cross-sections are shown as functions of
         $W$, $x$ and \qsq.
        }\label{fig:chap7_27}
\end{center}
\end{figure}
%
 A similar analysis has been performed by the OPAL experiment.
 Preliminary results were presented in Ref.~\cite{PRY-9901}
 based on data at $\ssee=189$~\gev, for an integrated luminosity of
 about $170$~\invpb, with average photon
 virtualities of about $10$~\gevsq, and for $W>5$~\gev.
 The differential cross-section as functions of $W$, $x$ and \qsq,
 corrected to the phase space defined by $\eonep,\etwop>65$~\gev,
 $34<\tonep,\ttwop<55$~mrad, and $W>5$~\gev, are shown in
 Figure~\ref{fig:chap7_27}, taken from Ref.~\cite{PRY-9901}.
 Due to the larger electron energies required in the OPAL analysis
 compared to the L3 result, the reach in $W$ for OPAL is only
 about $W=35$~\gev. This means the smallest value of $x$ reached is
 only about $8\cdot 10^{-3}$.
 The measured cross-section in the selected phase space is
 $0.32 \pm 0.05 ({\rm stat})\,^{+0.04}_{-0.05}({\rm sys})$~\pb,
 compared to the predicted cross-sections of $0.17$~pb for PHOJET and
 $2.2 / 0.26$~pb based on the BFKL calculation in leading/higher order.
 Also for the OPAL analysis, the data at $\ssee=189$~\gev are perfectly
 described by the PHOJET model and there is no room for large additional
 contributions.
 The precision of the results on the differential cross-sections are
 limited by the low statistics and they can considerably be improved
 by using the full statistics of the LEP2 programme.
%
%


%
%
\section{Future of structure function measurements}
\label{sec:future}
 As discussed in the previous sections, the QED and the hadronic
 structure of the photon have been measured up to average photon
 virtualities of $\qzm = 130$ and $\qzm = 400$~\gevsq respectively.
 In the future, the analysis of the photon structure can be extended
 to higher photon virtualities firstly, by exploring the high luminosity
 of the complete LEP2 programme and secondly, by using the full potential of
 the planned linear collider project.
 The prospects of these two parts of the future of structure function
 measurements are discussed briefly in this section.
%
%
\subsection{LEP2 programme}
\label{sec:futlep}
 So far the measurements of the QED structure of the photon are based
 on data taken at LEP1 energies for integrated luminosities of about
 $100$~\invpb per experiment and the results are mainly limited by
 statistics.
 Therefore, exploring the full integrated luminosity of $500$~\invpb
 expected at LEP2 energies, a reduction of the error by about a
 factor of two can be expected.
 \par
 For the case of the hadronic structure of the photon the situation is
 more difficult.
 Certainly the measurement can be extended to higher values of \qsq.
 The preliminary data from DELPHI already reach $\qzm=400$~\gevsq
 and, using the full data sample at LEP2 energies, decent statistics
 up to $\qzm=1000$~\gevsq can be reached.
 For the measurement of the hadronic structure at low values of \qsq
 the situation is different.
 Also in future this measurement suffers from theoretical
 uncertainties and
 considerable improvement in the description of the hadronic final state
 by the Monte Carlo models is needed first, before the experimental
 measurements can get more precise.
 \par
 The measurement of the hadronic structure for the exchange of two
 virtual photons suffers from low statistics, therefore
 using the expected $500$~\invpb at LEP2 energies will help to bring down
 the statistical errors, but it should be kept in mind that the results
 for the region $\qsq\approx\psq\gg\lamsq$
 are already based on about $200$~\invpb of data.
%
%
\subsection{A future linear collider}
\label{sec:nlc}
 At a future linear collider the measurements of the
 photon structure can be extended to larger photon virtualities
 and larger photon-photon centre-of-mass energies.
 A recent discussion of the prospects of these measurements
 can be found in Ref.~\cite{NIS-9802}.
 In particular the measurements of the photon structure functions
 can be performed at much higher values of \qsq. This is the subject
 of this section.
 \par
 The linear collider is an extension of the existing
 \epem colliders LEP and SLC.
 Table~\ref{tab:chap8_01} shows the improvements on several machine
 parameters which have to be achieved to arrive at a luminosity of the
 order of 10$^{34}/$cm$^2$s, which would lead to an integrated luminosity
 of about 100 fb$^{-1}$ per year of operation.
 \par
\begin{figure}[tbp]
\begin{center}
{\includegraphics[width=0.9\linewidth]{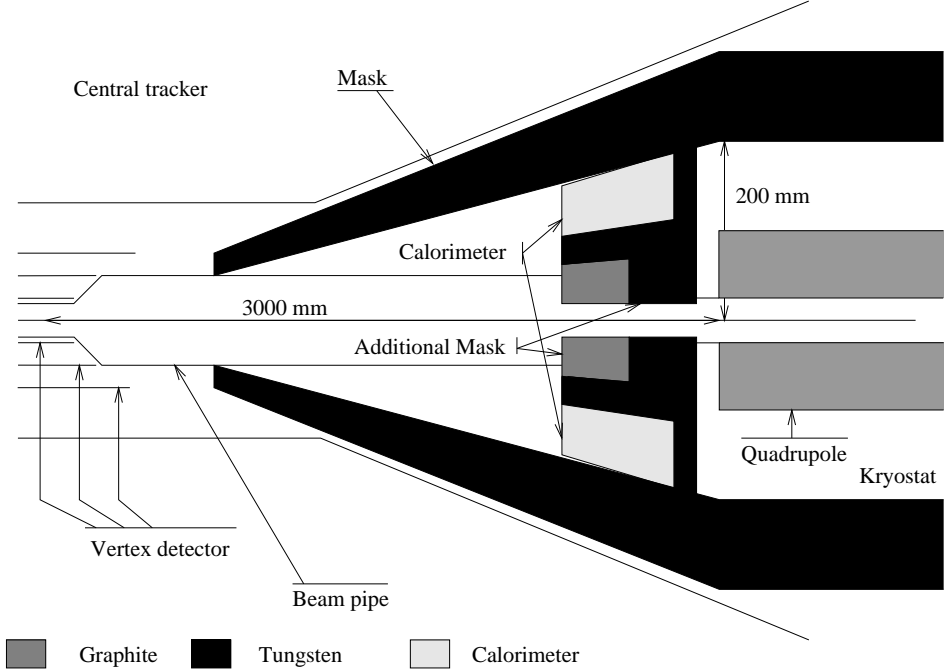}}
\caption[
         The forward region of a future linear collider detector.
        ]
        {
         The forward region of a future linear collider detector.
         A sketch of the proposed mask to protect the detector from
         the background for the TESLA design.
        }\label{fig:chap8_01}
\end{center}
\end{figure}

%
\renewcommand{\arraystretch}{0.92}
\begin{table}[tbp]
\caption[
         Parameters for a future linear collider of the TESLA design.
        ]
        {
         Parameters for a future linear collider of the TESLA design.
         Some approximate values of parameters of the present LEP
         and SLC colliders are shown together with goals for a future
         linear collider of the TESLA design.\\
        }\label{tab:chap8_01}
\begin{center}\begin{tabular}{rrccc}
\hline
              &                       &  LEP
                                      &  SLC
                                      &  TESLA     \\
\hline
 total length & [km]                  &  26.7
                                      &  4
                                      & 33         \\
 gradient     & [MV/m]                &  6
                                      &  10
                                      &  25         \\
 beam size  $\sigma_x/
 \sigma_y$    & [$\mu$m/$\mu$m]       &  110 / 5
                                      &  1.4 / 0.5
                                      &  0.845/0.019 \\
 electron energy& [GeV]               &  100
                                      &  50
                                      &  250        \\
 luminosity   & [10$^{31}/$cm$^2$s]   &  7.4
                                      &  0.1
                                      &  5000-10000 \\
 \lumi        & [1/pb y]              &  200
                                      &  15
                                      &  20000      \\\hline\\
\end{tabular}
\end{center}\end{table}
%
 The linear collider, even when running under optimal conditions will
 produce a huge amount of background where many particles are produced
 especially in the forward regions of the detector.
 Detailed background studies for the linear collider were performed.
 The background sources will lead mainly to \epem pair creation and
 to hadronic background.
 For the \epem collision mode the background simulation of
 Ref.~\cite{SCU-9801} showed that the amount of background expected
 per bunch crossing for the TESLA design is about $10^5$ \epem pairs
 with a total energy of $1.5\cdot10^5$~\gev and about 0.13 events of
 the type $\GG\rightarrow {\rm hadrons}$ for hadronic masses $W>5$~\gev
 with an average visible energy of 10~\gev.
 To accommodate this background the main detector has to be
 shielded with a massive mask as shown, for example for the TESLA
 design, in Figure~\ref{fig:chap8_01}, taken from Ref.~\cite{SCU-9801}.
 In addition, the photon radiation from the beam electrons will also lead
 to a significant energy smearing for the electrons of the beams.
 The prospects of structure function measurements have to be discussed
 in the context of this expected machine parameter dependent
 'soft' underlying background, and the energy spread of the beam
 electrons.
 \par
 There is also a strong interest in the construction of a photon linear
 collider which would in several aspects be complementary to an
 electron linear collider.
 For several reactions the cross sections for incoming
 photons are larger than for incoming electrons of the same energy.
 In addition, some reactions, for example, the very important process
 $\GG\rightarrow H$ only have sufficiently large event rates, when using the
 large flux of high energetic incoming photons from a photon linear collider.
 The linear collider, when operated in the electron-photon mode,
 would also be an ideal source of high energetic photons for structure
 function measurements, because the energy of the incoming real photons
 would be known rather precisely.
 \par
%
\begin{figure}[tbp]
\begin{center}
{\includegraphics[width=0.9\linewidth]{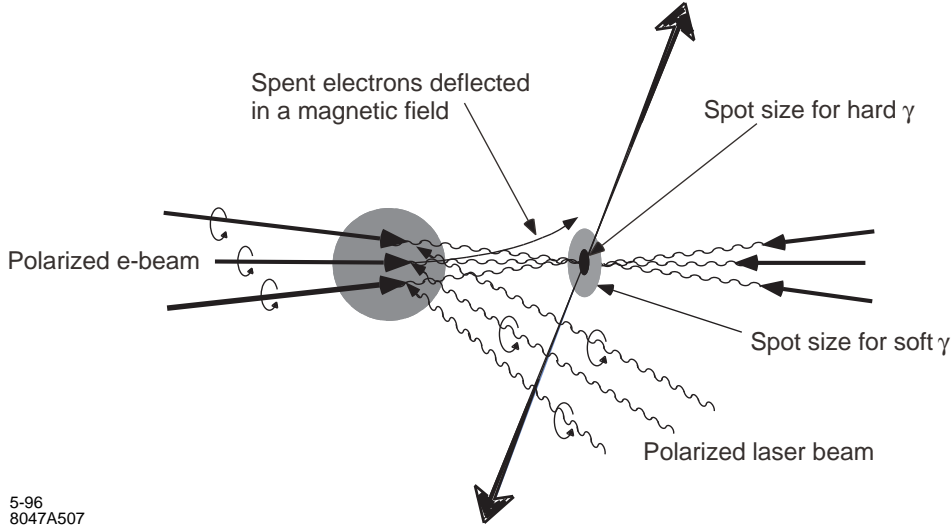}}
\caption[
         The production mechanism for high energetic photons.
        ]
        {
         The production mechanism for high energetic photons.
         A sketch of the creation of the photon beam by Compton
         backscattering of laser photons off the beam electrons.
        }\label{fig:chap8_02}
\end{center}
\end{figure}

%
 The construction of a photon linear collider is very demanding and
 only general concepts are available so far.
 The method to produce a beam of high energetic photons from
 an electron beam by means of the Compton backscattering process is
 shown in Figure~\ref{fig:chap8_02}, taken from Ref.~\cite{ADO-9601}.
 The photons are produced by a high intensity laser and brought into
 collision with the electron beams at distances of about $0.1-1$~cm
 from the interaction point.
 The photons are scattered into a small cone around the initial electron
 direction and receive a large fraction of the electron energy.
 By properly adjusting the machine parameters, like the
 distance along the beam line between of the production
 of the backscattered photons and the interaction region,
 by selecting the polarisations of the laser and the electron beams,
 and by magnetic reflection of the spent beam,
 the energy spectrum of the photons can be selected.
 As a result of this a typical distribution of \eG luminosity as a
 function of the invariant mass of the electron photon system, $\sega$,
 is expected to peak at the maximum reachable invariant mass of around
 $0.8\ssee$ with a width of 5$\%$, as described in Ref.~\cite{TEL-9801}.
 \par
%
\begin{figure}[tbp]
\begin{center}
{\includegraphics[width=1.0\linewidth]{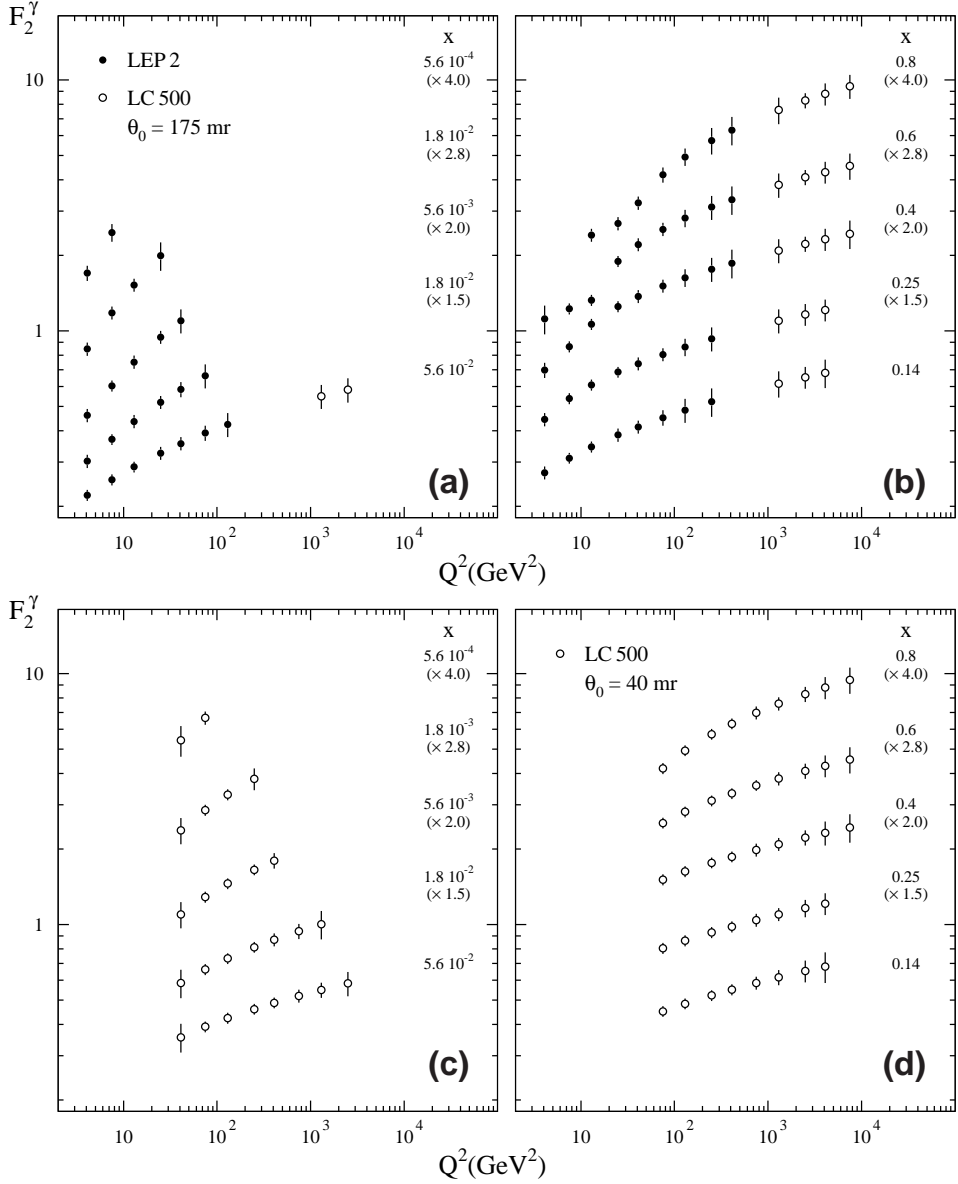}}\\
\caption[
         Prospects for structure function measurements at a future linear
         collider.
        ]
        {
         Prospects for structure function measurements at a future linear
         collider. Shown are hypothetical LEP data and linear collider data
         for different minimal detection angles of the deeply inelastically
         scattered electrons, \tonep. In (a,b) $\tonep > 175$~mrad is assumed
         and (c,d) are based on $\tonep > 40$~mrad.
        }\label{fig:chap8_03}
\end{center}
\end{figure}

%
 The two main questions concerning the photon structure function \ft
 addressed at LEP, namely the low-$x$ behaviour of \ft and the
 \qsq evolution of \ft can be studied at a future
 linear collider but stringent requirements have to be imposed on
 the detector design.
 The region of high values of \qsq and $x$ can already be studied with an
 electromagnetic calorimeter, located outside the shielding mask, and
 covering polar angles of the observed electrons of $\tonep > 175$ mrad,
 which is able to detect electrons with energies above 50$\%$ of the
 energy of the beam electrons, as shown in Figure~\ref{fig:chap8_03}(a,b).
 The errors assumed in Figure~\ref{fig:chap8_03}, taken from
 Ref.~\cite{ACC-9801}, are the quadratic sum
 of the statistical and the systematic components.
 The statistical error is calculated based on the leading order GRV
 structure function \ft for an integrated luminosity of 10 fb$^{-1}$.
 The systematic error is assumed to be equal to the statistical error
 but amounts to at least 5$\%$.
 Therefore, the precision indicated in Figure~\ref{fig:chap8_03} has to
 be taken with care, as the systematic errors shown do not reflect the
 present level of precision of the LEP data, as detailed in
 Section~\ref{sec:qcdres}.
 To achieve overlap in \qsq with the LEP data the electron detection has
 to be possible down to $\tonep > 40$~mrad, shown in
 Figure~\ref{fig:chap8_03}(c,d), which means the mask has to be
 instrumented, and the calorimeter has to be able to detect
 electrons which carry 50$\%$ of the energy of the beam electrons
 in the huge but flat background of electron pairs discussed above,
 certainly a non-trivial task.
 \par
%
\begin{figure}[tbp]
\begin{center}
{\includegraphics[width=1.0\linewidth]{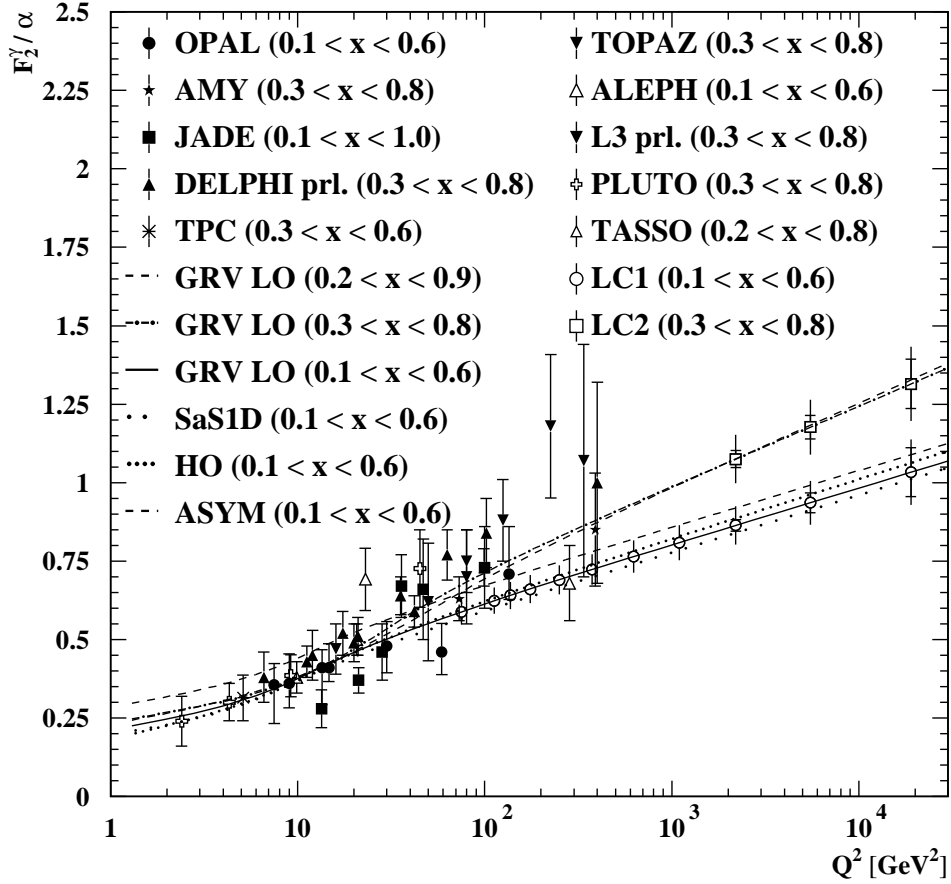}}
\caption[
         Prospects for the measurement of the \qsq evolution of \ft
         at a future linear collider.
        ]
        {
         Prospects for the measurement of the \qsq evolution of \ft
         at a future linear collider.
         The measured \qsq evolution of \ft is shown, together with
         the possible extensions at a future linear collider,
         denoted with LC1 and LC2.
         For the hypothetical data, the inner error bars indicate the
         statistical and the outer error bars the
         quadratic sum of the assumed statistical and systematic errors.
        }\label{fig:chap8_04}
\end{center}
\end{figure}

%
 The measurement of the \qsq evolution of the structure function \ft
 constitutes a fundamental test of QCD.
 In Figure~\ref{fig:chap8_04}
 the prospects of the extension of this measurement at a future linear
 collider with $\sqrt{s_{\ee}}= 500$~\gev are shown for two scenarios.
 This figure has been taken from Ref.~\cite{NIS-9802} and the
 measurements from PLUTO and TASSO have been added.
 As above, the scenarios assume that electrons can be observed
 for energies above 50$\%$ of the energy of the beam electrons,
 and for angles of $\tonep > 40$~mrad (LC1) and $\tonep >175$~mrad (LC2).
 It is further assumed that the measured structure function is equal
 to the prediction of the leading order GRV photon structure
 function \ft in the respective range in $x$.
 The statistical errors of the hypothetical
 measurements are calculated from the number of events,
 for an integrated luminosity of 10 fb$^{-1}$, predicted by
 the HERWIG Monte Carlo for the leading order GRV photon structure
 function \ft in bins of \qsq using the ranges in $x$ as indicated
 in Figure~\ref{fig:chap8_04}.
 The systematic error is taken to be 6.7$\%$ and to be independent of \qsq.
 This assumption is based on the systematic error of the published
 OPAL~\cite{OPALPR207} result at $\qsq=135$~\gevsq.
 The symmetrised value of the published systematic error is 13.4$\%$.
 It is assumed that this error can be improved by a factor of two.
 With these assumptions the error on the measurement is dominated
 by the systematic error up to the highest values of \qsq.
 It is clear from Figure~\ref{fig:chap8_04}
 that overlap in \qsq with the existing data can only be achieved if
 electron detection with $\tonep > 40$~mrad is possible. For
 $\tonep > 175$~mrad sufficient statistics is only available for
 \qsq above around 1000~\gevsq.
 \par
 In summary with the data from the linear collider the measurement of
 the \qsq evolution of the structure function \ft, can be extended to about
 $\qsq = 10000$~\gevsq, and the low-$x$ behaviour can be investigated
 down to $x\approx 5\cdot 10^{-2}$ ($x\approx 5\cdot 10^{-4}$) for an
 electron acceptance of $\tonep > 175$ mrad ($\tonep > 40$ mrad).
 \par
%
%
 At the largest values in \qsq also the contributions from \zn exchange
 in deep inelastic electron-photon scattering could be measured and the
 charged-current process ${\mathrm e}\gamma\to\nu X$ could be used to
 study the weak structure of the photon.
 Together these measurements would allow for a separate measurement of
 the parton distribution functions for up and down-type quark species,
 as has been discussed in Ref.~\cite{GEH-9901}.
 \par
 For a photon linear collider operating in the electron-photon mode
 a completely new scenario for photon structure function measurements
 would be opened.
 For the first time measurements could be performed with beams of high
 energetic photons of known energy with a rather small energy spread,
 instead of measurements using the broad bremsstrahlungs spectrum of
 photons radiated by electrons.
 With this the measurements of the photon structure function would be on a
 similar ground than the measurement of the proton structure function at
 HERA, which would probably also result in a strong reduction of the
 systematic error.
 \par
 Another very important improvement for structure function measurements
 would be the detection of the electron that radiates the quasi-real photon
 and is scattered under almost zero angle.
 The possibility of such very low angle tagging is presently under study,
 but it is not yet clear whether it can be realised.
 If this could be achieved the
 precision of structure function measurements may significantly be improved,
 because $x$ could be calculated from the two detected electrons.
 However, it should be kept in mind that this requires a very good
 resolution on the measured electron energy, despite the large background
 discussed above, a very demanding requirement.
 If this could be reached the measured electron energy would be used
 to determine the much smaller photon energy, which would allow for a
 measurement of $W$ independently of the hadronic final state.
 Given that the dominant systematic error of the structure function
 measurement comes from the imperfect description of the hadronic final
 state by the Monte Carlo models, this would be a important step to reduce
 the systematic error of structure function measurements.
 \par
 From the above it is clear that the investigation of the structure
 of the photon would greatly profit from the measurements performed
 at a future linear collider.
 However, it has to be kept in mind that at the moment it is not clear
 whether several of the desired features of the detector, like zero-angle
 tagging and excellent calorimetry inside the shielding mask, can be
 achieved.
%
%

%
%
\section{Probing the structure of the photon apart from DIS}
\label{sec:probe}
 In addition to the results on the structure of the photon from deep
 inelastic electron-photon scattering the photon structure has been
 studied in the scattering of two quasi-real photons at \epem colliders,
 and in photoproduction and deep inelastic electron-proton scattering
 at HERA.
 These two rich fields of investigations of the photon structure
 cannot be covered in all details here. Only the most important
 topics in the context of this review will be discussed below,
 focusing on the general ideas and the main results.
 For the important details, which are not given here the reader is
 referred to the most recent publications and
 to summaries of the LEP and HERA results, which can be
 found in Refs.~\cite{ERD-9601,SOL-9701,NIS-9903,KIE-9901}.
%
%
\subsection{Photon-photon scattering at \epem colliders}
\label{sec:gamgam}
 The scattering of two quasi-real photons has been studied in
 detail at LEP.
 The photon-photon scattering reaction has the largest hadronic cross-section
 at LEP2 energies and therefore, in most cases, the results are mainly
 limited by systematic uncertainties.
 Results have been derived
 on general properties of the hadronic final states
 in Refs.~\cite{ALE-9301,DEL-9401,L3C-9704},
 on the total hadronic photon-photon cross-section
 in Refs.~\cite{L3C-9903,OPALPR278,CSI-9901},
 on hadron production
 in Ref.~\cite{OPALPR241},
 on jet cross-sections
 in Refs.~\cite{OPALPR175,OPALPR250},
 on heavy quark production
 in Refs.~\cite{ALE-9501,L3C-9902,L3C-9903,L3C-9905,NEI-9901,PAT-9901},
 on lepton pair production
 in Ref.~\cite{L3C-9701}
 and on resonances in
 Refs.~\cite{L3C-9301,L3C-9501,L3C-9702,L3C-9703,L3C-9802,L3C-9901,OPALPR248}.
 The selected topics discussed below are the total hadronic
 cross-section for photon-photon scattering, \siggg,
 and more exclusively, hadron production, jet cross-sections
 and the production of heavy quarks.
%
%
\subsubsection{Total hadronic cross-section for photon-photon scattering}
\label{sec:ggtot}
 The measurement of \siggg is both, interesting and challenging.
 It is interesting, because in the framework of Regge theory \siggg can be
 related to the total hadronic cross-sections for photon-proton and
 hadron-hadron scattering, \siggp and \sighh, and a slow rise with the
 photon-photon centre-of-mass energy squared, $\wsq$, is predicted.
 It is challenging, firstly because experimentally
 the determination of $W$ is very difficult
 due to limited acceptance and resolution for the hadrons created in the
 reaction and secondly, because the composition of different event classes,
 for example, diffractive and quasi-elastic processes, is rather uncertain,
 which affects the overall acceptance of the events.
 The first problem is dealt with by determining $W$ from the
 visible hadronic invariant mass using unfolding programs, similarly
 to the measurements of the hadronic structure function.
 The second uncertainty is taken into account by using two models, namely
 PHOJET and PYTHIA, for the description of the hadronic final
 state and for the correction from the accepted cross-section to \siggg,
 leading to the largest uncertainty of the result.
 \par
%
\begin{figure}[tbp]
\begin{center}
{\includegraphics[width=0.8\linewidth]{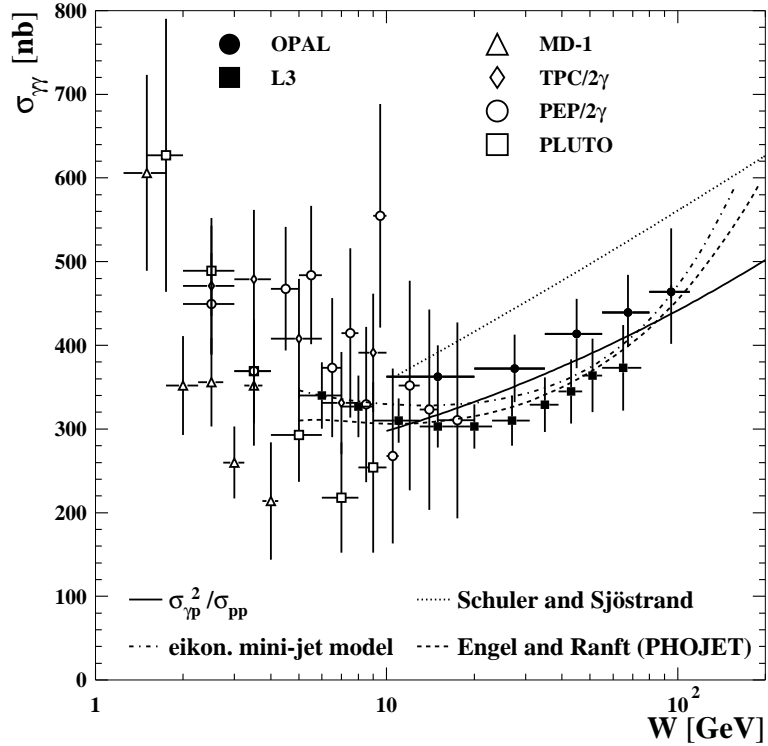}}
\caption[
         Published results on \siggg as a function of $W$.
        ]
        {
         Published results on \siggg as a function of $W$.
        }\label{fig:chap9_01}
\end{center}
\end{figure}
\begin{figure}[tbp]
\begin{center}
{\includegraphics[width=0.7\linewidth]{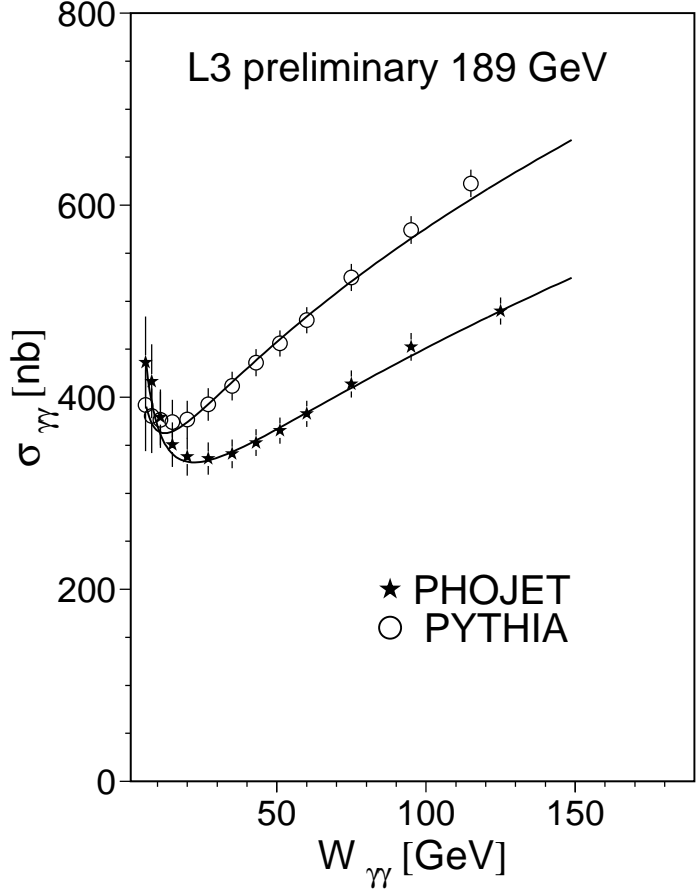}}
\caption[
         Preliminary results on \siggg as a function of $W$.
        ]
        {
         Preliminary results on \siggg as a function of $W$.
         The total hadronic cross-section for photon-photon scattering as a
         function of $W$ is shown for L3 data at $\ssee=189$~\gev and using
         two different Monte Carlo models for correcting the data.
        }\label{fig:chap9_02}
\end{center}
\end{figure}

%
 The published measurements of \siggg by L3 in Ref.~\cite{L3C-9704}
 and by OPAL in Ref.~\cite{OPALPR278} are shown in Figure~\ref{fig:chap9_01},
 and preliminary measurements by L3 presented in Ref.~\cite{CSI-9901}
 are shown in Figure~\ref{fig:chap9_02}.
 All results show a clear rise as a function of $W$.
 \par
%
%
 The cross-section \siggg is interpreted within the framework of Regge
 theory, motivated by the fact that \siggp and \sighh are well described
 by Regge parametrisations using terms to account for pomeron and reggeon
 exchanges.
 The originally proposed form of the Regge parametrisations for
 \siggg is
%
 \begin{equation}
 \siggg(\wsq)= X_{1\GG} (\wsq)^{\epsilon_1}+ Y_{1\GG} (\wsq)^{-\eta_1}\, ,
 \end{equation}
%
 where \wsq is taken in units of \gevsq.
 The first term in the equation is due to soft pomeron exchange
 and the second term is due to reggeon exchange.
 The exponents $\epsilon_1$ and $\eta_1$ are assumed to be universal.
 The presently used values of $\epsilon_1=0.095\pm0.002$ and
 $\eta_1=0.034\pm0.02$ are taken from Ref.~\cite{PDG-9801}.
 The parameters were obtained by a fit to the total hadronic cross-sections
 of pp, ${\rm p}\bar{{\rm p}}$, $\pi^{\pm}$p, K$^{\pm}$p,
 $\gamma$p and $\GG$ scattering reactions.
 The coefficients $X_{1\GG}$ and $Y_{1\GG}$ have to
 be extracted from the $\GG$ data.
 The values obtained in Ref.~\cite{PDG-9801} by a fit to previous
 $\GG$ data, including those of L3 from Ref.~\cite{L3C-9704},
 are $X_{1\GG}=(156 \pm 18)$~nb and $Y_{1\GG}=(320 \pm 130)$~nb.
 Recently an additional hard pomeron component has been suggested in
 Ref.~\cite{DON-9801} leading to
%
 \begin{equation}
 \siggg(\wsq)= X_{1\GG} (\wsq)^{\epsilon_1}+
 X_{2\GG} (\wsq)^{\epsilon_2}+Y_{1\GG} (\wsq)^{-\eta_1}\, ,
 \label{eqn:sigma}
 \end{equation}
%
 with a proposed value of $\epsilon_2=0.418$ and an expected
 uncertainty of $\epsilon_2$ of about $\pm 0.05$.
 \par
%
\begin{figure}[tbp]
\begin{center}
{\includegraphics[width=1.0\linewidth]{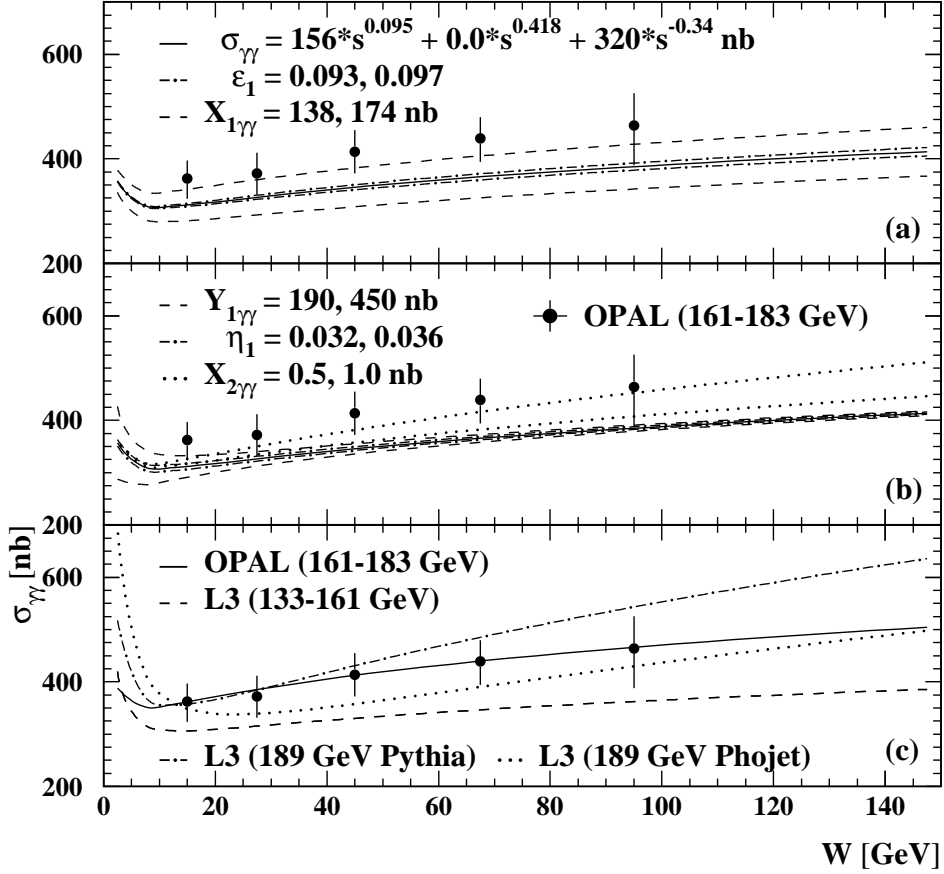}}
\caption[
         Fits to \siggg using various data and fit assumptions.
        ]
        {
         Fits to \siggg using various data and fit assumptions.
         In (a) and (b) the present theoretical predictions are shown
         using the central values and errors quoted in
         Ref.~\protect\cite{PDG-9801}.
         In (c) results for fits to various data as explained in the
         text are shown.
         In addition, the measurement from Ref.~\protect\cite{OPALPR278}
         is shown to illustrate the size of the experimental
         uncertainties.
         }\label{fig:chap9_03}
\end{center}
\end{figure}
%
 Different fits to the data have been performed by the experiments.
 The interpretation of the results is very difficult, because, firstly,
 the parameters are highly correlated, secondly, the main region
 of sensitivity to the reggeon term is not covered by the OPAL measurement
 and thirdly, different assumptions have been made when performing the fits.
 The correlation of the parameters of Eq.~(\ref{eqn:sigma}) can be
 clearly seen in Figure~\ref{fig:chap9_03}(a,b), where the theoretical
 predictions are shown, exploring the uncertainties
 for the soft pomeron term in (a) and for the reggeon as well as for
 the hard pomeron term in (b), using the central values and
 errors quoted in Ref.~\cite{PDG-9801}.
 It is clear from Figure~\ref{fig:chap9_03}(a,b) that by
 changing different parameters in (a) and (b) a very similar
 effect on the rise of the total cross-section can be achieved.
 Figure~\ref{fig:chap9_03}(c) shows the spread of the best fit curves
 for various data and various fit assumptions, explained  below.
 In Figure~\ref{fig:chap9_03}(a-c) in addition the results from
 Ref.~\cite{OPALPR278} are shown
 to illustrate the size of the experimental uncertainties.
 \par
 The different fits performed by the experiments yield the
 following results:
\begin{Enumerate}
 \item\underline{\emph{OPAL:}}
 The OPAL data taken at $\ssee=161-183$~\gev,
 within the present range of $W$, can be accounted for without
 the presence of the hard pomeron term. When fixing all exponents
 and $Y_{1\GG}$ to the values listed above the fit yields
 $X_{2\GG}=(0.5\pm0.2^{+1.5}_{-1.0})$~nb, which is not significantly
 different from zero, and $X_{1\GG}=(182 \pm 3 \pm 22)$~nb, which is
 consistent with the values from Ref.~\cite{PDG-9801}.
 Using $X_{2\GG}=0$ and leaving only $\epsilon_1$ and
 $X_{1\GG}$ as free parameters results in
 $\epsilon_1=0.101\pm0.004^{+0.025}_{-0.019}$ and
 $X_{1\gamma\gamma}=(180 \pm 5^{+30}_{-32})$~nb,
 Figure~\ref{fig:chap9_03}(c, full),
 again consistent with Ref.~\cite{PDG-9801}.
 \item\underline{\emph{L3:}}
 In all fits performed by L3 the hard pomeron term is set to zero.
 The L3 data from Ref.~\cite{L3C-9704} can be fitted using the old
 values for the exponents of $\epsilon_1=0.0790\pm0.0011$ and
 $\eta_1=0.4678\pm0.0059$ from  Ref.~\cite{PDG-9601} leading to
 $X_{1\GG}=(173 \pm 7)$~nb and $Y_{1\GG}=(519 \pm 125)$~nb,
 Figure~\ref{fig:chap9_03}(c, dash).
 The L3 data at $\ssee=189$~\gev indicate a faster rise with energy.
 Using $\epsilon_1=0.95$ and $\eta_1=0.34$, and the PHOJET Monte Carlo
 for correcting the data, leads to $X_{1\GG}=(172\pm 3)$~nb and
 $Y_{1\GG}=(325 \pm 65)$~nb, but the confidence level of the fit is only
 0.000034.
 Fixing only the reggeon exponent to $\eta_1=0.34$ leads to
 $\epsilon_1 = 0.222 \pm 0.019 / 0.206 \pm 0.013$,
 $X_{1\GG}=(50   \pm   9)\, /\, (78  \pm  10)$~nb and
 $Y_{1\GG}=(1153 \pm 114)\, /\, (753 \pm 116)$~nb, when using
 PHOJET/PYTHIA, Figure~\ref{fig:chap9_03}(c, dot/dot-dash).
 \end{Enumerate}
%
 In summary, the situation is unclear at the moment with OPAL being
 consistent with the universal Regge prediction, whereas L3
 indicating a faster rise with $W$ in connection with a very large
 reggeon component for the data at $\ssee=189$~\gev.
 In addition, the L3 data taken at different centre-of-mass energies
 show a different behaviour of the measured cross-section,
 with the data taken at $\ssee=133-161$~\gev being lower,
 especially for $W<30$~\gev.
%
%
\subsubsection{Production of charged hadrons}
\label{sec:gghad}
\begin{figure}[tbp]
\begin{center}
{\includegraphics[width=0.8\linewidth]{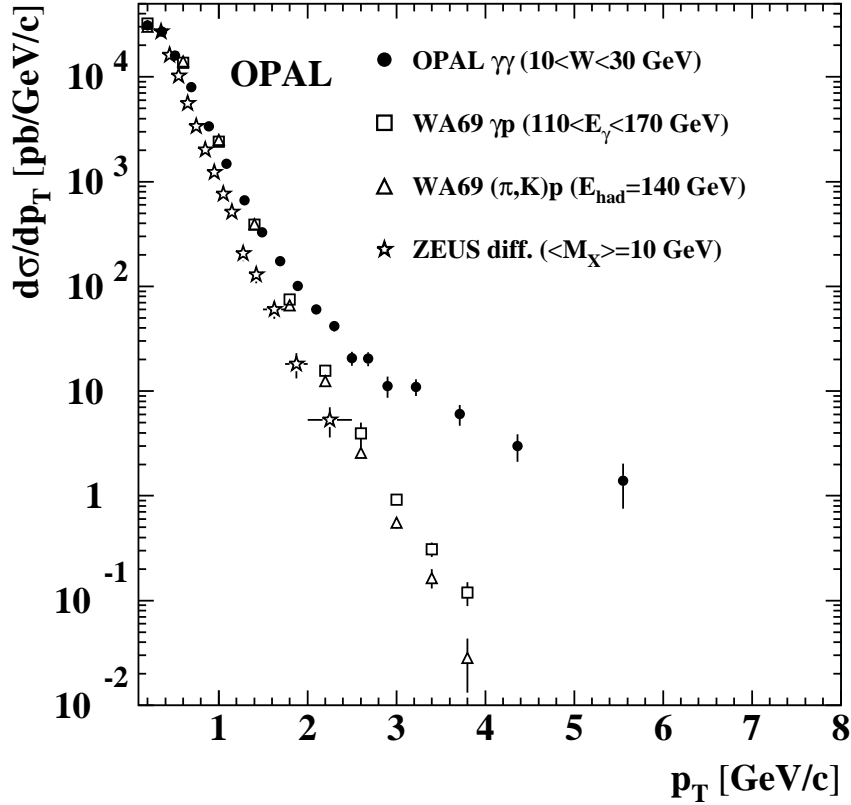}}
\caption[
         Transverse momentum distribution
         ${\mathrm{d}}\sigma/{\mathrm{d}}p_{\rm T}$ for hadron production
         in photon-photon scattering compared to other experiments.
        ]
        {
         Transverse momentum distribution
         ${\mathrm{d}}\sigma/{\mathrm{d}}p_{\rm T}$ for hadron production
         in photon-photon scattering compared to other experiments.
         The photon-photon scattering data taken
         at $\ssee=161-172$~\gev are compared to other experiments
         for $10<W<30$~\gev.
        }\label{fig:chap9_04}
\end{center}
\end{figure}

\begin{figure}[tbp]
\begin{center}
{\includegraphics[width=0.8\linewidth]{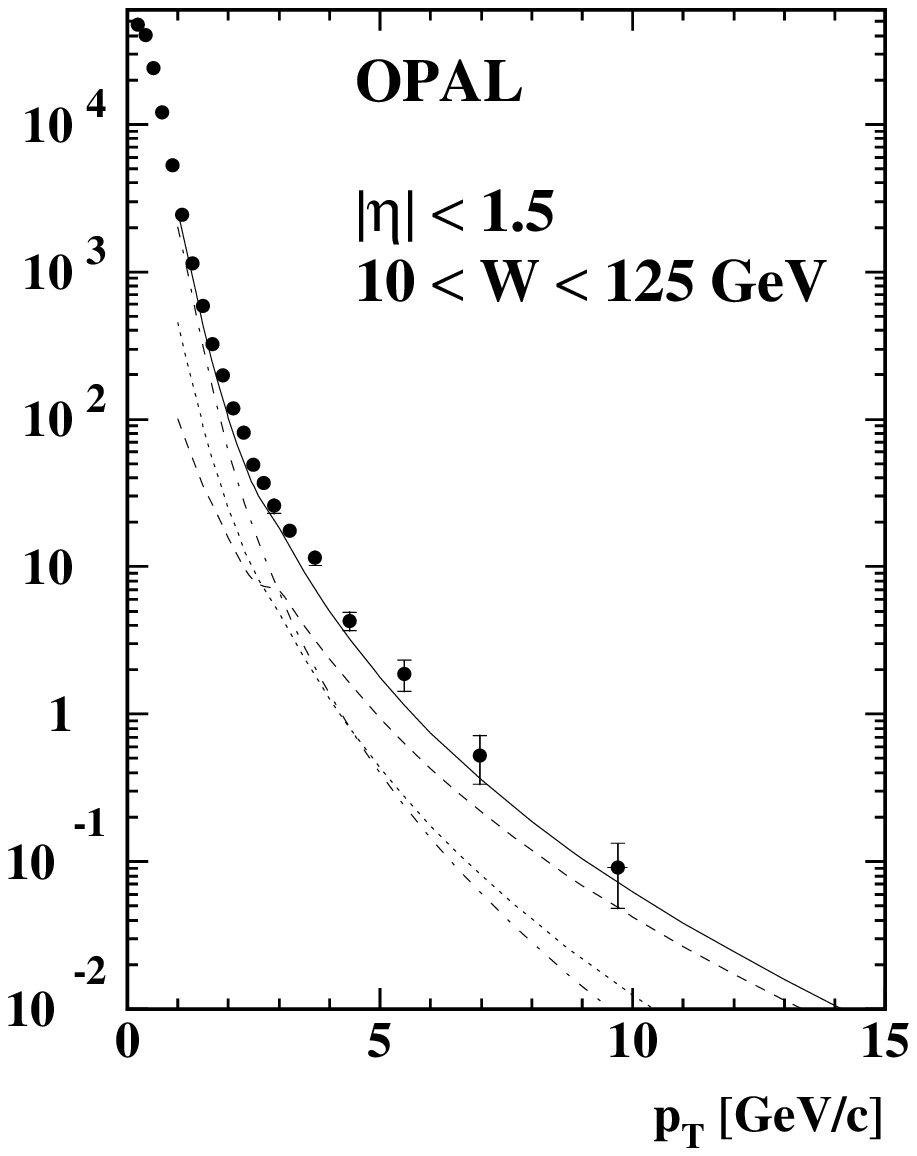}}
\caption[
         Transverse momentum distribution
         ${\mathrm{d}}\sigma/{\mathrm{d}}p_{\rm T}$ for hadron production
         in photon-photon scattering compared to next-to-leading order
         calculations.
        ]
        {
         Transverse momentum distribution
         ${\mathrm{d}}\sigma/{\mathrm{d}}p_{\rm T}$ for hadron production
         in photon-photon scattering compared to next-to-leading order
         calculations.
         The photon-photon scattering data taken
         at $\ssee=161-172$~\gev are compared to next-to-leading order
         calculations for $10<W<125$~\gev.
        }\label{fig:chap9_05}
\end{center}
\end{figure}

 The production of charged hadrons is sensitive to the structure of
 the photon-photon interactions without theoretical and experimental
 problems related to the definition and reconstruction of jets.
 The two main results from the study of hadron production at
 LEP are shown in Figures~\ref{fig:chap9_04} and~\ref{fig:chap9_05}.
 \par
%
%
%
%
 In Figure~\ref{fig:chap9_04} the differential single particle inclusive
 cross-section ${\mathrm{d}}\sigma/{\mathrm{d}}p_{\rm T}$
 for charged hadrons for $\GG$ scattering from
 Ref.~\cite{OPALPR241}, with $10<W<30$~\gev, is shown, together with
 results from $\gamma {\rm p}$, $\pi {\rm p}$ and $ {\rm K}{\rm p}$
 scattering from WA69 with a hadronic invariant mass of $16$~\gev
 from Ref.~\cite{W69-8901}. The WA69 data are normalised to the $\GG$
 data at $p_{\rm T}\approx 0.2$~\gev.
 In addition, ZEUS data on charged particle production in $\gamma {\rm p}$
 scattering with a diffractively dissociated photon
 from Ref.~\cite{ZEU-9502} are shown.
 These data have an average invariant mass of the diffractive system of
 $10$~\gev, and again they are normalised to the OPAL data.
 In Figure~\ref{fig:chap9_05} the differential single particle inclusive
 cross-section for $10<W<125$~\gev is compared to next-to-leading order
 QCD predictions.
 \par
 The main findings are:
%
\begin{Enumerate}
 \item
 The spectrum of transverse momentum of charged hadrons in photon-photon
 scattering is much harder than in the case of photon-proton, hadron-proton
 and `photon-Pomeron' interactions. This can be attributed to the direct
 component of the photon-photon interactions.
 \item
 The production of charged hadrons is found to be described by the
 next-to-leading order QCD predictions from Ref.~\cite{BIN-9601} over a wide
 range of $W$.
 These next-to-leading order calculations are based on the QCD partonic
 cross-sections, the next-to-leading order GRV parametrisation of the
 parton distribution functions for the photon and on fragmentation
 functions fitted to e$^+$e$^-$ data.
 The renormalisation and factorisation scales are set equal to $p_{\rm T}$.
 \end{Enumerate}
%
%
%
\subsubsection{Jet production}
\label{sec:ggjet}
 Jet production is the classical way to study the partonic structure
 of particle interactions.
 At LEP the di-jet cross-section in $\GG$ scattering was studied
 in Ref.~\cite{OPALPR250} at $\ssee=161-172$~\gev using
 the cone jet finding algorithm with $R=1$.
 Three event classes are defined, direct, single-resolved and
 double-resolved interactions.
 As explained in Section~\ref{sec:parton},
 direct means that the photon as a whole takes part in the
 hard interaction, as shown in Figure~\ref{fig:chap1_02}(a),
 whereas resolved means that a parton of a hadronic fluctuation of
 the photon participates in the hard scattering reaction,
 as shown in Figure~\ref{fig:chap1_02}(b,c).
 Experimentally, direct and double-resolved interactions can be
 clearly separated using the quantity
%
 \begin{equation}
 \xgpm = \frac{\sum_{\rm jets=1,2}(E \pm p_z)}
              {\sum_{\rm hadrons}(E \pm p_z)},
 \label{eqn:xgamma}
 \end{equation}
%
 whereas a selection of single-resolved events cannot be achieved
 with high purity.
 Here $E$ and $p_z$ are the energy and longitudinal momentum of a
 hadron, and the sum either runs over all hadrons in the two hard jets
 or over all observed hadrons.
 Ideally, in leading order, direct interactions have $\xgpm = 1$.
 However, due to resolution and higher order corrections the
 measured values of \xgpm are smaller.
 Experimentally, samples containing large fractions of direct events
 can be selected by requiring $\xgpm>0.8$, and samples containing
 large fractions of double-resolved events by using $\xgpm<0.8$.
 \par
%
\begin{figure}[tbp]
\begin{center}
{\includegraphics[width=0.9\linewidth]{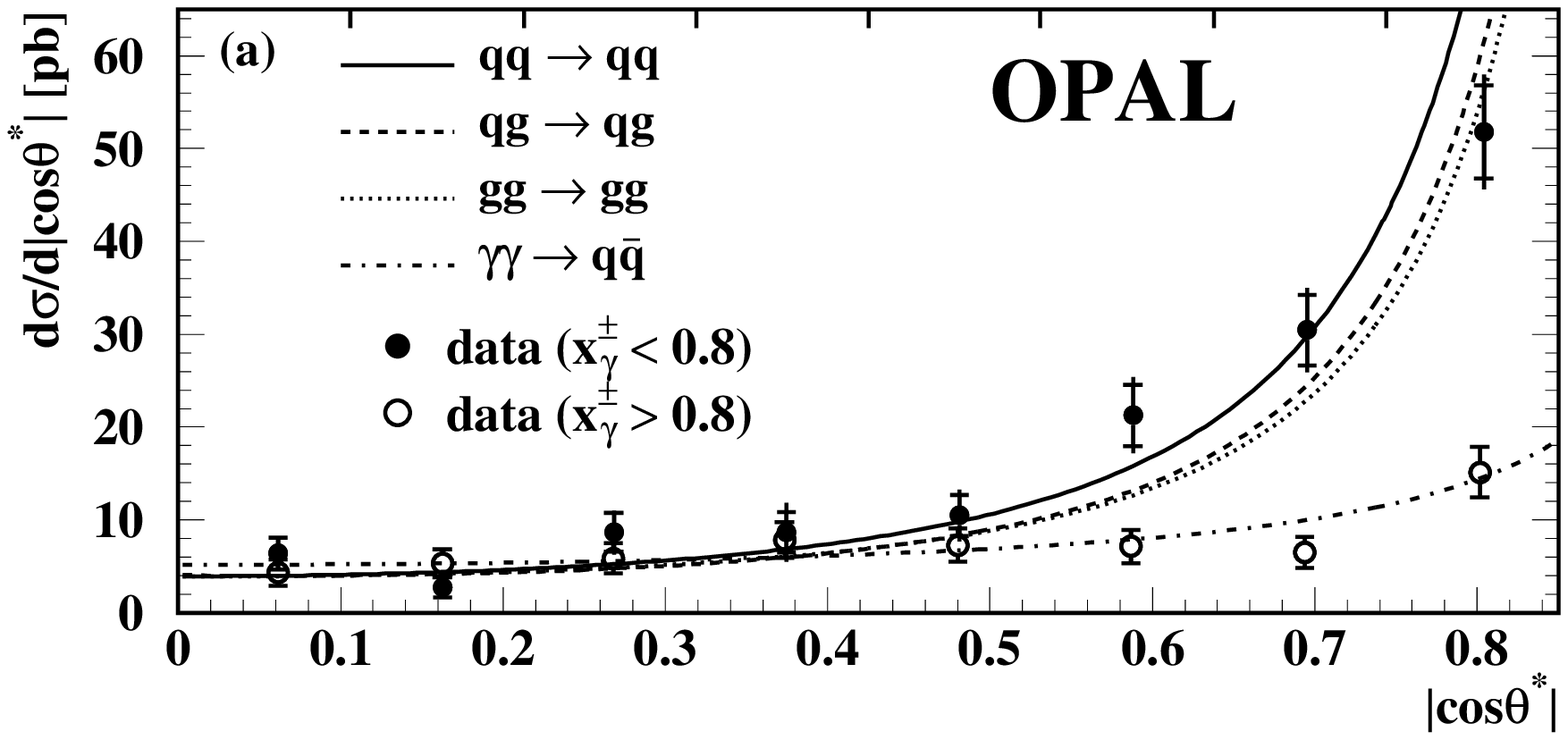}}
{\includegraphics[width=0.9\linewidth]{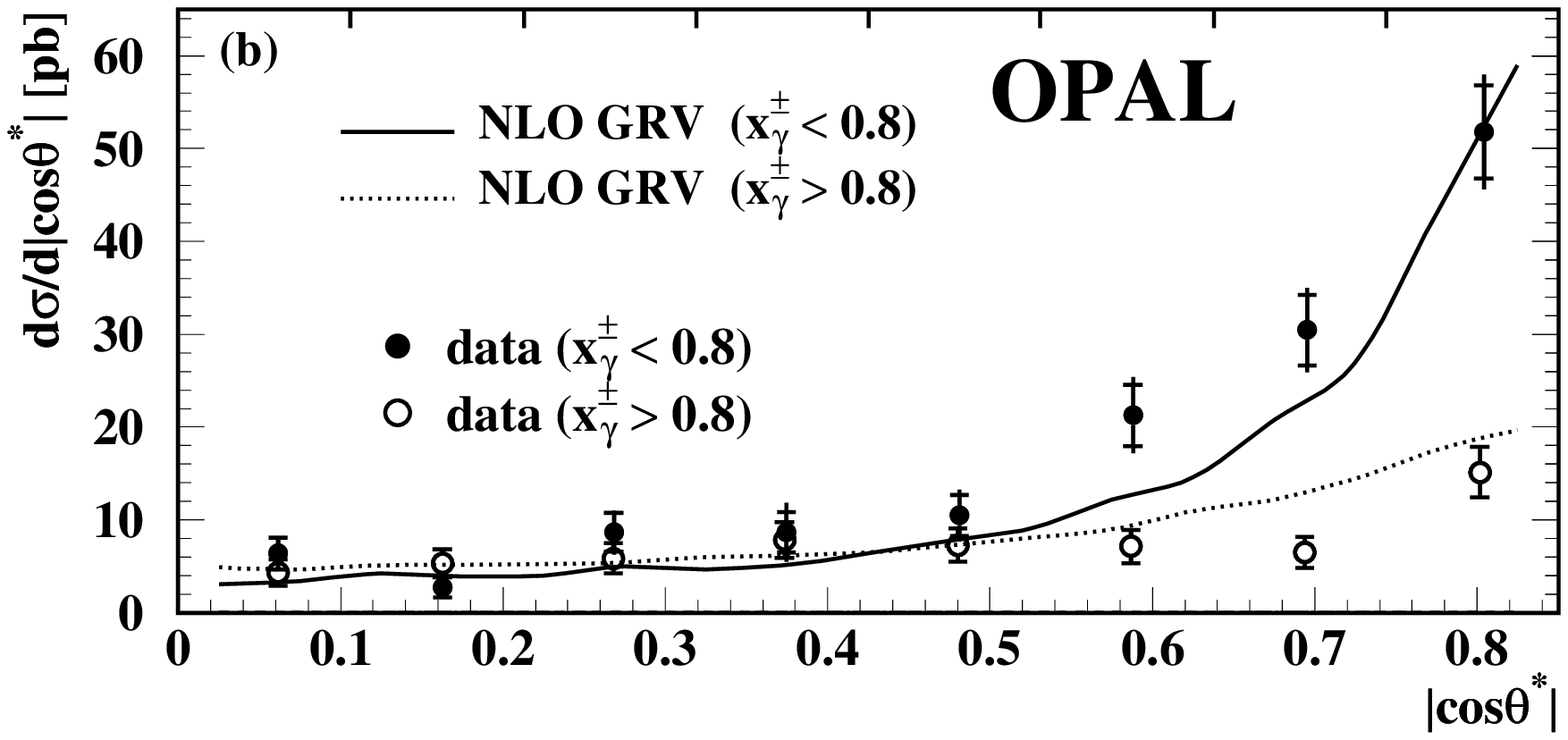}}
\caption[
         Angular dependence of di-jet production in photon-photon scattering.
        ]
        {
         Angular dependence of di-jet production in photon-photon scattering.
         The data at $\ssee=161-172$~\gev are compared to leading order
         matrix elements in (a) and to next-to-leading order (NLO)
         predictions in (b).
        }\label{fig:chap9_06}
\end{center}
\end{figure}
%
 The measurement of the distribution of \cts the cosine of the scattering
 angle in the photon-photon centre-of-mass system, allows for a test
 of the different matrix elements contributing to the reaction.
 The scattering angle is calculated from the jet rapidities in the
 laboratory frame using
%
 \begin{equation}
 \cts = \tanh \frac{\etajeto-\etajeto}{2}.
 \end{equation}
%
 In leading order the direct contribution $\GG\rightarrow q\bar{q}$
 leads to an angular dependence of the form $(1-\ctsq)^{-1}$,
 whereas double-resolved events, which are dominated by gluon induced
 reactions, are expected to behave approximately as
 $(1-\ctsq)^{-2}$.
 The steeper angular dependence of the double-resolved interactions
 can be clearly seen in Figure~\ref{fig:chap9_06}(a), where the shape
 of the di-jet cross-section, for events with di-jet masses above
 $12$~\gev and average rapidities of $\vert(\etajeto+\etajett)/2\vert<1$,
 is compared to leading order predictions.
 In addition, the shape of the angular distribution observed in the data
 is roughly described by the next-to-leading order prediction
 from Refs.~\cite{KLA-9801}, as shown in Figure~\ref{fig:chap9_06}(b).
 In both cases the theoretical predictions are normalised to the data
 in the first three bins.
 \par
%
\begin{figure}[tbp]
\begin{center}
{\includegraphics[width=0.8\linewidth]{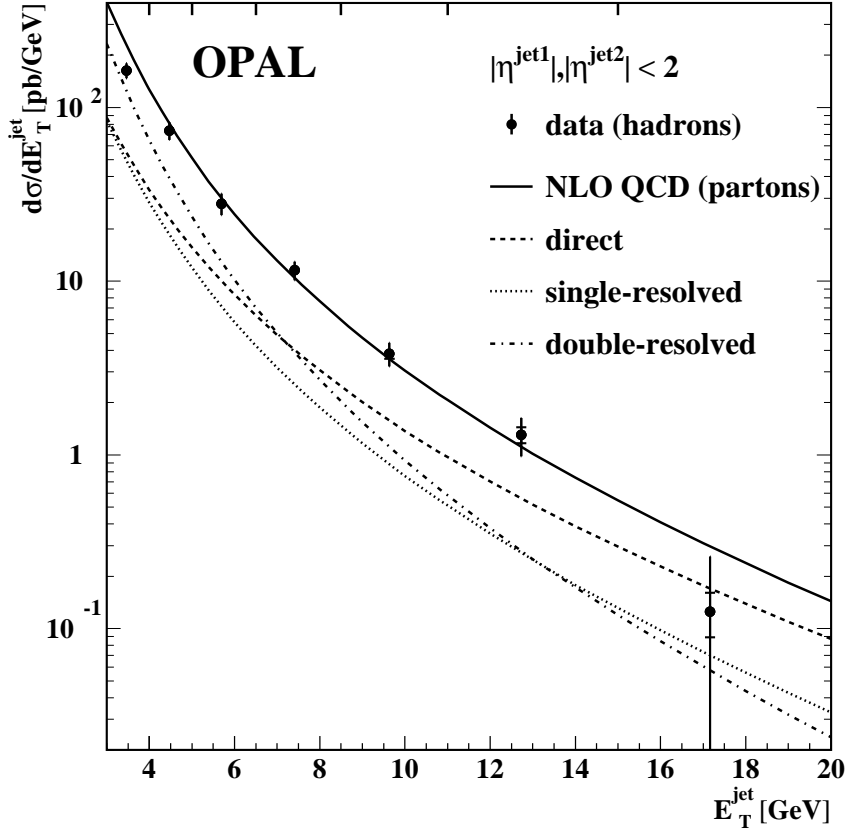}}
\caption[
         Transverse energy distribution
         ${\rm d}\sigma/{\rm d}E^{\rm jet}_{\rm T}$ for jet production
         in photon-photon scattering compared to next-to-leading order
         calculations.
        ]
        {
         Transverse energy distribution
         ${\rm d}\sigma/{\rm d}E^{\rm jet}_{\rm T}$ for jet production
         in photon-photon scattering compared to next-to-leading order
         calculations.
         The measured di-jet production at $\ssee=161-172$~\gev
         is compared to next-to-leading order (NLO) predictions for
         different event classes.
        }\label{fig:chap9_07}
\end{center}
\end{figure}
%
 These next-to-leading order calculations well account for the observed
 inclusive differential di-jet cross-section,
 ${\rm d}\sigma/{\rm d}\etjet$,
 as a function of jet transverse energy, \etjet,
 for di-jet events with pseudorapidities $\vert\etajet\vert<2$.
%
\begin{figure}[tbp]
\begin{center}
{\includegraphics[width=0.8\linewidth]{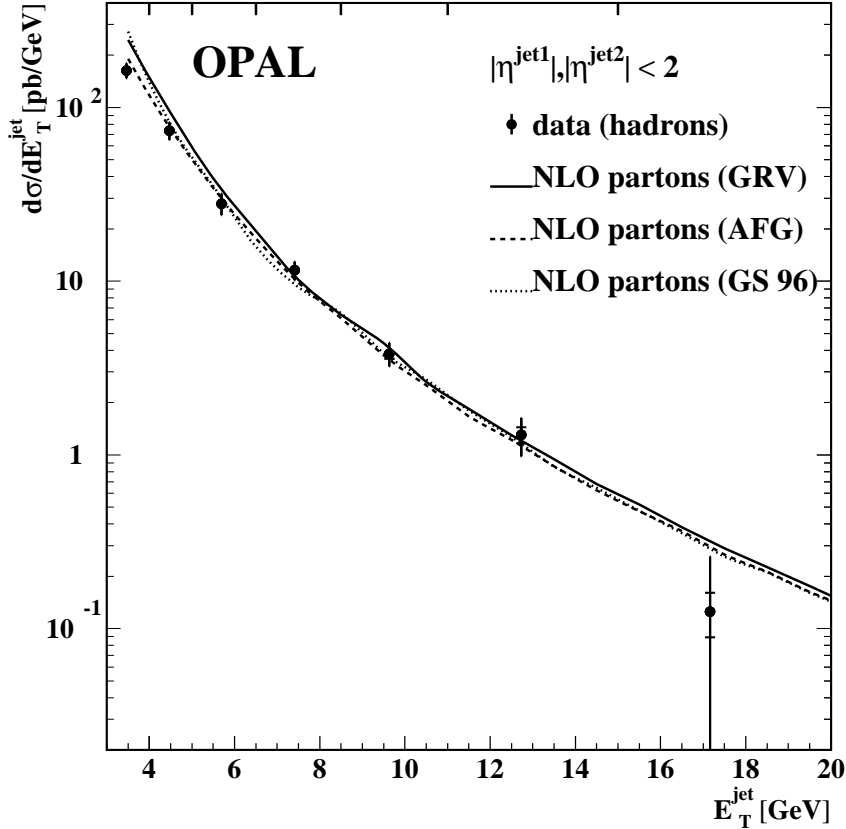}}
\caption[
         Transverse energy distribution
         ${\rm d}\sigma/{\rm d}E^{\rm jet}_{\rm T}$ for jet production
         in photon-photon scattering compared to
         predictions for different parton distribution functions.
        ]
        {
         Transverse energy distribution
         ${\rm d}\sigma/{\rm d}E^{\rm jet}_{\rm T}$ for jet production
         in photon-photon scattering compared to
         predictions for different parton distribution functions.
         The measured di-jet production at $\ssee=161-172$~\gev
         is compared to next-to-leading order (NLO) predictions for the
         GRV, GRS and GS parton distribution functions of the photon.
        }\label{fig:chap9_08}
\end{center}
\end{figure}
%
 As expected, the direct component can account for most of the cross-section
 at large \etjet, whereas the region of low
 \etjet is dominated by the double-resolved contribution,
 shown in Figure~\ref{fig:chap9_07}.
 The calculation for three different
 next-to-leading order parametrisations of the parton distribution
 functions of the photon are in good agreement with
 the data shown in Figure~\ref{fig:chap9_08}, except in the first bin,
 where theoretical as well as experimental uncertainties are large.
 Unfortunately, this is the region which shows the largest sensitivity to
 the differences of the parton distribution functions of the photon.
%
%
\subsubsection{Heavy quark production}
\label{sec:ggheavy}
 Similarly to the case of deep inelastic electron photon scattering
 discussed in Section~\ref{sec:QCD},
 in photon-photon scattering the production of heavy quarks is
 dominated by charm quark production, because the bottom quarks are
 much heavier and have a smaller electric charge.
 Due to the large scale of the process provided by the charm quark mass,
 the production of charm quarks can be predicted in next-to-leading order
 perturbative QCD.
 In QCD the production of charm quarks at LEP2 energies receives sizeable
 contributions from the direct and the single-resolved process.
 In contrast, the double-resolved contribution is expected to be
 very small, as discussed in Ref.~\cite{FRI-9901}.
 The direct contribution allows to test a pure QCD prediction and
 the single-resolved contribution is sensitive to the gluon
 distribution function of the photon.
 \par
 In photon-photon collisions the charm quarks have been tagged
 using standard techniques, either based on the observation of
 semileptonic decays of charm quarks using identified electrons and muons
 in Ref.~\cite{L3C-9902}, or by the measurement of $D^\star$
 production in Refs.~\cite{ALE-9501,PAT-9901,L3C-9905},
 using the decay $D^\star\rightarrow D^0\pi$, where the pion has
 very low energy, followed by the D$^0$ decay observed in one of the
 decay channels, $D^0\rightarrow K\pi, K\pi\pi^0, K\pi\pi\pi$.
 The leptons as well as the $D^\star$ can be clearly separated from
 background processes.  However, due to the small branching ratios and
 selection inefficiencies the selected event samples are small and the
 measurements are limited mainly by the statistical error.
 \par
%
\begin{figure}[tbp]
\begin{center}
{\includegraphics[width=0.8\linewidth]{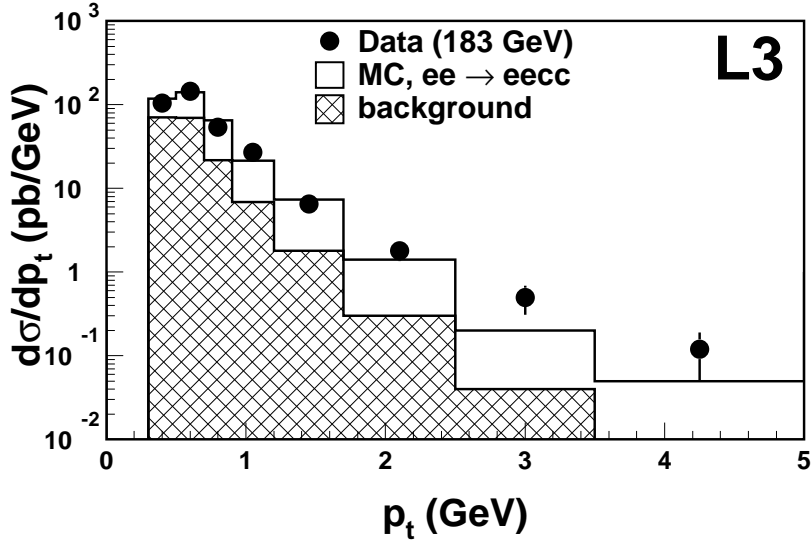}}
\caption[
         Differential cross-sections for charm quark production with
         semileptonic decays into electrons.
        ]
        {
         Differential cross-sections for charm quark production with
         semileptonic decays into electrons.
         The data with electrons fulfilling
         $\vert\cos\theta_{\rm e}\vert<0.9$ and $E_{\rm e}>0.6$~\gev
         and for invariant masses $W>3$~\gev are compared to the
         prediction of the PYTHIA Monte Carlo model, normalised to the
         number of data events observed.
        }\label{fig:chap9_09}
\end{center}
\end{figure}
%
%
\begin{figure}[tbp]
\begin{center}
{\includegraphics[width=0.8\linewidth]{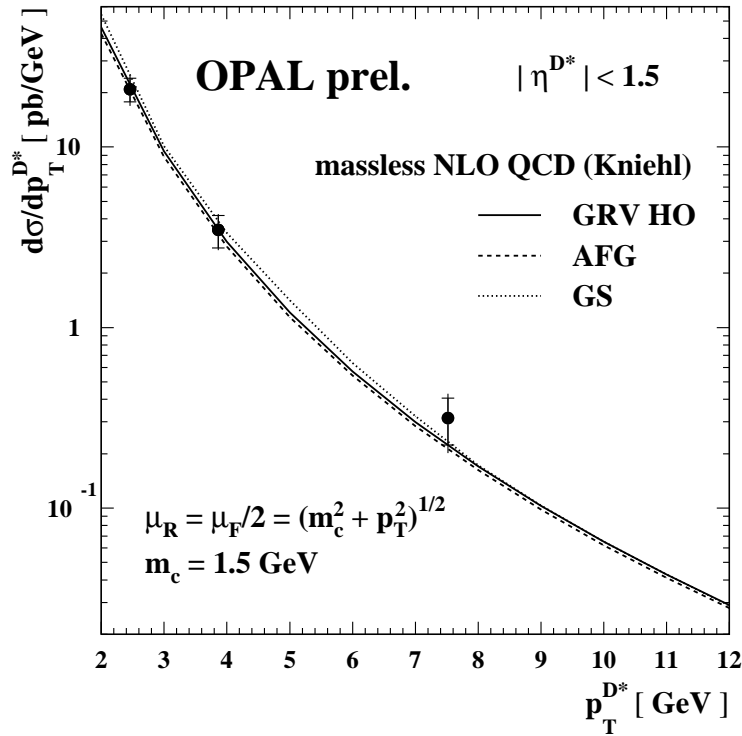}}
\caption[
         Differential cross-sections for charm quark production with
         tagged $D^\star$ mesons.
        ]
        {
         Differential cross-sections for charm quark production with
         tagged $D^\star$ mesons.
         The measured differential cross-sections for $D^\star$ mesons
         with $\vert\eta^{D^\star}\vert<1.5$ is compared to
         a next-to-leading order perturbative QCD predictions using
         the massless scheme.
        }\label{fig:chap9_10}
\end{center}
\end{figure}
%
 Based on these tagging methods differential cross-sections for charm quark
 production and $D^\star$ production in restricted kinematical regions
 have been obtained, examples of which are shown in
 Figure~\ref{fig:chap9_09} and~\ref{fig:chap9_10}.
 Figure~\ref{fig:chap9_09}, taken from Ref.~\cite{L3C-9905}, shows the
 differential cross-section for charm quark production, with semileptonic
 decays into electrons fulfilling $\vert\cos\theta_{\rm e}\vert<0.9$
 and $E_{\rm e}>0.6$~\gev and for $W>3$~\gev.
 The data are compared to the leading order prediction from PYTHIA,
 normalised to the number of data events observed. The shape of the
 distribution is well reproduced by the leading order prediction.
 Figure~\ref{fig:chap9_10}, taken from Ref.~\cite{PAT-9901}, shows the
 differential cross-sections for $D^\star$ production as a function of
 the transverse momentum of the $D^\star$, for
 $\vert\eta^{D^\star}\vert<1.5$ compared to the next-to-leading order
 predictions from Ref.~\cite{BIN-9801} calculated in the massless approach.
 The differential cross-sections as functions of the transverse
 momentum and rapidity of the $D^\star$ are well reproduced by the
 next-to-leading order perturbative QCD predictions, both for the OPAL
 results presented in Ref.~\cite{PAT-9901} and for the L3
 results from Ref.~\cite{L3C-9905}.
 The shape of the OPAL data can be reproduced by the NLO calculations
 from Ref.~\cite{FRI-9901}, however, the theoretical predictions
 are somewhat lower than the data, especially at low values of
 transverse momentum of the $D^\star$.
 \par
\begin{figure}[tbp]
\begin{center}
{\includegraphics[width=0.8\linewidth]{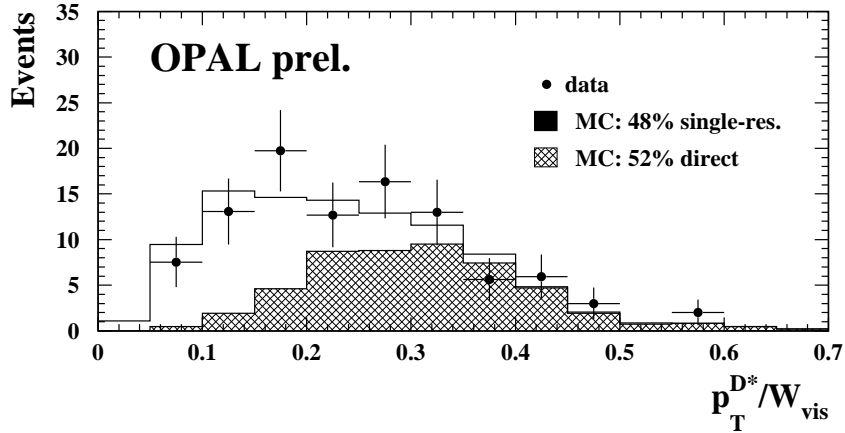}}
\caption[
         The separation of the $D^\star$ production into direct and
         single-resolved contributions.
        ]
        {
         The separation of the $D^\star$ production into direct and
         single-resolved contributions.
         The measured distribution of the transverse momentum of the
         $D^\star$ meson, $p_T^{D^\star}$, normalised to the visible
         hadronic invariant mass, \Wvis, is fitted by a superposition
         of the predicted distributions for direct and resolved events
         based on the PYTHIA Monte Carlo.
        }\label{fig:chap9_11}
\end{center}
\end{figure}

 Based on the observed cross-sections in the restricted ranges in phase
 space the
 total charm quark production cross-section is derived, very much relying
 on the Monte Carlo predictions for the unseen part of the cross-section.
 Two issues are addressed, firstly the relative contribution of
 the direct and single-resolved processes, and secondly the total
 charm quark production cross-section.
 The direct and single-resolved events, for example, as predicted by
 the PYTHIA Monte Carlo, show a different distribution as a function
 of the transverse momentum of the $D^\star$ meson, $p_T^{D^\star}$,
 normalised to the visible hadronic invariant mass, \Wvis, as can
 be seen in Figure~\ref{fig:chap9_11} from Ref.~\cite{PAT-9901}.
 This feature has been used to experimentally determine the relative
 contribution of direct and single-resolved events, which are found
 to contribute about equally to the cross-section.
 \par
\begin{figure}[tbp]
\begin{center}
{\includegraphics[width=0.8\linewidth]{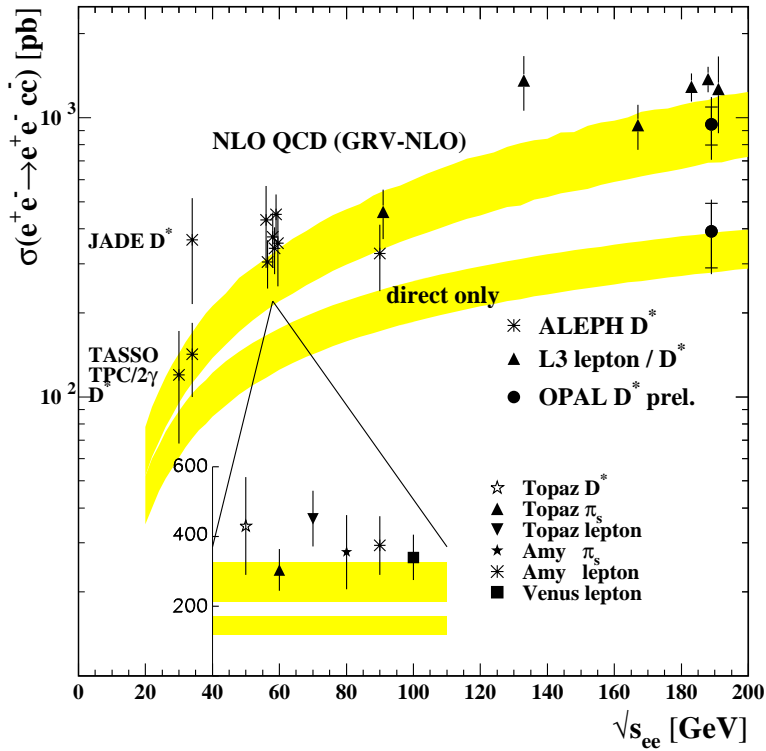}}
\caption[
         Total cross-section for charm quark production.
        ]
        {
         Total cross-section for charm quark production.
         The measured cross-sections from LEP using lepton tags and
         $D^\star$ tags are compared to previous measurements.
        }\label{fig:chap9_12}
\end{center}
\end{figure}

 The total cross-section for the production of charm quarks is shown in
 Figure~\ref{fig:chap9_12} together with previous results summarised
 in Ref.~\cite{AUR-9601}.
 The figure, taken from  Ref.~\cite{PAT-9901}, has been extended by
 additional L3 measurements presented in Refs.~\cite{NEI-9901},
 by the author of Ref.~\cite{PAT-9901}.
 The LEP results are consistent with each other and the theoretical
 predictions are in agreement with the data, both for the NLO
 prediction of the full cross-section based on the GRV parametrisation
 and for the leading order prediction of the direct component.
 The results suffer from additional errors due to the assumptions
 made in the extrapolation from the accepted to the total cross-section,
 which are avoided by measuring only cross-sections in restricted
 ranges in phase space.
 It has been shown in Ref.~\cite{FRI-9901}
 that the NLO calculations are flexible enough to account for the
 phase space restrictions of the experimental analyses and that the
 predicted cross-sections in restricted ranges in phase space are less
 sensitive to variations of the charm quark mass and to alterations of the
 renormalisation as well as the factorisation scale.
 Given this, more insight into several aspects of charm quark production
 may be gained by comparing experimental results and theoretical
 predictions for cross-sections in restricted ranges in phase space.
 \par
%
%
%
%
 In addition to the measurements of the charm quark production
 cross-sections, a preliminary measurement of the cross-section for
 bottom quark production has been reported in Ref.~\cite{NEI-9901}.
%
%
\subsection{Photon structure from HERA}
\label{sec:gamp}
 At HERA the photon structure is investigated mainly by measurements
 of the total photon-proton cross-section in
 Refs.~\cite{H1C-9301,H1C-9514,ZEU-9401}
 and by measurements of the production of charged particles and jets in
 Refs.~\cite{%
 H1C-9506,H1C-9702,H1C-9801,H1C-9802,H1C-9803,CVA-9901,%
 ZEU-9901,ZEU-9902,VOS-9901}.
 The most important results in the context of this review are the
 ones which try to extract information on the partonic content of
 quasi-real and also of virtual photons from photon-proton
 scattering and from deep inelastic electron-proton scattering.
 \par
 At \epem colliders the partonic structure of the quasi-real or
 virtual target photon, $\gamma(\psq,z)$, is probed by the
 highly virtual photon $\gamma^\star(\qsq)$,
 in the region $\qsq\gg\psq$, as shown in Figure~\ref{fig:chap2_01}.
 This leads to a measurement of \ftxq and, by using Eq.~(\ref{eqn:F2def}),
 also to a measurement of the parton distribution functions of the target
 photon probed at the factorisation scale \qsq.
 At HERA the partonic structure of the quasi-real or virtual target
 photon, $\gamma(\psq,z)$, is probed by a parton from the
 proton, producing a pair of partons of large transverse momentum
 squared \ptq, with $\ptq\gg\psq$, leading, for example, to a
 measurement of the effective parton distribution function
 of the target photon, introduced in Eq.~(\ref{eqn:pdfeff}),
 probed at the factorisation scale \ptq.
 Due to the large cross-section at HERA the structure of the photon can
 be probed at even larger factorisation scales than at LEP.
 However, this is only possible if the factorisation scale is
 identified with the transverse momentum of the partons with respect to
 the photon-proton axis in the photon-proton centre-of-mass system,
 $\qsq\equiv\ptq$, which experimentally is approximated, for example,
 by the transverse energy squared of observed jets.
 \par
 Unfortunately the variables used at HERA and LEP are denoted
 differently, for example at HERA the virtuality of the target photon
 is called \qsq instead of \psq and this should not be confused with
 the factorisation scale \qsq, which at HERA is usually taken as \ptq.
 For consistency, and to avoid confusion,
 in this review the factorisation scale
 will be denoted with \qsq, the photon virtuality with \psq,
 and the fractional energy of the photon from the electron,
 $E_\gamma/E$, with $z$,
 and the identifications will be made explicit in the figure captions.
 \par
 The photon-proton scattering reaction as well as deep inelastic
 electron-proton scattering also depends on the structure of the proton.
 Therefore, the investigations of the photon structure are restricted
 to phase space regions where the parton
 distribution functions of the proton are well constrained such that
 the dependence on the proton structure is removed as much as possible.
 \par
 There is one conceptual difference between the results
 obtained by ZEUS and those derived by H1.
 In the case of ZEUS all results are given at the hadron level, which
 means, the data are corrected for detector effects only.
 The phase space regions are selected such that hadronisation
 corrections, as predicted by Monte Carlo models, are expected to
 be small. However, the data are not corrected for these effects.
 The results at the hadron level
 are then compared to NLO calculations which are valid at
 the partonic level and do not contain hadronisation corrections.
 H1 also measures cross-sections corrected for detector effects.
 However, based on these cross-sections leading order partonic
 quantities are reconstructed, which can directly be compared to
 perturbative calculations at the parton level.
 This approach certainly reconstructs more fundamental quantities.
 However, they have additional uncertainties compared
 to the hadron level cross-sections stemming from
 the differences in the hadronisation procedures as assumed
 in the Monte Carlo programs used for the unfolding.
 In some cases these additional uncertainties even contribute the
 dominant error, as explained, for example, in Ref.~\cite{H1C-9803}.
 \par
 Based on measurements of jet production, and charged particle
 production, information on the partonic structure of
 quasi-real and also of virtual photons have been derived.
 These results will be discussed in Section~\ref{sec:ggreal}
 and~\ref{sec:ggvirt} respectively.
%
%
\subsubsection{Structure of quasi-real photons}
\label{sec:ggreal}
 Jet cross-sections for photoproduction reactions have been measured
 by H1 and ZEUS.
%
\begin{figure}[tbp]
\begin{center}
{\includegraphics[width=1.0\linewidth]{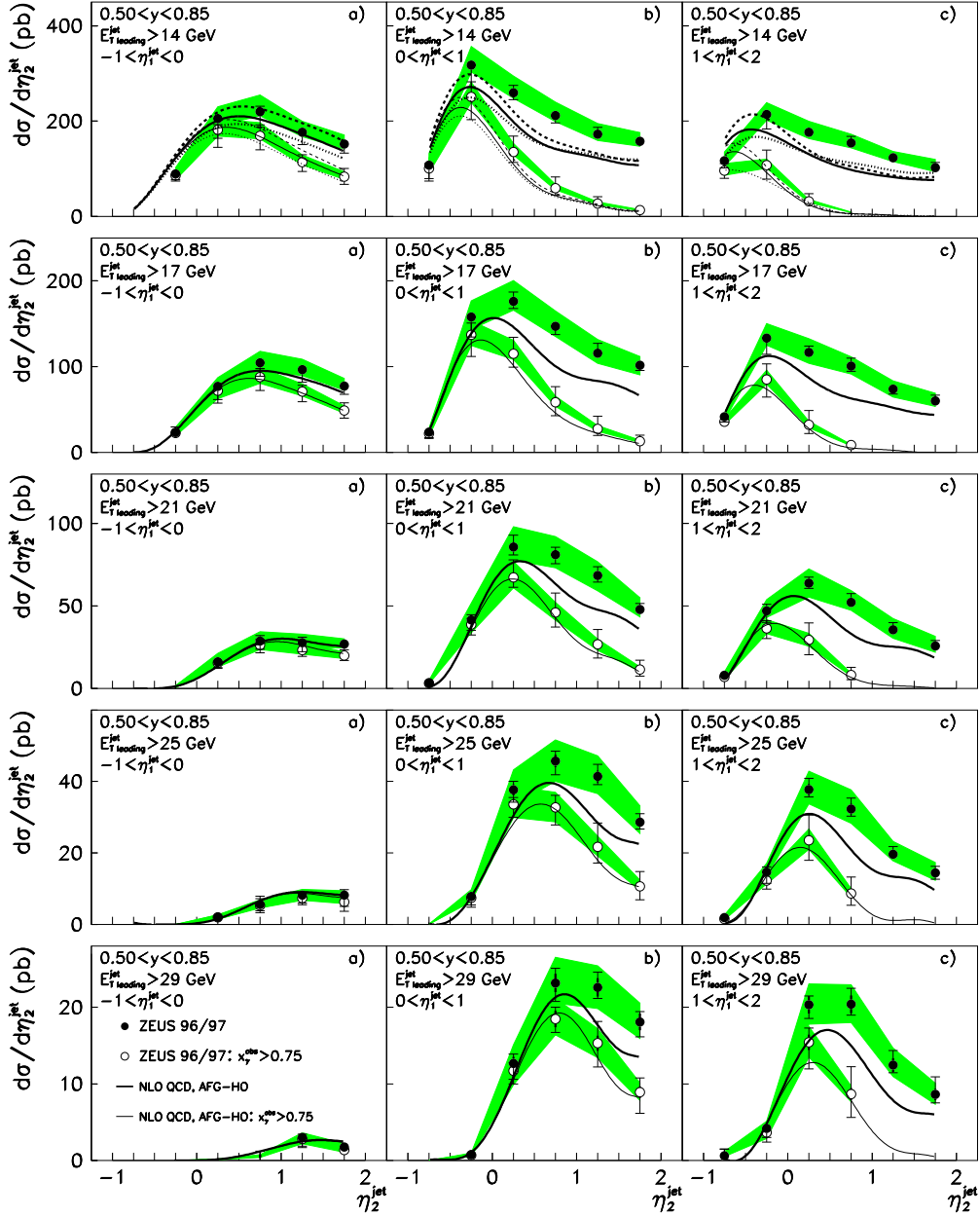}}
\caption[
         Di-jet cross-section in photon-proton scattering compared to
         next-to-leading order QCD predictions.
        ]
        {
         Di-jet cross-section in photon-proton scattering compared to
         next-to-leading order QCD predictions.
         In (a)$-$(c) the data are shown as a function of \etajett in bins
         of \etajeto, for $0.50<z=y<0.85$, for all selected
         events and for events with an observed value of $\xg > 0.75$.
         The rows correspond to different values for the minimum
         required \etjeto.
         The inner error bar indicates the statistical and the outer
         error bar the full error. The shaded band shows the uncertainty
         related to the energy scale. The curves denote the theoretical
         predictions based on the AFG (full), GRV (dash) and GS (dot)
         parametrisations
         of the parton distribution functions of the photon.
        }\label{fig:chap9_13}
\end{center}
\end{figure}
%
 Figure~\ref{fig:chap9_13}, taken from Ref.~\cite{VOS-9901},
 shows an example of a measured di-jet cross-section in
 electron-proton scattering from the ZEUS experiment.
 In this figure, additional theoretical predictions presented in
 Ref.~\cite{ZEU-9901} have been added by the author of Ref.~\cite{VOS-9901}.
 The di-jet cross-section is corrected for detector effects and
 displayed for different values for the minimum required \etjeto.
 The jets are found using the \kt-clustering algorithm in the inclusive
 mode from Ref.~\cite{ELL-9301,CAT-9301}, and the minimum required
 transverse energies of the jets are $\etjeto>14$~\gev
 and $\etjett>11$~\gev for jets with $-1<\etajet<2$.
 This ensures good stability of the next-to-leading order QCD
 predictions due to the asymmetric cuts on \etjet.
 The di-jet cross-section is corrected for detector effects
 using a bin-by-bin correction procedure.
 In addition, the photon virtuality is restricted to be smaller than
 1~\gevsq and the scaled photon energy is required to be in the range
 $0.5<z<0.85$.
 The restriction in the photon energy enhances the
 sensitivity to the parton distribution functions of the photon.
 \par
 The influence of the hadronisation on the di-jet cross-section
 has been studied based on the Monte Carlo programs HERWIG and PYTHIA.
 It was found that the di-jet cross-section at the parton level is 10$\%$
 to 50$\%$ higher than the di-jet cross-section at the hadron level and the
 largest corrections are predicted for the configuration where both jets
 have $\etajet<0$, which means the jets go in the same hemisphere as
 the incoming electron.
 In addition, based on the investigation of the transverse energy
 flow around the jets, it has been concluded that no inclusion of
 soft interactions in addition to the primary hard parton-parton
 scattering reaction is needed to describe the observed jet profiles at
 this large values of \etjet.
 This means, the jet energy profiles can be described without inclusion
 of the so-called soft-underlying event, a method to describe additional
 soft interactions between the photon and proton remnants.
 However, it should be noted that the predicted cross-sections of
 the HERWIG and PYTHIA Monte Carlo programs had to be scaled by large
 factors to account for the measured \xg distribution.
 The numbers used are 1.28/1.27 and 1.83/1.72 for the direct/resolved
 reactions, when using the PYTHIA and HERWIG Monte Carlo programs
 respectively.
 \par
 The measured jet cross-section is higher than the predictions from
 several groups of authors, especially for the region $\etajet>0$.
 It has been shown in Ref.~\cite{ZEU-9901} that
 the various predictions are in good agreement with each other.
 The cross-section shows some sensitivity to the choice of the parton
 distribution functions of the photon, given by the spread of the
 predictions seen in the first row of Figure~\ref{fig:chap9_13}.
 \par
 The different regions in \etajet correspond to different regions in
 \xg
%
 \begin{equation}
 \xg = \frac{\etjeto\,e^{-\etajeto}+\etjett\,e^{-\etajett}}{2zE}.
 \label{eqn:xgammaHERA}
 \end{equation}
%
 The region of large \xg is easiest identified by the region in
 \etajett where the curves for $\xg>0.75$ approach those for
 the full range in \xg.
 It is in this part where the parton distribution functions have
 the largest spread because they are only mildly constraint by
 the measurements of \ft shown in Figure~\ref{fig:chap7_15}.
 In this region the cross-section can be described by the
 perturbative calculations.
 However, for smaller values of \xg, shown in
 Figure~\ref{fig:chap9_13}(b,c) for increasing values of \etajett,
 none of the used sets of the parton distribution functions of the
 photon is able to describe the data, which is taken as an indication
 that they may underestimate the parton content of the photon.
 This region corresponds to about 0.1 to 0.6 in \xg, and here the
 constraints from the measurements of \ft are rather tight, as
 indicated by the small spread of the curves.
 It remains an interesting, but still open question, whether the
 parton distribution functions of the photon can be changed to
 describe the jet data and still being consistent with the measured \ft.
 However, it has to be taken into account that the data are only
 corrected for detector effects, and thus a hadron level quantity
 is compared to theoretical predictions at the parton level.
 \par
\begin{figure}[tbp]
\begin{center}
{\includegraphics[width=0.7\linewidth]{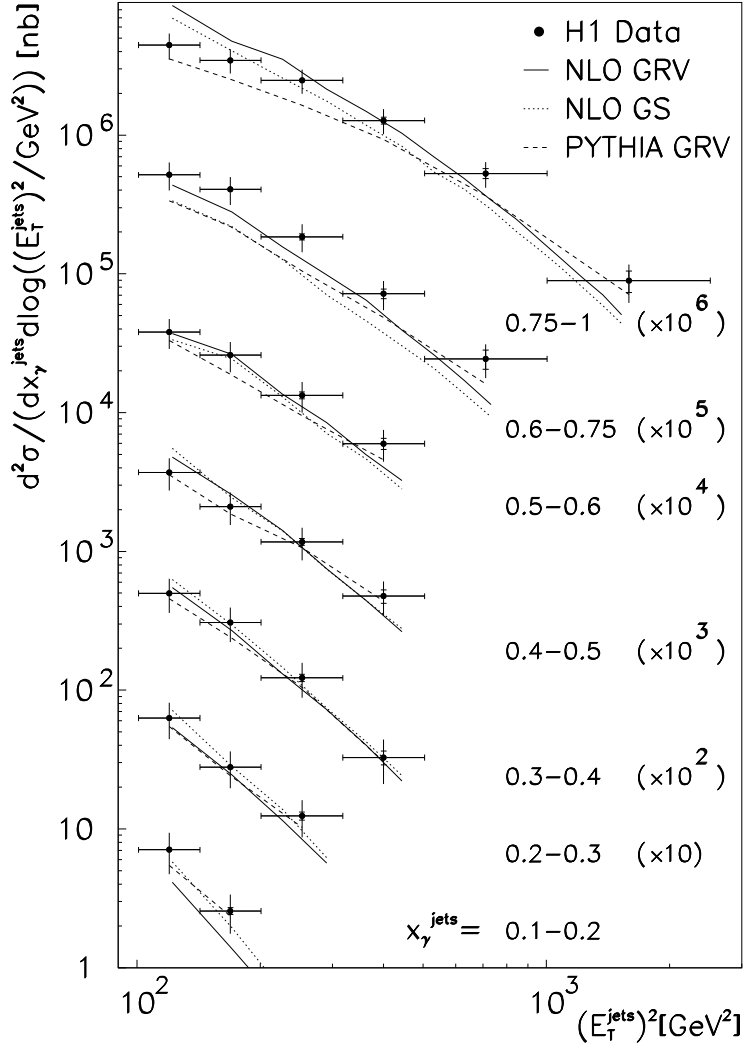}}
\caption[
         Inclusive di-jet cross-section for photon-proton
         scattering from H1.
        ]
        {
         Inclusive di-jet cross-section for photon-proton
         scattering from H1.
         The data are shown as a function of the average transverse
         energy squared of the jets for several bins in \xg.
         The inner error bars represent the statistical errors, the
         outer error bars the full errors.
         The data are compared to the leading order prediction
         from the PYTHIA generator (dash),
         and to analytical next-to-leading calculations using the GRV
         (full) and the GS (dot) parton distribution
         functions of the photon.
        }\label{fig:chap9_14}
\end{center}
\end{figure}

 Similarly, in the case of H1, the measured di-jet cross-section as a
 function of the average transverse energy squared of the jets in bins of \xg,
 is the starting point to investigate the partonic structure of the
 photon in Ref.~\cite{H1C-9801}.
 Again jets are found using the \kt-clustering algorithm in the inclusive
 mode from Ref.~\cite{ELL-9301,CAT-9301}.
 The average of the transverse energies of the two jets with the highest
 transverse energies is required to be above 10~\gev and the difference
 of the transverse energies should be less than 50$\%$ of their average.
 The average jet rapidity is constrained in the region 0-2 and the
 absolute difference to be smaller than unity.
 This ensures $\etjet>7.5$~\gev and, as above for the ZEUS measurement,
 good stability of the next-to-leading order QCD predictions.
 The cross-section is integrated for $\psq<4$~\gevsq and
 $0.2<z<0.83$.
 The measured differential electron-proton di-jet cross-section is
 shown in Figure~\ref{fig:chap9_14}, taken from Ref.~\cite{H1C-9801}.
 The di-jet cross-section is corrected for detector effects only,
 and compared to the predictions of the leading order PYTHIA Monte Carlo
 and to the next-to-leading order parton level predictions using the GRV
 and the GS parton distribution functions of the photon.
 The predictions well account for the observed jet cross-section with
 the exceptions of low and large values of \xg.
 For $\xg<0.2$ the next-to-leading order prediction based on
 the GRV parametrisations is lower than the data, whereas the GS
 parametrisations are able to describe the observed cross-section.
 As in the case of ZEUS, for large values of \xg the GRV and GS
 predictions tend to lie below the data, with GS predicting a smaller
 cross-section due to the strongly suppressed quark distribution
 functions at large values of $x$ discussed in Section~\ref{sec:PDF}.
 However, for H1 the disagreement seems to be less pronounced
 and also to be more concentrated at larger values of \xg,
 $0.6<\xg<0.75$.
 In the highest bin in \xg and at low values of \etjet both
 parametrisations predict too large cross-sections.
 \par
%
%
 This cross-section is then used to determine an effective
 parton distribution function of the photon.
 The reaction is factorised into the radiation of the photon
 off the incoming electron, followed by a subsequent photon-proton
 scattering reaction.
 The flux of transverse photons is described using the Weizs\"acker-Williams
 approximation, Eq.~(\ref{eqn:weiz}), discussed in Section~\ref{sec:EPA}.
 The photon-proton cross-section is approximated  using the concept of the
 single effective subprocess matrix element, \Mses, from Ref.~\cite{COM-8401}.
 This leading order approach relies on the fact that the angular
 dependence of the matrix elements of the most important contributions to
 the process is very similar, as can be seen from  Figure~\ref{fig:chap9_06}.
 Therefore the contributions to the photon-proton cross-section can
 be approximated by the product of the effective parton distribution
 functions and the \Mses, leading to
%
\begin{equation}
 \frac{\der^4\sigma}{\der z\der\xg\der\xp\der\cts}
 \propto \frac{1}{z}\Ngot
 \, \frac{\tilde{f}_{\gamma}(\xg,\qsq)}{\xg}
 \, \frac{\tilde{f}_{\rm p}(\xp,\qsq)}{\xp}
 \, \vert \Mses(\cts) \vert^2,
 \label{eqn:ses}
\end{equation}
%
 with effective parton distribution functions defined as
%
 \begin{eqnarray}
 \pdfexq & \equiv &
 \sum_{k=1}^{\nf}
 \left[q_k^{\gamma}(\xg,\qsq)+\bar{q}_k^{\gamma}(\xg,\qsq)\right] +
 \frac{9}{4} g^{\gamma}(\xg,\qsq), \label{eqn:pdfeff}\\
 \tilde{f}_{\rm p}(\xp,\qsq) & \equiv &
 \sum_{k=1}^{\nf}
 \left[q_k^{\rm p}(\xp,\qsq)+\bar{q}_k^{\rm p}(\xp,\qsq)\right] +
 \frac{9}{4} g^{\rm p}(\xp,\qsq).
 \end{eqnarray}
%
 Similarly to deep inelastic electron-photon scattering, the factorisation
 and renormalisation scales are taken to be equal in the analysis by H1.
 The factorisation scale \qsq is identified with \ptq,
 the transverse momentum squared of the partons with respect to the
 photon-proton axis in the photon-proton centre-of-mass system.
 It has been verified that the factorisation of the process into the
 photon flux and the two effective parton distribution
 functions is meaningful.
 This has been done by showing that within the experimental uncertainty,
 the observed \xp distribution is independent of the measured values of
 $z$ and \xg for a fixed product $z\xg$,
 which means a fixed energy entering the hard parton-parton scattering
 from the photon side.
 In addition, in the region of the H1 analysis it is found that the
 relative amount of quark and gluon initiated di-jet events agree
 with the weight 9/4 to better than 5$\%$.
 \par
%
\begin{figure}[tbp]
\begin{center}
{\includegraphics[width=0.49\linewidth]{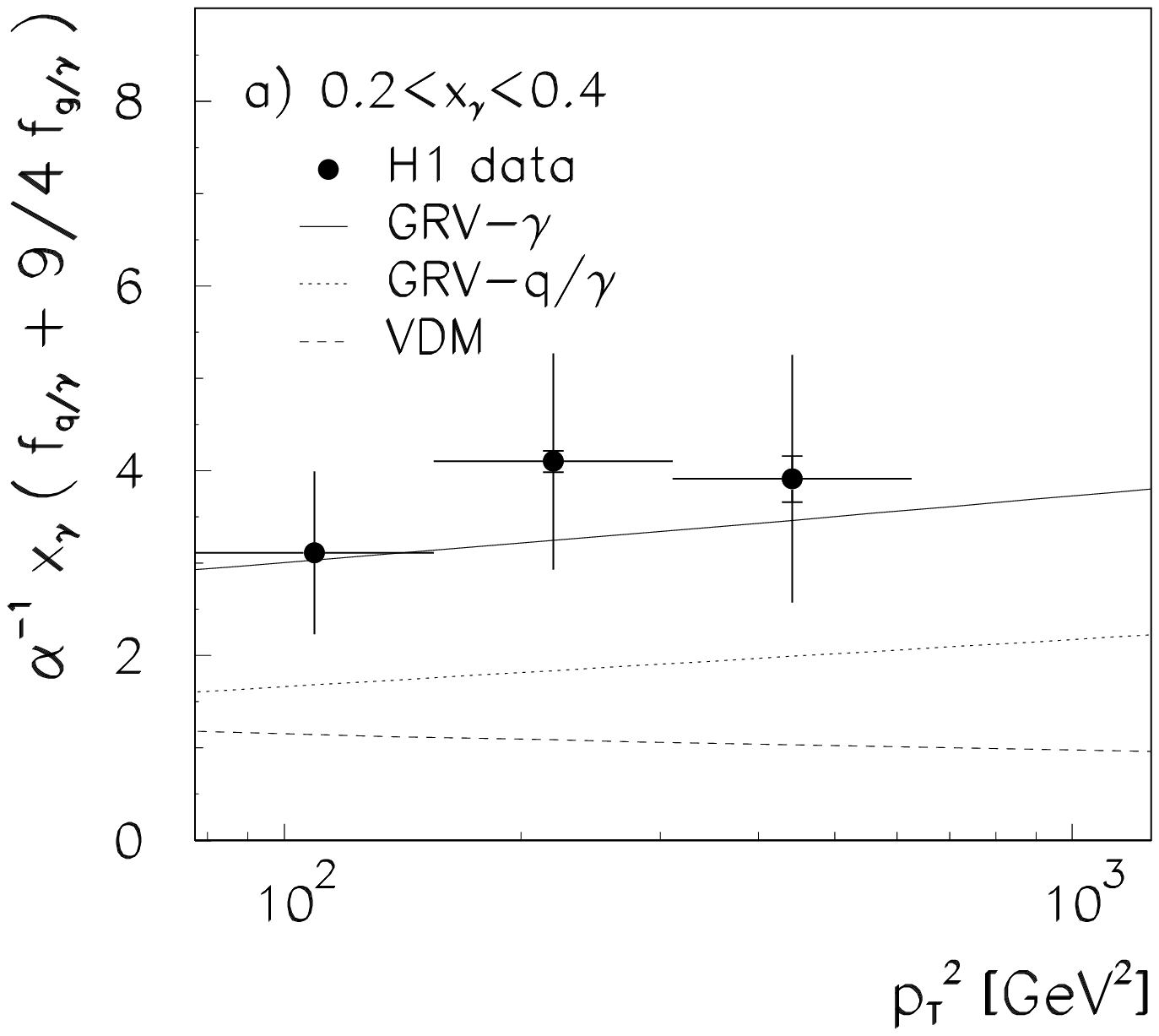}}
{\includegraphics[width=0.49\linewidth]{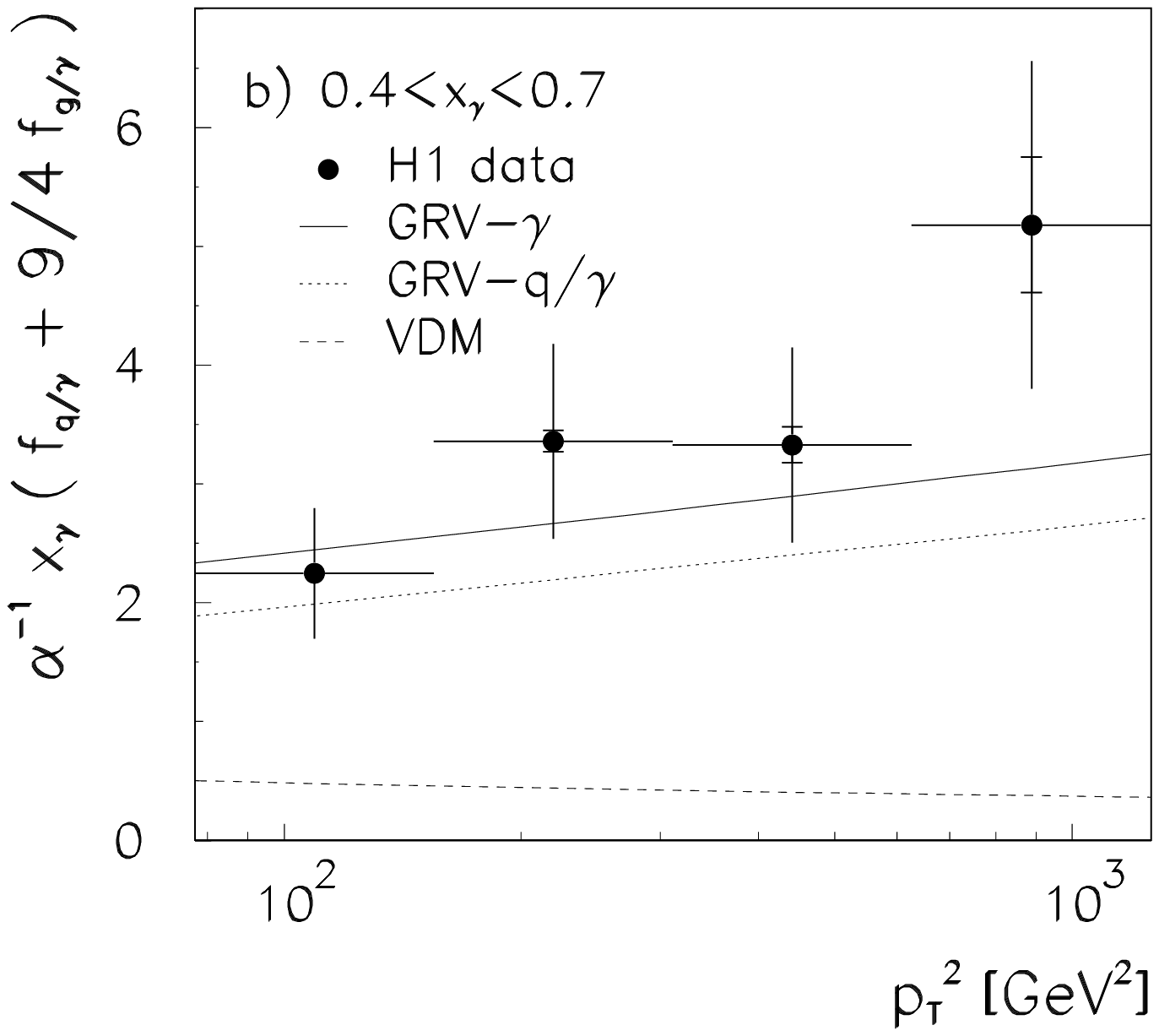}}
\caption[
         Leading order effective parton distribution function
         of the photon from H1.
        ]
        {
         Leading order effective parton distribution function
         of the photon from H1.
         The data are shown as a function of the factorisation scale
         $\qsq=\pthq$, averaged over \xg in the ranges a) $0.2<\xg<0.4$ and
         b) $0.4<\xg<0.7$.
         The inner error bar indicates the statistical and the outer
         error bar the full error. The data are compared to several
         theoretical predictions explained in the text.
        }\label{fig:chap9_15}
\end{center}
\end{figure}
%
 The evolution of the extracted leading order effective parton
 distribution function $\xg\pdfe/\aem$ as a function of the
 factorisation scale \qsq is shown in
 Figure~\ref{fig:chap9_15} taken from Ref.~\cite{H1C-9801} for two
 regions of \xg, $0.2<\xg<0.4$ and $0.4<\xg<0.7$.
 The data are compared to three predictions based on the GRV
 parametrisation of the parton distribution functions of the photon.
 The predictions shown are the effective parton
 distribution function (full), the quark component of \pdfe (dot),
 and the VMD contribution to \pdfe (dash), based on the VMD prediction
 of \ft, explained in Section~\ref{sec:VMD}.
 The full prediction describes the measurement.
 As expected, the VMD contribution is not able to account for the data, and
 the importance of the gluon part increases for decreasing value of \xg.
 Under the assumption that the factorisation and renormalisation scales
 can be identified with the transverse momentum of the partons,
 and within the uncertainties of the concept of the single effective
 subprocess matrix element and the effective parton distribution functions,
 the measurement shows the universality of the parton distribution
 functions of the photon, which are able to describe both photon-proton
 scattering and deep inelastic electron-photon scattering reactions.
 In addition this analysis extends the measurement
 of the photon structure to factorisation scales of the order of
 $\qzm=\langle\pthq\rangle\approx 900$~\gevsq.
 \par
 The sensitivity to the gluon distribution function of the photon
 seen in Figure~\ref{fig:chap9_15} can be explored in the measurement
 of the di-jet cross-section and by using the production of charged
 particles to obtain the gluon distribution function of the photon.
 Both methods have been used by H1 in Refs.~\cite{H1C-9506,CVA-9901}
 and Ref.~\cite{H1C-9802} respectively.
 The two methods are complementary.
 The jet cross-section receives error contributions, for example,
 from the accuracy of the knowledge of the energy scales of the
 calorimeters and from the jet definition, which are absent when
 using charged particles.
 In addition, for sufficiently large transverse momenta of the particles,
 the dependence on the soft underlying event is also reduced, since most
 of the particles produced in the soft underlying event have momenta of
 the order of 0.3~\gev.
 In contrast, the distribution of charged particles are more sensitive to
 details of the hadronisation, which are integrated over when using jets.
 The variable \xg is either obtained from the jets using
 Eq.~(\ref{eqn:xgamma}), or similarly, from the sum
 over all charged particles with transverse momentum
 with respect to the beam axis of more than 2~\gev using the relation
 $\xg\approx1/E_\gamma\,\sum\pt e^{-\eta}$.
 Since the incoming partons of the hard scattering process cannot
 be identified, no distinction can be made between quark and gluon
 initiated processes.
 Therefore, both methods yield only an indirect measurement of the
 gluon distribution function, because the quark initiated contribution
 has to be subtracted based on existing parton distribution functions
 obtained from measurements of \ft at \epem colliders.
 \par
%
\begin{figure}[tbp]
\begin{center}
{\includegraphics[width=0.45\linewidth]{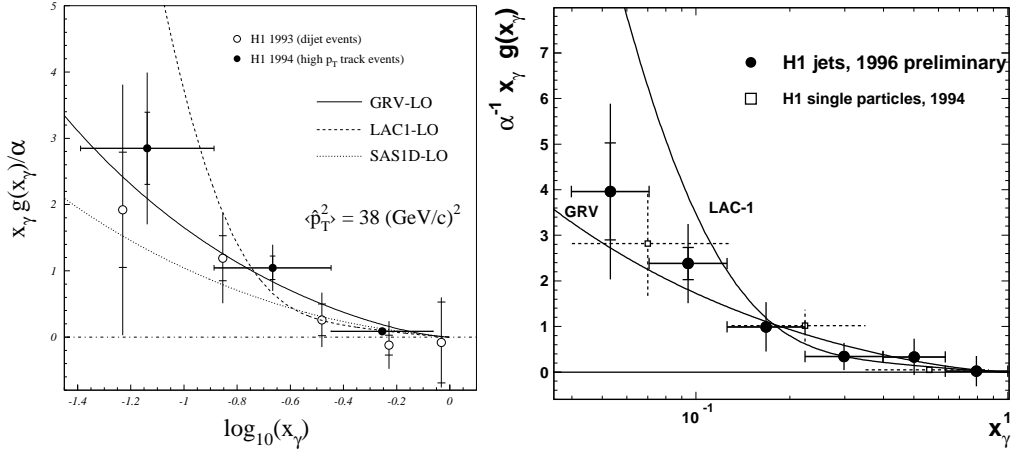}}
{\includegraphics[width=0.50\linewidth]{./eps/chap9_16b}}
\caption[
         Measurements of the gluon distribution function of the photon from H1.
        ]
        {
         Measurements of the gluon distribution function of the photon from H1.
         The leading order gluon distribution function as a
         function of \xg, as obtained from the charged particle
         cross-section is shown twice
         (left full circles and right open squares) for an average
         factorisation scale $\qzm=\langle\pthq\rangle=38$~\gevsq,
         together measurements obtained from di-jet
         cross-sections (left open circles from Ref.~\cite{H1C-9506}
         and right closed circles from Ref.~\cite{CVA-9901})
         for $\qzm=75$~\gevsq.
         The inner error bar indicates the statistical and the outer
         error bar the full error. The data are compared to several
         theoretical predictions explained in the text.
        }\label{fig:chap9_16}
\end{center}
\end{figure}
%
 The results from the two methods as a function of \xg are shown in
 Figure~\ref{fig:chap9_16}(left,right) taken from Refs.~\cite{H1C-9802}
 and Refs.~\cite{CVA-9901} respectively.
 Both figures are shown because the older figure contains more
 comparisons to theoretical predictions.
 The two results, obtained from single particles and jets respectively,
 are consistent and the gluon distribution function is found to be
 small at large values of $x$ and to rise towards small values of $x$.
 The measured leading order gluon distribution function is consistent
 with the existing parametrisation from GRV.
 As in the case of the structure function measurements discussed
 in Section~\ref{sec:qcdresf2}, the measurements disfavour the
 strongly rising gluon distribution functions of the photon,
 for example, the LAC1 gluon distribution function.
 In addition, the recent H1 measurements are above the SaS1D prediction
 which is slightly disfavoured.
 This is similar to the measurements of \ft shown in
 Figures~\ref{fig:chap7_10} and~\ref{fig:chap7_12} which tend to be
 above the SaS1D prediction.
 However, the preliminary update of the \ft measurement at low
 values of \qsq from OPAL, Figure~\ref{fig:chap7_12}, seems to
 be consistent with the SaS1D prediction.
 Certainly more precise data is needed to draw definite conclusions, but,
 again these measurements shows the universality of the parton distribution
 functions of the photon.
 \par
 This concludes the discussion of measurements of the structure
 of quasi-real photons and the remaining part is devoted to virtual
 photons.
%
%
\subsubsection{Structure of virtual photons}
\label{sec:ggvirt}
 As has been discussed in Section~\ref{sec:qcdresvirt} the structure
 of the virtual photon can be studied via the measurement of photon
 structure functions in the kinematical regime $\qsq\gg\psq\gg\lamsq$.
 The evolution of \ftxq and of the parton distribution functions of the
 photon with the factorisation scale \qsq is predicted by QCD, whereas
 several models exist for the expected suppression with the photon
 virtuality \psq.
 In the limit $\qsq\gg\psq\gg\lamsq$ these models approach
 the perturbative QCD results of Refs.~\cite{UEM-8101,UEM-8201}.
 A discussion of the predictions can be found in Section~\ref{sec:PDF}.
 The measurements of \feff from Refs.~\cite{PLU-8405,ERN-9901}
 can be described by next-to-leading order QCD predictions,
 as explained in Section~\ref{sec:qcdresvirt}.
 However, the precision of the \epem data is very limited due
 to low statistics and further information on the structure of the
 virtual photon is certainly needed.
 \par
 At HERA both the evolution with \qsq and the suppression with \psq can
 be investigated in deep inelastic electron-proton scattering.
 The largest part of the cross-section is due to the direct coupling
 of the virtual photon to the partons in the proton, but there is a
 small region of phase space where $\ptq\gg\psq$, there the structure of
 the virtual photon can be resolved.
 \par
\begin{figure}[tbp]
\begin{center}
{\includegraphics[width=1.0\linewidth]{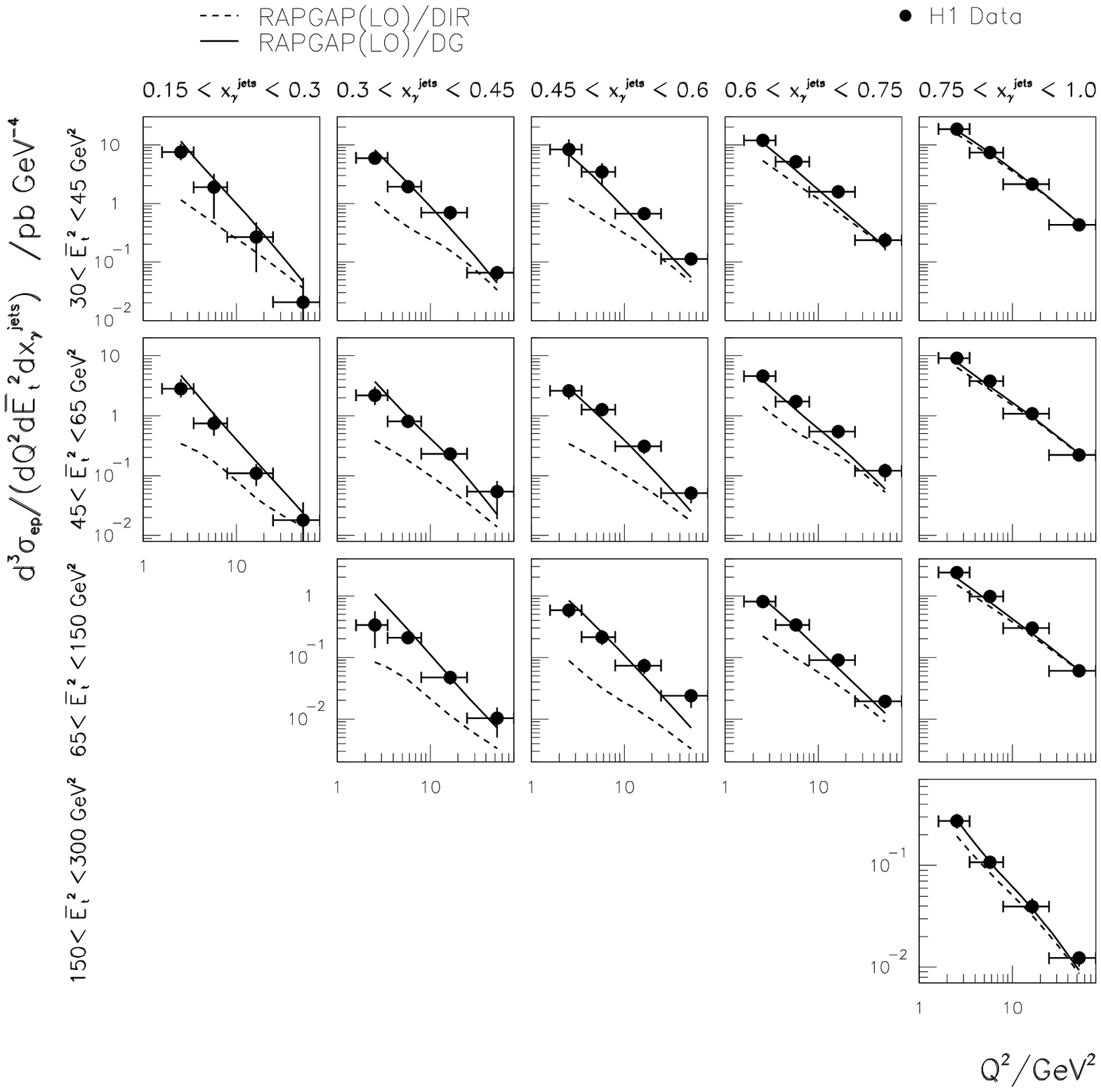}}
\caption[
         Triple differential jet cross-section
         $\der\sigma/\der\psq\der\etjavq\der\xg$
         for virtual photons from H1.
        ]
        {
         Triple differential jet cross-section
         $\der\sigma/\der\psq\der\etjavq\der\xg$
         for virtual photons from H1.
         The data are shown as a function of the photon virtuality
         $\psq=\qsq$ in the ranges of \xg and using several
         bins in the factorisation scale $\qsq=\etjavq$.
         The inner error bar indicates the statistical and the outer
         error bar the full error. The data are compared to several
         theoretical predictions explained in the text.
        }\label{fig:chap9_17}
\end{center}
\end{figure}

 The dependence of the triple differential jet cross-section
 $\der\sigma/\der\psq\der\etjavq\der\xg$ as a function of
 the photon virtuality \psq, and in bins of \xg and \etjet,
 is shown in Figure~\ref{fig:chap9_17} taken from Ref.~\cite{H1C-9803}.
 The data exhibits a strong decrease as a function of \psq and a Monte
 Carlo prediction using the RAPGAP program from Ref.~\cite{JUN-9501},
 based on the direct coupling of the virtual photon alone (dash) is
 only able to account for the data in the region of large values of \xg.
 For low values of \xg the data are much higher than the prediction
 from the direct coupling, and this difference is attributed to resolved
 interaction due to the hadronic structure of the virtual photon.
 A much better description of the data by the RAPGAP prediction is
 achieved when also the partonic structure of the virtual photon is
 taken into account by using the GRV parton distribution functions of the
 quasi-real photon, suppressed by the Drees Godbole scheme (full),
 Eq.~(\ref{eqn:DGsup}), with $\pcsq=\omega^2=0.04$~\gevsq.
 \par
%
%
 Based on this observation the leading order effective parton
 distribution function of virtual photons is extracted very similar to
 the case of quasi-real photons, discussed above.
 Again the photon-proton cross-section is approximated by the product
 of the effective parton distribution function and the \Mses.
 Due to the non-zero virtuality \psq  of the photon the situation is
 more complex.
 Firstly, also the flux of longitudinal photons has to be taken into
 account and secondly, parton distribution functions of longitudinal
 virtual photons are needed, which have not yet been determined.
 Given the known ratio of the flux of transverse and longitudinal
 photons, $\epsilon(z)$, defined in Eq.~(\ref{eqn:eparat}),
 only the flux of transverse photons, Eq.~(\ref{eqn:epa}), is needed
 and the cross-section can be expressed in a factorised form by
%
\begin{eqnarray}
 \frac{\der^5\sigma}{\der z\der\xg\der\xp\der\cts\der\psq}
 &\propto& \frac{1}{z}\Ngtt
 \, \frac{\tilde{f}_{\gamma}(\xg,\qsq,\psq)}{\xg}
 \nonumber\\&&
 \, \frac{\tilde{f}_{\rm p}(\xp,\qsq)}{\xp}
 \, \vert \Mses(\cts) \vert^2,
 \label{eqn:sesvir}
\end{eqnarray}
%
 with effective parton distribution functions defined as
%
 \begin{eqnarray}
 \pdfexqp & \equiv &
 \sum_{k=1}^{\nf}
 \left[q_k^{\gamma}(\xg,\qsq,\psq)+\bar{q}_k^{\gamma}(\xg,\qsq,\psq)\right]
 \nonumber\\&&\quad\quad
 + \frac{9}{4} g^{\gamma}(\xg,\qsq,\psq),
 \label{eqn:pdfeffvir}
 \end{eqnarray}
%
 and $\tilde{f}_{\rm p}(\xp,\qsq)$ as above.
 Here \pdfe is to be understood as the polarisation averaged effective
 parton distribution function $\pdfe = \pdfet + \epsilon(z)\pdfel$.
 According to Refs.~\cite{BZS-9301,FRI-9301,GLU-9601} \pdfel is expected
 to be small in most of the kinematical range of the H1 analysis.
 If this is the case \pdfe reduces to the purely transverse effective
 parton distribution function for virtual photons.
 \par
%
\begin{figure}[tbp]
\begin{center}
{\includegraphics[width=1.0\linewidth]{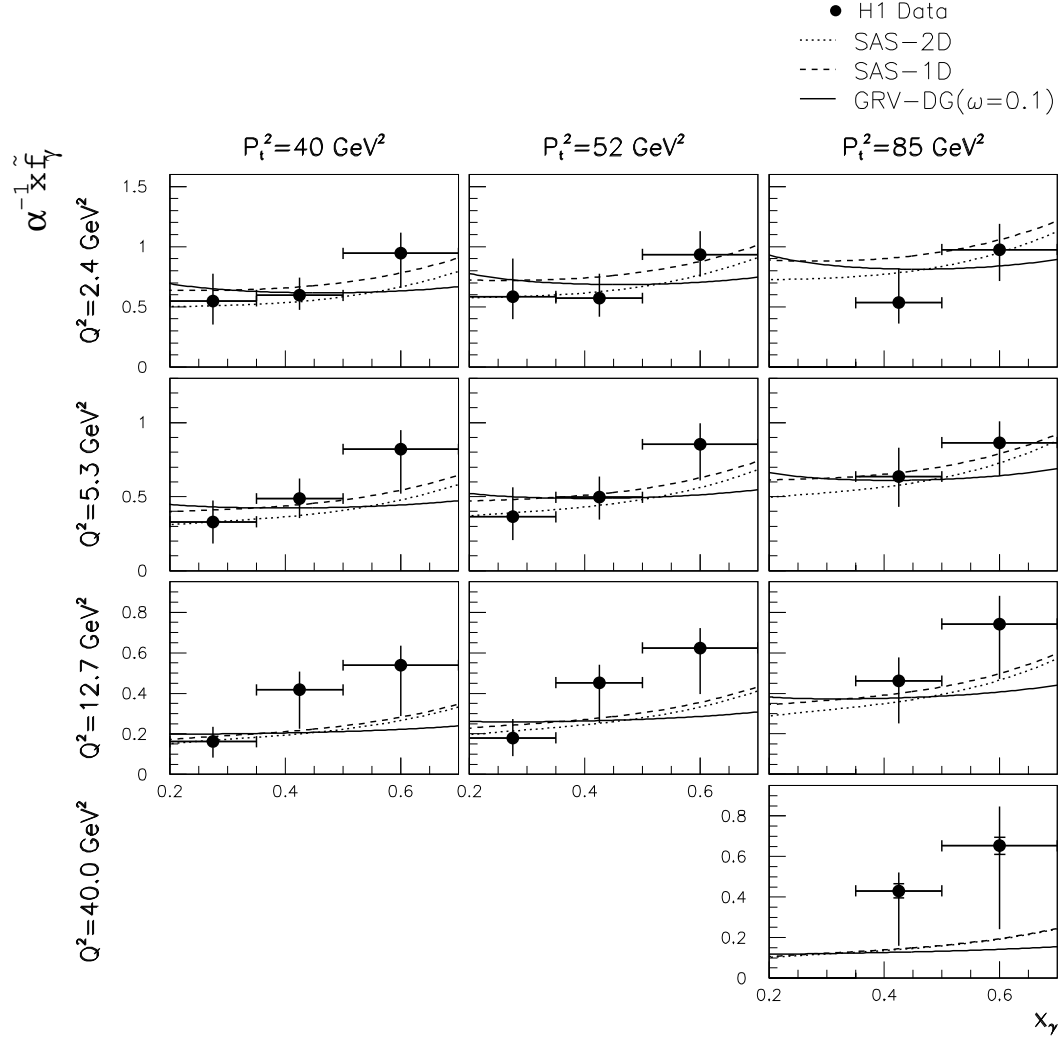}}
\caption[
         Leading order effective parton distribution function
         of virtual photons from H1.
        ]
        {
         Leading order effective parton distribution function
         of virtual photons from H1.
         The effective parton distribution function is shown as a function
         of \xg for several bins for the factorisation scale $\qsq=\ptq$
         and for the photon virtuality $\psq=\qsq$
         The inner error bar indicates the statistical and the outer
         error bar the full error. The data are compared to several
         theoretical predictions explained in the text.
        }\label{fig:chap9_18}
\end{center}
\end{figure}
%
 In Figure~\ref{fig:chap9_18} the measured leading order effective
 parton distribution function $\xg\pdfe/\aem$
 is shown as a function of \xg in the region $0.2<\xg<0.7$ and in bins of
 the factorisation scale \qsq and the photon virtuality \psq.
 A slow increase for increasing \xg is observed for all values
 of the factorisation scale and the photon virtuality.
 The lever arm in the factorisation scale \qsq
 is too small to observe the predicted logarithmic growth.
 The observed decrease with the photon virtuality \psq is much
 stronger at low values than at large values of \xg.
 For example at $\xg=0.275$, $\xg\pdfe/\aem$ decreases from 0.55 to 0.16 when
 \psq changes from 2.4 to 12.7~\gevsq, whereas at $\xg=0.6$ and for the
 same range in \psq the decrease is only from 0.95 to 0.54.
 \par
%
\begin{figure}[tbp]
\begin{center}
{\includegraphics[width=1.0\linewidth]{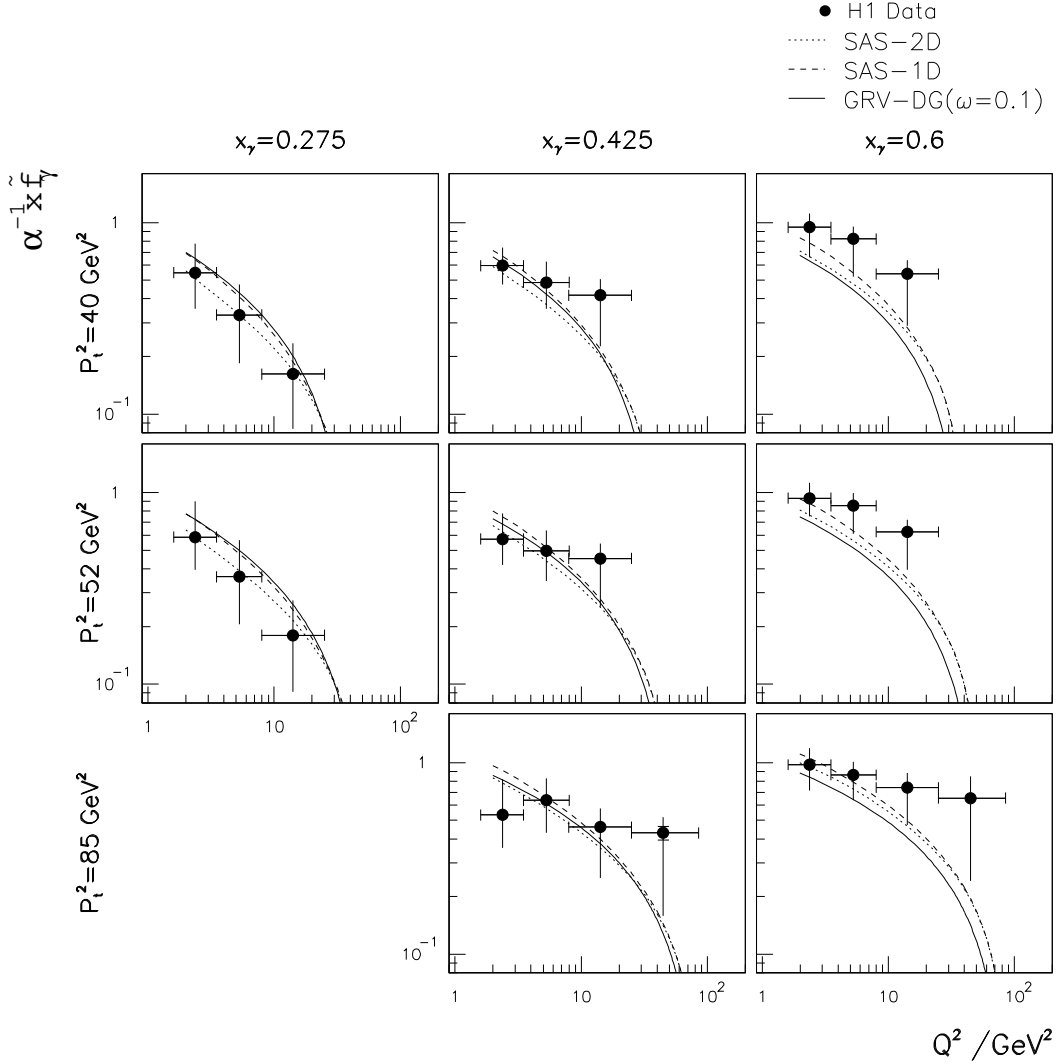}}
\caption[
         The \psq suppression of the leading order effective parton
         distribution function of virtual photon from H1.
        ]
        {
         The \psq suppression of the leading order effective parton
         distribution function of virtual photon from H1.
         The effective parton distribution function is shown as a function of
         the photon virtuality $\psq=\qsq$ for several bins in \xg
         and using several bins for the factorisation scale $\qsq=\ptq$.
         The inner error bar indicates the statistical and the outer
         error bar the full error. The data are compared to several
         theoretical predictions explained in the text.
        }\label{fig:chap9_19}
\end{center}
\end{figure}
%
 In most of the phase space the data presented in Figure~\ref{fig:chap9_18}
 can be described by the predictions based on
 the SaS1D (dash) and SaS2D (dot) parton distribution functions of the
 virtual photon using the default suppression with the
 photon virtuality $\mbox{IP2}=0$, see Section~\ref{sec:PDF} for details.
 In addition, also the GRV parton distribution functions of the
 quasi-real photon suppressed by the Drees Godbole scheme (full),
 Eq.~(\ref{eqn:DGsup}) with $\pcsq=\omega^2=0.01$~\gevsq, is
 in agreement with the data.
 The only exception is the region of large values of \xg and large photon
 virtualities, where the predictions tend to lie below the data,
 however, in this region the data suffer from large errors.
 \par
 In Figure~\ref{fig:chap9_19} the effective parton distribution function
 is shown as a function of the photon virtuality \psq in bins of \xg and
 of the factorisation scale \qsq.
 Again the observed decrease with \psq is described for
 the region $\qsq\gg\psq$ by the same model predictions as above.
 When the photon virtuality approaches the factorisation scale,
 $\psq\approx\qsq$, the models predict a
 faster decrease with the photon virtuality, than is seen in the data.
 However, it is clear that the structure function picture has to break down
 when the target virtuality and the probing scale approach each other,
 and the method to extract \pdfe can no longer be applied.
 This is the same situation as in the case of the \siggsgs
 cross-section discussed in Section~\ref{sec:qcdresvirt}.
 \par
 This completes the discussion about the experimental results on the
 measurements of the photon structure. It is evident from the discussion
 above that this field of research receives complementary information
 from different reactions.
 A rather consistent picture emerges and the general features of the
 photon structure can be accounted for by the theoretical predictions.
 However, there is still a long way to go until we reach precise
 measurements of all features of the structure of the photon.
%
%

 \ack
 I very much enjoy to participate in this active field of research
 dealing with the structure of the photon.
 It is clear that the achievements in this area of research are due
 to the effort of many people.
 Given this, it is only natural that in writing this review I greatly
 profited from the discussions with my colleagues about experimental
 and theoretical aspects of the investigations of the photon structure.
 I also received strong support from various people who provided me
 with their software and valuable advice on how to use it.
 Without this I could not have performed the comparisons between
 the data and the theoretical predictions.
 The excellent working conditions within the CERN OPAL group
 were very helpful in various aspects of this research.
 \par
 I would like to thank very much the following persons:
 \newline
 V.P.~Andreev,
 P.~Aurenche,
 C.~Berger,
 V.~Blobel,
 A.~B{\"o}hrer,
 C.~Brew,
 A.~Buijs,
 J.~Chyla,
 A.~De Roeck,
 R.~Engel,
 M.~Erdmann,
 F.C.~Ern{\'e},
 A.~Finch,
 M.~Fontannaz,
 R.~Freudenreich,
 L.~Gordon,
 K.~Hagiwara,
 R.D.~Heuer,
 R.G.~Kellogg,
 E.~Laenen,
 J.A.~Lauber,
 C.H.~Lin,
 D.J.~Miller,
 J.~Patt,
 D.E.~Plane,
 I.~Schienbein,
 G.A.~Schuler,
 M.H.~Seymour,
 T.~Sj{\"o}strand,
 S.~S{\"o}ldner-Rembold,
 H.~Spiesberger,
 M.~Stratmann,
 B.~Surrow,
 G.~Susinno,
 I.~Tyapkin,
 A.~Vogt,
 J.H.~Vossebeld,
 M.~Wadhwa,
 A.~Wagner,
 P.M.~Zerwas
 and
 V.~Zintchenko.
 \par
 Even more important was the continuous encouragement and patience
 of my wife. Without her help, I would have never been able to write
 this review during a long period of illness, vielen Dank.
%

%
%
\appendix
%
%
\section{Connecting the cross-section and the structure function picture}
\label{sec:SIGTOSF}
 In this section the connection between the cross-section and the
 structure function picture for deep inelastic scattering,
 $\qsq \gg \psq \approx 0$, is made by deriving
 Eq.~(\ref{eqn:approxlt}) from Eq.~(\ref{eqn:truesing}).
 The limit $\psq\to 0$ is explored, using the fact that the scattering
 angle of the quasi-real photon also approaches zero.
 \par
 Starting point are the four vectors of the particles defined in
 Figure~\ref{fig:chap2_01}, using the notation $a=(E_{\rm a},\vec{a})$.
%
\begin{eqnarray}
 \pone &=& (E,0,0,E),\quad \ptwo = (E,0,0,-E),\quad p \approx (\eg,0,0,-\eg),
\end{eqnarray}
%
 with $ \ptwop= \ptwo - p$. This leads to the simplified equations
%
\begin{equation}
 r =      \frac{p\cdot q}{\ptwo\cdot q} =
          \frac{(\ptwo-\ptwop)\cdot q}{\ptwo\cdot q}
 \approx  1 - \frac{E-\eg}{E} = \frac{\eg}{E} = z
\label{eqn:simpr}
\end{equation}
%
 and
%
\begin{equation}
 (p\cdot q)^2 - \qsq\psq \approx (p\cdot q)^2.
\label{eqn:simpV}
\end{equation}
%
 In addition,
%
\begin{equation}
\pone\cdot p = 2E\eg
\label{eqn:p1p}
\end{equation}
%
 will be used.
 Inserting the relations between the structure functions and the
 cross-sections in the limit of Eq.~(\ref{eqn:simpV})
 as given in Eq.~(\ref{eqn:strucnull}) into Eq.~(\ref{eqn:truesing})
 yields
%
 \begin{eqnarray}
 \mathrm{d}^6\sigma
  &=&
  \frac{\der^3\ponep\der^3\ptwop}{\eonep\etwop}
  \frac{\aemsq}{16\pi^4\qsq\psq}
  \left[
  \frac{(p\cdot q)^2-\qsq\psq}{(\pone\cdot\ptwo)^2-
         m_{\mathrm e}^2m_{\mathrm e}^2}
  \right]^{1/2} 4\ropp\rtpp\cdot
  \nonumber  \\  &&
  \frac{4\pi^2\aem}{\qsq}
  \left[2x\fTxq +\frac{\ronn}{2\ropp}\flxq\right]\, .
 \label{eqn:truesingsf}
 \end{eqnarray}
%
 The limits for the individual terms are derived below.
 \par
 The term in the first square brackets reduces to
%
\begin{equation}
  \left[\frac{(p\cdot q)^2-\qsq\psq}{(\pone\cdot\ptwo)^2-
        m_{\mathrm e}^2m_{\mathrm e}^2}\right]^{1/2}
 = \frac{p\cdot q}{2E^2} = \frac{y}{2E^2}\pone\cdot p = yz,
\label{eqn:square}
\end{equation}
%
 where Eqs.~(\ref{eqn:simpV}), (\ref{eqn:p1p}) and~(\ref{eqn:see})
 have been used.
 The photon density matrix element \rtpp defined in Eq.~(\ref{eqn:rhos})
 can be written as
%
\begin{eqnarray}
 2\rtpp
 &=& \frac{\left(2\ptwo\cdot q-p\cdot q\right)^2}
     {(p\cdot q)^2 -\qsq\psq} + 1 - 4\frac{\me^2}{\psq}
  = \left(\frac{2\ptwo\cdot q}{p\cdot q} -1\right)^2+ 1 - 4\frac{\me^2}{\psq}
                                                                 \nonumber\\
 &=& \left(\frac{2}{r} - 1\right)^2 + 1 - 4\frac{\me^2}{\psq}
  = \frac{2}{z}\left[
    \frac{1+\left(1-z\right)^2}{z} - \frac{2\me^2z}{\psq}
    \right],
\label{eqn:rtppred}
\end{eqnarray}
%
 where Eqs.~(\ref{eqn:simpV}), (\ref{eqn:y}) and~(\ref{eqn:z})
 have been used.
 Similarly \ropp leads to
%
\begin{equation}
 2\ropp = \frac{2}{y}\left[\frac{1+\left(1-y\right)^2}{y}\right],
\label{eqn:roppred}
\end{equation}
%
 where, due to $\qsq\gg\me^2$, the mass term can be neglected.
 \par
 Using this and the relation $\ronn=2(\ropp-1)$,
 Eq.~(\ref{eqn:rhoslq}), the term in front of \fl simplifies to
%
\begin{equation}
 \epsilon_1=\frac{\ronn}{2\ropp} = \frac{\ropp-1}{\ropp}
 = 1-\frac{y^2}{1+\left(1-y\right)^2}
 = \frac{2\left(1-y\right)}{1+\left(1-y\right)^2}=\epsilon(y).
\label{eqn:ropprat}
\end{equation}
%
 This shows that all terms factorise into quantities which depend only
 on the quasi-real or on the virtual photon, but not on a combination
 of them. This is certainly not obvious from the original form
 of  Eq.~(\ref{eqn:truesing}).
 \par
 Next the change of variables is performed.
 Differentiating Eq.~(\ref{eqn:p2}) with respect to $\cos\ttwop$ yields
%
\begin{equation}
 \der\psq = - 2E\etwop\der\cos\ttwop
\end{equation}
%
 and by in addition using $-\der\etwop = \der\eg= E\der z$ one derives
%
\begin{equation}
\frac{\der^3\ptwop}{\etwop} = \etwop \der\etwop \der\phtwop\der\cos\ttwop
                            = - \pi \der z \der \psq.
\label{eqn:dif2}
\end{equation}
%
 Combining Eq.~(\ref{eqn:ye}) and Eq.~(\ref{eqn:q2}) for $\psq=0$,
 a relation between \eonep and $x$ is obtained
%
\begin{equation}
\eonep = \frac{\qsq}{4E} + E - \frac{\qsq E}{x\segaq},
\end{equation}
%
 and therefore
%
\begin{equation}
\der\eonep = \frac{\qsq E}{\segaq x^2} \der x.
\end{equation}
%
 Using in addition
%
\begin{equation}
 \der\qsq = - 2E\eonep\der\cos\tonep, \nonumber
\end{equation}
%
 derived from Eq.~(\ref{eqn:q2}) as above for \psq, the other
 differential reads
%
%
\begin{equation}
 \frac{\der^3\ponep}{\eonep} = \eonep \der\eonep \der\phonep\der\cos\tonep
                             = - \pi \frac{y}{x} \der x \der \qsq.
\end{equation}
%
 Inserting all pieces into Eq.~(\ref{eqn:truesingsf})
 recovers Eq.~(\ref{eqn:approxlt})
%
 \begin{eqnarray}
 \frac{\der^4\sigma}
 {\der x\,\der\qsq\,\der z\,\der\psq}
 &=& \frac{4\pi^2\aem y}{x\,\qsq} \cdot \nonumber\\
 && \frac{\aem}{2\pi}\left[\frac{1+(1-z)^2}{z}\frac{1}{\psq}
     -\frac{2\,\me^2\,z}{P^4}\right] \cdot \nonumber\\
 && \frac{\aem}{2\pi}\left[\frac{1+(1-y)^2}{y}\frac{1}{\qsq}\right]
                                                           \cdot \nonumber\\
 && \left[2x\fTxq +
          \frac{2\left(1-y\right)}{1+\left(1-y\right)^2}\flxq
    \right].
 \label{eqn:approxfinal}
 \end{eqnarray}
%
 In this form the individual pieces can be nicely identified.
 The second line is the flux of the transverse quasi-real target photons.
 The third line is the flux of the transverse virtual photons, where
 the mass term has been neglected.
 The term in front of \fl is the ratio of the flux of the
 transverse and longitudinal virtual photons.
 Finally the structure functions \fT and \fl contain the information
 on the structure of the transverse quasi-real target photons when probed
 by transverse and longitudinal virtual photons respectively.
 \par
 The most important approximation made in deriving Eq.~(\ref{eqn:approxlt})
 from Eq.~(\ref{eqn:truesing}) is Eq.~(\ref{eqn:simpV}).
 As can be seen from the functional form of \stt and \slt, listed in
 the appendix of Ref.~\cite{BUD-7501}, exactly this term also appears
 in these cross-sections and therefore, for example, in \ftqed.
 Given this Eq.~(\ref{eqn:approxlt}) should not be used when studying
 the \psq dependence of \ftqed, and Eq.~(\ref{eqn:truesing})
 should be used instead.

%
%
\section{General concepts for deriving the parton distribution functions}
\label{sec:PDFTH}
 In this section the procedure to derive the parton distribution
 functions by solving the full evolution equations is discussed.
 There are several groups using this approach.
 However, they differ in the choices made for \qnsq, for the
 factorisation scheme and for the assumptions on the input
 parton distribution functions at the starting scales.
 The general strategy is outlined following the discussion given
 in Ref.~\cite{VOG-9701}. The individual sets of parton distribution
 functions have been discussed in Section~\ref{sec:PDF}.
 \par
 The parton distribution functions of the photon obey the
 following inhomogeneous evolution equations
%
 \begin{eqnarray}
  \frac{\der q_i^{\gamma}}{\der\ln Q^{2}} & = &
  \frac{\alpha}{2\pi} P_{q_{i}\gamma} \otimes \Gamma^{\gamma}
  +\frac{\alpha_{s}}{2\pi}\bigg\{\sum_{k=1}^{\nf}\left[
    P_{q_{i}q_{k}}\otimes q_{k}^{\gamma}
  + P_{q_{i}\bar{q}_{k}}\otimes \bar{q}_{k}^{\gamma}\right]
  + P_{q_{i}g} \otimes g^{\gamma} \bigg\},\nonumber\\
  \frac{\der \bar{q}_i^{\gamma}}{\der\ln Q^{2}} & = &
  \frac{\alpha}{2\pi} P_{\bar{q}_{i}\gamma} \otimes \Gamma^{\gamma}
  +\frac{\alpha_{s}}{2\pi}\bigg\{\sum_{k=1}^{\nf}\left[
    P_{\bar{q}_{i}q_{k}}\otimes q_{k}^{\gamma}
  + P_{\bar{q}_{i}\bar{q}_{k}}\otimes \bar{q}_{k}^{\gamma}\right]
  + P_{\bar{q}_{i}g} \otimes g^{\gamma} \bigg\},\nonumber\\
  \frac{\der g^{\gamma}}{\der \ln Q^{2}} & = &
  \frac{\alpha}{2\pi} P_{g\gamma}\otimes \Gamma^{\gamma}
  +\frac{\alpha_{s}}{2\pi} \bigg\{\sum_{k=1}^{\nf}\left[
    P_{gq_{k}}\otimes q_{k}^{\gamma}
  + P_{g\bar{q}_{k}}\otimes \bar{q}_{k}^{\gamma}\right]
  +P_{gg}\otimes g^{\gamma} \bigg\},   \nonumber \\
  \frac{\der\Gamma^{\gamma}}{\der\ln Q^{2}} & = &
  \frac{\alpha}{2\pi} \bigg\{
  P_{\gamma\gamma}\otimes \Gamma^{\gamma}
  + \sum_{k=1}^{\nf}\left[
    P_{\gamma q_{k}}\otimes q_{k}^{\gamma}
  + P_{\gamma \bar{q}_{k}}\otimes \bar{q}_{k}^{\gamma}\right]
  + P_{\gamma g}\otimes g^{\gamma} \bigg\}.
  \label{eqn:evol1}
\end{eqnarray}
%
 The parton distribution functions for the quarks
 and antiquarks are denoted with
 $q_i^{\gamma}(x,Q^2)$ and $\bar{q}_i^{\gamma}(x,Q^2)$, the
 gluon distribution function with $g^{\gamma}(x,Q^2)$, and
 $\Gamma^{\gamma}(x,Q^2)$ is the photon distribution function.
 The symbol $\otimes$ represents the convolution integral, defined as
%
\begin{equation}
 \label{eqn:Mellin}
 P\otimes q^{\gamma}(x,\qsq) =
 \int_x^1 \frac{\der y}{y}\,P(\frac{x}{y})\cdot q^{\gamma}(y,\qsq).
\end{equation}
%
 The sum runs over all active quark flavours $k=1,\dots\, ,\nf$,
 and the $P_{ab}$ are the usual Altarelli-Parisi splitting kernels
%
\begin{equation}
 \label{eqn:pgen}
 P_{ab}(z,\aem,\al) = \sum_{l,m}^{\infty}
 \frac{\alpha^l\alpha_s^m}{(2\pi)^{l+m}}P_{ab}^{(l,m)}(z).
\end{equation}
%
 \par
 Since $\aem\approx 1/137$ is very small, the expansion of
 Eq.~(\ref{eqn:evol1}) in powers of \aem is cut at ${\mathcal{O}}(\aem)$.
 To this order the terms $P_{\gamma q_{k}}$, $P_{\gamma \bar{q}_{k}}$
 and $P_{\gamma g}$ do not contribute and the evolution
 equation for the photon inside the photon can be solved directly.
 Since photon radiation from photons starts at order \aemsq one can
 use ${P}_{\gamma \gamma}\propto \delta (1-x)$ to all orders in $\alpha_s$,
 and with this, the photon distribution function
 in leading order is given by
%
\begin{equation}
 \label{eqn:evol2}
 \Gamma_{\rm LO}^{\gamma}(x,Q^2) = \delta (1-x)\Big[1-\frac{\alpha}{\pi}
 \big(\sum_{k=1}^{\nf} \eqkt\ln\frac{Q^2}{Q_0^2} + c_1 \big) \Big],
\end{equation}
%
 where \qnsq is the starting scale of the evolution and $c_1$ is an unknown
 parameter.
 Since the photon coupling to quarks and anti-quarks is the same the
 quark distribution functions fulfill $q_i^\gamma=\bar{q}_i^\gamma$ and
 one of the first two equations from Eq.~(\ref{eqn:evol1}) can be removed.
 Since $ q_i^{\gamma}$ and $g^{\gamma}$ are already of order $\alpha$
 only the ${\mathcal{O}}(1)$ contribution from $\Gamma^{\gamma}$ has
 to be used in the remaining evolution equations which thereby reduce to
%
\begin{eqnarray}
 \frac{\der q_i^{\gamma}}{\der\ln Q^{2}} &= & \frac{\alpha}{2\pi}P_{q_i\gamma}
 + \frac{\alpha_{s}}{2\pi}\bigg\{\sum_{k=1}^{\nf}\left(
   P_{q_{i}q_{k}} + P_{q_{i}\bar{q}_{k}}\right)\otimes q_{k}^{\gamma}
 + P_{q_{i}g} \otimes g^{\gamma} \bigg\},\nonumber \\
 \frac{\der g^{\gamma}}{\der\ln Q^{2}} &=& \frac{\alpha}{2\pi} P_{g\gamma}
 +\frac{\alpha_{s}}{2\pi}\bigg\{\sum_{k=1}^{\nf}\left(
  P_{g q_{k}}+P_{g \bar{q}_{k}}\right) \otimes q_{k}^{\gamma}
 +P_{gg}\otimes g^{\gamma} \bigg\}.
\label{eqn:evol3}
\end{eqnarray}
%
 \par
%
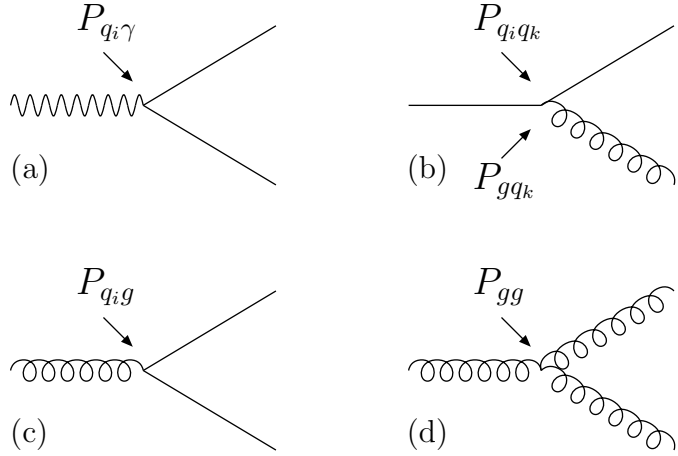
\begin{figure}[tbp]\unitlength 1pt
\begin{center}
\begin{picture}(300,200)(0,0)
 \Photon(30,150)(80,150){4}{8.5}
 \Line(80,150)(130,180)
 \Line(80,150)(130,120)
 \Text(55,175)[lb]{\mbox{\large $P_{q_i\gamma}$}}
 \LongArrow(65,170)(75,160)
 \Text(30,120)[lb]{(a)}
 \Line(180,150)(230,150)
 \Line(230,150)(280,180)
 \Gluon(230,150)(280,120){4}{6.5}
 \Text(205,175)[lb]{\mbox{\large $P_{q_{i}q_{k}}$}}
 \LongArrow(215,170)(225,160)
 \Text(205,115)[lb]{\mbox{\large $P_{gq_{k}}$}}
 \LongArrow(215,130)(225,140)
 \Text(180,120)[lb]{(b)}
 \Gluon(30,50)(80,50){4}{6.5}
 \Line(80,50)(130,80)
 \Line(80,50)(130,20)
 \Text(55,75)[lb]{\mbox{\large $P_{q_{i}g}$}}
 \LongArrow(65,70)(75,60)
 \Text(30,20)[lb]{(c)}
 \Gluon(180,50)(230,50){4}{6.5}
 \Gluon(230,50)(280,80){4}{6.5}
 \Gluon(230,50)(280,20){4}{6.5}
 \Text(205,75)[lb]{\mbox{\large $P_{gg}$}}
 \LongArrow(215,70)(225,60)
 \Text(180,20)[lb]{(d)}
\end{picture}
\caption[
         Feynman diagrams illustrating the Altarelli Parisi splitting kernels.
        ]
        {
         Feynman diagrams illustrating the Altarelli Parisi splitting kernels.
         Shown are
         (a) the branching of a photon into a quark pair,
         (b) the branching of a quark  into a quark and a gluon,
         (c) the branching of a gluon  into a quark pair,
         and
         (d) the branching of a gluon  into two gluons.
        }\label{fig:chap11_1}
\end{center}
\end{figure}
%
 The Altarelli-Parisi splitting kernels describe the parton branchings,
 as illustrated in Figure~\ref{fig:chap11_1}.
 In leading order they have the following form
%
\begin{eqnarray}
 P_{q_{i}\gamma}\left(z\right) &=&
 3\eqit\left[z^2+\left(1-z\right)^2\right]\,,\nonumber\\
 P_{q_{i}q_{k}}\left(z\right)  &=&
 \delta_{ik}\left[\frac{4}{3}\frac{1+z^2}{\left(1-z\right)_{+}}+
 2\delta\left(1-z\right)\right]\,,\nonumber\\
 P_{q_{i}\bar{q}_{k}}\left(z\right) &=& 0 \,,\nonumber\\
 P_{q_{i}g}\left(z\right) &=&
 \frac{1}{2}\left[z^2+\left(1-z\right)^2\right]\,,\nonumber\\
 P_{g\gamma}\left(z\right) &=& 0 \,,\nonumber\\
 P_{gq_{k}}\left(z\right)  &=&
 \frac{4}{3}\left[\frac{1+\left(1-z\right)^2}{z}\right]\,,\nonumber\\
 P_{g\bar{q}_{k}}\left(z\right) &=& P_{gq_{k}}\left(z\right)
 \,,\nonumber\\
 P_{gg}\left(z\right) &=&
 6\bigg[\frac{1-z}{z}+\frac{z}{\left(1-z\right)_+}+z\left(1-z\right)+
 \left(\frac{11}{12}-\frac{\nf}{18}\right)
 \delta\left(1-z\right)\bigg]
 \,.
\label{eqn:split}
\end{eqnarray}
%
 \par
 The evolution equations are inhomogeneous, because of the
 occurrence of the term $P_{q_{i}\gamma}$ describing the coupling of the
 photon to quarks. If it were not for this term, the evolution
 equations would be identical to the evolution equations for
 parton distribution functions of hadrons like the proton.
 This is why the solution of the homogeneous evolution equations
 can be identified with the hadron-like part of the photon structure function,
 and its $x$ and \qsq behaviour is just as in the hadron case.
 A particular solution to the inhomogeneous evolution equations can be
 identified with the point-like part of the photon structure function.
 \par
 The parton distribution functions are subject to a momentum sum rule,
 which can be expressed as
%
\begin{equation}
 \label{eqn:msr1}
 \int_0^1 \der x \bigg\{x \Big[ \Sigma^{\gamma}(x,Q^2) + g^{\gamma}(x,Q^2) +
 \Gamma^{\gamma}(x,Q^2) \Big]\bigg\} = 1,
\end{equation}
%
 where $\Sigma^{\gamma}$ is defined in Eq.~(\ref{eqn:Sigdef}).
 This momentum sum rule holds order by order in
 $\alpha$, thus by using Eq.~(\ref{eqn:evol2}) to order
 ${\mathcal{O}}(\aem)$ the momentum sum rule reads
%
\begin{equation}
 \label{eqn:msr2}
 \int_0^1\der x\bigg\{x\Big[\Sigma_{\rm LO}^{\gamma}(x,Q^2)+g^{\gamma}_{\rm
  LO}(x,Q^2) \Big]\bigg\} = \frac{\alpha}{\pi} \Big(\sum_{k=1}^{\nf}
                             \eqkt\ln \frac{Q^2}{Q_0^2} + c_1 \Big).
\end{equation}
%
 This means that the quark and gluon distribution functions of the
 photon do not obey a momentum sum rule which is independent of \qsq as,
 for example, the parton distribution functions of the proton.
 In contrast, the momentum carried by
 the partons of the photon rises logarithmically with \qsq, with an
 unknown parameter $c_1$, which has to be obtained from somewhere
 else, as performed, for example, in Refs.~\cite{SCH-9501,FRK-9601,FRK-9602}.
 This has been done by relating Eq.~(\ref{eqn:evol2}) to the hadronic
 \epem annihilation cross-section $\sigma(\epem\rightarrow\mbox{hadrons})$
 by means of a dispersion relation in the photon virtuality.
 For example, the value of $c_1/\pi$ obtained in Ref.~\cite{SCH-9501}
 is about 0.55 with an uncertainty of about 20$\%$
 for $\qnsq=0.36$~\gevsq.
 The main consequence of this is that for the photon the
 constraint on the gluon distribution function from the momentum sum
 rule together with the measurement of \ft is not as powerful
 as in the case of the proton.
 In addition, there is a theoretical debate on this issue, and the
 applicability of this sum rule has been questioned in Ref.~\cite{GLU-9801}.
 \par
 The general solution of the inhomogeneous evolution equations,
 Eqs.~(\ref{eqn:evol3}), for the flavour singlet part, which
 is given by
%
\begin{equation}
\label{eqn:sol1}
 \vec{q}^{\,\,\gamma}(x,Q^2) = \left(\begin{array}{c}\Sigma^{\,\gamma}(x,Q^2)
 \\ g^{\gamma}(x,Q^2)\end{array}\right)
 =\vec{q}_{\rm PL}^{\,\,\gamma}(x,Q^2)+\vec{q}_{\rm had}^{\,\,\gamma}(x,Q^2) ,
\end{equation}
%
 can be expressed in terms of
 the point-like part, $\vec{q}_{\rm PL}^{\,\,\gamma}(x,Q^2)$,
 and
 the hadron-like part, $\vec{q}_{\rm had}^{\,\,\gamma}(x,Q^2)$,
 taken as the solution of the homogeneous evolution equation.
 At next-to-leading order these solutions can be written as
%
\begin{eqnarray}
 \label{eqn:sol2}
 \vec{q}^{\,\,\gamma}_{\rm had}(x,Q^2) &=&\bigg(\Big[\frac{\alpha_s}{\alpha_0}
 \Big]^{\hat{d}(x)} + \frac{\alpha_s}{2\pi} \bigg\{ \hat{U}(x)\otimes\Big[\frac
 {\alpha_s} {\alpha_0} \Big]^{\hat{d}(x)}-\Big[\frac{\alpha_s}{\alpha_0} \Big]
 ^{\hat{d}(x)}\otimes\hat{U}(x)\bigg\}\bigg)\nonumber\\
 &&\quad\quad\quad\quad\otimes\vec{q}^{\,\,\gamma}(Q_0^2),
\end{eqnarray}
%
 and
%
\begin{eqnarray}
 \label{eqn:sol3}
 \vec{q}^{\,\,\gamma}_{\rm PL}(x,Q^2)&=&\bigg\{\frac{2\pi}{\alpha_s}\!+
 \hat{U}(x)
 \bigg\} \otimes \bigg\{ 1\! -\Big[ \frac{\alpha_s}{\alpha_0} \Big]
 ^{1+\hat{d}(x)} \bigg\}\otimes\frac{1}{1\! +\hat{d}(x)} \otimes \vec{a}(x)
 \nonumber\\&&
 + \bigg\{1-\Big[\frac{\alpha_s}{\alpha_0} \Big]^{\hat{d}(x)} \bigg\}
 \otimes \frac{1}{\hat{d}(x)} \otimes \vec{b}(x),
\end{eqnarray}
%
 where $\alpha_s = \alpha_s(Q^2)$ and $\alpha_0 = \alpha_s(Q_0^2)$.
 The quantities $\vec{a}(x)$, $\vec{b}(x)$, $\hat{d}(x)$ and $\hat{U}(x)$
 abbreviate combinations of the splitting functions and the QCD
 $\beta$-function.
%
 The solutions to the leading order evolution equations are contained in
 Eqs.~(\ref{eqn:sol2}) and~(\ref{eqn:sol3}), and are obtained
 for $\hat{U}(x) = 0$ and $\vec{b}(x)=\vec{0}$.
 The asymptotic point-like solution is obtained from Eq.~(\ref{eqn:sol3})
 in next-to-leading order for $\hat{U}(x)=0$ and by dropping the
 terms proportional to $\al/\alpha_0$, which vanish for
 $\qsq\to\infty$ leading to
%
\begin{equation}
 \label{eqn:sol4}
 \vec{q}^{\,\,\gamma}_{\rm asy}(x,\qsq) = \frac{2\pi}{\alpha_s(\qsq)}\otimes
 \frac{1}{1+\hat{d}(x)}\otimes\vec{a}(x)+\frac{1}{\hat{d}(x)}\otimes\vec{b}(x).
\end{equation}
%
 The leading order result is obtained by in addition setting
 $\vec{b}(x)=\vec{0}$.
 The pole at $\hat{d}(x) = -1$ is responsible for the divergence
 of the asymptotic solution.
 The most recent parametrisation of the leading order asymptotic
 solution is given in Ref.~\cite{GOR-9201}.
 \par
 The next-to-leading order structure function \ft for light quarks
 is given by
%
\begin{equation}
 \label{eqn:f2ph}
  F_{2}^{\,\gamma} = 2x \sum_{k=1}^{3} \eqkt \Big\{ q_k^{\gamma}
  + \frac{\alpha_{s}}{2\pi} \big(C_{2,q} \otimes q_k^{\gamma}+ C_{2,g} \otimes
  g^{\gamma} \big) + \frac{\alpha}{2\pi} \eqkt C_{2,\gamma}\Big\},
\end{equation}
%
 where the $C_{2,{\rm i}}$ are the next-to-leading order coefficient
 functions.
 In next-to-leading order there exist the factorisation scheme
 ambiguity, which means a freedom in the definition of the terms
 belonging to the parton density functions and the terms
 which are included in the hard scattering matrix elements.
 The different choices are known as factorisation schemes.
 A physics quantity like \ft, which is a combination of
 parton density functions and hard scattering matrix elements,
 is invariant under this choice, if calculated to all orders in
 perturbation theory, but in fixed order the results from
 different factorisation schemes can differ by finite terms,
 as discussed, for example, in Ref.~\cite{AUR-9402}.
 Commonly used factorisation schemes for the proton structure function
 are the \msbl scheme and the DIS scheme.
 For the photon structure function the DIS$_\gamma$ scheme, as
 introduced in Ref.~\cite{GLU-9201}, is motivated by the DIS
 scheme\footnote{
 The DIS scheme absorbs all higher order terms into the definition of the
 quark distribution functions, such that \ftp is proportional solely to
 the quark distribution functions in all orders in \al.
 In contrast to this, the \disg scheme in next-to-leading order
 absorbs only the purely photonic part, $C_{2,\gamma}$,
 into the quark distribution functions.}.
 The photonic, non-universal part, $C_{2,\gamma}$, in next-to-leading
 order is given by
%
\begin{equation}
 \label{eqn:c2ph}
  C_{2,\gamma}^{\overline{\rm MS}}(x)=3\Big\{\big[x^2+(1-x^2)\big]
 \ln\frac{1-x}{x}-1+8x(1-x)\Big\}.
\end{equation}
%
 In the \disg scheme this term is absorbed into the quark
 distribution functions by using the definitions
%
\begin{equation}
 \label{eqn:disg}
  q^{\gamma}_{\,{\rm k,DIS}_{\gamma}} = q^{\gamma}_{k,\overline{\rm MS}}
  + \frac{\alpha}{2\pi} \eqkt\, C_{2,\gamma}^{\,\overline{\rm MS}}
  \: , \:\:\:\:  C_{\, 2,\gamma}^{\,{\rm DIS}_{\gamma}} = 0.
\end{equation}
%
 If the calculation is performed in the DIS$_\gamma$ factorisation
 scheme, then there is a good stability of the perturbative prediction
 when comparing the leading order and next-to-leading order
 results, as can be seen from Ref.~\cite{GLU-9201}.
 By applying the \msbl factorisation scheme, there are much larger
 differences between the leading order and next-to-leading order
 results of \ft, stemming from the large negative contribution
 to the point-like part of the
 purely photonic part $C_{2,\gamma}$ in next-to-leading order
 at large values of $x$, as can be seen from
 Eq.~(\ref{eqn:c2ph}) in the limit $x\to 1$.
 This results even in a negative structure function \ft at large values of
 $x$, if it is not compensated for by carefully choosing also the point-like
 input distribution functions, as done in Refs.~\cite{GOR-9701,AUR-9402}.
 The negative structure function \ft at large values of
 $x$ can by avoided by using a \emph{technical\/} next-to-leading order
 input distribution function for the point-like part of the
 following form
%
\begin{equation}
 \label{eqn:msbg}
 q_{{\rm k,}\overline{\rm PL}}^{\gamma}(x,Q_{0}^{2}) = - \frac{\alpha}{2\pi}
 \eqkt \, C_{2,\gamma}^{\prime}(x) \: , \:\:\:\:
 g_{\overline{\rm PL}}^{\gamma}(x,Q_{0}^{2})=0.
\end{equation}
%
 Suitable expression for the term $C_{2,\gamma}^{\prime}$ are either
 $C_{2,\gamma}^{\overline{\rm MS}}(x)$ as defined in Eq.~(\ref{eqn:c2ph})
 or
\begin{equation}
 \label{eqn:c2pr}
 C^{\prime}_{\gamma}(x)=3\Big\{\big[x^2+(1-x^2)\big]\ln(1-x)+2x(1-x)\Big\}.
\end{equation}
%
 The first solution is chosen in Refs.~\cite{GOR-9701} for the
 construction of the parametrisations from Gordon and Storrow, and
 Eq.~(\ref{eqn:c2pr}) was developed for the parametrisations from
 Aurenche, Fontannaz and Guillet in Refs.~\cite{AUR-9402}, based
 on an analysis of the momentum integration of the box diagram.
 As stated above these are purely technical questions on how to deal
 with the factorisation scheme ambiguity.
 The predictions for the photon structure function \ft calculated
 in the different schemes are different.
 However, the differences in \ft are much smaller than the differences
 in the parton distribution functions, as can be seen from
 Ref.~\cite{VOG-9701}.
 This also shows that although \ft is a well defined quantity its
 interpretation in terms of parton distribution functions of the photon
 is a delicate issue.
%
%

%
%
\section{Collection of results on the QED structure of the photon}
\label{sec:tabqed}
 This section contains a summary of the available results on the
 QED structure of the photon obtained either from measurements of
 deep inelastic electron-photon scattering or by exploring
 the exchange of two highly virtual photons.
 The numbers listed in the tables are the basis of the summary plots
 shown in Section~\ref{sec:qedres}.
 The sources of information used to obtain the numbers are always
 given in the caption of the respective table.
 If the total error $\sigma_{\rm tot}$ has been obtained in this review,
 it was calculated from the statistical error $\sigma_{\rm stat}$
 and the systematic error $\sigma_{\rm sys}$ using the relation
%
\begin{equation}
\sigma_{\rm tot} = \sqrt{\sigma^2_{\rm stat} + \sigma^2_{\rm sys}}.
\label{eqn:error}
\end{equation}
%
 First the measurements of \ftqed are listed, followed by results on
 \faqed and \fbqed, and by the results for the exchange of two highly
 virtual photons.
 \par
%
\renewcommand{\arraystretch}{1.0}
\begin{table}[htb]
\caption[
         Results on the average photon structure function
         $\langle\ftqed\rangle$ from the CELLO experiment.
        ]
        {
         Results on the average photon structure function
         $\langle\ftqed\rangle$ from the CELLO experiment.
         The numbers are read off the published
         figure, which probably contains only the statistical error.
         The additional quoted systematic error of 8$\%$ is added in
         quadrature.
         The measured \ftqed is averaged over the \qsq range
         $1.2-39$~\gevsq, with an average value of $\qzm=9.5$~\gevsq.
         No information is available to which value of \pzm the result
         corresponds.\\
        }\label{tab:chap12_01}
\begin{center}
\begin{tabular}{cccccc}\hline
\multicolumn{6}{c}{\textbf{CELLO}}\\\hline
 $\qsq\,\left[\gevsq\right]$ & $x$ & $\langle\ftqed\rangle$ &
 $\sigma_{\rm stat}$ & $\sigma_{\rm tot}$ & Ref. \\\hline
 $1.2-39$
     & 0.00$-$0.10& 0.222& 0.077& 0.079&\protect\cite{CEL-8301}\\
     & 0.10$-$0.20& 0.426& 0.128& 0.132&\\
     & 0.20$-$0.30& 0.562& 0.162& 0.168&\\
     & 0.30$-$0.40& 0.511& 0.153& 0.158&\\
     & 0.40$-$0.50& 0.597& 0.170& 0.177&\\
     & 0.50$-$0.60& 0.571& 0.170& 0.176&\\
     & 0.60$-$0.70& 0.545& 0.170& 0.176&\\
     & 0.70$-$0.80& 1.202& 0.256& 0.273&\\
     & 0.80$-$0.90& 1.057& 0.273& 0.286&\\
     & 0.90$-$1.00& 1.185& 0.528& 0.536&\\
\hline
\end{tabular}
\end{center}\end{table}
%
\renewcommand{\arraystretch}{1.0}
\begin{table}[htb]
\caption[
         Results on the average photon structure function
         $\langle\ftqed\rangle$ from the DELPHI experiment.
        ]
        {
         Results on the average photon structure function
         $\langle\ftqed\rangle$ from the DELPHI experiment.
         The numbers are provided by V.~Podznyakov.
         Only statistical errors are available.
         The measured \ftqed is averaged over the \qsq range
         $4-30$~\gevsq, with an average of $\qzm=12$~\gevsq.
         The best fit of the QED prediction to the data is
         obtained for $\pzm=0.04$~\gevsq.
         In addition, there exist preliminary results, which are listed
         in Table\protect~\ref{tab:chap12_08}.\\
        }\label{tab:chap12_02}
\begin{center}
\begin{tabular}{cccccc}\hline
\multicolumn{6}{c}{\textbf{DELPHI}}\\\hline
 $\qsq\,\left[\gevsq\right]$ & $x$ & $\langle\ftqed\rangle$ &
 $\sigma_{\rm stat}$ & $\sigma_{\rm tot}$ & Ref. \\\hline
 $4-30$
     & 0.00$-$0.08& 0.077& 0.013& &\protect\cite{DEL-9601}\\
     & 0.08$-$0.18& 0.193& 0.016& &\\
     & 0.18$-$0.31& 0.327& 0.026& &\\
     & 0.31$-$0.48& 0.513& 0.032& &\\
     & 0.48$-$0.70& 0.719& 0.051& &\\
     & 0.70$-$1.00& 0.969& 0.109& &\\
\hline
\end{tabular}
\end{center}\end{table}
%
\renewcommand{\arraystretch}{1.0}
\begin{table}[htb]
\caption[
         Results on the average photon structure function
         $\langle\ftqed\rangle$ from the L3 experiment.
        ]
        {
         Results on the average photon structure function
         $\langle\ftqed\rangle$ from the L3 experiment.
         The numbers for $\langle\ftqed\rangle$ are read off the
         published figure, and
         the average value of \qsq is provided by G.~Susinno.
         The measured \ftqed is averaged over the \qsq range
         $1.4-7.6$~\gevsq, with an average of $\qzm=3.25$~\gevsq.
         The best fit of the QED prediction to the data
         is obtained for $\pzm=0.033$~\gevsq.\\
        }\label{tab:chap12_03}
\begin{center}
\begin{tabular}{cccccc}\hline
\multicolumn{6}{c}{\textbf{L3}}\\ \hline
 $\qsq\,\left[\gevsq\right]$ & $x$ & $\langle\ftqed\rangle$ &
 $\sigma_{\rm stat}$ & $\sigma_{\rm tot}$ & Ref. \\\hline
  $1.4-7.6$
      & 0.00$-$0.10& 0.062& 0.060& 0.063&\protect\cite{L3C-9801}\\
      & 0.10$-$0.20& 0.216& 0.015& 0.018&\\
      & 0.20$-$0.30& 0.326& 0.022& 0.025&\\
      & 0.30$-$0.40& 0.391& 0.026& 0.030&\\
      & 0.40$-$0.50& 0.477& 0.028& 0.032&\\
      & 0.50$-$0.60& 0.534& 0.029& 0.035&\\
      & 0.60$-$0.70& 0.654& 0.035& 0.039&\\
      & 0.70$-$0.80& 0.709& 0.037& 0.044&\\
      & 0.80$-$0.90& 0.775& 0.046& 0.052&\\
      & 0.90$-$1.00& 0.549& 0.069& 0.072&\\
\hline
\end{tabular}
\end{center}\end{table}
%
\renewcommand{\arraystretch}{1.0}
\begin{table}[htb]
\caption[
         Results on the photon structure function \ftqed from
         the OPAL experiment.
        ]
        {
         Results on the photon structure function \ftqed from
         the OPAL experiment.
         For explanations see Table~\ref{tab:chap12_05}.\\
        }\label{tab:chap12_04}
\begin{center}
\begin{tabular}{cccccc}\hline
\multicolumn{6}{c}{\textbf{OPAL}}\\ \hline
 $\qzm\,\left[\gevsq\right]$ & $x$ & \ftqed &
 $\sigma_{\rm stat}$ & $\sigma_{\rm tot}$ & Ref. \\\hline
  2.2& 0.00$-$0.10& 0.115& 0.007& 0.009&\protect\cite{OPALPR271}\\
     & 0.10$-$0.20& 0.219& 0.010& 0.013&\\
     & 0.20$-$0.30& 0.282& 0.012& 0.016&\\
     & 0.30$-$0.40& 0.347& 0.015& 0.019&\\
     & 0.40$-$0.50& 0.356& 0.017& 0.020&\\
     & 0.50$-$0.60& 0.400& 0.020& 0.023&\\
     & 0.60$-$0.70& 0.483& 0.025& 0.030&\\
     & 0.70$-$0.80& 0.491& 0.031& 0.033&\\
     & 0.80$-$0.90& 0.532& 0.034& 0.036&\\
     & 0.90$-$0.97& 0.308& 0.032& 0.078&\\
\hline
  4.2& 0.00$-$0.10& 0.108& 0.010& 0.019&\protect\cite{OPALPR271}\\
     & 0.10$-$0.20& 0.237& 0.014& 0.017&\\
     & 0.20$-$0.30& 0.320& 0.018& 0.022&\\
     & 0.30$-$0.40& 0.378& 0.020& 0.023&\\
     & 0.40$-$0.50& 0.373& 0.020& 0.022&\\
     & 0.50$-$0.60& 0.421& 0.025& 0.028&\\
     & 0.60$-$0.70& 0.519& 0.029& 0.032&\\
     & 0.70$-$0.80& 0.556& 0.034& 0.036&\\
     & 0.80$-$0.90& 0.601& 0.040& 0.042&\\
     & 0.90$-$0.97& 0.470& 0.041& 0.065&\\
\hline
  8.4& 0.00$-$0.10& 0.090& 0.012& 0.014&\protect\cite{OPALPR271}\\
     & 0.10$-$0.20& 0.271& 0.022& 0.029&\\
     & 0.20$-$0.30& 0.334& 0.029& 0.035&\\
     & 0.30$-$0.40& 0.409& 0.033& 0.040&\\
     & 0.40$-$0.50& 0.496& 0.038& 0.046&\\
     & 0.50$-$0.60& 0.563& 0.043& 0.050&\\
     & 0.60$-$0.70& 0.596& 0.049& 0.054&\\
     & 0.70$-$0.80& 0.687& 0.056& 0.061&\\
     & 0.80$-$0.90& 0.891& 0.072& 0.084&\\
     & 0.90$-$0.97& 0.761& 0.074& 0.089&\\
\hline
 12.4& 0.00$-$0.15& 0.151& 0.022& 0.025&\protect\cite{OPALPR271}\\
     & 0.15$-$0.30& 0.297& 0.033& 0.037&\\
     & 0.30$-$0.45& 0.402& 0.041& 0.048&\\
     & 0.45$-$0.60& 0.434& 0.044& 0.056&\\
     & 0.60$-$0.75& 0.758& 0.062& 0.074&\\
     & 0.75$-$0.90& 0.723& 0.072& 0.077&\\
     & 0.90$-$0.97& 0.714& 0.085& 0.090&\\
\hline
\end{tabular}
\end{center}\end{table}
%
\renewcommand{\arraystretch}{1.0}
\begin{table}[htb]
\caption[
         Results on the photon structure function \ftqed from
         the OPAL experiment, continued.
        ]
        {
         Results on the photon structure function \ftqed from
         the OPAL experiment continued.
         The values are taken from the published Tables of
         Ref.\protect~\cite{OPALPR271}.
         These results on \ftqed superseed the results published in
         Ref.\protect~\cite{OPALPR088}.
         The results given at $\qzm=2.2,4.2,8.4,12.4,21.0$ and 130~\gevsq
         are all statistically independent and are
         unfolded from data in the \qsq ranges $1.5-3$,
         $3-7$, $6-10$, $10-15$, $15-30$, and $70-400$~\gevsq.
         The results given at $\qzm=3.0$~\gevsq are unfolded from data in the
         \qsq ranges $1.5-7$~\gevsq, which means they are not independent
         results, but contain the data at $\qzm=2.2$ and 4.2~\gevsq.
         This data is
         used for the comparisons made in Figure\protect~\ref{fig:chap6_04}.
         All results are unfolded for $\pzm=0.05$~\gevsq.\\
        }\label{tab:chap12_05}
\begin{center}
\begin{tabular}{cccccc}\hline
\multicolumn{6}{c}{\textbf{OPAL}}\\ \hline
 $\qzm\,\left[\gevsq\right]$ & $x$ & \ftqed &
 $\sigma_{\rm stat}$ & $\sigma_{\rm tot}$ & Ref. \\\hline
 21.0& 0.00$-$0.15& 0.117& 0.028& 0.030&\protect\cite{OPALPR271}\\
     & 0.15$-$0.30& 0.302& 0.039& 0.044&\\
     & 0.30$-$0.45& 0.403& 0.051& 0.059&\\
     & 0.45$-$0.60& 0.559& 0.058& 0.065&\\
     & 0.60$-$0.75& 0.782& 0.070& 0.078&\\
     & 0.75$-$0.90& 0.907& 0.080& 0.087&\\
     & 0.90$-$0.97& 0.802& 0.103& 0.108&\\
\hline
  130& 0.10$-$0.40& 0.343& 0.094& 0.100&\protect\cite{OPALPR271}\\
     & 0.40$-$0.60& 0.578& 0.079& 0.095&\\
     & 0.60$-$0.80& 0.936& 0.109& 0.126&\\
     & 0.80$-$0.90& 1.125& 0.130& 0.142&\\
\hline
\hline
  3.0& 0.00$-$0.10& 0.113& 0.006& 0.011&\protect\cite{OPALPR271}\\
     & 0.10$-$0.20& 0.230& 0.009& 0.012&\\
     & 0.20$-$0.30& 0.300& 0.011& 0.015&\\
     & 0.30$-$0.40& 0.363& 0.013& 0.018&\\
     & 0.40$-$0.50& 0.364& 0.014& 0.016&\\
     & 0.50$-$0.60& 0.409& 0.017& 0.020&\\
     & 0.60$-$0.70& 0.507& 0.021& 0.029&\\
     & 0.70$-$0.80& 0.516& 0.025& 0.018&\\
     & 0.80$-$0.90& 0.574& 0.029& 0.045&\\
     & 0.90$-$0.97& 0.397& 0.029& 0.066&\\
\hline
\end{tabular}
\end{center}\end{table}
%
\renewcommand{\arraystretch}{1.0}
\begin{table}[htb]
\caption[
         Results on the photon structure function \ftqed from
         the PLUTO experiment.
        ]
        {
         Results on the photon structure function \ftqed from
         the PLUTO experiment.
         The numbers are read off the published
         figure, which probably contains the full error.
         The results given at $\qzm=5.5$ and 40~\gevsq
         are unfolded from data in the \qsq ranges $1-16$
         and $10-160$~\gevsq.
         No information is available to which value of \pzm the result
         corresponds.\\
        }\label{tab:chap12_06}
\begin{center}
\begin{tabular}{cccccc}\hline
\multicolumn{6}{c}{\textbf{PLUTO}}\\ \hline
 $\qzm\,\left[\gevsq\right]$ & $x$ & \ftqed &
 $\sigma_{\rm stat}$ & $\sigma_{\rm tot}$ & Ref. \\\hline
  5.5& 0.00$-$0.10& 0.081&    & 0.040&\protect\cite{PLU-8501}\\
     & 0.10$-$0.20& 0.177&    & 0.048&\\
     & 0.20$-$0.30& 0.532&    & 0.089&\\
     & 0.30$-$0.40& 0.403&    & 0.105&\\
     & 0.40$-$0.50& 0.532&    & 0.113&\\
     & 0.50$-$0.60& 0.597&    & 0.161&\\
     & 0.60$-$0.70& 0.952&    & 0.322&\\
     & 0.70$-$0.80& 0.887&    & 0.444&\\
\hline
   40& 0.00$-$0.20& 0.177&    & 0.113&\protect\cite{PLU-8501}\\
     & 0.20$-$0.40& 0.565&    & 0.210&\\
     & 0.40$-$0.60& 0.532&    & 0.241&\\
     & 0.60$-$0.80& 1.532&    & 0.468&\\
     & 0.80$-$1.00& 0.807&    & 0.581&\\
\hline
\end{tabular}
\end{center}\end{table}
%
\renewcommand{\arraystretch}{1.0}
\begin{table}[htb]
\caption[
         Results on the average photon structure function
         $\langle\ftqed\rangle$ from the TPC/2$\gamma$ experiment.
        ]
        {
         Results on the average photon structure function
         $\langle\ftqed\rangle$ from the TPC/2$\gamma$ experiment.
         The numbers are read off the published figure.
         The measured \ftqed is averaged over the approximate range
         in \qsq of $0.14-1.28$~\gevsq, with an average of about
         $\qzm=0.45$~\gevsq, which was estimated from the GALUGA Monte Carlo
         using the experimental requirements of the TPC/2$\gamma$
         analysis.
         No information is available to which value of \pzm the result
         corresponds.\\
        }\label{tab:chap12_07}
\begin{center}
\begin{tabular}{cccccc}\hline
\multicolumn{6}{c}{\textbf{TPC/2$\gamma$}}\\ \hline
 $\qzm\,\left[\gevsq\right]$  & $x$ & $\langle\ftqed\rangle$ &
 $\sigma_{\rm stat}$ & $\sigma_{\rm tot}$ & Ref. \\\hline
  $0.14-1.28$
     & 0.00$-$0.05& 0.038& 0.010& 0.012&\protect\cite{TPC-8401}\\
     & 0.05$-$0.10& 0.104& 0.010& 0.020&\\
     & 0.10$-$0.15& 0.135& 0.017& 0.027&\\
     & 0.15$-$0.20& 0.172& 0.021& 0.035&\\
     & 0.20$-$0.25& 0.219& 0.031& 0.047&\\
     & 0.25$-$0.30& 0.281& 0.042& 0.061&\\
     & 0.30$-$0.35& 0.320& 0.042& 0.066&\\
     & 0.35$-$0.40& 0.344& 0.037& 0.066&\\
     & 0.40$-$0.50& 0.370& 0.042& 0.072&\\
     & 0.50$-$0.60& 0.373& 0.037& 0.070&\\
     & 0.60$-$0.70& 0.357& 0.037& 0.068&\\
     & 0.70$-$0.80& 0.354& 0.037& 0.068&\\
     & 0.80$-$0.90& 0.291& 0.031& 0.056&\\
     & 0.90$-$1.00& 0.323& 0.068& 0.085&\\
\hline
\end{tabular}
\end{center}\end{table}
%
\renewcommand{\arraystretch}{1.0}
\begin{table}[htb]
\caption[
         Preliminary results on the photon structure function \ftqed from
         the DELPHI experiment.
        ]
        {
         Preliminary results on the photon structure function \ftqed from
         the DELPHI experiment.
         The numbers are provided by A.~Zintchenko.
         The results given at $\qzm=12.5$ and 120~\gevsq
         are unfolded from data in the \qsq ranges $2.4-51.2$
         and $45.9-752.8$~\gevsq.
         The best fit of the QED prediction to the data
         is obtained for $\pzm= 0.025$ and 0.066~\gevsq for
         $\qzm=12.5$ and 120~\gevsq.\\
        }\label{tab:chap12_08}
\begin{center}
\begin{tabular}{cccccc}\hline
\multicolumn{6}{c}{\textbf{DELPHI preliminary}}\\ \hline
$\qzm\,\left[\gevsq\right]$ & $x$ & \ftqed &
 $\sigma_{\rm stat}$ & $\sigma_{\rm tot}$ & Ref. \\\hline
 12.5& 0.00$-$0.10& 0.106 & 0.008 & 0.024 &\protect\cite{ZIN-9901}\\
     & 0.10$-$0.20& 0.273 & 0.012 & 0.017&\\
     & 0.20$-$0.30& 0.426 & 0.017 & 0.021&\\
     & 0.30$-$0.40& 0.515 & 0.021 & 0.024&\\
     & 0.40$-$0.50& 0.573 & 0.024 & 0.024&\\
     & 0.50$-$0.60& 0.645 & 0.029 & 0.029&\\
     & 0.60$-$0.70& 0.743 & 0.038 & 0.043&\\
     & 0.70$-$0.80& 0.942 & 0.060 & 0.080&\\
     & 0.80$-$1.00& 1.152 & 0.112 & 0.146&\\
\hline
 120 & 0.00$-$0.20& 0.426 & 0.291 & 0.291&\protect\cite{ZIN-9901}\\
     & 0.20$-$0.40& 0.436 & 0.134 & 0.144&\\
     & 0.40$-$0.60& 0.678 & 0.143 & 0.153&\\
     & 0.60$-$0.80& 1.039 & 0.170 & 0.176&\\
     & 0.80$-$1.00& 1.524 & 0.247 & 0.257&\\
\hline
\end{tabular}
\end{center}\end{table}
%
\renewcommand{\arraystretch}{1.0}
\begin{table}[htb]
\caption[
         Results on the photon structure functions
         \faqed and \fbqed from the L3 experiment.
        ]
        {
         Results on the photon structure functions
         \faqed and \fbqed from the L3 experiment.
         The numbers are taken from the published table of
         Ref.~\protect\cite{L3C-9801}. The original numbers for the
         measurement of \faqed are multiplied by -1/2 to account for
         the different definitions of \faqed as detailed in
         Section~\protect\ref{sec:kinem}.
         For the measurements of \faoft and 1/2\fboft
         the first error is statistical and the second systematic.
         The measured structure functions are averaged over the \qsq range
         $1.4-7.6$~\gevsq, with an average of $\qzm=3.25$~\gevsq.
         The values of \ftqed are not corrected for the effect of
         non-zero \psq in the data.\\
        }\label{tab:chap12_09}
\begin{center}
\begin{tabular}{cccc}
 \cline{1-3}
 \textbf{L3}&
 \multicolumn{2}{c}{$\qzm = 3.25$~\gevsq, Ref.\protect\cite{L3C-9801}}\\
 \cline{1-3}
 $x$         & \faoft               &     1/2\fboft  \\
 \cline{1-3}
 0.00$-$0.25 & \pz \XZ{0.159}{0.040}{0.034} & \XZ{0.046}{0.012}{0.012}&
 \multicolumn{1}{c}{}\\
 0.25$-$0.50 & \pz \XZ{0.087}{0.071}{0.056} & \XZ{0.111}{0.019}{0.038}&
 \multicolumn{1}{c}{}\\
 0.50$-$0.75 & $-$ \XZ{0.210}{0.102}{0.057} & \XZ{0.141}{0.026}{0.048}&
 \multicolumn{1}{c}{}\\
 0.75$-$1.00 & $-$ \XZ{0.236}{0.091}{0.079} & \XZ{0.061}{0.019}{0.030}&
 \multicolumn{1}{c}{}\\\cline{1-3}
 \multicolumn{4}{c}{}\\
 \hline
 \textbf{L3}&
 \multicolumn{3}{c}{$\qzm = 3.25$~\gevsq, Ref.\protect\cite{L3C-9801}}\\
 \hline
 $x$         &          \ftqed &               \faqed &        \fbqed  \\
 \hline
 0.00$-$0.25 & \X{0.090}{0.008}& \pz \X{0.014}{0.024} & \X{0.008}{0.010}\\
 0.25$-$0.50 & \X{0.404}{0.016}& \pz \X{0.036}{0.032} & \X{0.090}{0.021}\\
 0.50$-$0.75 & \X{0.597}{0.020}& $-$ \X{0.126}{0.052} & \X{0.168}{0.040}\\
 0.75$-$1.00 & \X{0.731}{0.032}& $-$ \X{0.174}{0.062} & \X{0.089}{0.045}\\
\hline
\end{tabular}
\end{center}\end{table}
%
\renewcommand{\arraystretch}{1.0}
\begin{table}[htb]
\caption[
         Results on the photon structure functions
         \faqed and \fbqed from the OPAL experiment.
        ]
        {
         Results on the photon structure functions
         \faqed and \fbqed from the OPAL experiment.
         The numbers are taken from the published table of
         Ref.~\protect\cite{OPALPR271}.
         The first error is statistical and the second systematic.
         The results given at $\qzm=5.4$~\gevsq
         are unfolded from data in the \qsq range $1.5-30$~\gevsq.
         The values of \ftqed are corrected for the \psq effect
         and correspond to \ftqed for $\psq=0$.\\
        }\label{tab:chap12_10}
\begin{center}
\begin{tabular}{cccc}
\cline{1-3}
\textbf{OPAL}&
\multicolumn{2}{c}{$\qzm = 5.4$~\gevsq, Ref.\protect\cite{OPALPR271}}\\
\cline{1-3}
 $x$         & \faoft                     & 1/2\fboft\\
\cline{1-3}
 $x<0.25$    & \pz\XZ{0.176}{0.031}{0.010}& \XZ{0.075}{0.025}{0.008}&
 \multicolumn{1}{c}{}\\
 0.25$-$0.50 & \pz\XZ{0.018}{0.028}{0.008}& \XZ{0.099}{0.024}{0.010}&
 \multicolumn{1}{c}{}\\
 0.50$-$0.75 & $-$\XZ{0.171}{0.029}{0.007}& \XZ{0.081}{0.027}{0.011}&
 \multicolumn{1}{c}{}\\
 $x>0.75$    & $-$\XZ{0.228}{0.037}{0.014}& \XZ{0.037}{0.033}{0.011}&
 \multicolumn{1}{c}{}\\
\cline{1-3}
\multicolumn{4}{c}{}\\
\hline
\textbf{OPAL}&
\multicolumn{3}{c}{$\qzm = 5.4$~\gevsq, Ref.\protect\cite{OPALPR271}}\\
 \hline
 $x$& \ftqed        & \faqed             & \fbqed\\\hline
 $x<0.25$    & \XZ{0.249}{0.006}{0.008}
             & \pz\XZ{0.039}{0.007}{0.003} & \XZ{0.029}{0.010}{0.003}\\
 0.25$-$0.50 & \XZ{0.523}{0.011}{0.014}
             & \pz\XZ{0.011}{0.016}{0.004} & \XZ{0.101}{0.025}{0.011}\\
 0.50$-$0.75 & \XZ{0.738}{0.017}{0.019}
             & $-$\XZ{0.122}{0.021}{0.006} & \XZ{0.121}{0.041}{0.017}\\
 $x>0.75$    & \XZ{0.871}{0.027}{0.021}
             & $-$\XZ{0.201}{0.033}{0.013} & \XZ{0.063}{0.056}{0.018}\\
\hline
\end{tabular}
\end{center}\end{table}
%
\renewcommand{\arraystretch}{1.0}
\begin{table}[htb]
\caption[
         Preliminary results on the photon structure function ratios
         \faoft and 1/2\fboft from the DELPHI experiment.
        ]
        {
         Preliminary results on the photon structure function ratios
         \faoft and 1/2\fboft from the DELPHI experiment.
         The numbers are provided by A.~Zintchenko.
         The original numbers for the measurement of \faoft
         are multiplied by -1 to account
         for the different definitions of \faqed as detailed in
         Section~\protect\ref{sec:kinem}.
         The first error is statistical and the second systematic.
         The results given at $\qzm=12.5$~\gevsq
         are unfolded from data in the \qsq range $2.4-51.2$~\gevsq.\\
        }\label{tab:chap12_11}
\begin{center}
\begin{tabular}{ccc}
\hline
{\textbf{DELPHI preliminary}}&
\multicolumn{2}{c}{$\qzm = 12.5$~\gevsq, Ref.~\protect\cite{ZIN-9901}}\\
\hline
 $x$         & \faoft                     & 1/2\fboft\\
\hline
 0.0$-$0.2 & \pz\XZ{0.135}{0.037}{0.016}& \XZ{0.004}{0.026}{0.009}\\
 0.2$-$0.4 & \pz\XZ{0.140}{0.034}{0.012}& \XZ{0.077}{0.024}{0.009}\\
 0.6$-$0.6 & \pz\XZ{0.038}{0.034}{0.023}& \XZ{0.099}{0.028}{0.015}\\
 0.6$-$1.0 & $-$\XZ{0.263}{0.049}{0.035}& \XZ{0.182}{0.035}{0.022}\\
\hline
\end{tabular}
\end{center}\end{table}
%
\renewcommand{\arraystretch}{1.0}
\begin{table}[htb]
\caption[
         Differential QED cross-section \dsigdx for two exchanged 
         virtual photons.
        ]
        {
         Differential QED cross-section \dsigdx for two exchanged 
         virtual photons.
         The results on the differential cross-section \dsigdx from
         the OPAL experiment are taken from the published Table of
         Ref.~\protect\cite{OPALPR271}.
         The results given at $\qzm=3.6$~\gevsq and $\pzm=2.3$~\gevsq
         are unfolded from data in the \qsq and \psq range $1.5-6$~\gevsq,
         and the results given at $\qzm=14.0$~\gevsq and $\pzm=5.0$~\gevsq
         are unfolded from data in the \qsq range $6-30$~\gevsq and
         the \psq range $1.5-20$~\gevsq.\\
        }\label{tab:chap12_12}
\begin{center}
\begin{tabular}{ccccccc}\hline
\multicolumn{7}{c}{\textbf{OPAL}}\\ \hline
$\qzm\,\left[\gevsq\right]$&
$\pzm\,\left[\gevsq\right]$& $x$ & \dsigdx &
            $\sigma_{\rm stat}$ & $\sigma_{\rm tot}$ & Ref. \\\hline
 3.6 & 2.3  & 0.00$-$0.20& $\pz$9.77 & 1.62 & 1.80&\protect\cite{OPALPR271}\\
     &      & 0.20$-$0.40&     10.45 & 1.26 & 1.39&\\
     &      & 0.40$-$0.65& $\pz$4.34 & 1.07 & 1.09&\\
\hline
14.0 & 5.0 & 0.00$-$0.25&      5.26 & 0.82 & 1.29 &\protect\cite{OPALPR271}\\
     &     & 0.25$-$0.50&      6.87 & 0.78 & 1.08 &\\
     &     & 0.50$-$0.75&      2.75 & 0.60 & 0.63 &\\
\hline
\end{tabular}
\end{center}\end{table}
%

%
%
\section{Collection of results on the hadronic structure of the photon}
\label{sec:tabqcd}
 Because in some cases it is not easy to correctly derive the errors
 of several of the measurements, a detailed survey of the available
 results has been performed, the outcome of which is presented below.
 In some of the cases it is rather difficult to obtain the central values
 and the errors of the measurements, because, especially in older
 publications, it is not always clear which errors are contained in
 the figures. The strategy taken to obtain the results is the following.
 If possible the values are taken from the published numbers.
 If no numbers are published, the values are obtained from inspecting the
 figures, either based on the RAL database or by the author himself.
 In some cases it is unclear whether the errors shown in the figures are
 only statistical, or whether they include also the systematic error.
 If only statistical errors are given in the figures, the total
 error is obtained using  Eq.~(\ref{eqn:error}), which means
 by adding in quadrature the statistical errors
 and the global systematic errors given in the publications.
 The prescription on how the central values and the errors are
 evaluated can always be found in the corresponding tables.
 The numbers listed in the tables are the basis of the summary plots
 shown in Section~\ref{sec:qcdres}.
 \par
%
\renewcommand{\arraystretch}{1.0}
\begin{table}[htb]
\caption[
         Results on the photon structure function \ft from
         the ALEPH experiment.
        ]
        {
         Results on the photon structure function \ft from
         the ALEPH experiment.
         The results at $\qzm=9.9$, 20.7 and 284~\gevsq
         are unfolded from data in the \qsq ranges
         6$-$13, 13$-$44 and 35$-$3000~\gevsq.
         All numbers are taken from the published tables.
         In addition, there exist preliminary results, which are listed
         in Table\protect~\ref{tab:chap13_11}.\\
        }\label{tab:chap13_01}
\begin{center}
\begin{tabular}{ccccccc}\hline
\multicolumn{7}{c}{\textbf{ALEPH}}\\ \hline
 $\qzm\,\left[\gevsq\right]$& \nf & $x$ & \ft &
 $\sigma_{\rm stat}$ & $\sigma_{\rm tot}$ & Ref. \\\hline
  9.9& 4& 0.005$-$0.080&  0.30&  0.02&  0.03&\protect\cite{ALE-9901}\\
     &  & 0.080$-$0.200&  0.40&  0.03&  0.07&\\
     &  & 0.200$-$0.400&  0.41&  0.05&  0.10&\\
     &  & 0.400$-$0.800&  0.27&  0.13&  0.16&\\
\hline
 20.7& 4& 0.009$-$0.120&  0.36&  0.02&  0.05&\protect\cite{ALE-9901}\\
     &  & 0.120$-$0.270&  0.34&  0.03&  0.12&\\
     &  & 0.270$-$0.500&  0.56&  0.05&  0.11&\\
     &  & 0.500$-$0.890&  0.45&  0.11&  0.12&\\
\hline
  284& 4&   0.03$-$0.35&  0.65&  0.10&  0.14&\protect\cite{ALE-9901}\\
     &  &   0.35$-$0.65&  0.70&  0.16&  0.25&\\
     &  &   0.65$-$0.97&  1.28&  0.26&  0.37&\\
\hline
\end{tabular}
\end{center}\end{table}
%
\renewcommand{\arraystretch}{1.0}
\begin{table}[htb]
\caption[
         Results on the photon structure function \ft from
         the AMY experiment.
        ]
        {
         Results on the photon structure function \ft from
         the AMY experiment.
         The result from Ref.\protect~\cite{AMY-9501} at $\qzm=73$~\gevsq
         is an update of the measurement from Ref.\protect~\cite{AMY-9002}
         at the same value of \qzm, where the previous measurement is no
         more included in this review.
         The results at $\qzm=6.8$, 73 and 390~\gevsq
         are unfolded from data in the \qsq ranges
         3.5$-$12, 25$-$220~\gevsq and for $\qsq>110$~\gevsq.
         The total error has been calculated in this review from the
         quadratic sum of the statistical error and the systematic error
         using the errors given in the published tables.\\
        }\label{tab:chap13_02}
\begin{center}
\begin{tabular}{ccccccc}\hline
\multicolumn{7}{c}{\textbf{AMY}}\\ \hline
 $\qzm\,\left[\gevsq\right]$ & \nf & $x$ & \ft &
 $\sigma_{\rm stat}$ & $\sigma_{\rm tot}$ & Ref. \\\hline
  6.8& 4& 0.015$-$0.125& 0.337& 0.030& 0.053&\protect\cite{AMY-9701}\\
     &  & 0.125$-$0.375& 0.302& 0.040& 0.049&\\
     &  & 0.373$-$0.620& 0.322& 0.049& 0.097&\\
\hline
   73& 4& 0.125$-$0.375&  0.65&  0.08&  0.10&\protect\cite{AMY-9501}\\
     &  & 0.375$-$0.625&  0.60&  0.16&  0.16&\\
     &  & 0.625$-$0.875&  0.65&  0.11&  0.14&\\
\hline
  390& 4& 0.120$-$0.500&  0.94&  0.23&  0.25&\protect\cite{AMY-9501}\\
     &  & 0.500$-$0.800&  0.82&  0.16&  0.19&\\
\hline
\end{tabular}
\end{center}\end{table}
%
\renewcommand{\arraystretch}{1.0}
\begin{table}[htb]
\caption[
         Results on the photon structure function \ft from
         the DELPHI experiment.
        ]
        {
         Results on the photon structure function \ft from
         the DELPHI experiment.
         The two results are not independent, but use the same data, which
         is unfolded for four bins on a linear scale in $x$, and also
         for three bins on a logarithmic $x$ scale for $x<0.35$.
         The results at $\qzm=12$~\gevsq
         are unfolded from data in the \qsq range 4$-$30~\gevsq.
         The total error has been calculated in this review from the
         quadratic sum of the statistical error and the systematic error
         using the errors given in the published table
         of Ref.\protect~\cite{DEL-9601}.
         In addition, there exist preliminary results, which are listed
         in Table\protect~\ref{tab:chap13_12}.\\
        }\label{tab:chap13_03}
\begin{center}
\begin{tabular}{ccccccc}\hline
\multicolumn{7}{c}{\textbf{DELPHI}}\\ \hline
 $\qzm\,\left[\gevsq\right]$& \nf & $x$ & \ft &
 $\sigma_{\rm stat}$ & $\sigma_{\rm tot}$ & Ref. \\\hline
   12& 4& 0.001$-$0.080&  0.21&  0.03&  0.05&\protect\cite{DEL-9601}\\
     &  & 0.080$-$0.213&  0.41&  0.04&  0.06&\\
     &  & 0.213$-$0.428&  0.45&  0.05&  0.07&\\
     &  & 0.428$-$0.847&  0.45&  0.11&  0.15&\\
\hline
   12& 4& 0.001$-$0.046&  0.24&  0.03&  0.06&\protect\cite{DEL-9601}\\
     &  & 0.046$-$0.117&  0.41&  0.05&  0.09&\\
     &  & 0.117$-$0.350&  0.46&  0.17&  0.19&\\
\hline
\end{tabular}
\end{center}\end{table}
%
\renewcommand{\arraystretch}{1.0}
\begin{table}[htb]
\caption[
         Results on the photon structure function \ft from
         the JADE experiment.
        ]
        {
         Results on the photon structure function \ft from
         the JADE experiment.
         The results at $\qzm=24$ and 100~\gevsq
         are unfolded from data in the \qsq ranges 10$-$55
         and 30$-$220~\gevsq.
         The full errors are obtained by the RAL database from
         the figures of Ref.\protect~\cite{JAD-8401} which contain only
         the full errors, and the statistical errors are not available.\\
        }\label{tab:chap13_04}
\begin{center}
\begin{tabular}{ccccccc}\hline
\multicolumn{7}{c}{\textbf{JADE}}\\ \hline
 $\qzm\,\left[\gevsq\right]$& \nf & $x$ & \ft &
 $\sigma_{\rm stat}$ & $\sigma_{\rm tot}$ & Ref. \\\hline
   24& 4& 0.000$-$0.100&  0.51&    &  0.15&\protect\cite{JAD-8401}\\
     &  & 0.100$-$0.200&  0.29&    &  0.12&\\
     &  & 0.200$-$0.400&  0.34&    &  0.10&\\
     &  & 0.400$-$0.600&  0.59&    &  0.12&\\
     &  & 0.600$-$0.900&  0.23&    &  0.12&\\
\hline
  100& 4& 0.100$-$0.300&  0.52&    &  0.23&\protect\cite{JAD-8401}\\
     &  & 0.300$-$0.600&  0.75&    &  0.22&\\
     &  & 0.600$-$0.900&  0.90&    &  0.27&\\
\hline
\end{tabular}
\end{center}\end{table}
%
\renewcommand{\arraystretch}{1.0}
\begin{table}[htb]
\caption[
         The results on the photon structure function \ft from
         the L3 experiment.
        ]
        {
         Results on the photon structure function \ft from
         the L3 experiment.
         The results given at $\qzm=1.9$ and 5.0~\gevsq
         are unfolded from data in the \qsq ranges 1.2$-$3.0,
         3.0$-$9.0~\gevsq. For these measurements two results are
         given in Ref.\protect~\cite{L3C-9803}, one is obtained by an
         unfolding based
         on the PHOJET Monte Carlo, the other is based on the TWOGAM Monte
         Carlo. The numbers given here use the results based on PHOJET
         as the central values with the corresponding statistical error.
         The systematic error is calculated from the quadratic sum of the
         systematic error for the result based on PHOJET and the
         difference between the results obtained from PHOJET and TWOGAM.
         The results given at $\qzm=10.8,15.3$ and 23.1~\gevsq
         are unfolded from data in the \qsq ranges 9$-$13,
         13$-$18 and 13$-$30~\gevsq. For these results
         the central values with the corresponding statistical errors
         are taken from the published table in Ref.\protect~\cite{L3C-9804}.
         The systematic error is calculated from the quadratic sum of the
         systematic error and the additional systematic error
         due to the dependence on the Monte Carlo model.
         In addition, there exist preliminary results, which are listed
         in Table\protect~\ref{tab:chap13_13}.\\
        }\label{tab:chap13_05}
\begin{center}
\begin{tabular}{ccccccc}\hline
\multicolumn{7}{c}{\textbf{L3}}\\ \hline
 $\qzm\,\left[\gevsq\right]$& \nf & $x$ & \ft &
 $\sigma_{\rm stat}$ & $\sigma_{\rm tot}$ & Ref. \\\hline
  1.9& 4& 0.002$-$0.005& 0.184& 0.009& 0.050&\protect\cite{L3C-9803}\\
     &  & 0.005$-$0.010& 0.179& 0.007& 0.023&\\
     &  & 0.010$-$0.020& 0.176& 0.006& 0.017&\\
     &  & 0.020$-$0.030& 0.191& 0.008& 0.009&\\
     &  & 0.030$-$0.050& 0.193& 0.008& 0.012&\\
     &  & 0.050$-$0.100& 0.185& 0.007& 0.027&\\
\hline
  5.0& 4& 0.005$-$0.010& 0.307& 0.021& 0.096&\protect\cite{L3C-9803}\\
     &  & 0.010$-$0.020& 0.282& 0.014& 0.047&\\
     &  & 0.020$-$0.040& 0.263& 0.011& 0.023&\\
     &  & 0.040$-$0.060& 0.278& 0.013& 0.015&\\
     &  & 0.060$-$0.100& 0.270& 0.012& 0.023&\\
     &  & 0.100$-$0.200& 0.252& 0.011& 0.047&\\
\hline
 10.8& 4&   0.01$-$0.10&  0.30&  0.02&  0.04&\protect\cite{L3C-9804}\\
     &  &   0.10$-$0.20&  0.35&  0.03&  0.04&\\
     &  &   0.20$-$0.30&  0.30&  0.04&  0.11&\\
\hline
 15.3& 4&   0.01$-$0.10&  0.37&  0.02&  0.04&\protect\cite{L3C-9804}\\
     &  &   0.10$-$0.20&  0.42&  0.04&  0.05&\\
     &  &   0.20$-$0.30&  0.42&  0.05&  0.09&\\
     &  &   0.30$-$0.50&  0.35&  0.05&  0.16&\\
\hline
 23.1& 4&   0.01$-$0.10&  0.40&  0.03&  0.05&\protect\cite{L3C-9804}\\
     &  &   0.10$-$0.20&  0.44&  0.04&  0.06&\\
     &  &   0.20$-$0.30&  0.47&  0.05&  0.06&\\
     &  &   0.30$-$0.50&  0.44&  0.05&  0.13&\\
\hline
\end{tabular}
\end{center}\end{table}
%
\renewcommand{\arraystretch}{1.0}
\begin{table}[htb]
\caption[
         Results on the photon structure function \ft from
         the OPAL experiment.
        ]
        {
         Results on the photon structure function \ft from
         the OPAL experiment.
         The results at
         $\qzm=7.5, 14.7, 135, 9.0, 14.5, 30.0, 59.0, 1.86$ and 3.76~\gevsq
         are unfolded from data in the \qsq ranges
         $6-8$, $8-30$, $60-400$, $6-11$, $11-20$, $20-40$, $40-100$,
         $1.1-2.5$ and $2.5-6.6$~\gevsq.
         The numbers are taken from the published tables in
         Refs.\protect~\cite{OPALPR185,OPALPR207,OPALPR213}.
         Only the statistically independent results are listed here, the
         original publication also contains results on combined \qsq
         ranges. The asymmetric errors are listed using the
         positive / negative values.\\
        }\label{tab:chap13_06}
\begin{center}
\begin{tabular}{ccccccc}\hline
\multicolumn{7}{c}{\textbf{OPAL}}\\ \hline
 $\qzm\,\left[\gevsq\right]$ & \nf & $x$ & \ft &
 $\sigma_{\rm stat}$ & $\sigma_{\rm tot}$ & Ref. \\\hline
  7.5& 4& 0.001$-$0.091&  0.28&  0.02& 0.04 / 0.10&\protect\cite{OPALPR185}\\
     &  & 0.091$-$0.283&  0.32&  0.02& 0.08 / 0.14&\\
     &  & 0.283$-$0.649&  0.38&  0.04& 0.07 / 0.22&\\
\hline
 14.7& 4& 0.006$-$0.137&  0.38&  0.01& 0.06 / 0.13&\protect\cite{OPALPR185}\\
     &  & 0.137$-$0.324&  0.41&  0.02& 0.06 / 0.04&\\
     &  & 0.324$-$0.522&  0.41&  0.03& 0.09 / 0.12&\\
     &  & 0.522$-$0.836&  0.54&  0.05& 0.31 / 0.14&\\
\hline
  135& 4& 0.100$-$0.300&  0.65&  0.09& 0.35 / 0.11&\protect\cite{OPALPR185}\\
     &  & 0.300$-$0.600&  0.73&  0.08& 0.09 / 0.11&\\
     &  & 0.600$-$0.800&  0.72&  0.10& 0.81 / 0.12&\\
\hline
  9.0& 4& 0.020$-$0.100&  0.33&  0.03& 0.07 / 0.07&\protect\cite{OPALPR207}\\
     &  & 0.100$-$0.250&  0.29&  0.04& 0.06 / 0.06&\\
     &  & 0.250$-$0.600&  0.39&  0.08& 0.13 / 0.31&\\
\hline
 14.5& 4& 0.020$-$0.100&  0.37&  0.03& 0.16 / 0.03&\protect\cite{OPALPR207}\\
     &  & 0.100$-$0.250&  0.42&  0.05& 0.06 / 0.15&\\
     &  & 0.250$-$0.600&  0.39&  0.06& 0.12 / 0.13&\\
\hline
 30.0& 4& 0.050$-$0.100&  0.32&  0.04& 0.12 / 0.05&\protect\cite{OPALPR207}\\
     &  & 0.100$-$0.230&  0.52&  0.05& 0.08 / 0.14&\\
     &  & 0.230$-$0.600&  0.41&  0.09& 0.22 / 0.10&\\
     &  & 0.600$-$0.800&  0.46&  0.15& 0.42 / 0.21&\\
\hline
 59.0& 4& 0.050$-$0.100&  0.37&  0.06& 0.29 / 0.09&\protect\cite{OPALPR207}\\
     &  & 0.100$-$0.230&  0.44&  0.07& 0.11 / 0.10&\\
     &  & 0.230$-$0.600&  0.48&  0.09& 0.19 / 0.14&\\
     &  & 0.600$-$0.800&  0.51&  0.14& 0.50 / 0.14&\\
\hline
 1.86& 4& 0.0025$-$0.0063&  0.27&  0.03& 0.06 / 0.08&\protect\cite{OPALPR213}\\
     &  & 0.0063$-$0.0200&  0.22&  0.02& 0.03 / 0.05&\\
     &  & 0.0200$-$0.0400&  0.20&  0.02& 0.09 / 0.03&\\
     &  & 0.0400$-$0.1000&  0.23&  0.02& 0.04 / 0.05&\\
\hline
 3.76& 4& 0.0063$-$0.0200&  0.35&  0.03& 0.09 / 0.09&\protect\cite{OPALPR213}\\
     &  & 0.0200$-$0.0400&  0.29&  0.03& 0.07 / 0.07&\\
     &  & 0.0400$-$0.1000&  0.32&  0.02& 0.07 / 0.05&\\
     &  & 0.1000$-$0.2000&  0.32&  0.03& 0.09 / 0.05&\\
\hline
\end{tabular}
\end{center}\end{table}
%
\renewcommand{\arraystretch}{1.0}
\begin{table}[htb]
\caption[
         Results on the photon structure function \ft from
         the PLUTO experiment.
        ]
         {
         Results on the photon structure function \ft from
         the PLUTO experiment.
         The results from Ref.\protect~\cite{PLU-8401} at $\qzm=2.4,4.3$ and
         9.2~\gevsq are unfolded from data in the \qsq ranges 1.5$-$3,
         3$-$6 and 6$-$16~\gevsq.
         The measurement at $\qzm=5.3$~\gevsq contains all data and is not
         an independent measurement.
         The total error has been calculated in this review from the
         quadratic sum of the statistical error and the systematic error
         using the statistical error and the systematic error
         of 15/25$\%$ for data above/below $x=0.2$ of the measured values
         for $\nf=4$, as given in Ref.\protect~\cite{PLU-8401}.
         The data from Ref.\protect~\cite{PLU-8701} at $\qzm=45$~\gevsq
         are unfolded from data in the \qsq range 18$-$100~\gevsq.
         The statistical errors are obtained by the RAL database
         from Figure 5 of Ref.\protect~\cite{PLU-8701} which contains only
         statistical errors, and the systematic error has been
         added according to the quoted systematic error of 10$\%$.
         For both publications, the charm subtraction has been performed
         by PLUTO.\\
        }\label{tab:chap13_07}
\begin{center}
\begin{tabular}{ccccccc}\hline
\multicolumn{7}{c}{\textbf{PLUTO}}\\ \hline
 $\qzm\,\left[\gevsq\right]$ & \nf & $x$ & \ft &
 $\sigma_{\rm stat}$ & $\sigma_{\rm tot}$ & Ref. \\\hline
  2.4& 3/4& 0.016$-$0.110& 0.183/0.204& 0.014& 0.053&\protect\cite{PLU-8401}\\
     &    & 0.110$-$0.370& 0.263/0.272& 0.026& 0.049&\\
     &    & 0.370$-$0.700& 0.222/0.222& 0.064& 0.072&\\
\hline
  4.3& 3/4& 0.030$-$0.170& 0.218/0.256& 0.014& 0.066&\protect\cite{PLU-8401}\\
     &    & 0.170$-$0.440& 0.273/0.295& 0.020& 0.048&\\
     &    & 0.440$-$0.800& 0.336/0.336& 0.044& 0.067&\\
\hline
  9.2& 3/4& 0.060$-$0.230& 0.300/0.354& 0.027& 0.093&\protect\cite{PLU-8401}\\
     &    & 0.230$-$0.540& 0.340/0.402& 0.029& 0.067&\\
     &    & 0.540$-$0.900& 0.492/0.492& 0.069& 0.101&\\
\hline
   45& 3/4& 0.100$-$0.250& 0.360/0.480& 0.170& 0.177&\protect\cite{PLU-8701}\\
     &    & 0.250$-$0.500& 0.400/0.550& 0.120& 0.132&\\
     &    & 0.500$-$0.750& 0.770/0.890& 0.160& 0.183&\\
     &    & 0.750$-$0.900& 0.840/0.870& 0.260& 0.274&\\
\hline
\hline
  5.3& 3/4& 0.035$-$0.072& 0.216/0.245& 0.015& 0.063&\protect\cite{PLU-8401}\\
     &    & 0.072$-$0.174& 0.258/0.307& 0.010& 0.078&\\
     &    & 0.174$-$0.319& 0.222/0.277& 0.025& 0.049&\\
     &    & 0.319$-$0.490& 0.329/0.329& 0.037& 0.061&\\
     &    & 0.490$-$0.650& 0.439/0.439& 0.052& 0.084&\\
     &    & 0.650$-$0.840& 0.361/0.361& 0.076& 0.093&\\
\hline
\end{tabular}
\end{center}\end{table}
%
\renewcommand{\arraystretch}{1.0}
\begin{table}[htb]
\caption[
         Results on the photon structure function \ft from
         the TASSO experiment.
        ]
        {
         Results on the photon structure function \ft from
         the TASSO experiment.
         The results at $\qzm=23$~\gevsq
         are unfolded from data in the \qsq range 7$-$70~\gevsq.
         The statistical errors are obtained by the RAL database from
         Figure 7 of Ref.\protect~\cite{TAS-8601}, which probably
         contains only statistical errors, and
         the systematic error has been added according to the
         quoted systematic error of 19$\%$ on the \xvis distribution,
         which in several cases is larger than the full error given in
         Figure 7 of Ref.\protect~\cite{TAS-8601}.\\
        }\label{tab:chap13_08}
\begin{center}
\begin{tabular}{ccccccc}\hline
\multicolumn{7}{c}{\textbf{TASSO}}\\ \hline
 $\qzm\,\left[\gevsq\right]$ & \nf & $x$ & \ft &
 $\sigma_{\rm stat}$ & $\sigma_{\rm tot}$ & Ref. \\\hline
   23& 4& 0.020$-$0.200& 0.366& 0.089& 0.112&\protect\cite{TAS-8601}\\
     &  & 0.200$-$0.400& 0.670& 0.086& 0.153&\\
     &  & 0.400$-$0.600& 0.722& 0.104& 0.172&\\
     &  & 0.600$-$0.800& 0.693& 0.116& 0.176&\\
     &  & 0.800$-$0.980& 0.407& 0.222& 0.235&\\\cline{2-6}
     & 3& 0.020$-$0.200& 0.281& 0.087& 0.111&\\
     &  & 0.200$-$0.400& 0.441& 0.085& 0.153&\\
     &  & 0.400$-$0.600& 0.469& 0.104& 0.172&\\
     &  & 0.600$-$0.800& 0.549& 0.115& 0.175&\\
     &  & 0.800$-$0.980& 0.422& 0.224& 0.237&\\
\hline
\end{tabular}
\end{center}\end{table}
%
\renewcommand{\arraystretch}{1.0}
\begin{table}[htb]
\caption[
         Results on the photon structure function \ft from
         the TPC/2$\gamma$ experiment.
        ]
        {
         Results on the photon structure function \ft from
         the TPC/2$\gamma$ experiment.
         The results at
         $\qzm=0.24, 0.38, 0.71, 1.3, 2.8$ and 5.1~\gevsq
         are unfolded from data in the \qsq ranges
         $0.2-0.3$, $0.3-0.5$, $0.5-1.0$, $1.0-1.6$, $1.8-4.0$
         and $4.0-6.6$~\gevsq.
         The statistical errors are obtained by the RAL database.
         The quoted systematic errors from Ref.\protect~\cite{TPC-8701}
         have been added. They amount to 11$\%$ for the regions
         $0.2<\qsq<1$~\gevsq with $x<0.1$, and $1<\qsq<7$~\gevsq with
         $x<0.2$, and to 14$\%$ for the regions $0.2<\qsq<1$~\gevsq with
         $x>0.1$, and $1<\qsq<7$~\gevsq with $x>0.2$.\\
        }\label{tab:chap13_09}
\begin{center}
\begin{tabular}{ccccccc}\hline
\multicolumn{7}{c}{\textbf{TPC/2$\gamma$}}\\ \hline
 $\qzm\,\left[\gevsq\right]$& \nf & $x$ & \ft &
 $\sigma_{\rm stat}$ & $\sigma_{\rm tot}$ & Ref. \\\hline
 0.24& 4& 0.000$-$0.020&  0.084&  0.005&  0.011&\protect\cite{TPC-8701}\\
     &  & 0.020$-$0.060&  0.074&  0.008&  0.012&\\
     &  & 0.060$-$0.180&  0.062&  0.013&  0.015&\\
\hline
 0.38& 4& 0.000$-$0.020&  0.113&  0.007&  0.014&\protect\cite{TPC-8701}\\
     &  & 0.020$-$0.055&  0.118&  0.011&  0.017&\\
     &  & 0.055$-$0.111&  0.171&  0.021&  0.023&\\
     &  & 0.111$-$0.243&  0.151&  0.028&  0.044&\\
\hline
 0.71& 4& 0.000$-$0.028&  0.117&  0.006&  0.014&\protect\cite{TPC-8701}\\
     &  & 0.028$-$0.065&  0.130&  0.010&  0.018&\\
     &  & 0.065$-$0.121&  0.170&  0.017&  0.025&\\
     &  & 0.121$-$0.340&  0.133&  0.013&  0.023&\\
\hline
  1.3& 4& 0.000$-$0.050&  0.107&  0.013&  0.017&\protect\cite{TPC-8701}\\
     &  & 0.050$-$0.126&  0.184&  0.021&  0.029&\\
     &  & 0.126$-$0.215&  0.215&  0.034&  0.041&\\
     &  & 0.215$-$0.507&  0.102&  0.031&  0.034&\\
\hline
  2.8& 4& 0.000$-$0.080&  0.134&  0.018&  0.023&\protect\cite{TPC-8701}\\
     &  & 0.080$-$0.156&  0.234&  0.031&  0.040&\\
     &  & 0.156$-$0.303&  0.198&  0.042&  0.050&\\
     &  & 0.303$-$0.600&  0.160&  0.033&  0.040&\\
\hline
  5.1& 4& 0.021$-$0.199&  0.224&  0.034&  0.042&\protect\cite{TPC-8701}\\
     &  & 0.199$-$0.359&  0.373&  0.057&  0.077&\\
     &  & 0.359$-$0.740&  0.300&  0.044&  0.061&\\
\hline
\end{tabular}
\end{center}\end{table}
%
\renewcommand{\arraystretch}{1.0}
\begin{table}[htb]
\caption[
         Results on the photon structure function \ft from
         the TOPAZ experiment.
        ]
        {
         Results on the photon structure function \ft from
         the TOPAZ experiment.
         The results at $\qzm=5.1,16,80$~\gevsq
         are unfolded from data in the \qsq ranges 3$-$10,
         10$-$30 and 45$-$130~\gevsq.
         The total error has been calculated in this review from the
         quadratic sum of the statistical error and the systematic error
         using the errors given in the published table.\\
        }\label{tab:chap13_10}
\begin{center}
\begin{tabular}{ccccccc}\hline
\multicolumn{7}{c}{\textbf{TOPAZ}}\\ \hline
 $\qzm\,\left[\gevsq\right]$ & \nf & $x$ & \ft &
 $\sigma_{\rm stat}$ & $\sigma_{\rm tot}$ & Ref. \\\hline
  5.1& 4& 0.010$-$0.076&  0.33&  0.02&  0.05&\protect\cite{TOP-9402}\\
     &  & 0.076$-$0.200&  0.29&  0.03&  0.04&\\
\hline
   16& 4& 0.020$-$0.150&  0.60&  0.08&  0.10&\protect\cite{TOP-9402}\\
     &  & 0.150$-$0.330&  0.56&  0.09&  0.10&\\
     &  & 0.330$-$0.780&  0.46&  0.15&  0.16&\\
\hline
   80& 4& 0.060$-$0.320&  0.68&  0.26&  0.27&\protect\cite{TOP-9402}\\
     &  & 0.320$-$0.590&  0.83&  0.22&  0.23&\\
     &  & 0.590$-$0.980&  0.53&  0.21&  0.22&\\
\hline
\end{tabular}
\end{center}\end{table}
%
\renewcommand{\arraystretch}{1.0}
\begin{table}[htb]
\caption[
         Additional preliminary results on the photon structure function
         \ft from the ALEPH experiment.
        ]
        {
         Additional preliminary results on the photon structure function
         \ft from the ALEPH experiment.
         The results at $\qzm=13.7$ and 56.6~\gevsq
         are unfolded from data in the \qsq ranges 7$-$24,
         and 17$-$200~\gevsq.
         The numbers for $\qzm=13.7$~\gevsq are taken from the published
         table and the numbers for $\qzm=56.6$~\gevsq are read off the
         figures presented in Ref.\protect~\cite{BOE-9901}.\\
        }\label{tab:chap13_11}
\begin{center}
\begin{tabular}{ccccccc}\hline
\multicolumn{7}{c}{\textbf{ALEPH preliminary}}\\ \hline
 $\qzm\,\left[\gevsq\right]$& \nf & $x$ & \ft &
 $\sigma_{\rm stat}$ & $\sigma_{\rm tot}$ & Ref. \\\hline
  13.7& 4& 0.002$-$0.065&  0.32&  0.02&  0.04&\protect\cite{BOE-9901}\\
      &  & 0.130$-$0.343&  0.41&  0.03&  0.03&\\
      &  & 0.343$-$0.560&  0.53&  0.04&  0.06&\\
      &  & 0.560$-$0.900&  0.37&  0.07&  0.12&\\
\hline
  56.5& 4& 0.003$-$0.05&  0.48&  0.04&  0.05&\protect\cite{BOE-9901}\\
      &  & 0.05 $-$0.25&  0.41&  0.04&  0.10&\\
      &  & 0.25 $-$0.48&  0.38&  0.06&  0.09&\\
      &  & 0.48 $-$0.98&  0.54&  0.07&  0.16&\\
\hline
\end{tabular}
\end{center}\end{table}
%
\renewcommand{\arraystretch}{1.0}
\begin{table}[htb]
\caption[
         Additional preliminary results on the photon structure function
         \ft from the DELPHI experiment.
        ]
        {
         Additional preliminary results on the photon structure function
         \ft from the DELPHI experiment.
         For the results reported in Ref.\protect~\cite{TIA-9701}
         the numbers have been provided by I.~Tiapkin, whereas the
         results from Ref.\protect~\cite{TIA-9801} have been taken from
         the published tables.
         No information is available which ranges of \qsq have been used
         for the results.\\
        }\label{tab:chap13_12}
\begin{center}
\begin{tabular}{cccccccc}\hline
\multicolumn{7}{c}{\textbf{DELPHI preliminary}}\\ \hline
 $\qzm\,\left[\gevsq\right]$ & \nf & $x$ & \ft &
 $\sigma_{\rm stat}$ & $\sigma_{\rm tot}$ & Ref. \\\hline
  6.3& 4& 0.002$-$0.020& 0.204&      &  0.03&\protect\cite{TIA-9701}\\
     &  & 0.020$-$0.070& 0.261&      &  0.03&\\
     &  & 0.070$-$0.200& 0.303&      &  0.04&\\
     &  & 0.200$-$0.700& 0.377&      &  0.11&\protect\cite{TIA-9701}\\
\hline
   13& 4& 0.002$-$0.020& 0.266&      &  0.03&\\
     &  & 0.023$-$0.140& 0.316&      &  0.03&\\
     &  & 0.140$-$0.280& 0.366&      &  0.04&\\
     &  & 0.280$-$0.750& 0.424&      &  0.05&\\
\hline
   21& 4&  0.01$-$0.10 & 0.33 & 0.01 &  0.03&\protect\cite{TIA-9801}\\
     &  &  0.10$-$0.30 & 0.41 & 0.03 &  0.04&\\
     &  &  0.30$-$0.80 & 0.51 & 0.05 &  0.06&\\
\hline
   42& 4&  0.01$-$0.10 & 0.41 & 0.01 &  0.03&\protect\cite{TIA-9801}\\
     &  &  0.10$-$0.30 & 0.48 & 0.02 &  0.03&\\
     &  &  0.30$-$0.80 & 0.59 & 0.03 &  0.05&\\
\hline
   99& 4&  0.01$-$0.10 & 0.45 & 0.06 &  0.06&\protect\cite{TIA-9801}\\
     &  &  0.10$-$0.30 & 0.52 & 0.05 &  0.06&\\
     &  &  0.30$-$0.80 & 0.73 & 0.05 &  0.06&\\
\hline
  400& 4&  0.01$-$0.10 & 0.5  & 0.3  &  0.3 &\protect\cite{TIA-9801}\\
     &  &  0.10$-$0.30 & 0.7  & 0.2  &  0.3 &\\
     &  &  0.30$-$0.80 & 1.0  & 0.1  &  0.3 &\\
\hline
\end{tabular}
\end{center}\end{table}
%
\renewcommand{\arraystretch}{1.0}
\begin{table}[htb]
\caption[
         Additional preliminary results on the photon structure function
         \ft from the L3 experiment.
        ]
        {
         Additional preliminary results on the photon structure function
         \ft from the L3 experiment.
         The results at $\qzm=120$~\gevsq
         are unfolded from data in the \qsq range 40$-$500~\gevsq.
         The numbers have been provided by F.C.~Ern{\'e}.\\
        }\label{tab:chap13_13}
\begin{center}
\begin{tabular}{cccccccc}\hline
\multicolumn{7}{c}{\textbf{L3 preliminary}}\\ \hline
 $\qzm\,\left[\gevsq\right]$ & \nf & $x$ & \ft &
 $\sigma_{\rm stat}$ & $\sigma_{\rm tot}$ & Ref. \\\hline
  120& 4& 0.05$-$0.20&  0.66&  0.06&  0.08&\protect\cite{ERN-9901}\\
     &  & 0.20$-$0.40&  0.81&  0.04&  0.08&\\
     &  & 0.40$-$0.60&  0.76&  0.10&  0.12&\\
     &  & 0.60$-$0.80&  0.85&  0.12&  0.14&\\
     &  & 0.80$-$0.98&  0.91&  0.18&  0.19&\\
\hline
\end{tabular}
\end{center}\end{table}
%
\renewcommand{\arraystretch}{1.1}
\begin{table}[htb]
\caption[
         Results on the \qsq evolution of \ft for three active flavours.
        ]
        {
         Results on the \qsq evolution of \ft for three active flavours.
         The values for the AMY and the TOPAZ experiments are the
         published numbers from Refs.\protect~\cite{AMY-9501}
         and\protect~\cite{TOP-9402}.
         The values for the PLUTO experiment are obtained by the RAL
         database from the published figures of Ref.\protect~\cite{PLU-8701}.
         The TASSO result is the sum of the three middle bins
         of Table\protect~\ref{tab:chap13_08}.\\
        }\label{tab:chap13_14}
\begin{center}
\begin{tabular}{ccccc}\hline
 Exp.          & $\qzm\,\left[\gevsq\right]$
                       &     $x$   & $\ft\pm\sigma_{\rm tot}$& Ref. \\\hline
 \textbf{AMY}  &  73   & 0.3$-$0.8 & $0.42 \pm 0.08$&\protect\cite{AMY-9501}\\
               & 390   &           & $0.50 \pm 0.18$&\\\hline
 \textbf{PLUTO}&   2.4 & 0.3$-$0.8 & $0.24 \pm 0.08$&\protect\cite{PLU-8701}\\
               &   4.3 &           & $0.30 \pm 0.06$&\\
               &   9.2 &           & $0.36 \pm 0.07$&\\
               &  45   &           & $0.55 \pm 0.12$&\\\hline
 \textbf{TASSO}&  23   & 0.2$-$0.8 & $0.48 \pm 0.10$&\protect\cite{TAS-8601}\\
                                                                        \hline
 \textbf{TOPAZ}&  16   & 0.3$-$0.8 & $0.38 \pm 0.08$&\protect\cite{TOP-9402}\\
               &  80   &           & $0.49 \pm 0.15$&\\
               & 338   &           & $0.72 \pm 0.37$&\\\hline
\end{tabular}
\end{center}\end{table}
%
\renewcommand{\arraystretch}{1.1}
\begin{table}[htb]
\caption[
         Results on the \qsq evolution of \ft for four active flavours.
        ]
        {
         Results on the \qsq evolution of \ft for four active flavours.
         The values for the AMY, ALEPH, DELPHI, OPAL and TOPAZ
         experiments are the published numbers from
         Refs.\protect~\cite{AMY-9501},\protect~\cite{ALE-9901}
         ,\protect~\cite{DEL-9601}
         ,\protect~\cite{OPALPR207} and\protect~\cite{TOP-9402}
         respectively.
         The values for the JADE and PLUTO experiments are obtained by
         the RAL database from the published figures of
         Refs.\protect~\cite{JAD-8401} and\protect~\cite{PLU-8701}.
         For the PLUTO result the contribution from charm quarks has been
         added as explained in Section~\ref{sec:qcdresf2}.
         The TASSO result is the sum of the three middle bins
         of Table\protect~\ref{tab:chap13_08}.
         For the TPC/2$\gamma$ the statistical errors are obtained by
         the RAL database. The quoted systematic error from
         Ref.\protect~\cite{TPC-8701} has been added.
         In addition, there exist preliminary results from the DELPHI
         and L3 experiments, which are listed in
         Table\protect~\ref{tab:chap13_16}.\\
        }\label{tab:chap13_15}
\begin{center}
\begin{tabular}{ccccc}\hline
 Exp.          & $\qzm\,\left[\gevsq\right]$
                       &    $x$    & $\ft\pm\sigma_{\rm tot}$& Ref. \\\hline
 \textbf{ALEPH}&   9.9 & 0.1$-$0.6 & $0.38 \pm 0.05$&\protect\cite{ALE-9901}\\
               &  20.7 &           & $0.50 \pm 0.05$&\\
               & 284   &           & $0.68 \pm 0.12$&\\\hline
 \textbf{AMY}  &  73   & 0.3$-$0.8 & $0.63 \pm 0.07$&\protect\cite{AMY-9501}\\
               & 390   &           & $0.85 \pm 0.18$&\\\hline
 \textbf{DELPHI}
               &  12.0 & 0.3$-$0.8 & $0.45 \pm 0.08$&\protect\cite{DEL-9601}\\
\hline
 \textbf{JADE} &  13.4 &  $>0.1$   & $0.28 \pm 0.06$&\protect\cite{JAD-8401}\\
               &  21.2 &           & $0.37 \pm 0.04$&\\
               &  28.3 &           & $0.46 \pm 0.09$&\\
               &  35.8 &           & $0.67 \pm 0.10$&\\
               &  46.9 &           & $0.66 \pm 0.16$&\\
               & 100   &           & $0.73 \pm 0.13$&\\\hline
 \textbf{OPAL} &   7.5 & 0.1$-$0.6 &$\Y{0.36}{0.07}{0.12}$&
                                                     \protect\cite{OPALPR207}\\
               &   9   &           &$\Y{0.36}{0.09}{0.08}$&\\
               &  13.5 &           &$\Y{0.41}{0.06}{0.13}$&\\
               &  14.7 &           &$\Y{0.41}{0.08}{0.05}$&\\
               &  30   &           &$\Y{0.48}{0.08}{0.09}$&\\
               &  59   &           &$\Y{0.46}{0.09}{0.07}$&\\
               & 135   &           &$\Y{0.71}{0.15}{0.08}$&\\\hline
 \textbf{PLUTO}&   2.4 & 0.3$-$0.8 & $0.24 \pm 0.08$&\protect\cite{PLU-8701}\\
               &   4.3 &           & $0.30 \pm 0.06$&\\
               &   9.2 &           & $0.39 \pm 0.07$&\\
               &  45   &           & $0.73 \pm 0.12$&\\\hline
 \textbf{TASSO}&  23   & 0.2$-$0.8 & $0.69 \pm 0.10$&\protect\cite{TAS-8601}\\
                                                                \hline
 \textbf{TPC/2$\gamma$}
               &   5.1 & 0.3$-$0.6 & $0.31 \pm 0.07$&\protect\cite{TPC-8701}\\
                                                                \hline
 \textbf{TOPAZ}&  16   & 0.3$-$0.8 & $0.47 \pm 0.08$&\protect\cite{TOP-9402}\\
               &  80   &           & $0.70 \pm 0.15$&\\
               & 338   &           & $1.07 \pm 0.37$&\\\hline
\end{tabular}
\end{center}\end{table}
%
\renewcommand{\arraystretch}{1.1}
\begin{table}[htb]
\caption[
         Preliminary results on the \qsq evolution of \ft.
        ]
        {
         Preliminary results on the \qsq evolution of \ft.
         For the DELPHI results reported in Ref.\protect~\cite{TIA-9701}
         the numbers have been provided by I.~Tiapkin, whereas the
         results from Ref.\protect~\cite{TIA-9801} have been taken from
         the published tables.
         The numbers from the L3 experiment have been provided by
         F.C.~Ern{\'e}.\\
        }\label{tab:chap13_16}
\begin{center}
\begin{tabular}{ccccc}\hline
 Exp.          & $\qzm\,\left[\gevsq\right]$
                       &    $x$   & $\ft\pm\sigma_{\rm tot}$& Ref. \\\hline
 \textbf{DELPHI prel.}
               &   6.6 & 0.3$-$0.8& $ 0.38 \pm 0.08 $&\protect\cite{TIA-9701}\\
               &  11.2 &          & $ 0.43 \pm 0.05 $\\
               &  17.4 &          & $ 0.52 \pm 0.07 $\\
               &  20   &          & $ 0.49 \pm 0.06 $\\
               &  35.5 &          & $ 0.64 \pm 0.06 $\\
               &  63.0 &          & $ 0.77 \pm 0.08 $\\
               & 102.0 &          & $ 0.84 \pm 0.11 $\\
\hline
               &  21   & 0.3$-$0.8& $ 0.51 \pm 0.06 $&\protect\cite{TIA-9801}\\
               &  42   &          & $ 0.59 \pm 0.05 $\\
               &  99   &          & $ 0.73 \pm 0.06 $\\
               & 400   &          & $ 1.00 \pm 0.32 $\\
\hline
 \textbf{L3 prel.}
               &  50   & 0.3$-$0.8& $ 0.62 \pm 0.19 $&\protect\cite{ERN-9901}\\
               &  80   &          & $ 0.75 \pm 0.10 $&\\
               & 125   &          & $ 0.88 \pm 0.13 $&\\
               & 225   &          & $ 1.18 \pm 0.23 $&\\\hline
\end{tabular}
\end{center}\end{table}
%

%
%

\end{document}